\documentclass[12pt,twoside,numbychap]{thesis}
\usepackage{thesis}
\pdfoutput=1 
\versiondate{August 1, 2009}
\version{}
\begin{document}
\frontmatter
\title{Modified Gravity}
\subtitle{and the phantom of dark matter}
\author{Joel Richard Brownstein} 
\date{\today}
\maketitle
\begin{abstract}
Astrophysical data analysis of the weak-field predictions support the claim that modified gravity (MOG) theories provide a self-consistent, scale-invariant, universal description of galaxy rotation curves, without the need of non-baryonic dark matter.  Comparison to the predictions of Milgrom's modified dynamics (MOND) provide a best-fit and experimentally determined universal value of the MOND acceleration parameter.  The predictions of the modified gravity theories are compared to the predictions of cold non-baryonic dark matter (CDM), including a constant density core-modified fitting formula, which produces excellent fits to galaxy rotation curves including the low surface brightness and dwarf galaxies.

Upon analysing the mass profiles of clusters of galaxies inferred from X-ray luminosity measurements, from the smallest nearby clusters to the largest of the clusters of galaxies, it is shown that while MOG provides consistent fits, MOND does not fit the observed shape of cluster mass profiles for any value of the MOND acceleration parameter.  Comparison to the predictions of CDM confirm that whereas the Navarro-Frenk-White (NFW) fitting formula does not fit the observed shape of galaxy cluster mass profiles, the core-modified dark matter fitting formula provides excellent best-fits, supporting the hypothesis that baryons are dynamically important in the distribution of dark matter halos.

\end{abstract}
\clearpage
\chapter{Acknowledgements}

This work was supported by the Natural Sciences and Engineering Research Council of Canada (NSERC) and the University of Waterloo.  Research at Perimeter Institute for Theoretical Physics is supported in part by the Government of Canada through NSERC and by the Province of Ontario through the Ministry of Research and Innovation (MRI).

It was a privilege to be supervised by John W. Moffat of the Universities of Toronto and Waterloo.  I thank John for illuminating so many subjects in gravity theory, cosmology and high energy physics with his insight and rational perspective. His scientific expertise and originality continue to inspire me to seek an ever clearer picture of theories of space-time and their connection to physical reality. I am grateful to all my mentors -- David J. Rowe of the University of Toronto, James B. Hartle of the University of California at Santa Barbara, Jess K. Brewer of the University of British Columbia -- for supplying brilliantly challenging and enormously satisfying physics.

\clearpage
\newcommand{\cellphy}[1]{\textcolor{cayenne}{\large\sf #1}}
\chapter{Dedications}

\noindent The most stimulating and rewarding process of modelling gravity is entirely due to the physicists who have illuminated the physical universe, in their unique and visionary ways.  I dedicate my dissertation to each of the world's greatest  {\it classical} and {\it quantum} physicists, each of whom have strongly affected my journey:
\vspace{9mm}

\begin{center} 
\begin{tabular}{ccc}  \hline 
\cellphy{Albert Einstein} &\phantom{thanks}& \cellphy{Sir Isaac Newton} \\ [1ex]
\cellphy{Steven Weinberg} && \cellphy{James C. Maxwell} \\ [1ex]
\cellphy{Wolfgang Pauli} && \cellphy{Paul Dirac} \\ [1ex]
\cellphy{Richard P. Feynman} && \cellphy{Erwin Schr\"odinger} \\ [1ex]
\cellphy{Brian D. Josephson} && \cellphy{Kenneth G. Wilson} \\ [1ex]\hline
\end{tabular}
\end{center}
\vspace{9mm}

\tableofcontents
\listoffigures
\listoftables
\mainmatter
\chapterquote{Whoever undertakes to set himself up as a judge of Truth and Knowledge is shipwrecked by the laughter of the gods.}{Albert Einstein}
\chapter{\label{chapter.introduction}Introduction}

\noindent As Einstein observed, one's perspective into the nature of physical reality determines the degree to which one can understand physics within existing notions, or whether a shift of paradigm is needed.

\section{\label{section.introduction.objective}Motivation and Objectives}

Motivated by ongoing advances in gravity theory, the basic objective of this thesis is to present the current state of the art in astrophysical models of the astronomically observed data and to explore the frontier of cosmology by presenting a model universe without the necessity of dominant non-baryonic dark matter. 

\index{Dark matter!Missing mass problem}The central computation -- in the question of dark matter --  is the spatial distribution of the unseen component.  Application of the Newtonian \(1/r^2\) gravitational force law inevitably points to dark matter halos dominating disks of visible baryons within galaxies and clusters of galaxies throughout the cosmos~\citep{Oort:1932}.  

\index{Modified gravity!Phantom of dark matter}\index{Fifth force!Geometric origin}\index{Equivalence principle!Universality of free fall} Alternatively, a modified gravity field theory, which is sourced only by baryons of the standard model of particle physics, may provide the solution to the unseen component.  Regardless of how minuscule the physical effect must be at planetary and solar scales, the inclusion requires a fundamental modification of the known interactions, resulting in additional configuration variables and new couplings.  The addition of scalar-vector-tensor fields and their couplings to the action for pure gravity will be identified as a measurable fifth force, that does not vanish in a local frame, but without necessarily violating the weak equivalence principle and the {\it universality of free fall}.  Although each modified gravity theory may generate a fifth force through dissimilar means, the necessary violation of the equivalence principle may be quantified by measuring the dynamical mass factor as the ratio of the dynamic mass to the observed baryonic mass in each of the theories, depending on its distance from the astrophysical center of the system.\index{Dynamic mass factor, \(\Gamma\)}

Motivations for each perspective are presented in greater detail, according to whether there exists physical dark matter candidates corresponding to as yet undetected massive particle fields, as in \sref{chapter.introduction.objective.darkmatter}, or the hypothesis that modified gravity is responsible for the phantom of dark matter, as in \sref{chapter.introduction.objective.mog}, according to some violation of the strong equivalence, or relativity principle. The full hypothesis is detailed in \sref{chapter.introduction.objective.theory}, defining the candidate theories and the scope of astronomical case studies. The consequences for astrophysics and cosmology are identified in \sref{chapter.introduction.objective.astroph} and \sref{chapter.introduction.objective.cosmo}, respectively.

Whereas each of the main ideas are theoretically developed in \pref{part.theory}, the specific objective of this thesis, documented in \pref{part.astroph}, is to provide the results of a computational astrophysical data analysis, across a range of scales, in order to determine whether dark matter is real, or a phantom of a modified gravity theory, according to the method:
\begin{enumerate}
\item To measure the visible baryon distributions in galaxies and clusters of galaxies within Newton's theory, Milgrom's modified dynamics (MOND) and Moffat's modified gravity theories (MOG), and to measure the dark matter distributions in a sample of galaxies and clusters of galaxies. 
\item To compute the dynamic mass factors in each of the sample's galaxies and clusters of galaxies, which provide the extra gravity and phantom dark matter in MOND and MOG; and to compare the dark matter factor with cosmological values.
\item To observe whether the results may be universally understood across the range of kiloparsec to megaparsec scales within the sample.
\end{enumerate}
Conclusions are supplied in \pref{part.conclusions}.

\subsection{\label{chapter.introduction.objective.darkmatter}Non-Baryonic Dark Matter}

\index{Dark matter!Particle physics}
Despite over 20 years of focussed experimental effort, no direct evidence of dark matter particles has ever been found, and no annihilation radiation from any non-baryonic, dark matter candidate has ever been detected.  In fact, no experiment has ever supported any physics beyond the standard model 
of particle physics on which the dark matter hypothesis depends.  One of the goals of CERN's new Large Hadron Collider (LHC) is to understand what constitutes dark matter.  Capturing the notion that dark matter and dark energy are emergent gravitational phenomena due to terms beyond the Einstein-Hilbert action is a primary objective for this thesis.  Just as the Michelson-Morley experiment gives a null result and falsifies the prediction of the luminiferous {\ae}ther -- the medium for the propagation of light as it was thought until the late 19th century -- the LHC and the dark matter project (and future experiments) may never confirm any candidate for the dark matter particle.  The scientific community was strongly resistant to accept the falsification of the luminiferous {\ae}ther, and experiments continue to this day in search of this classically motivated, but expendable substance.  The search for dark matter continues, as Kipling's poem {\it IF} reasons, regardless.

The most popular candidates for non-baryonic dark matter, and the experimental status of their respective searches, are listed in \tref{table.introduction.objective.theory.pp.darkmatter}.  In all cases, a modification to the standard model of particle physics is required to explain why the \(\Lambda\)-CDM predicted density of dark matter dominates the baryon component by a factor of \(5.7 \pm 0.4\)~\protect\citep{Spergel:ApJS:2007}.

\begin{table}
\caption[Standard model extensions for non-baryonic dark matter]{\label{table.introduction.objective.theory.pp.darkmatter}{\sf Standard model extensions for non-baryonic dark matter}}
\bigskip\begin{center}
\parbox{5.0in}{Some of the primary candidates for non-baryonic dark matter are listed~\protect\citep{Munoz:2004IJMPA..19.3093M},~\protect\citep{Taoso.JCAP.2008.0803}.\index{Dark matter!Particle physics}}
\begin{tabular}{cc} \\ \hline 
{\tt Particle} & {\tt Conjecture} \\[1pt] \hline\hline
Axion & \fcolorbox{white}{white}{\parbox{4.2in}{Neutral scalars associated with the spontaneous symmetry breaking of global \(U(1)\) Peccei-Quinn symmetry as a mechanism to solve the strong CP problem in QCD.  These light -- 10 \(\mu\)eV -- non-thermally produced species are viable CDM candidates with relic abundances matching the cosmological density.  Reviewed in \citet{Sikivie.ArXiv:hep-ph/0509198} and \citet{Raffelt.JPA.2007.40}. }} \\ \hline 
WIMP relics & \fcolorbox{white}{white}{\parbox{4.2in}{Weakly interacting massive particles whose annihilation rate fell below the cosmological expansion rate, freezing out at primordial density, are thermal relics.  The required small annihilation cross-section introduces new physics at the weak scale.}}  \\ \hline
Neutralino & \fcolorbox{white}{white}{\parbox{4.2in}{The lightest supersymmetric WIMP may be stable and may act as a heavy thermal relic.  The near collisionless aspect of the neutralino makes it a prototype for cold dark matter.   Supersymmetry is badly broken in nature, and requires new physics at the weak scale.}}  \\ \hline
Neutrino & \fcolorbox{white}{white}{\parbox{4.2in}{The first proposed dark matter candidate was a fourth generation heavy neutrino, but collider experiments have ruled this out up to 1 TeV.}} \\ \hline 
sNeutrino & \fcolorbox{white}{white}{\parbox{4.2in}{The supersymmetric partner to the neutrino has been ruled out by LEP up to 1 TeV, which is too massive to be the lightest stable supersymmetric particle.  The left-handed sNeutrino has further been excluded in the minimal supersymmetric standard model (MSSM), but the right-handed sNeutrino is viable in extensions to the MSSM.}} \\ \hline 
Gravitino & \fcolorbox{white}{white}{\parbox{4.2in}{The supersymmetric partner to the graviton may have a mass of order keV, and is a candidate for non-baryonic dark matter in the absence of the inflationary paradigm.  However, relic gravitinos at big bang nucleosynthesis would lead to a higher reheating temperature than permitted by thermal leptogenesis. }} \\ \hline 
\end{tabular} \end{center}
\end{table}

\index{Dark matter!Dwarf galaxy problem}\index{Dark matter!Missing mass problem}
The status of dark matter is entirely reversed in the astrophysics of galaxies and clusters of galaxies, where the baryon fraction in certain critical systems can be entirely neglected, and has led some astrophysicists to claim detection of dark galaxies devoid of stars~\protect\citep{Minchin.APJ.2007.670}.  The search for dark dwarf galaxies is an important prediction of the \(\Lambda\)-CDM cosmological model of structure formation.  However, after years of intense search, all of the candidate dark dwarf galaxies have turned out to have been dwarf galaxies with a luminous stellar component, observed with high-powered optical telescopes.  In fact, for decades, dark matter simulations have predicted many times more companion galaxies than are actually observed~\protect\citep{Merrifield:2005idm..conf...49M}.  Large galaxies like the Milky Way are accompanied by a local group including many small satellites.  However, a robust prediction of dark matter simulations is that the Milky Way should be accompanied by \(\sim 500\) satellites, whereas only 35 have been observed~\protect\citep{Moore:1999ApJ...524L..19M}.  \citet{Klypin.APJ.1999.522} used the circular velocity distribution of the galaxy satellites to conclude that unless a large fraction of the Local Group satellites has been missed in observations, there is a dramatic discrepancy between observations and hierarchical models, regardless of the model parameters. Furthermore, \citet{Somerville:2004ApJ...600L.135S} showed that the \(\Lambda\)-CDM hierarchical scenario predicts that the largest most massive galaxies should form last, yet high redshift observations are beginning to indicate that some fraction of 
very luminous galaxies were present quite early in the process of structure formation.

The dark matter paradigm has spawned an entire industry of computer simulations which attempts to model the formation of structure in the universe through the gravitational collapse of dark matter dominated clumps.  As the resolution of very large computer simulations continues to improve, definite and robust results for the dark matter distribution within individual galaxies show a central power-law cusp, \(\rho(r) \propto r^{-\gamma}\).  The steepness of the cusp is a topic of debate~\protect\citep{Navarro:2004MNRAS.349.1039N}, particularly in regard to the inclusion of baryons in the simulations.

\citet{Bullock.APJ.2001.555}  considered the angular momentum of dark matter halos, and provided high-resolution N-body simulations of the \(\Lambda\)CDM cosmological structure formation of galactic halos, showing that the HI (and He) gas components cool at early times into small mass halos, leading to massive low-angular momentum cores in conflict with the observed exponential disks. The simulated \(\Lambda\)CDM galaxies have profiles which are too dense at small radii, and with tails extending too far. A possible solution is to associate the central excesses with bulge components and the outer regions with extended gaseous disks.\index{Dark matter!Cusp problem}

In the cores of spiral galaxies that have been fitted, the power-law index, \(1 \lesssim \gamma \lesssim 1.5\), and the general consensus is that at large radii, the  profile steepens approaching a power-law index, \(\gamma = 3\).  At intermediate distance, the fits reproduce the observed flat rotation curves with \(\gamma = 2\).  However, the large amount of luminous stellar material in the core of spiral galaxies means that such cusps will have negligible effects on the rotation curves of spiral galaxies.  Fortunately, a class of galaxies exists in which this issue is not the case.  These low surface brightness galaxies contain a very low density of luminous material, even in the core, so that the observed dynamics should be dominated by the gravitational forces of the dark halo.  Unlike in high surface brightness spiral galaxies, the small amount of luminous matter should not be efficient in redistributing the dark matter, so the central cusp should remain. \protect\citet{McGaugh:2001AJ....122.2396D} showed that the dark matter fits to a sample of 30 low surface brightness galaxies do not fit the data, showing systematic deviations in the galaxy cores.  Although the low surface galaxy rotation curves do tend to flatten off to the constant rotation velocity characteristic of the dark matter halo, they rise significantly more slowly out to several kiloparsecs than the best-fit cold dark matter prediction.  The rotation curves of low surface brightness galaxies prefer dark matter distributions with constant density modified cores.

\index{Dark matter!Occam's razor}
Just as the luminiferous {\ae}ther is considered an unnecessary addition to physics that violates the principle of Occam's razor, non-baryonic dark matter may be considered a superfluous component to be removed from our description of the cosmos.  The application of Occam's razor here would be to decrease the computed masses of galaxies and clusters of galaxies by an order of magnitude, and several orders of magnitude in the case of the critical systems.  Although the evidence we currently have for dark matter is concordant, it is unconfirmed, so it is vital to investigate the alternatives.  In some cases, modified gravity may seem to fit the observations better than dark matter.

\index{Dark matter!Cusp problem}\index{Dark matter!Missing mass problem}\index{Dark matter!NFW formula}The conflict between the cuspy dark matter halos predicted by hydrodynamical N-body simulations and the constant density cores preferred by dwarf and low surface brightness galaxies made it impossible to \(\chi^2\)-fit some of the galaxies in \pref{part.astroph} with the NFW fitting formula of \citet{Navarro.APJ.1996.462,Navarro.APJ.1997.490} with a nonvanishing stellar mass-to-light ratio.  Regardless, a cure to the cusp problem was found in \cref{chapter.galaxy} by implementing a core-modified fitting formula of \eref{eqn.newton.darkmatter.coremodified}\index{Dark matter!Core-modified}:
\begin{equation} \label{eqn.introduction.objective.darkmatter} 
\rho(r) = \frac{\rho_{0} r_{s}^3}{r^3+r_{s}^3},
\end{equation}
which provided excellent best-fits across the sample of high and low surface brightness galaxies including dwarf galaxies; and supports the existing \(\Lambda\)-CDM cosmology based on logarithmically divergent dark matter halos.  For each rotation curve in the Ursa Major sample of \sref{section.galaxy.uma}, the power-law index was found to asymptotically rise (\(\gamma \rightarrow 3\)) at large radii, whereas at small radii, the baryon-dominated constant density dark matter core (\(\gamma \rightarrow 0\)) appeared, as shown in \fref{figure.galaxy.powerlaw}.  Including the baryons, the total power-law indices oscillate around \(\gamma \rightarrow 1\) at the small radii and asymptotically rise to \(\gamma \rightarrow 3\) at large radii, where the dark matter core dominates and the baryonic data runs out.

Although cuspy dark matter halos are robust predictions from hydrodynamical N-body simulations of clusters of galaxies which ignore the baryon components, it is a feature which has not been observed.  Whereas it was impossible to \(\chi^2\)-fit any of the clusters of galaxies in \pref{part.astroph} with the NFW fitting formula of \citet{Navarro.APJ.1996.462,Navarro.APJ.1997.490}, the core-modified profile of \erefs{eqn.introduction.objective.darkmatter}{eqn.newton.darkmatter.coremodified}  that solved the cusp problem for galaxy rotation curves was successfully applied to clusters of galaxies in \cref{chapter.cluster}, providing excellent best-fits across the sample from the smallest X-ray cluster in nearby Virgo to the largest and most radiant {\bc} main cluster.  The core-modified dark matter halos described in \sref{section.cluster.models.darkmatter} do not dispute the existing \(\Lambda\)-CDM cosmology based on logarithmically divergent dark matter halos, but challenge dark matter simulations to reinsert the baryons, and make predictions that are consistent with astrophysical observations.

\subsection{\label{chapter.introduction.objective.mog}Modified gravity}
\index{Modified gravity!History}

Aristotle's notion of the motion of bodies was that a constant force maintains a body in uniform motion, and that force could only be applied by contact so action at a distance was considered impossible.  Aristotle produced a number of arguments why the heavens revolved around the Earth, and denied the possibility that the Earth rotated on its axis.  This geocentric model stood the test of time for over eighteen hundred years until Copernicus postulated that the sun was the center of the universe and the earth revolved around it at a distance related to the size of the orbit.  This idea, although controversial, initiated a scientific revolution that allowed a mathematical description of the force of gravity.  Kepler's empirical three laws of motion determined the elliptical orbits of planets orbiting the sun and opened an era of precision astronomy.  Galileo continued the revolution with a series of experiments on projectile motion, including the legendary Tower of Pisa experiment on the universality of free fall, and developed the mathematical theory of falling bodies.  \protect\citet{Newton:Principia:1687} set the foundation for classical mechanics and introduced the law of universal gravitation and a derivation of  Kepler's laws of planetary motion.  Newton's gravity theory describes action at a distance through Poisson's equation, in which any change in the matter distribution is instantaneously communicated to bodies in motion through the gravitational potential field.\index{Equivalence principle!Universality of free fall}

\subsubsection{Scalar modified gravity}
\index{Newton's constant!Varying in Jordan-Brans-Dicke}\index{Modified gravity!Jordan-Brans-Dicke gravity}\index{Equivalence principle!Violations}
The idea that Newton's constant varies from one point in space-time to another was first considered by \citet{Jordan.ZfP.1959.157} and first implemented in \citet{BransDicke:PR:1961}, which attempted to modify general relativity to be compatible with Mach's principle. A non-geometric, scalar field, \(\phi(x)\), was coupled to the Ricci curvature scalar in the action for gravity. This field is massless, but self-interacting as it couples to its own kinetic term through the coupling constant, \(\omega_{BD}\), which must be determined experimentally.  This  self-consistent and energetically stable modification to general relativity leads to the result that the locally measured value of Newton's constant varies spatially, and depends on the expectation value of the inverse of the Jordan-Brans-Dicke field, \(G(x) = \mean{1/\phi(x)}\).  However, this scalar-tensor modification to general relativity violates the strong equivalence principle and leads to a variation in Kepler's third law -- which is locally measurable through the \(\gamma\) parameter of the parametrized post-Newtonian formalism.  Solar system tests and to a lesser degree, PSR1913+16, have constrained Jordan-Brans-Dicke theory~\protect\citep{Will:LLR:2006}.

\subsubsection{Renormalized gravity}\index{Newton's constant!Renormalized}Even without the presence of auxiliary scalar fields which carry energy and momentum, the evidence from quantum field theory suggests that coupling constants like the fine structure constant, \(\alpha\), in quantum electrodynamics (QED) are not really universal, but are scale dependent ``running'' quantities.  The value measured in the lab for the ``fundamental'' charge of the electron depends on the renormalization scale, \(k\).  The physical mechanism behind the running fine structure constant is the appearance of a sea of virtual electron-positron pairs which are in a constant state of creation/annihilation.  Offshell photons surround the test charge, and contribute to the polarization of the bare charge, screening it at large distances.  Experimental measurements of the fine structure constant, at sufficiently large renormalization scale \(k\), show values \(\alpha(k) \gg \tfrac{1}{137}\).  Just as the Coulomb force law in classical electromagnetism is modified by the quantum corrected Ueling potential, the Newtonian force law may be subject to a running Newton's constant, leading to quantum corrected modified gravity~\protect\citep{Reuter:PRD:1998,Reuter:hep-th/0012069}.

\subsection{\label{chapter.introduction.objective.theory}Hypothesis}

\index{Dark matter!Missing mass problem}\index{Modified gravity!Phantom of dark matter}The solution to the missing mass problem in galaxy rotation curves and clusters of galaxies may be one of the candidates:
\begin{enumerate}
\item Cold non-baryonic dark matter (CDM),
\item Milgrom's modified Newtonian dynamics (MOND),\index{MOND}
\item Moffat's metric skew-tensor gravity theory (MSTG),\index{Modified gravity!Metric skew-tensor gravity}
\item Moffat's scalar tensor vector gravity theory (STVG).\index{Modified gravity!Scalar-tensor-vector gravity}
\end
{enumerate}

\index{Dark matter!Missing mass problem}\index{Modified gravity!Phantom of dark matter}\index{Dynamic mass factor, \(\Gamma\)}\index{Dark matter!Dynamic mass factor \(\Gamma\)}For the case of CDM, the total mass interior  to a spherical region divided by the integrated baryonic mass of the combined, visible, components is a measure of how much dark matter is required -- this is the dark matter factor.  For the case of MOND, the dynamic mass interior to a region divided by the integrated baryonic mass on the same region is a measure of how much MOND is required -- this dynamic mass factor is the inverse of the MOND interpolating function, \(\mu\).  For the case of Moffat's modified gravity (MOG), the dynamic mass factor is the ratio of a running Newtonian constant, \(G(r)\) divided by the bare Newtonian constant, \(G_N=6.67428 \times 10^{-11}\ \mbox{m}^3\mbox{kg}^{-1}\mbox{s}^2\).  

The hypothesis will be tested, performing a best-fit to the data for each of the objects in \tref{table.introduction.organization.astroph} and measuring the dynamic mass factor for each of the candidate theories, completely determining the predictions that are distinct from Newton's theory.  The radial profile of these dynamic mass factors across the Ursa Major sample of \sref{section.galaxy.uma} are shown in \fref{figure.galaxy.Gamma}.  The radial profile of the stellar mass-to-light ratios, for the same sample, are shown in \fref{figure.galaxy.masslight}, providing the means for the best-fits to the galaxy rotation curves of \cref{chapter.galaxy}.  The high resolution properties of the solutions are used to study orphan features in \sref{section.galaxy.halos.orphans}, and the integrated results are used to study the theoretical Tully-Fisher relations, shown in \fref{figure.galaxy.halos.tfr.diskbary}.  In \cref{chapter.cluster}, the radial profile of these dynamic mass factors across a sample of best-fit clusters models, of \sref{section.cluster.models}, is shown in \fref{figure.cluster.models.Gamma}, for each theory.  Consistency with solar system experimental constraints are examined in \cref{chapter.solar}, using data from \sref{section.solar.pioneer} on the Pioneer 10/11 Anomaly, to set an upper bound on dynamic mass factors, plotted in \fref{figure.solar.pioneer.Gamma}.  The planetary predictions and observational

\newcommand{\cellc}[1]{\parbox[t]{3.4in}{#1\vspace{2pt}}}
\begin{table}[ht]
\caption[Catalogue of astronomical case studies]{\label{table.introduction.organization.astroph}
				{\sf Catalogue of astronomical case studies}}
\begin{picture}(460,240)(0,0)
	\put(0,110){
	\begin{tabular}{c|l} \hline 
		{\tt Case study} & {\tt Astronomical objects}\\ \hline\hline
	{\creft{chapter.galaxy}} & 	\sreft{section.galaxy.uma} \\
High surface brightness galaxies & \cellc{NGC~3726, NGC~3769, NGC~3877, NGC~3893, NGC~3949, NGC~3953, NGC~3972, NGC~3992, NGC~4013, NGC~4051, NGC~4085, NGC~4088, NGC~4100, NGC~4138, NGC~4157, NGC~4217, NGC~4389, UGC~6399, UGC~6973} \\ 
	 Low surface brightness galaxies & \cellc{NGC~3917, NGC~4010, NGC~4183, UGC~6446, UGC~6667, UGC~6818, UGC~6917, UGC~6923, UGC~6983, UGC~7089} \\ \hline
		\creft{chapter.cluster}&  \sreft{section.cluster.models} \\ 
		& \cellc{{\bul}, Abell 2142, Coma, Abell 2255, Perseus, Norma, Hydra-A, Centaurus, Abell 400, Fornax, Messier 49} \\ 
		&  \sreft{section.cluster.bullet} \\ \hline
 \end{tabular}		}\end{picture}
\end{table}

\noindent limits are provided in \tref{table.solar.pioneer.predictions} and the mean ephemerides of planetary orbits is provided in \tref{table.solar.pioneer.ephemerides}.  Kepler's laws of motion are explored in \sref{section.solar.pioneer.kepler}, and the anomalous perihelion advance is studied in   \sref{section.solar.pioneer.perihelion}, with constraints listed in \tref{table.solar.pioneer.perihelion}.

As regards CDM halos, testing the hypothesis in \cref{chapter.galaxy} and \cref{chapter.cluster} will explore the fine details of the cusp problem and too many dwarf problem, discussed in \cref{chapter.darkmatter}, and may indicate a consistent cure based upon the model of \sref{subsection.galaxy.dynamics.coremodified}. Dynamic mass measurements, according to \sref{section.darkmatter.astroph}, are used to establish a fundamental connection to the Tully-Fisher relation, demonstrated in \sref{section.galaxy.tullyfisher}, in which the total mass vs. velocity relation, including dark matter, is plotted in \fref{figure.galaxy.halos.tfr.darkmatter}, and confirming the importance of including the baryons in curve-fitting dark matter halos.  The core-modified dark matter model, as indicated in \sref{section.cluster.models.darkmatter}, also cures the cusp problem for the sample of clusters of galaxies studied in \cref{chapter.cluster}, whereas the well established NFW model of \sref{subsection.newton.darkmatter.nfw} does not allow \(\chi^2\)-fits in these systems.

As regards MOND, testing the hypothesis will provide a better measurement of the MOND acceleration, \(a_{0}\), and provide a more detailed understanding of the MOND interpolating function, \(\mu\), in the context of Milgrom's acceleration law presented in \sref{section.mog.mond.action}.  Dynamical mass measurements, according to \sref{section.mog.mond.dynamic}, are used to establish the scales at which the theory fits observations; and where MOND potentially falls short, considering both the possibilities of MOND without dark matter, in \sref{section.cluster.models.mond}, and the possibility of neutrino halos in \sref{section.cluster.bullet.neutrino}.  Covariant theoretical foundations of MOND are explored in \sref{section.mog.mond.dynamic}.

As regards Moffat's MOG theories, with MSTG presented in \sref{section.mog.mstg} and STVG presented in \sref{section.mog.mstg}, the respective point source modified acceleration laws of \erefs{eqn.mog.mstg.eom.accelerationlaw3}{eqn.mog.stvg.eom.accelerationlaw3} are derived from an action principle in which a Yukawa fifth force combines with a Newtonian gravitational force, leading to modified Poisson equations, derived in \sref{section.mog.mstg.mog} and \sref{section.mog.stvg.mog}, respectively.  Whereas the MSTG theory has phenomenological parameters derived from the Tully-Fisher relation, the STVG theory has parameters which emerge as integration constants from integrated field equations.  In either case, testing the hypothesis will provide a better measurement of the MOG parameters.  Dynamical mass measurements, according to \sref{section.mog.mstg.dynamic} and \sref{section.mog.stvg.dynamic}, are used to provide support for the conjecture that the combination of a weak fifth force and a renormalization of Newton's constant will dynamically provide stronger gravity at astrophysical scales, providing a consistent solution to the missing mass problem without the addition of baryonic or non-baryonic dark matter.\index{Newton's constant!Renormalized}

Although not specifically tested in the list of theories, Moffat's NGT is considered a candidate for halos of phantom dark matter, with overlapping predictions with MSTG and possibly STVG.  It is natural that the source of the fifth force in either of the MOG theories is due to the fundamental NGT field excitations.

\subsection{\label{chapter.introduction.objective.astroph}Consequences for Astrophysics}
\index{Newton's central potential}

We know from general relativity that the Newtonian force is an emergent phenomenon due to the laws of physics in a curved space-time.  Although we no longer treat gravity as a simple force, we do make such a simplification to perform astrophysics by means of the central potential, 
\begin{equation}\label{eqn.introduction.objective.potential}
\Phi(t,\vect{x}) = - \int d^3 x^{\prime} \frac{G_{N}\rho(t,\vectprime{x})}{\norm{\vect{x}-\vectprime{x}}},
\end{equation}
where \(G_{N} = 6.67428(67) \times 10^{-11}\,\mbox{m}^{3}\mbox{kg}^{-1}\mbox{s}^{-2}\) is Newton's constant measured experimentally\footnote{\href{http://www.physics.nist.gov/cgi-bin/cuu/Value?bg}{NIST 2006 CODATA value.}}. Newton's central potential, where \(\Phi\) is given by \eref{eqn.introduction.objective.potential}, is an unshakeable foundation of modern physics, and appears upon identification of the constant of integration in the static spherically symmetric Schwarzschild solution to general relativity, 
\begin{equation}\label{eqn.introduction.objective.schwarzschild}
ds^2 = \left(1+\frac{2\Phi}{c^2}\right)dt^2 - \left(1+\frac{2\Phi}{c^2}\right)^{-1}\left(dx^2+dy^2+dz^2\right).
\end{equation}
Astrophysics assumes that the measured velocity dispersions and temperature profiles are determined to a good approximation by the Newtonian acceleration, 
\begin{equation}\label{eqn.introduction.objective.poisson}
\mathbf{a}(t,\vect{x}) = -\vect{\nabla} \Phi(t,\vect{x}) ,
\end{equation}
which neglects relativistic effects.  However, for realistic distributions of matter in galaxies, we have neither analytic, nor numerical solutions to general relativity from which orbits can be predicted.  For realistic distributions of matter in clusters of galaxies, the high degree of symmetry improves the situation in general relativity, and we may well approximate the intracluster medium by the interior solution of the rotating, axially symmetric Kerr solution to general relativity, from which cluster masses and temperatures may be predicted with precision.  However these interior solutions contain gravitomagnetic components due to the rotational energy of the system, which are not easily measured astronomically.  Newton's universal law of motion will emerge from these other solutions with relativistic corrections.  Unlike the simplest case of the Schwarzschild metric, the familiar distance-squared law may be modified, in effect.  Any such modifications would have impact on the computations of the mass-to-light ratio in galaxy rotation curves,
the temperature to mass relationship in X-ray clusters of galaxies,
and gravitational lensing in galaxies and clusters of galaxies.

\subsection{\label{chapter.introduction.objective.cosmo}Cosmological Models}

When matter sources are dominated by radiation, as in the early universe, the formation of local inhomogeneities is suppressed and the universe expands homogeneously and isotropically, without the growth of structures such as stars, galaxies or clusters of galaxies.  In the idealized case that the dominant radiation density is constant, the Newtonian central potential vanishes from the solution to the Einstein equations, and the universe is perfectly described by the Friedmann Robertson Walker solution to general relativity, 
\begin{equation}\label{eqn.introduction.objective.frwMetric}
ds^{2} = dt^{2} - a^{2}(t) \left[ \frac{dr^{2}}{1-k r^{2}} + r^{2}(d\theta^{2} + sin^{2}\theta d\phi^{2}) \right],
\end{equation}
where \(k = \{0,\pm 1\}\) is the curvature index and \(a(t)\) is the expansion factor.  The  expansion in time, or redshift \(z\), is governed by nonlinear Friedmann equations of motion,
\begin{equation}\label{eqn.introduction.objective.frwEquation}
{\dot a(z)}^{2} = \frac{k c^{2}}{\Omega(z) -1},
\end{equation}
where 
\begin{equation}\label{eqn.introduction.objective.densityParameter}
\Omega(z) = \frac{8\pi}{3} \frac{G_{N} \rho(z)}{H(z)^2}
\end{equation}
is the cosmological density parameter, \(\rho(z)\) is the mean cosmological mass density. \begin{equation}\label{eqn.introduction.objective.Hubble}
H(z) = \frac{{\dot a}(z)}{a(z)}
\end{equation}
 is the Hubble parameter, measured experimentally as a fraction of today's value.
 
\citet{Moffat.ArXiv:0710.0364} explored the cosmological consequences of MOG, and found that it provides, using a minimal number of parameters, good fits to the data, including the cosmic microwave background temperature anisotropy, the galaxy power spectrum, and the supernova luminosity-distance observations without the necessity of dark matter. 
 
 As astrophysics cannot simply do away with the Newtonian central potential, cosmology cannot simply do away with Newton's constant, but neither does gravity theory predict its value.  However, in both astrophysics and cosmology, it can always be arranged that Newton's constant, \(G_{N}\), and the mass density, \(\rho\), appear together in the combination \(G_{N}\rho\) as in \erefs{eqn.introduction.objective.potential}{eqn.introduction.objective.densityParameter}.  This argument is   due to dimensional reasoning, and leads to an ambiguity between the necessity of dark matter, \(\rho_{m} > \rho_{b}\), and a running Newton's constant, \(G > G_{N}\), or the existence of a MOND regime, \(\mu < 1\).  In the case of the dark matter paradigm, the density of matter \(\rho_m\) exceeds the density of baryons \(\rho_b\) and the combination \(G_N \rho_m > G_N \rho_b\).  In the case of a running Newton's constant,  the combination \(G \rho_b > G_{N} \rho_b\), and the visible baryon distribution provides ``extra gravity'' without non-baryonic dark matter.  However, the apparent degeneracy between dark matter and a running Newton's constant may be broken by calculations which involve a spatial integral or derivative of the combination \(G \rho\).  Such is the case for galaxy and cluster lensing experiments and cosmological models.

\section{\label{section.introduction.publications}Citations to published results}

Large portions of \preft{part.astroph} have been published:

\begin{publications}
\citet{Brownstein:ApJ:2006}, ``Galaxy rotation curves without non-baryonic dark matter'', {\apj} {\bf 636} 721--741. \eprint{astro-ph/0506370} \\
\citet{Brownstein:MNRAS:2006}, ``Galaxy cluster masses without non-baryonic dark matter'', {\mnras} {\bf 367} 527--540. \eprint{astro-ph/0507222} \\
\citet{Brownstein:CQG:2006}, ``Gravitational solution to the Pioneer 10/11 anomaly'', {\cqg} {\bf 23} 3427--3436. \eprint{gr-qc/0511230} \\
\citet{Brownstein.MNRAS.2007.382}, ``The Bullet Cluster 1E0657-558 evidence shows Modified Gravity in the absence of Dark Matter'', {\mnras} {\bf 382} 29--47. \eprint{astro-ph/0702146}.  \href{http://ras.joelbrownstein.com}{\tt Roy.\,Astron.\,Soc.\,Press Note 07/44}
\end{publications}

\noindent Some sections of \cref{chapter.mog}, particularly \sref{section.mog.equivalence} on violations of the strong equivalence principle, and \sref{section.mog.mstg} on the geometric origin of a fifth force,  and \cref{chapter.solar}, on Solar system tests, are motivated from my master's thesis, which is published:

\begin{publications}
\citet{Moffat.PRD.1990.41}, ``Spinning test particles and the motion of a 
gyroscope in the nonsymmetric theory of gravitation'', {Phys.\,Rev.} {\bf D41} 3111--3117.\\
\end{publications}

\section{\label{section.introduction.organization}Organization of the thesis}

\cref{chapter.introduction} is an introduction to the problem of non-baryonic dark matter in astrophysics, and an overview of the solution in which a modified gravity phantom of dark matter appears at astrophysical distances, followed by a summary of motivations and objectives in \sref{section.introduction.objective}, with citations to published results in \sref{section.introduction.publications}.

\pref{part.theory} of the thesis is divided into two chapters:  Dark matter halo fitting formulae are provided in \cref{chapter.darkmatter}, and the derivations of the modified acceleration laws are provided in \cref{chapter.mog}.  The two dark matter profiles used in curve-fitting, including baryons, are the Navarro-Frenk-White (NFW) profile, described in \sref{subsection.newton.darkmatter.nfw}, and the core-modified halo, derived in \sref{subsection.galaxy.dynamics.coremodified}.  The three modified gravity theories used in curve-fitting, using only baryons, are Milgrom's modified Newtonian dynamics, described in \sref{section.mog.mond}, and Moffat's metric skew-tensor gravity (MSTG) and scalar-tensor-vector gravity (STVG), in \sref{section.mog.mstg}  and \sref{section.mog.stvg}, respectively, which produce a finite range, Yukawa-type, fifth force~\protect\citep{Yukawa:PRMSJ:1935}.

The relativistic field theoretical versions of Milgrom's modified Newtonian dynamics (MOND), including Bekenstein's TEVES theory, and the general family of Einstein-{\ae}ther gravity models that may provide a weak-field MOND-like phantom of dark matter are documented in \sref{section.mog.mond.aether}. 

\pref{part.astroph} covers a survey of astronomical observations across a tremendous range of astrophysical scales.  The data used in the dissertation, and range of astrophysical phenomenon are organized in a catalogue of astronomical case studies, in \tref{table.introduction.organization.astroph}. The investigation into the available data starts with galaxy rotation curves, in \cref{chapter.galaxy}, and is concerned with dynamics in the weak field in \sref{section.galaxy.dynamics}, and uses the Ursa Major filament of galaxies, in \sref{section.galaxy.uma}, as the primary experimental observations between the 1 kiloparsec to 50 kiloparsec range.

The best-fitting core-modified dark matter model of \eref{eqn.galaxy.dynamics.dm.coremodified} provide excellent fits, including the dwarf galaxies, consistent with the large distance power law behaviour of cold collisionless non-baryonic dark matter (CDM).  All of the galaxy fits include the best-fitting Newtonian core model of \sref{section.galaxy.halos.core}, provided for comparison.  The theoretical underpinning and the experimental status of the Tully-Fisher relation  are reviewed in \sref{section.galaxy.tullyfisher}. \index{Newton's central potential!Galaxy core}

\cref{chapter.cluster} continues the investigation with X-ray clusters, in \sref{section.cluster.xraymass}, as the primary experimental observations between the 50 kiloparsec to 1000 kiloparsec range to the largest range of virialized matter, which compares the observed X-ray luminosities with the temperature profiles of the best-fit isothermal gas spheres in \sref{section.cluster.models}.  The {\bc} provides a laboratory to distinguish the {\it direct} gravitational lensing evidence for CDM with the modified gravity solution, in \sref{section.cluster.bullet}.

The search for the phantom of dark matter within the solar system in \cref{chapter.solar}, at ranges between 1 \(\au\) to 50 \(\au\), is primarily concerned with the Pioneer 10/11 Anomaly in \sref{section.solar.pioneer}, and experimental bounds.

Conclusions are presented in \pref{part.conclusions}, which supplies a summary of contributions in \cref{chapter.summary}, and a list of some possible future astrophysical tests in \cref{chapter.future}.  Lessons learned from CDM halos and modified gravity theories are supplied in \sref{section.summary.lessons}.  Specific conclusions on galactic astrophysics are summarized in \sref{section.summary.galaxy} with future tests in \sref{section.future.galaxy}, and specific conclusions on cluster-scale astrophysics are summarized in \sref{section.summary.cluster} with future tests in \sref{section.future.cluster}.

\part{\label{part.theory}Theory}
\chapterquote{No great discovery was ever made without a bold guess.}{Sir Isaac Newton}
\chapter{\label{chapter.darkmatter}Non-baryonic dark matter}

\noindent For three centuries, Newton's theory has proven to be remarkably successful, but is limited to weak gravitational fields. As a classical nonrenormalizable effective theory, Einstein's theory has proven to be 
remarkably successful, and together with Newton's theory, is believed to fully describe the measurable gravitational physics in astrophysical systems and cosmology.  The fact that these theories predict the necessary existence of non-baryonic dark matter which dominates the visible matter in the universe will create a new era for precision astrophysics -- provided the dark matter candidate is identified and experimentally confirmed.  Otherwise a modification of gravity, as in \cref{chapter.mog}, may solve the missing mass question, provided there are gravitational degrees of freedom in nature that are not captured by Newton's or Einstein's theory.

Whereas flat cosmological models with a mixture of radiation, ordinary baryonic matter, cold collisionless dark matter and cosmological constant (or quintessence) and a nearly scale-invariant adiabatic spectrum of density fluctuations provide good fits to large scale (\(\gg\) 1 Mpc) observations, there remains a large amount of data on galactic and sub-galactic scales (\(\ll\) 100 kpc) which may be in conflict with the \(\Lambda\)CDM halo structure -- or support a core-modified dark matter fitting formula which retains the large scale \(\Lambda\)CDM halo structure~\citep{Zhao.MNRAS.1996.278}.\index{Dark matter!Core-modified}


\section{\label{section.newton.darkmatter}Dark matter halos}

\index{Dark matter!Missing mass problem}Based on three rotation curves, \citet{Roberts.AAP.1973.26} concluded that spiral galaxies must be larger than indicated by the usual photometric measurements, and suggested the existence of an unseen massive halo beyond the last measured point -- to explain the slower than Keplerian decline at large radii.  This view challenged the notion of a constant mass-to-light ratio, with radius, and suggested a mass-to-light ratio which increases with distance from the center.

\citet{Ostriker.APJL.1974.193} argued that the masses of ordinary galaxies -- found by assuming a constant mass-to-light ratio --  may have been underestimated by a factor of 10; but that the galaxy rotation curve in the inner region provides almost no information about the exterior halo mass.  Upon application of a Newtonian force law,
\begin{equation}\label{eqn.newton.darkmatter.newton}
a(r) = -\frac{G_{N} M_{N}(r)}{r^2} 
\end{equation}
one may obtain the Newtonian dynamic mass, \(M_{N}(r)\), which is the mass interior to the sphere of radius, \(r\), needed to support the galaxy rotation curve.  \citet{Ostriker.APJL.1974.193} observed that although the surface luminosity profiles, \(L(r)\), do appear to be convergent, \(M_{N}(r)\), diverges with r either weakly (logarithmic) or strongly (linear) depending on the method of measurement, and concluded that within local giant spiral galaxies,
\begin{equation}\label{eqn.newton.darkmatter.divergent}
M_{N}(r) \propto r\,\quad\quad \mbox{for}\ 20\ \mbox{kpc}\ \le r \le 500\ \mbox{kpc}.
\end{equation}
This divergent mass-to-light ratio necessitates the existence of giant halos surrounding ordinary galaxies of dark matter -- the implied density distribution similar to isothermal gas spheres in the outer parts,~\citep{Begeman.MNRAS.1991.249}
\begin{equation}\label{eqn.newton.darkmatter.isothermal}
\rho(r) = \frac{\rho_0 r_{c}^2}{r^2+r_{c}^2}, 
\end{equation}
where \(r_{c}\) is the core radius and \(\rho_{0}\) is the central dark matter density.  In the limit of small \(r \ll r_{c}\), the isothermal sphere model approaches a constant density core. Spherically integrating the constant density core model of \eref{eqn.newton.darkmatter.isothermal} one obtains a simple formula for the mass of dark matter,
\begin{equation} \label{eqn.newton.darkmatter.isothermal.mass} 
M(r) =  4\pi\rho_{0} r_{c}^3 \left\{\frac{r}{r_{c}} -\tan^{-1}(r/r_{c})\right\},
\end{equation}
which diverges with the behaviour of \eref{eqn.newton.darkmatter.divergent}, for  \(r \gg r_{c}\).

\citet{Einasto.NAT.1974.250} studied the distribution of missing mass, as it relates to galactic morphology, concluding that the distribution is suggestive of a corona (surrounding the luminous disk), increasing the total mass of the galaxy by an order of magnitude.  \citet{Rubin.APJL.1978.225} considered extended rotation curves of 10 high-luminosity galaxies, and reproduced observed velocities using mass distributions from disk or spherical models; and suggested that the {\it flat} rotation velocity, \(v_{out}\), is not correlated with luminosity or with radius, but with extended dark matter.  However, the observations did not suggest whether spherical or disk models were favoured.  \citet{Rubin.APJL.1978.225} concluded that the total mass-to-light ratio is higher for early-type galaxies leading to a large intrinsic scatter in the Tully-Fisher relation.

On larger than galaxy scales, \citet{Fillmore.APJ.1984.281} considered the self-similar gravitational collapse of collisionless dark matter in a perturbed Einstein-de Sitter universe, and suggested that spherically averaged solutions prefer similar halo mass profiles which may be approximated by a power-law in the distance from center of symmetry.  In the case of structure evolving hierarchically from a scale-free Gaussian field of a given power spectrum, \citet{Hoffman.APJ.1985.297,Hoffman.APJ.1988.328} suggested that the final virialized halo should have an asymptotic density profile given by
\index{Dark matter!Power law, \(\gamma\)}
\begin{equation} \label{eqn.newton.darkmatter.powerlaw} 
\rho(r) \propto r^{-\gamma},
\end{equation}
where \(\gamma = 2\) assuming that the CDM power spectrum, on galactic scales, is effectively, 
\begin{equation} \label{eqn.newton.darkmatter.powerspectrum} 
P(k) \propto k^{n_{\rm eff}},\ \mbox{where}\ n_{\rm eff} = 2.
\end{equation}

\subsection{\label{subsection.newton.darkmatter.nfw}Navarro-Frenk-White profile}
\index{Dark matter!NFW formula}
In search of a universal description of collisionless dark matter, \citet{Navarro.APJ.1996.462,Navarro.APJ.1997.490} provided power-law fits to halo density profiles using N-body simulations, showing that halo profiles are shallower than \(r^{-2}\) near the center and steeper than \(r^{-2}\) near the virial radius.  The NFW profile is then a simple fitting formula to \eref{eqn.newton.darkmatter.powerlaw}, with a radially {\it varying} powerlaw \(1 \le \gamma(r) \le 3\), to describe spherically averaged density profiles:
\begin{equation} \label{eqn.newton.darkmatter.nfw} 
\rho(r) = \frac{\rho_{0} r_{s}^3}{r(r+r_{s})^2}.
\end{equation}
Spherically integrating the NFW profile of \eref{eqn.newton.darkmatter.nfw} one obtains a simple formula for the mass of dark matter,
\begin{equation} \label{eqn.newton.darkmatter.nfw.mass} 
M(r) = 4\pi \rho_{0} r_{s}^3 \left\{\ln(r+r_{s})  - \ln(r_{s}) - \frac{r}{r+r_{s}}\right\},
\end{equation} 
which diverges logarithmically, for  \(r \gg r_{s}\).  In the limit of small \(r \ll r_{s}\), the NFW fitting formula of \eref{eqn.newton.darkmatter.nfw} approaches the power-law with \(\gamma \rightarrow 1\); and in the limit of large \(r \gg r_{s}\) approaches the power-law with \(\gamma \rightarrow 3\) --  which does not approximate isothermal spheres.  \index{Dark matter!NFW formula}\citet{Navarro.APJ.1996.462} reported that rotation curves from galaxies ranging in size from giant to dwarf, satellites and gaseous atmospheres are compatible with the NFW halo structure of \eref{eqn.newton.darkmatter.nfw} provided the mass-to-light ratio increases with luminosity.  \citet{Navarro.APJ.1996.462} determined that the central regions of the NFW distribution have densities comparable to the luminous parts of galaxies.

\index{Dark matter!Power law, \(\gamma\)}\index{Dark matter!Cusp problem}Although the N-body problem can easily be defined, and numerically simulated in the world's best computers, the problem defies any rigorous analytic treatment.  \citet{Zait.APJ.2008.682} reported that N-body numerical simulations are providing conflicting evidence regarding the asymptotic behaviours of the density slope, \(\gamma\) of the profile at small radii (in the inner region of the halo).

\subsection{\label{subsection.galaxy.dynamics.generals}Generalized profile}
\index{Dark matter!Dwarf galaxy problem}\citet{Burkert.APJL.1995.447} fitted a sample of several dark matter dominated dwarf galaxies employing a phenomenologically modified universal fitting formula, 
\begin{equation} \label{eqn.newton.darkmatter.burkert} 
\rho(r) = \frac{\rho_{0} r_{s}^3}{(r+r_{s})(r^2+r_{s}^2)}.
\end{equation}
which, as in the case of the isothermal sphere of \eref{eqn.newton.darkmatter.isothermal},  approximates a constant density core, \(\gamma \rightarrow 0\) at \(r \ll r_{s}\) -- instead of a divergent \(\gamma=1\) core -- but otherwise agrees with the NFW profile, with \(\gamma \rightarrow 3\) at \(r >> r_{s}\).  Spherically integrating the Burkert model of \eref{eqn.newton.darkmatter.burkert}, one obtains an analytic formula for the mass of dark matter,
\begin{equation} \label{eqn.newton.darkmatter.burkert.mass} 
M(r) = \pi\rho_{0} r_{s}^3 \left\{\ln(r^{2}+r_{s}^{2}) + 2\ln(r+r_{s})  - 4 \ln(r_{s}) -2 \tan^{-1}(r/r_{s})\right\},
\end{equation}
which diverges logarithmically, for  \(r \gg r_{s}\).

\index{Dark matter!Core-modified}\citet{Zhao.MNRAS.1996.278} hypothesized that the NFW fitting formula must be broadened to account for the basic observed features of galactic dynamics, including less cuspy cores:
\begin{equation} \label{eqn.newton.darkmatter.zhao} 
\rho(r) = \frac{\rho_{0} r_{s}^{b}}{r^{c}(r^{a}+r_{s}^{a})^{(b-c)/a}},
\end{equation}
where \((a,b,c)\) are free parameters.  The NFW fitting formula of \erefs{eqn.newton.darkmatter.nfw}{eqn.newton.darkmatter.nfw.mass} correspond to \eref{eqn.newton.darkmatter.zhao} with an inner cusp with logarithmic slope \(c=1\), an outer corona with logarithmic slope \(b = 3\), and a ``turnover'' exponent of \(a=1\).  ~\citet{Syer.MNRAS.1998.293} argued that the existence of a \(\gamma \ll 1\) core is inconsistent with the hierarchical formation scenario of dark halos, which are much more likely to result in cuspy central density distributions.  The least cuspy fitting formula, the isothermal spheres of \erefs{eqn.newton.darkmatter.isothermal}{eqn.newton.darkmatter.isothermal.mass} correspond to \eref{eqn.newton.darkmatter.zhao} with a constant density inner core with logarithmic slope \(c = 0\), an outer corona with logarithmic slope \(b = 2\), and a ``turnover'' exponent of \(a=2\).  Although Burkert's fitting formula of \eref{eqn.newton.darkmatter.burkert} cannot be expressed in the core-modified form of \eref{eqn.newton.darkmatter.zhao}, it does bridge the constant density, \(\gamma \rightarrow 0\), core behaviour of the isothermal sphere with the \(\gamma \rightarrow 3\) large \(r\) behaviour of the NFW profile. 

\citet{Kravtsov.APJ.1998.502} used the rotation curves of a sample of dark matter dominated dwarf and low surface brightness (LSB) galaxies, employing the modified universal fitting formula of \eref{eqn.newton.darkmatter.zhao} with a shallow cusp, \((a,b,c) = (2,3,0.2)\), and computed that a dominant fraction (\(\sim 95\%\)) of the dynamical mass is due to dark matter at the last measured point in the rotation curve; but with \(0.2 < \gamma < 0.4\) in the inner region, \(r \ll r_{s}\), of every galaxy in the sample.

\citet{McGaugh.APJ.1998.499} enforced the claim that the severity of the mass discrepancy in spiral galaxies is strongly correlated with the central surface brightness of the disk. Progressively lower surface brightness galaxies have ever larger mass discrepancies. No other parameter (luminosity, size, velocity, morphology) is so well correlated with the magnitude of the mass deficit. 

\index{Dark matter!Core-modified}Regardless of the galactic and sub-galactic data, collisionless dark matter N-body simulations continue to predict steep inner cusps~\citep{Moore.APJL.1998.499}.  \citet{Moore.MNRAS.1999.310} argued that a universe dominated by cold dark matter fails to reproduce the rotation curves of dark matter dominated dwarf and LSB galaxies; and instead provided fits employing the modified universal fitting formula of \eref{eqn.newton.darkmatter.zhao} with a steeper cusp, \((a,b,c) = (1.5,3,1.5)\).  However, these fits purposely ignored the contribution from the HI gas and the stellar disk to maximize the dark matter halo in the core.  In contrast, the stellar mass-to-light ratio, \(\Upsilon\), is critical to the study of galaxy rotation curves; and the requirement that the stellar mass-to-light ratio, \(\Upsilon=0\), is too strong and therefore should be suspect as the reason for \citet{Moore.MNRAS.1999.310} good core-modified best-fits.

\index{Dark matter!Dwarf galaxy problem}\index{Dark matter!NFW formula}\citet{Navarro.ArXiv:astro-ph/9807084} remarked that a subset of spiral galaxies have flat rotation curves, and suggested that disagreement with the rotation curves of a few dwarf galaxies may signal systematic departures from the NFW shape, and that the rotation curves for LSB galaxies may be better described by shallower central density profiles, than presumed in the NFW fitting formula.  \citet{vandenBosch.AJ.2000.119} argued that the spatial resolution of LSB rotation curves is not sufficient to put any meaningful constraints on the dark matter density profiles, but conceded that the rotation curves of dark matter dominated dwarf galaxies are inconsistent with steeply cusped dark halos.  ~\citet{Kleyna.APJL.2003.588} demonstrated that the most dark matter dominated dwarf galaxies in the Local Group have constant density \(\gamma \rightarrow 0\) halo cores, and suggested that CDM disagrees with observations in that end of the galaxy mass spectrum.    Based on a series of high-resolution N-body simulations designed to examine whether the density profiles of dark matter halos are universal, \citet{Jing.APJL.2000.529} found that the dark matter density depends on the total halo mass, making it difficult to link the inner slope with the primordial index of the fluctuation spectrum. 

The unexplained behaviour of the computed dark matter distribution in the core is known as the {\it cusp problem} and casts doubt on the choice of the NFW fitting formula which presupposes the core behaviour.  These discrepancies at the galactic and sub-galactic scale have stimulated a number of alternative proposals.  \citet{Spergel.PRL.2000.84,Ostriker.Science.2003.300} reviewed the situation for collisionless dark matter predictions -- overly dense cores in the centers of galaxies and clusters and an overly large number of halos with the Local Group compared to actual observations --  and suggested the alternative of {\it self-interacting} dark matter produces distinctive modifications on small scales that can be tested through improved astronomical observations.  \citet{Stoehr.MNRAS.2002.335} commented that these self-interacting dark matter modifications either may fail to reproduce the large observed velocity dispersions in the Local Group dwarf galaxies; or may suffer from a fine-tuning problem.  Modifying the microscopic physics of the dark matter particles may work to reduce the concentration in the central regions of galaxies and to reduce the abundance of halo substructure (unseen dwarf galaxies).

\citet{Binney.MNRAS.2001.327} claimed that the Milky Way has considerably less dark matter in the luminous disk than expected, particularly near the galactic center, and concluded that cuspy halos favoured by the cold dark matter cosmology (and its variants) are inconsistent with the observational data.  \citet{Dave.APJ.2001.547} presented a comparison of halo properties in cosmological simulations, confirming that collisionless dark matter yields cuspy halos that are too centrally concentrated, as compared to observations.  \citet{deBlok.APJL.2001.552} found that, at small radii, the mass density distribution is dominated by a nearly constant density core with a core radius of a few kiloparsecs, and found no clear evidence for a cuspy halo in any of the low surface brightness galaxies. \citet{Swaters.APJ.2003.583} presented a sample of 15 dwarf and low surface brightness galaxies, showing that most are equally well or better explained by constant density cores, and none require halos with steep cusps.  \citet{Gentile.MNRAS.2004.351} confirmed that the distribution of dark matter in spiral galaxies is consistent with constant density cores.

\subsection{\label{subsection.galaxy.dynamics.coremodified}Core-modified profile}\index{Dark matter!Core-modified}
Consider a fitting formula of the form of \eref{eqn.newton.darkmatter.zhao} with a constant density inner core with logarithmic slope \(c = 0\), an outer corona with logarithmic slope \(b = 3\), and a ``turnover'' exponent of \(a=3\):
\begin{equation} \label{eqn.newton.darkmatter.coremodified} 
\rho(r) = \frac{\rho_{0} r_{s}^3}{r^3+r_{s}^3}.
\end{equation}
which, as in the case of the isothermal sphere of \eref{eqn.newton.darkmatter.isothermal} and the Burkert model of \eref{eqn.newton.darkmatter.burkert},  approximates a constant density core, \(\gamma \rightarrow 0\) at \(r \ll r_{s}\) -- instead of a divergent \(\gamma=1\) core -- but otherwise agrees with the NFW profile, with \(\gamma \rightarrow 3\) at \(r >> r_{s}\).  Spherically integrating this core-modified model of \eref{eqn.newton.darkmatter.coremodified}, one obtains a new analytic formula for the mass of dark matter,
\begin{equation} \label{eqn.newton.darkmatter.coremodified.mass} 
M(r) = \frac{4}{3}\pi\rho_{0} r_{s}^3 \left\{\ln(r^3+r_{s}^3) - \ln(r_{s}^3) \right\},
\end{equation}
which diverges logarithmically, for  \(r \gg r_{s}\).

Utilizing the form of the power-law of \eref{eqn.newton.darkmatter.powerlaw}, the power-law index of the profile of \eref{eqn.newton.darkmatter.coremodified} is minus the logarithm slope
\begin{equation} 
\label{eqn.newton.darkmatter.coremodified.gamma} \gamma(r) = -\frac{d \ln \rho(r)}{d \ln r} = \frac{3r^3}{r^3+r_{s}^3}.
\end{equation}
The central density, \(\rho(0) = \rho_{0}\), is finite and may be written in terms of the cosmological critical density, \(\rho_{c}(z)\), and the concentration parameter, \(\delta_c\),
\begin{equation} \label{eqn.newton.darkmatter.coremodified.concentration} 
\rho_{0} = \rho_{c}(z) \delta_{c},
\end{equation}
where \(z\) is the redshift.  Moreover, the dark matter density at \(r=r_{s}\) is one-half the central density,
\begin{equation} \label{eqn.newton.darkmatter.coremodified.rho.rs} 
\rho(r_{s}) = \frac{1}{2}\rho_{0},
\end{equation}
and the power-law index of \eref{eqn.newton.darkmatter.coremodified.gamma} is
\begin{equation} \label{eqn.newton.darkmatter.coremodified.gamma.rs} 
\gamma(r_{s}) = 3/2,
\end{equation}
which is the intermediate value between the inner core with logarithmic slope \(\gamma \rightarrow 0\), and outer corona with logarithmic slope \(\gamma\rightarrow 3\).  This means that the halo's constant density core is limited to the region \(r<r_{s}\), where baryons dominate the galaxy, which is important for N-body simulations.

\section{\label{section.darkmatter.astroph}Dynamic mass}

\index{Modified gravity!Mass profile}Theoretical predictions of dynamical quantities such as galaxy rotation curves and cluster masses of galaxies, as in \cref{chapter.galaxy} and \cref{chapter.cluster}, respectively, are either difference calculations as in the case of dark matter, or divisive ones as in the case of the modified gravity models of \cref{chapter.mog}, and the preferred frame gravity models of \sref{section.mog.mond.aether}, including those with modified dynamics at small accelerations, as in \sref{section.mog.mond.dynamic}.

Each of the modified acceleration laws applied in the astrophysics computations of \pref{part.astroph} determine the acceleration felt by a test particle, at a distance \(r\) from the center.  This acceleration is proportional to the mass enclosed within a spherical region of radius, \(r\), so that
\begin{equation}\label{eqn.darkmatter.astrophbirkhoff}
a(r) \propto M(r),\quad\quad \mbox{where}\ M(r) = \int_0^r \rho({r^\prime}) dV,
\end{equation}
where \(\rho(r)\) is the density at the position, \(r\), and \(dV\) denotes the spherical volume element.  
 
\index{Dynamic mass factor, \(\Gamma\)}In this case, the cells of data can be related by a factor,
\begin{equation}\label{eqn.darkmatter.astrophGamma}
M_N({\bf r}) = M({\bf r}) \Gamma({\bf r})
\end{equation}
where \(M_N({\bf r})\) is the dynamical mass of the integrated cells of data within a spherical region or separation, \({\bf r}\), and \(M({\bf r})\) is the observed baryonic mass of the same region.  \(\Gamma({\bf r})\) is the dynamical mass factor and is related to the dark matter ratio, whereby 
\index{Dark matter!Dynamic mass factor \(\Gamma\)}
\begin{equation}\label{eqn.darkmatter.astrophGamma.darkmatter}
\Gamma({\bf r}) = 1 + M_{\rm dark~matter}({\bf r})/M({\bf r}),
\end{equation}
where \(M_{\rm dark~matter}({\bf r})\) is the integrated mass of dark matter inside the common spherical region.  At cosmological scales, where \({\bf r}\) is course grained away, the dark matter factor of \citet{Spergel:ApJS:2007} is 
\begin{equation}\label{eqn.darkmatter.astrophGamma.dmbaryonfraction}
\Gamma = \Omega_{\rm matter}/\Omega_{\rm baryon} = 5.73 \pm 0.40
\end{equation}

The dark matter fits to the Ursa Major sample of \sref{section.galaxy.uma} confirm that \(\Gamma \leqslant 10\) across the galaxies, therefore the dark matter factor is consistent with the \(\Lambda\)CDM scenario.

\index{Dynamic mass factor, \(\Gamma\)}
Alternatively, the dynamic mass factor predicted by MSTG and STVG, as in \sref{section.mog.mstg} and \sref{section.mog.stvg}, respectively, is effectively due to a renormalized gravitation coupling of \eref{eqn.mog.mstg.mog.Gforce} with\index{Newton's constant!Renormalized}
\begin{equation}\label{eqn.darkmatter.astrophGamma.mog}
\Gamma({\bf r}) = M_N({\bf r})/M({\bf r}) = G({\bf r})/G_N,
\end{equation}
where \(G({\bf r})\) is the best-fitted gravitational coupling to the dynamical data at the separation \({\bf r}\), and  \(G_{N} = 6.67428(67) \times 10^{-11}\,\mbox{m}^{3}\mbox{kg}^{-1}\mbox{s}^{-2}\) is Newton's constant measured experimentally\footnote{\href{http://www.physics.nist.gov/cgi-bin/cuu/Value?bg}{NIST 2006 CODATA value.}}. 

Dynamic mass factors, constructed from galaxy rotation curves in the Ursa Major filament of galaxies, are provided in \fref{figure.galaxy.Gamma}, and those constructed from a sample of clusters of galaxies are provided in \fref{figure.cluster.models.Gamma}.  Conclusions drawn from the astrophysics on CDM halos may be found in \sref{section.summary.darkmatter}.
\chapterquote{A hundred times every day I remind myself that my inner and outer life depend on the labors of other men, living and dead, and that I must exert myself in order to give in the same measure as I have received and am still receiving.}{Albert Einstein}
\chapter{\label{chapter.mog}Modified gravitation theory}

In the absence of a fifth force in nature, either the dark matter paradigm ensues at astrophysical and cosmological scales, or the relativity principle may come into question.  Local SO(3,1) invariance is a foundation of relativistic gravity theories, and is made manifest by general covariance.  However, the equivalence principle, as in \sref{section.mog.equivalence}, may be violated by fifth-force fields or preferred space-time frames.   Milgrom's modified dynamics (MOND), as in \sref{section.mog.mond}, is phenomenologically derived from observations of galaxy rotation curves and the Tully-Fisher relation, relativistic theories with a preferred frame, as in \sref{section.mog.mond.aether}, are manifestly covariant, but violate SO(3,1) Lorentz covariance by means of a constraint.  Moffat's metric skew-tensor gravity, as in \sref{section.mog.mstg}, is a relativistic metric gravity theory, with massive fifth-force fields, Moffat's scalar-tensor-vector gravity, as in \sref{section.mog.stvg}, is furthermore without phenomenological input from the Tully-Fisher relation.
\section{\label{section.mog.equivalence}Equivalence principle}\index{Equivalence principle|(}

\subsection{\label{section.mog.equivalence.so31}Local SO(3,1) theory} 

At the turn of the last century, Lorentz conjectured that Newton's universal gravitation law needed to be modified so that changes in the gravitation field propagate with the speed of light. Days before the \protect\citet{Einstein:SR:1905} paper on special relativity, \protect\citet{Poincare:1904,Poincare:SR:1905} suggested that all forces, including gravity, should transform according to Lorentz transformations.  Einstein set himself the task of modifying Newton's gravity theory to accommodate the principles of special relativity, and  proposed the equivalence principle based on the empirically observed universality of free fall:\index{Equivalence principle!Universality of free fall}

\subsubsection{Einstein equivalence principle}\begin{quote}

As far as we know, the physical laws with respect to an accelerated system do not differ from those with respect to a system at rest; this is based on the fact that all bodies are equally accelerated in the gravitational field. At our present state of experience we have thus no reason to assume that the accelerating and inertial systems differ from each other in any respect, and in the discussion that follows, we shall therefore assume the complete physical equivalence of a gravitational field and a corresponding acceleration of the reference system~\protect\citep{Einstein:EP:1907}.
\end{quote}
\protect\citet{Einstein:GR:1916} formulated his gravity theory geometrically so that particles travel along geodesics in a curved space-time. Observables are invariant under local Lorentz transformations, generalizing the global Lorentz invariance of special relativity.  The Newtonian gravitational attraction is the effect outside of a test particle's rest frame, modulo {\it relativistic corrections}.  It is the curvature of the pseudo-Riemannian manifold which is fundamental, and the space-time metric is a dynamical solution to the Einstein equations.

\subsection{Strong equivalence principle}\index{Equivalence principle!Universality of free fall}

The demand that the laws of nature, in a sufficiently small region of a given space-time point, take the same form as they do in special relativity is stronger than the {\it universality of free fall} as it means that there are no fields 
unified with the metric.

The is contrary to the case in which the unified field is associated with Maxwell's electromagnetism -- which obviously does not vanish locally.  However, the strong equivalence principle holds for pure gravity, where the unified field is constrained to vanish by the metric-connection compatibility equations.  Conversely, the  dynamic nature of the connection field does not imply a vanishing torsion trace,  and there are two degrees of violation of the strong equivalence principle in the general hermitian theory.

Even though Einstein's theory may be written formally as a gauge theory, with field variables suitably chosen, it does not predict the form of the Newtonian universal force law.   \citet{Bianchi.CQG.2006.23} considered the graviton propagator within background independent, nonperturbative quantum gravity, yielding results that are consistent with Newton's universal law, but the renormalized interaction remains to be calculated.\index{Newton's constant!Renormalized}

\subsection{\label{section.mog.equivalence.violations}Violations of the strong equivalence principle}

The two mechanisms for potential violations of the equivalence principle are:

\index{Equivalence principle!Violations}
\begin{description}
\item[Charge violations] The possibility that fermions possess quantum numbers related to conserved fifth force charges leads to direct violations of the weak equivalence principle and  severely constrain modified gravity theory~\protect\citep{Will:LLR:2006}.  The non-abelian gauge theory for gravity necessitates the consideration that the nonmetric degrees of freedom that are associated with the larger symmetry group carry their charged quantum numbers.  This would cause violations of {\it universality of free fall}, although Cavendish and Eotv\"os lunar laser ranging experiments tightly constraint any charge associated with the hermitian theory. The possibility of weak equivalence principle violations due to the Earth's rotation have been tightly constrained~\protect\citep{Moffat.PRD.1990.41}, and are not expected to be measurable by the Gravity Probe B in Earth's orbit. 
\item[Scalar-vector-tensor violations] Unlike the local SO(3,1) theory in \sref{section.mog.equivalence.so31},   local gravity in the general theory cannot be removed due to the presence of dynamical scalar/vector fields which are not determined by the metric.  This is also a general property of scalar-vector-tensor modifications, including Brans-Dicke gravity theory.  These strong equivalence principle violations do not of themselves imply any violation of the weak equivalence principle: Scalar-vector-tensor gravity preserves the {\it universality of free fall}.  This is important for the consideration of astrophysical phenomena, for which the {\it universality of free fall} is assumed.\index{Equivalence principle!Universality of free fall}\index{Modified gravity!Jordan-Brans-Dicke gravity}
\end{description}

Each of the modified gravity theories, including Modified Newtonian dynamics, in \sref{section.mog.mond}, Metric skew-tensor gravity, in \sref{section.mog.mstg}, and Scalar tensor vector gravity, in \sref{section.mog.stvg}, violate the strong equivalence principle, but maintain the universality of free fall for bodies in motion.

\index{Equivalence principle|)}

\section{\label{section.mog.mond}Modified Newtonian dynamics}\index{MOND|(}

\index{MOND!Equivalence principle violation}\index{Equivalence principle!Violations}Milgrom's modified Newtonian dynamics (MOND) is a nonrelativistic {\it small acceleration} model which softens the Newtonian \(1/r^{2}\) force law to the \(1/r\) behaviour preferred by galaxy rotation curves, introduced by \citet{Milgrom.APJ.1983.270.365,Milgrom.APJ.1983.270.371}.  MOND violates the strong equivalence principle since, at sufficiently low accelerations, the gravitational mass of a test particle exceeds the inertial mass.

\citet[Appendix B]{Bekenstein.APJ.1984.286} showed that a modified Newtonian potential may emerge, in the case of spherical symmetry, from a covariant Lagrangian formalism in which a cosmological scalar field, sourced by ordinary baryons, is added to the Einstein-Hilbert action.  This relativistic, metric-scalar gravity theory is a modification of \citet{Jordan.ZfP.1959.157,BransDicke:PR:1961} theory and similarly leads to violations of the strong equivalence principle, as in \sref{section.mog.equivalence.violations}.\index{Modified gravity!Jordan-Brans-Dicke gravity}
\begin{equation}\label{eqn.mog.mond.aether.dynamic.bransdicke}
S[g,\phi] =  \int d^4 x \sqrt{-g} \left[\frac{c^4}{16 \pi G} \left(\phi R-2\Lambda -\omega \frac{\phi_{,\nu}\phi^{,\nu}}{\phi}\right)\right],
\end{equation}
where the Jordan-Brans-Dicke parameter, \(\omega\), is not treated as a universal constant, but instead is treated as a function of the magnitude of the scalar gradient~\protect\citep{Sanders.MNRAS.1986.223}:
\begin{equation}\label{eqn.mog.mond.aether.dynamic.omega}
\omega =  \left(\omega_0 + \frac{3}{2}\right) \frac{f(x)}{x},
\end{equation}
where
\begin{equation}\label{eqn.mog.mond.aether.dynamic.x}
x =  \frac{c^4}{4} \frac{\phi_{,\nu}\phi^{,\nu}}{(2\omega_0+4)^{2}a_{0}^{2}\phi^{3}},
\end{equation}
and \(a_{0}\) is the Milgrom universal acceleration parameter, and 
\begin{equation}\label{eqn.mog.mond.aether.dynamic.mu}
\mu(x) =  \frac{df(x)}{dx},
\end{equation}
is the MOND interpolating function.\index{MOND!Interpolating function, \(\mu\)}.   \citet{Sanders.MNRAS.1986.223} extended the Bekenstein-Milgrom modification to include a fixed Yukawa-type length scale~\protect\citep{Yukawa:PRMSJ:1935}, which fits the galaxy rotation curves studied by \citet{Sanders.AAP.1986.154} so that at cosmic distances from the source, the gravitationally strong MOND force would vanish entirely.

To address the hypothesis stated in \sref{chapter.introduction.objective.theory}, of fitting galaxy rotation curves and galaxy cluster masses without dominant dark matter, Milgrom's acceleration law is presented in \sref{section.mog.mond.action}, and the resulting modified dynamics are considered in \sref{section.mog.mond.dynamic}.  In addition, the notion of building a relativistic, metric-scalar version of MOND is presented, and theories with dynamical preferred frames including Bekenstein's TEVES theory and the generalized Einstein-{\ae}ther theory are presented in \sref{section.mog.mond.aether}.

\subsection{\label{section.mog.mond.action}Milgrom's acceleration law}\index{Modified gravity!Phantom of dark matter|(}

\citet{Milgrom.APJ.1983.270.365} challenged the hidden mass hypothesis and introduced a nonrelativistic modification of Newtonian dynamics (MOND) at small accelerations, \(a < a_0\), whereupon the gravitational acceleration of a test particle is given by
\begin{equation}
\label{eqn.mog.mond.milgromacc} a \mu\biggl(\frac{a}{a_0}\biggr)=a_{\rm N},
\end{equation}
where $\mu(x)$ is a function that interpolates between the Newtonian regime, $\mu(x)=1$, when $x\gg 1$ and the MOND
regime, $\mu(x)=x$, when $x\ll 1$.  \citet{Milgrom.APJ.1983.270.371} introduced the interpolating function normally used for galaxy fitting,
\begin{equation}\label{eqn.mog.mond.interpolating}\index{MOND!Interpolating function, \(\mu\)}
\mu(x)=\frac{x}{\sqrt{1+x^2}},
\end{equation}
where
\begin{equation}\label{eqn.mog.mond.aether.dynamic.xr}
x \equiv x(r) = \left|\frac{\nabla \Phi(r)}{a_0}\right| = \left|\frac{a(r)}{a_0}\right|,
\end{equation}
and determined that the MOND acceleration was of the order \(a_{0} \approx cH_{0}/6\), and proportional to the Hubble constant, implying a cosmological connection to the modified dynamics.  

Substituting \erefs{eqn.mog.mond.interpolating}{eqn.mog.mond.aether.dynamic.xr} into \eref{eqn.mog.mond.milgromacc} gives,
\begin{equation}
\label{eqn.mog.mond.milgromseqn} \frac{a(r)^2}{\sqrt{a(r)^2+a_0^2}}=a_{\rm N}(r),
\end{equation}
which has the solution,
\begin{equation}
\label{eqn.mog.mond.milgromlaw} a(r) = a_0 \sqrt{\frac{1}{2}\left(\frac{a_N(r)}{a_0}\right)^2+\sqrt{\frac{1}{4}\left(\frac{a_N(r)}{a_0}\right)^4 + \left(\frac{a_N(r)}{a_0}\right)^2}},
\end{equation}
written in terms of the Newtonian acceleration of a test particle at a separation, \(r\),
\begin{equation}
\label{eqn.mog.mond.newtonianacc} a_N(r) = \frac{G_{N}M(r)}{r^2},
\end{equation}
where \(M(r)\) is the baryonic mass integrated within a sphere of radius, \(r\).  

Milgrom's acceleration law, given by \eref{eqn.mog.mond.milgromlaw}, is applied to galaxy rotation curves in \cref{chapter.galaxy}, in \eref{eqn.galaxy.mond.milgromlaw}.  The galaxy rotation curves, plotted in \fref{figure.galaxy.velocity}, are {\bf one parameter} best-fits by the stellar mass-to-light ratio, \(\Upsilon\), applying the MOND acceleration, \(a_0\) of \eref{eqn.galaxy.mond.a0}, universally.  Milgrom's acceleration law is applied to clusters of galaxies in \cref{chapter.cluster}, according to \sref{section.cluster.models.mond}.  In \sref{section.cluster.models}, the MOND mass is best-fitted to the X-ray gas mass of a sample of 11 clusters of galaxies, and plotted in \fref{figure.cluster.models.mass} according to the best-fit cluster model parameters tabulated in Panel (b) of \tref{table.cluster.models.bestfit}, for Milgrom's MOND.\index{Modified gravity!Phantom of dark matter|)}
\subsection{\label{section.mog.mond.dynamic}Modified dynamics at small acceleration}

Substituting \eref{eqn.mog.mond.newtonianacc} into \eref{eqn.mog.mond.milgromacc}, the MOND acceleration law can be written,
\begin{equation}
\label{eqn.mog.mond.mondacc} a(r) = \frac{1}{\mu(r)}\frac{G_{N}M(r)}{r^2},
\end{equation}
and therefore MOND can be interpreted as gravity theory with a varying gravitational coupling
\begin{eqnarray}
\label{eqn.mog.mond.mondacc} a(r) &=& \frac{G(r)M(r)}{r^2},\\
\label{eqn.mog.mond.gravitational} G(r) &=& \frac{G_{N}}{\mu(r)},
\end{eqnarray}
and \(G(r) \sim G_N\) in the Newtonian regime and \(G(r) > G_{N}\) in the MOND regime.  It is important to note that MOND has a classical instability in the deep MOND regime corresponding to \(\mu \rightarrow 0\) which leads to a divergent gravitational coupling of \eref{eqn.mog.mond.gravitational}, and that MOND violates the strong equivalence principle for all \(\mu \ne 1\).

For gravity fields interior to galaxies and clusters of galaxies, the accelerations are sufficiently small that the MOND interpolating function, \(\mu(x) \gg 1\), so that the Newtonian dynamic mass determined by MOND is much larger than the actual mass visible in the system.

\citet{Angus.APJL.2007.654} clarified the central issue in regards to gravitational lensing and the modified dynamics at small acceleration, since the total mass of baryons enclosed in a sphere of radius, \(r\), is determined from the divergence theorem,\index{Gravitational lensing!Modified dynamics}
\begin{equation}\label{eqn.mog.mond.aether.dynamic.mass}
M(r) =  \int \frac{\sin \theta d\theta d\phi}{4\pi G(r)}\frac{\partial\Phi(r,\theta,\phi)}{\partial r},
\end{equation}
where \(\Phi\) is the modified gravitational potential, and \(G(r)\) is given by \eref{eqn.mog.mond.gravitational}.  Therefore, the MOND dynamic mass factor is precisely the inverse of the MOND interpolating function,\index{Dynamic mass factor, \(\Gamma\)}
\begin{equation}\label{eqn.mog.mond.aether.dynamic.Gamma}
\Gamma(r) = \frac{G(r)}{G_N} = 1/\mu(x(r)),
\end{equation}
plotted in \fref{figure.galaxy.Gamma} for the Ursa Major filament of galaxies, and in \fref{figure.cluster.models.Gamma} for the sample of X-ray clusters of galaxies.

\citet{Bekenstein.PRD.2006.73} considered the behaviour of the MOND interpolating function in the deep MOND regime signalled by the small gradient of the dynamical scalar field, \(\phi\) of \sref{section.mog.mond.aether.bekenstein}, where \(\mu(x) \approx x\) and \eref{eqn.mog.mond.aether.dynamic.xr} implies
\begin{equation}\label{eqn.mog.mond.aether.dynamic.mur}
\mu(r) \approx  \left|\frac{\nabla \Phi(r)}{a_0}\right| = \left|\frac{a(r)}{a_0}\right|.
\end{equation}
In this regime, say far outside a spherically symmetric point source of mass, \(M\),  the \citet{Milgrom.APJ.1983.270.365} acceleration law,
\begin{equation}
\label{eqn.mog.mond.aether.dynamic.milgromacc} a \mu\biggl(\frac{a}{a_0}\biggr)=\frac{G_{N} M}{r^2},
\end{equation}
simplifies upon substitution of \eref{eqn.mog.mond.aether.dynamic.mur}:
\begin{equation}
\label{eqn.mog.mond.aether.dynamic.milgromacc.r} a(r)=\sqrt{\frac{a_{0}G_{N} M}{r^{2}}} = \frac{\sqrt{a_{0}G_N M}}{r},
\end{equation}
and thus the modified dynamics, at small acceleration scales, yields the gravitational field as \(1/r\) instead of the Newtonian \(1/r^{2}\) law.  Substitution of \eref{eqn.mog.mond.aether.dynamic.milgromacc.r} into \eref{eqn.mog.mond.aether.dynamic.mur} gives
\begin{equation}\label{eqn.mog.mond.aether.dynamic.mur.deep}
\mu(r) \approx  r^{-1}\sqrt{\frac{G_{N}M}{a_0}},
\end{equation}
which is valid in the deep MOND regime.  Substitution of \eref{eqn.mog.mond.aether.dynamic.mur.deep} into the dynamic mass factor of \eref{eqn.mog.mond.aether.dynamic.Gamma} gives
\begin{equation}\label{eqn.mog.mond.aether.dynamic.Gamma.deep}
\Gamma(r) = r\sqrt{\frac{a_0}{G_{N}M}}
\end{equation}
in the deep MOND regime, which shows a linear dependence with \(r\) at large distances. \citet{Milgrom.APJ.2008.678} defined a transition radius in MOND,
\begin{equation}\label{eqn.mog.mond.aether.dynamic.rt}
r_{t} = \sqrt{\frac{G_{N}M}{a_0}},
\end{equation}
so that the dynamic mass factor of \eref{eqn.mog.mond.aether.dynamic.Gamma.deep} in the deep MOND regime, can be written,
\begin{equation}\label{eqn.mog.mond.aether.dynamic.Gamma.deep.rt}
\Gamma(r) = \frac{r}{r_{t}},
\end{equation}

The dynamical mass factors plotted in \fref{figure.galaxy.Gamma}, in \cref{chapter.galaxy}, do indeed show a monotonically near-linear increasing \(\Gamma(r) < 10\), reaching the maximum value at the outermost observed data point, \(r_{\rm out}\), where \(r \sim 10 \cdot r_{t}\), typically.  This may imply that the MOND interpolating function is bounded from below, \(\mu > 0.1\).\index{MOND!Interpolating function, \(\mu\)}  Otherwise, as the gradient of the scalar field approaches zero, and the MOND interpolating function of \eref{eqn.mog.mond.aether.dynamic.mur} approaches zero, the dynamic mass factor of \eref{eqn.mog.mond.aether.dynamic.Gamma} grows without bound, \(\Gamma(r) \rightarrow \infty\) indicating a classical instability.

\index{MOND!Interpolating function, \(\mu\)}All of the modified gravity models examined in this dissertation provide the needed phantom dark matter, which is quantified by the dynamic mass factor, \(\Gamma>1\).  For MOND this corresponds to \(\mu(a/a_N)<1\), although it is not known if the MOND interpolating function approaches 0, this would correspond to the ultra-deep MOND regime and if MOND's dynamic mass factor is not bounded, \(\Gamma \rightarrow 0\), would effectively renormalize gravity's coupling \(G\rightarrow \infty\).  Conversely, if the inverse of the MOND interpolating function approaches a finite value, so \(\mu_\infty < \mu < 1\), then the instability of the theory is made regular (finite), and instead gravity's coupling approaches an asymptotic value, \(G\rightarrow G_\infty\).  This is consistent with the Ursa major sample of \sref{section.galaxy.uma}, from which it is clear that \(\mu_\infty < 10\).  Such a cutoff applied to clusters of galaxies could potentially cure MOND's unfortunate prediction of requiring dominant dark matter to fit clusters of galaxies, as in \cref{chapter.cluster}.  However, the final form of MOND's interpolating function should be dynamically determined from -- or at least correlated with -- the action of the covariant field theory from which it is derived. \index{MOND|)}
\subsection{\label{section.mog.mond.aether}Dynamical preferred frames}

The antithesis of Einstein's theory of special relativity, with local Lorentz SO(3,1) invariance, is the {\ae}ther theory in which the symmetry is broken.  Named after the luminiferous {\ae}ther -- the medium for the propagation of light as it was thought until the late 19th century -- the {\ae}ther theory is a generally covariant extension of general relativity by the addition of a unit timelike vector field.  The {\ae}ther has a preferred rest frame, and thus breaks local Lorentz SO(3,1) invariance.  In an address delivered on May 5, 1920, at the University of Leyden, Einstein commented,
\begin{quote}
How does it come about that alongside of the idea of ponderable matter, which is derived by abstraction from everyday life, the physicists set the idea of the existence of another kind of matter, the {\ae}ther? The explanation is probably to be sought in those phenomena which have given rise to the theory of action at a distance, and in the properties of light which have led to the undulatory theory.\end{quote}
It is the space-time components of the Maxwell field, \(F_{\mu \nu}\), which undulate; whereas the {\ae}ther vector field is not free to undulate because it is constrained to spacelike oscillations, and the vacuum cannot be empty of {\ae}ther excitations.

\subsubsection{\label{section.mog.mond.aether.bekenstein}Bekenstein's TEVES theory}

\citet{Bekenstein:PRD:2004} introduced the tensor-vector-scalar (TEVES) theory as a relativistic implementation of Milgrom's modified Newtonian dynamics (MOND), as in \sref{section.mog.mond}, with an additional scalar field, \(\phi\), and also a non-dynamical scalar field, \(\sigma\).  The vector field in TEVES, \(A_\mu\), has timelike unit norm,
\begin{equation}\label{eqn.mog.mond.aether.bekenstein.vectorconstraint}
g^{\mu\nu}A_{\mu}A_{\nu} = -1,
\end{equation}
and dynamically selects a preferred reference frame, breaking local Lorentz SO(3,1) invariance.   \(g_{\mu\nu}\) is the Einstein metric, with a well defined inverse, \(g^{\lambda\mu}\), such that
\begin{equation}\label{eqn.mog.mond.aether.bekenstein.metric.inverse}
g^{\lambda\mu}g_{\mu\nu} = \delta^{\lambda}_{\nu}.
\end{equation}
However, all types of matter see the same physical metric
\begin{equation}\label{eqn.mog.mond.aether.bekenstein.metric.physical}
{\tilde g}_{\mu\nu} = e^{-2\phi}g_{\mu\nu}-2\mbox{sinh}(2\phi)A_{\mu}A_{\nu},
\end{equation}
with a well defined inverse,
\begin{equation}\label{eqn.mog.mond.aether.bekenstein.metric.physical.inverse}
{\tilde g}^{\lambda\mu} = e^{-2\phi}g^{\lambda\mu}+2\mbox{sinh}(2\phi)g^{\lambda\alpha}g^{\mu\beta}A_{\alpha}A_{\beta},
\end{equation}
so adding a preferred frame is not in conflict with the weak equivalence principle.
However, because TEVES is a relativistic, bimetric theory, it permits the computation of geodesics in the presence of matter sources, and makes predictions for lensing convergences, time-delays and other metric effects~\protect\citep{Zhao.astro-ph:0611777}.\index{Gravitational lensing!Modified dynamics}

In TEVES, the vector field action is taken to be that of a Maxwell vector field, \(A_{\mu}\), with an additional Lagrange multiplier, \(\lambda\), to enforce the timelike, unit norm constraint of the vector field of \eref{eqn.mog.mond.aether.bekenstein.vectorconstraint}.  The action for the pair of scalar fields, \(\phi\) and \(\sigma\), is a generalization of the \citet{Bekenstein.PLB.1988.202} phase coupling gravity (PCG) theory including a vector-scalar interaction.  The total action for TEVES is formed by combining the Einstein-Hilbert action of \eref{eqn.mog.stvg.action.lagrangian.EH} with the vector and scalar actions:
\begin{eqnarray}\nonumber
S[g,A,\phi,\sigma] &=&  \int d^4 x \sqrt{-g} \left\{\frac{c^4}{16 \pi G} \left[ R-2\Lambda - \frac{K}{2}F_{\mu\nu}F^{\mu\nu} + \lambda\left(A_{\mu}A^{\mu}+1\right)\right]\right.\\ 
\label{eqn.mog.mond.aether.bekenstein.action} && \left.\phantom{\frac{c^4}{G}} -\frac{1}{2}\left[\sigma^2\left(g^{\mu\nu} -A^{\mu}A^{\nu}\right)\phi_{,\mu}\phi_{,\nu} + \frac{1}{2}G\ell^{-2}\sigma^4 \mathcal{F}(kG\sigma^2)\right]\right\},
\end{eqnarray}
where \(F_{\mu\nu} = A_{\nu,\mu} - A_{\mu,\nu}\) is the Maxwell vector field strength, \(K\) and \(k\) are dimensionless couplings,  \(\ell\) is a positive constant with units of length, and \(\mathcal{F}\) is a free dimensionless function, similar to the PCG potential, whose behaviour is determined phenomenologically by requiring that the dynamics at slow accelerations correspond to MOND.

\citet{Bekenstein.PRD.2006.73} predicted a universal acceleration scale, in terms of the positive coupling constant, \(k\), and the length scale, \(\ell\), of the TEVES action of \eref{eqn.mog.mond.aether.bekenstein.action}
\begin{equation}\label{eqn.mog.mond.aether.dynamic.milgromacc}
a_{0} =  \frac{\sqrt{3k}}{4\pi\ell} \approx 10^{-8}\ \mbox{cm s}^{-2},
\end{equation}
consistent with the MOND acceleration of \eref{eqn.galaxy.mond.a0}.

\citet{Zhao.APJL.2006.638} argued that the \citet{Bekenstein:PRD:2004} model produces a MOND interpolating function with the wrong behaviour to accurately fit galaxy rotation curves; and suggested a refinement to the TEVES Lagrangian to accommodate the standard MOND interpolating function of \eref{eqn.mog.mond.interpolating}.\index{MOND!Interpolating function, \(\mu\)}

\subsubsection{\label{section.mog.mond.aether.jacobson}Einstein-{\ae}ther theory}

\citet{Jacobson.PRD.2001.64} proposed a generally covariant model in which local Lorentz invariance is broken by a dynamical unit timelike vector field, \(A_{\mu}\), which is nowhere vanishing.  The Einstein-{\ae}ther theory leads to gravity with a dynamical preferred frame, via the \citet{Jacobson.PRD.2004.70} action
\begin{equation}\label{eqn.mog.mond.aether.action}
S[g,A] = S[g] - \int d^4 x \sqrt{-g} \left[\frac{c^4}{16 \pi G} \left({K^{\alpha \beta}}_{\mu\nu} \nabla_{\alpha}A^{\mu}\nabla_{\beta}A^{\nu}+\lambda(A^{\alpha}A_{\alpha}-1)\right)\right],
\end{equation}
where \(S[g]\) is the Einstein-Hilbert action of \eref{eqn.mog.mstg.action.action.EH}, and 
\begin{equation}\label{eqn.mog.mond.aether.action.k}
{K^{\alpha \beta}}_{\mu\nu} = c_{1} g^{\alpha \beta}g_{\mu\nu} +  c_{2} {\delta^{\alpha}}_{\mu}{\delta^{\beta}}_{\nu} +  c_{3} {\delta^{\alpha}}_{\nu}{\delta^{\beta}}_{\mu} + c_{4}g_{\mu\nu}A^{\alpha}A^{\beta},
\end{equation}
is written in terms of four dimensionless coefficients, \(c_{i}\), and \(\lambda\) is a lagrange multiplier which enforces the unit timelike nature of the vector field.

\citet{Jacobson.ArXiv:0801.1547} reviewed the theory, phenomenology, and observational constraints on the coupling parameters of Einstein-{\ae}ther gravity, showing that the unit timelike vector field, which breaks the local Lorentz invariance, must be dynamical; and the preferred frame must therefore be dynamical.  

\citet{Jacobson.PRD.2001.64} showed that such a field carries a nonlinear representation of the local Lorentz group since the field does not take values in a vector space on the tangent space, but on the unit hyperboloid.  \citet{Jacobson.PRD.2004.70} developed the linearized Einstein-{\ae}ther theory, finding the speeds and polarizations of the wave modes, determining in addition to the usual two transverse traceless metric modes, three coupled {\ae}ther-metric modes.  \citet{Eling.CQG.2006.23} claimed that regular perfect fluid star solutions exist with static {\ae}ther exteriors, with the {\ae}ther field pointing in the direction of a timelike Killing vector, but there are no spherically-symmetric solutions constructed purely from the {\ae}ther without naked singularities.  \citet{Seifert.PRD.2007.76} applied the action of \eref{eqn.mog.mond.aether.action}, and found that the flat space solution and the static vacuum {\ae}ther solution of \citet{Eling.CQG.2006.23} is stable to linear perturbations, provided the coefficients \(c_{i}\) satisfy an auxiliary inequality relation.

\citet{Clayton.gr-qc:0104103} showed that Einstein-{\ae}ther theories of the type of \eref{eqn.mog.mond.aether.action} are energetically unstable, having a Hamiltonian, in Minkowski flat space-time, that is unbounded from below; and the linearized analyses about configurations with a vanishing {\ae}ther vector field are singular. \citet{Jacobson.ArXiv:0801.1547} pointed out that \citet{Clayton.gr-qc:0104103} considered the question of energy positivity, but examined a limited {\it Maxwellian} subclass of \eref{eqn.mog.mond.aether.action.k} in which \(c_{3}=-c_{1}\) and \(c_2=c_4=0\), and restricted to the case where the coupling to gravity is neglected.  \citet{Seifert.PRD.2007.76} confirmed that the subclass investigated by \citet{Clayton.gr-qc:0104103} has spherically symmetric static solutions which are unstable, likely related to the unbounded Hamiltonian, even though kinetic terms in the unit timelike vector for a range of coefficients, \(c_{i}\),  that were ignored by \citet{Clayton.gr-qc:0104103}, may stabilize the theory.  \citet{Jacobson.ArXiv:0801.1547} suggested that the linear perturbations all have positive energy for coefficients, \(c_{i}\), within a particular range, but the total nonlinear energy has not been shown to be positive in this range. It is an unsatisfactory situation that the theory requires special values of the \(c_{i}\).  This places too great a burden on phenomenology, limiting the theory's ability to make testable and falsifiable predictions, but there have been no successful attempts to identify a principle of symmetry to restrict the action.  \citet{Carroll.PRD.2009.79.A,Carroll.PRD.2009.79.B} found that a timelike vector field leads to an unbounded Hamiltonian, and generates instability, except provided the kinetic term in the action is in the form of a \(\sigma\)-model, and introduced a \(\sigma\)-model {\ae}ther modified gravity theory, with a timelike vector field. 

\citet{ArkaniHamed.JHEP.2005.7} studied the effects of direct couplings between the Goldstone boson (which appear due to the broken time diffeomorphism symmetry), and standard model fermions, which necessarily accompany Lorentz-violating terms in the theory, finding that the {\ae}ther field couples to spin in the non-relativistic limit. A spin moving relative to the {\ae}ther rest frame will emit Goldstone-Cerenkov radiation.  The Goldstone boson also induces a long-range inverse-square law force between spin sources.

\subsubsection{\label{section.mog.mond.aether.ae}Generalized Einstein-{\ae}ther theory}

\citet{Zlosnik.PRD.2006.74} interpreted TEVES as a special case of the Einstein-{\ae}ther theory of \sref{section.mog.mond.aether.jacobson} with non-canonical kinetic terms, and showed that there exists a tensor-vector-scalar theory equivalent to TEVES, without the additional scalar field, \(\phi\), but retains the non-dynamical scalar field, \(\sigma\).  The equivalent theory is cast as a single-metric theory, because the Einstein metric which satisfies the Einstein-Hilbert action couples minimally to the matter fields.  However, there would be modifications to gravity resulting from the metric coupling to the vector field as a direct consequence of the Lorentz violating, dynamical {\ae}ther.  \citet{Zlosnik.PRD.2007.75} generalized the Einstein-{\ae}ther theory of \citet{Eling.PRD.2004.69}, replacing 
\begin{equation}\label{section.mog.mond.aether.bekenstein.zlosnik}
\mathcal{K} = M^{-2} {K^{\alpha \beta}}_{\mu\nu}\nabla_{\alpha}A^{\mu}\nabla_{\beta}A^{\nu},
\end{equation}
by \(\mathcal{F}(\mathcal{K})\), where \({K^{\alpha \beta}}_{\mu\nu}\) is given by \eref{eqn.mog.mond.aether.action.k}, but restricted to a class of theories spanned by the first three coefficients \(c_{i}\), \(i=1,2,3\), and \(M\) has the dimension of mass in order to make \eref{section.mog.mond.aether.bekenstein.zlosnik} dimensionless.  Although the generalized Einstein-{\ae}ther theory of \citet{Zlosnik.PRD.2007.75} does not include the TEVES equivalent theory of \citet{Zlosnik.PRD.2006.74}, each of these theories are reducible to MOND in the weak-field limit due to the never vanishing vector field.  \citet{Carroll.PRD.2009.79.A,Carroll.PRD.2009.79.B} found that because the Lorentz violating timelike vector field has kinetic terms in the action that are not in the form of a \(\sigma\)-model, the theory leads to an unbounded Hamiltonian, and is not stable, whereas \(\sigma\)-model {\ae}ther modified gravity probably does not have a low acceleration MOND limit.

\citet{Seifert.PRD.2007.76} considered the stability of spherically symmetric solutions in TEVES, without matter fields, finding that the perturbational Hamiltonian arising from the variational principle has an indefinite kinetic term.  In the absence of a well-defined variational principle with a sensible inner product, \citet{Seifert.PRD.2007.76} applied a WKB analysis to measure the instability of the spherically symmetric {\it vacuum} solution, and predicted a timescale of \(10^6\) seconds -- two weeks -- before a solar mass object would collapse under the weight of the nonvanishing vector-scalar fields.

\citet{Contaldi.PRD.2008.78} confirmed that TEVES is a fully causal theory for positive values of the scalar field, and represents a relativistic modification of gravity which may depend on acceleration (since one must have a reference frame to measure the acceleration), but develops classical singularities which may prevent the weak acceleration limit from resembling MOND; and argued that caustic singularities are symptomatic of Einstein-{\ae}ther theory, in general.  However, \citet{Contaldi.PRD.2008.78} speculated that problems with the vector field dynamics may be rectified by choosing more general kinetic terms, which may also include MOND in the nonrelativistic limit.

\section{\label{section.mog.mstg}Metric skew-tensor gravity}\index{Modified gravity!Metric skew-tensor gravity|(}\index{Equivalence principle!Violations}\index{Fifth force!Yukawa meson|(}\index{Modified gravity!Phantom of dark matter}\index{Equivalence principle!Universality of free fall}\index{Modified gravity!History}

\citet{Nieuwenhuizen.NPB.1973.60} found that the only massive antisymmetric tensor fields free of ghosts, tachyons and higher-order poles in the propagator for linearized gravitation are the massive spin-1 Maxwell-Proca fields.  \citet{Isenberg.AP.1977.107} performed a Hamilton-Dirac analysis of vector fields, determining that only Maxwell fields, Proca-Maxwell fields, and purely longitudinal vector fields are free of instability when minimally coupled to gravity.

In light of the difficulty in obtaining physically consistent modified gravity theories, it is instructive to study the emergent Kalb-Ramond-Proca field, as in \sref{section.mog.mstg.krp}.  The action in \sref{section.mog.mstg.action} for the metric skew-tensor gravity (MSTG) theory, given by \eref{eqn.mog.mstg.action.lagrangian}, couples an Einstein metric background (the metric sector) to the Kalb-Ramond-Proca field (massive skew sector).  To address the hypothesis stated in \sref{chapter.introduction.objective.theory}, of fitting galaxy rotation curves and galaxy cluster masses without dominant dark matter, it is sufficient to work in the weak-field spherically symmetric limit of MSTG, where the test particle equations of motion, calculated in \sref{section.mog.mstg.eom}, are used to derive the point source acceleration law in \sref{section.mog.mstg.yukawa} and effective Poisson equations are deduced in \sref{section.mog.mstg.mog} for distributions of matter.  The quadratic equations for the MSTG dynamic mass are solved exactly in \sref{section.mog.mstg.dynamic} by \erefs{eqn.mog.mstg.mass.mstg.soln}{eqn.mog.mstg.mass.mstg.xi}.
\subsection{\label{section.mog.mstg.krp}Kalb-Ramond-Proca field}\index{Nonsymmetric gravitation theory!Kalb-Ramond-Proca field}

\citet{Clayton.JMP.1996.37} showed that the massive nonsymmetric gravity theory (NGT) becomes identical to a Kalb-Ramond-Proca field with an additional curvature coupling term when considered as a perturbation about a Ricci-flat background.  Since the Kalb-Ramond-Proca theory does not require a conserved current and yet has no negative energy ghost modes, higher order poles or tachyons, the additional terms in the action for massive NGT allow the linearized field equations to take on this form in the antisymmetric sector.

\citet{Moffat.JMP.1995.36,Moffat.PLB.1996.378}\index{Nonsymmetric gravitation theory!Yukawa force}\index{Fifth force!Yukawa meson} determined that in the weak-field approximation relevant to galaxy dynamics, a range dependent Yukawa-type, fifth force~\protect\citep{Yukawa:PRMSJ:1935} emerges in addition to the Newtonian \(1/r^2\) central force due to the exchange of the spin-1\(^{+}\) skewons between fermions; and asserted that this additional potential due to the interaction of the field structure with matter in the halos of galaxies can explain galaxy rotation curves, as in \cref{chapter.galaxy}, and is a candidate for phantom dark matter.  This hypothesis is studied in the dynamics of the weak-field, as in \sref{section.galaxy.dynamics}, using the modified gravity theory of \sref{section.galaxy.dynamics.mog}, and extended to clusters of galaxies, in \cref{chapter.cluster}, with running gravitational couplings, as in \sref{section.cluster.models.mog}.

Geodesic and path motion in the nonsymmetric gravitational theory (NGT) were shown in  \citet{Moffat.PLB.1995.355} to have similar weak-field limits.  The correction to the weak-field gravitational force was found to be due to a Yukawa potential, resulting in a renormalized gravitational coupling.  The Yukawa interaction, considered as an alternative to dominant dark matter, must account for the majority of astrophysical forces and meanwhile be completely undetected at terrestrial scales.  It is remarkable that the astrophysical studies in \pref{part.astroph} show that the dark matter to baryon ratio can be consistently explained using the same Yukawa meson theory, from the smallest dwarf galaxies to the clusters of galaxies.  Measurements in the weak-field, according to \sref{section.galaxy.dynamics}, provide support of the hypothesis that dark matter is a phantom of MSTG, with galaxy specific density distributions.   Whereas the best-fitting dark matter theory, according to \eref{eqn.introduction.objective.darkmatter} of \sref{chapter.introduction.objective.darkmatter}, requires at least two additional dark matter parameters, \(\rho_0, r_s\), per galaxy, MSTG provides low reduced-\(\chi^2\) best-fits with universal mean parameters across galaxy scales.  Clusters of galaxies, however, show significantly improved \(\chi^2\) best-fits with variable parameters. \index{Fifth force!Yukawa meson|)}

Violations of the strong equivalence principle, described in \sref{section.mog.equivalence.violations}, are the means by which scalar-vector-tensor modifications to the action for gravity result in a fifth force which preserves the {\it universality of free fall}.  The effect due to the scalar-vector-tensor fields on the motion of a test particle requires careful approximation, such as the weak-field limit of a static, spherically symmetric space-time.  At astrophysical scales, we neglect any contribution to the fifth force due to baryons with charged quantum numbers.  Alternatively, we seek {\it gravitationally strong} contributions to the fifth force from a Yukawa (range dependent) meson emerging from the spin-1\(^{+}\) massive vector skewon of the Kalb-Ramond-Proca field, as in \sref{section.mog.mstg.action}.

The first measurable predictions for galaxy dynamics in the NGT appeared in \citet{Moffat.astro-ph:9412095} and \citet{Moffat.PLB.1996.378}, where the appearance of a Yukawa-like potential produced by a new spin \(1^{+}\) boson interacting with fermions emerged.  In \citet{LegareMoffat.GRG.1996.28}, the effects of three new interactions were identified, and possible modifications to the geodesic and path motion were calculated in the weak-field limit.  It was recognized by these attempts to provide an alternate explanation to the dark matter paradigm that the static, weak-field, slow speed, spherically symmetric limit of NGT may provide an adequate solution to the missing mass problem through the nonvanishing skewon mass and the coupling to baryons.  In the static, spherically symmetric limit, the skewon field strength tensor,
\begin{equation}\label{eqn.galaxy.dynamics.mog.fieldstrength}
F_{[\mu \nu \lambda]} = \partial_{[\mu} g_{\nu \lambda]}
\end{equation}
has only one independent, non-zero component, \(F_{[\theta \phi r]}\).  The modifications to the radial orbit equations of motion were explicitly computed, and the surviving Yukawa contribution -- potentially attractive or repulsive -- added a new phenomenology to the dynamics of astrophysical scale measurements.

\index{Modified gravity!Phantom of dark matter}
For the case of a repulsive Yukawa potential added to the attractive Newtonian potential, \citet{Sanders.AAP.1984.136} provided a preliminary analysis of circular orbit velocities in which the combined potentials lead to a deviation from the inverse square-distance law and may produce rotation curves which are ``nearly flat from 10 to 100 kpc''.  \citet{Sanders.AAP.1984.136} speculated that ``a very low mass vector boson carries an effective antigravity force which on scales smaller than that of galaxies almost balances the normal attractive gravity force.''  

\index{Nonsymmetric gravitation theory!Yukawa force}\index{Fifth force!Yukawa meson}In principle, for each astrophysical experiment, the Yukawa coupling constant and the mass of the vector boson (Yukawa range) provide two additional parameters which may be modelled through the mass-to-light ratio.  \citet{Sanders.AAP.1986.154} provided a best-fit to six galaxy rotation curves ranging in size from 5 to 40 kpc to determine whether the modification to gravity is associated with a fixed length scale.  Using the overall best-fit Yukawa coupling and range parameters, \citet{Sanders.AAP.1986.154} computed mass-to-light ratios between 1 and 3, which are considered reasonable, showing no systematic variation with the size of the galaxy.  The observed infrared Tully-Fisher law is shown to be consistent with the predictions of the Yukawa modified gravity for large galaxies greater in size than 15 kpc, whereas the smaller galaxies under 10 kpc do not exhibit a {\it maximum} flat rotation velocity.  Admittedly, the sample is too small to statistically determine whether the parameters are universal constants, although best-fitting universal constants for the finite length-scale Yukawa repulsive gravity does lead to agreement with the data without introducing mass discrepancies.

\index{Nonsymmetric gravitation theory!Degrees of freedom}The issue of whether the Yukawa meson coupling and range are universal is not certain in the weak-field limit of NGT, where the Yukawa potential is emergent. \citet{Moffat.ArXiv.hepth:9512018} derived the mismatch between the six degrees of freedom in the full nonlinear theory, and the three degrees of freedom that survive in the symmetry reduced, and linear, weak-field limit due to a \citet{KalbRamond.PRD.1974.9} field, identified clearly as the skewon, \(h_{[\mu \nu]}\), to explain the effective, low energy coupling to the Yukawa meson.

\index{Nonsymmetric gravitation theory!Field equations}Whereas \citet{Moffat.ArXiv.astroph:0403266} developed the radial orbit equations of motion for the problem of galaxy rotation curves from the full NGT action,  \citet{Moffat.ArXiv.grqc:0404076} derived the linear weak-field approximation, from which the \citet{KalbRamond.PRD.1974.9} field emerges as the field strength of the massive skewon. The modified  acceleration law corresponds to the low energy, low speed limit of NGT, effectively suppressing the high energy contributions of the full theory.  Metric skew-tensor gravity (MSTG) is introduced in \citet{Moffat.JCAP05.2005}, where the modified acceleration law results from coupling the massive skew symmetric \(F_{\mu \nu \lambda}\) field to Einstein's metric.  At astrophysical scales, the emergent low energy Yukawa meson is the only feature of the full NGT left in  MSTG to explain galaxy rotation curves.
\subsection{\label{section.mog.mstg.action}Action}\index{Nonsymmetric gravitation theory!Kalb-Ramond-Proca field}\index{Modified gravity!Kalb-Ramond-Proca|(}


\citet{DDM:PhysRevD.47.1541} analysed a class of physically consistent and ghost-free nonsymmetric gravity models with finite range massive spin-\(1^{+}\) gauge boson described by a second rank skew symmetric tensor, \(A_{\mu \nu}\), with an action in which the skewon's field strength tensor is coupled to a conserved fermion current vector with a dimensionless coupling constant.  The similarity to Maxwell's field, but for a massive skewon (instead of a massless photon), is described by the massive \citet{KalbRamond.PRD.1974.9} action,
\begin{equation}\label{eqn.mog.mstg.action.action.F}
	S_{F}=\int d^4x\sqrt{-g}\big(\frac{1}{12}F_{\mu\nu\lambda}F^{\mu\nu\lambda}-\frac{1}{4}\mu^2A_{\mu\nu}A^{\mu\nu}\big),
\end{equation}
with
\begin{equation}\label{eqn.mog.mstg.action.F}
   F_{\mu\nu\lambda} = \partial_{\mu}A_{\nu\lambda} + \partial_{\nu}A_{\lambda\mu} + \partial_{\lambda}A_{\mu\nu}, 
\end{equation}
and \(\mu\) is the mass of the Kalb-Ramond-Proca field, \(A_{\mu\nu}\).   The action is invariant under diffeomorphisms, and invariant under the U(1) local gauge transformation,
\begin{equation}\label{eqn.mog.mstg.action.skewon.u1}
\delta_0 A_{\mu \nu} = \partial_\mu \epsilon_\nu - \partial_\nu \epsilon_\mu,
\end{equation} 
only in the massless case.  Therefore, the dependence of the action based on the Lagrangian of \eref{eqn.mog.mstg.action.action.F} must be quadratic in the Proca field's strength tensor, \(F_{\mu \nu \lambda}\) of \eref{eqn.mog.mstg.action.F}. 

\citet{DDM:PhysRevD.47.1541} showed that, although the Kalb-Ramond-Proca field leads to minuscule -- as yet unmeasured -- deviations from Newtonian gravity at terrestrial scales consistent with stringent bounds on possible violations of the weak equivalence principle, the field may acquire {\it gravitational} strength at sufficiently large astrophysical scales because the coupling is unbounded as the range increases, and that the magnitude of the field is proportional to the coupling.\index{Fifth force!Geometric origin}

\citet{Moffat.JCAP05.2005} introduced the metric skew-tensor gravity (MSTG) gravity theory by adding the Kalb-Ramond-Proca action, coupled to a conserved fermion current, to an Einstein-Hilbert action:
\begin{equation}\label{eqn.mog.mstg.action.lagrangian}
	S=S_{EH}+S_{F}+S_{FM}+S_M,
\end{equation}
where
\begin{equation}\label{eqn.mog.mstg.action.action.EH}
	S_{EH}=\frac{c^4}{16\pi G}\int d^4x\sqrt{-g}\big(R-2\Lambda\big),
\end{equation}
is the Einstein-Hilbert action, and \(S_{F}\) is the Kalb-Ramond-Proca action of \eref{eqn.mog.mstg.action.action.F}.  A possible action for the coupling between the Kalb-Ramond-Proca field and matter was suggested by \citet{DDM:PhysRevD.47.1541} in regards to the NGT, and applied to MSTG by \citet{Moffat.JCAP05.2005}, according to
\begin{equation}\label{eqn.mog.mstg.action.action.FM}
	S_{FM}=\int
d^4xF_{\lambda\mu\nu}J^{*\lambda\mu\nu} =-3\int
d^4x\epsilon^{\alpha\beta\mu\nu}A_{\alpha\beta}\partial_\mu J_\nu,
\end{equation}
where $J_\mu$ is a conserved vector current and
$J^{*\mu\nu\lambda}=\epsilon^{\mu\nu\lambda\alpha}J_\alpha$ is the
dual tensor current density.  \(S_{M}\) is the ordinary matter action.

Varying the action of \eref{eqn.mog.mstg.action.lagrangian} with respect to the metric,
\begin{equation}\label{eqn.mog.mstg.action.action.metricvariation}
\frac{1}{\sqrt{-g}}\frac{\delta S}{\delta
g^{\mu\nu}}=-\frac{1}{2}\left(T_{M\mu\nu}+T_{F\mu\nu}\right),
\end{equation}
gives the field equations,
\begin{equation}\label{eqn.mog.mstg.action.action.metriceom}
G_{\mu\nu}+\Lambda g_{\mu\nu}=\frac{8\pi G}{c^4}\left(T_{M\mu\nu}+T_{F\mu\nu}\right),
\end{equation}
where \(T_{M\mu\nu}\) and \(T_{F\mu\nu}\) are the energy-momentum tensors for matter and the Kalb-Ramond-Proca field, \(A_{\mu\nu}\), respectively.  The Bianchi identities satisfied by the Einstein tensor,
\begin{equation}\label{eqn.mog.mstg.action.einsteintensor}
G_{\mu\nu}=R_{\mu\nu}-\frac{1}{2}g_{\mu\nu}R
\end{equation}
lead to the conservation laws
\begin{equation}
\nabla^\nu (T_{M\mu\nu}+T_{F\mu\nu})=0.
\end{equation}

Varying the action of \eref{eqn.mog.mstg.action.lagrangian} with respect to the field, \(A_{\mu\nu}\), 
\begin{equation}\label{eqn.mog.mstg.action.action.krpvariation}
\frac{1}{\sqrt{-g}}\frac{\delta S}{\delta
A^{\mu\nu}}=-\frac{1}{\sqrt{-g}}J_{\mu\nu},
\end{equation}
gives the Kalb-Ramond-Proca field equations,
\begin{equation}\label{eqn.mog.mstg.action.action.krpeom}
\nabla^\sigma
F_{\mu\nu\sigma}+\mu^2A_{\mu\nu}=\frac{1}{\sqrt{-g}}J_{\mu\nu},
\end{equation}
where $J_{\mu\nu}$ is the tensor density source for the $A_{\mu\nu}$ field:
\begin{equation}\label{eqn.mog.mstg.action.action.krpsource}
J_{\mu\nu}=\epsilon_{\mu\nu\alpha\beta}\partial^\alpha J^\beta.
\end{equation}\index{Modified gravity!Kalb-Ramond-Proca|)}

\subsection{\label{section.mog.mstg.eom}Motion under the fifth force}

The equations of motion of a test particle are
\begin{equation}
\label{eqn.mog.mstg.eom.geodesic2} \frac{du^\mu}{d\tau}+ \left\{{\mu\atop
\alpha\beta}\right\}u^\alpha u^\beta
=g^{\mu\alpha}f_{\alpha\nu}u^\nu,
\end{equation}
where $\tau$ is the proper time along the path of the particle and
$u^\lambda=dx^\lambda/d\tau$ is the 4-velocity of the particle, and
\begin{equation}
\left\{{\lambda\atop \mu\nu}\right\}={1\over 2}g^{\lambda\rho}
\left(g_{\mu\rho,\nu}+g_{\rho\nu,\mu}-g_{\mu\nu,\rho}\right),
\end{equation}
is the Christoffel connection, and \(f_{\alpha\nu}\) is derived from the Euler-Lagrange equations for a test particle, of mass \(m\), and fifth force charge \(\lambda m\), where $\lambda$ couples the skew field to the test particle and is assumed constant for the universality of free fall.  Such a test particle has a point particle action~\protect\citep{LegareMoffat.gr-qc:9412074,Moffat.JCAP05.2005},
\begin{equation} \label{eqn.mog.mstg.eom.testparticleaction} 
S_{\rm TP} = -m \int d\tau \sqrt{g_{\alpha\beta}u^\alpha u^\beta} - \lambda m \int d\tau \frac{\epsilon^{\alpha\sigma\eta\lambda}}{\sqrt{-g}}
F_{\sigma\eta\lambda}g_{\alpha\beta}u^\beta.
\end{equation}
Variation of \eref{eqn.mog.mstg.eom.testparticleaction} yields \eref{eqn.mog.mstg.eom.geodesic2} with 
\begin{equation}\label{eqn.mog.mstg.eom.testparticle} 
f_{\alpha\nu}=\frac{1}{3}\lambda\partial_{[\alpha}
\biggl(\frac{\epsilon^{\mu\sigma\eta\lambda}}{\sqrt{-g}}
F_{\sigma\eta\lambda}g_{\nu]\mu}\biggr).
\end{equation}

For a spherically symmetric, static skew symmetric potential field
$A_{\mu\nu}$ there are two non-vanishing components, the
``magnetic'' field potential $A_{0r}(r)=w(r)$ and the ``electric''
potential field $A_{\theta\phi}(r)=f(r)\sin\theta$.  According to \citet{Moffat.JCAP05.2005}, we may 
assume that there are no static magnetic poles, so only the electric field contribution $f(r)\sin\theta$
is non-zero, Therefore, $F_{\mu\nu\lambda}$ has only one non-vanishing component:
\begin{equation}\label{eqn.mog.mstg.eom.fsintheta}
F_{\theta\phi r}=\partial_rA_{\theta\phi}=f'\sin\theta,
\end{equation}
where the prime notation implies differentiation with respect to \(r\), so $f'=df/dr$. Substituting \eref{eqn.mog.mstg.eom.fsintheta} into \eref{eqn.mog.mstg.eom.testparticle},
\begin{equation}\label{eqn.mog.mstg.eom.lineelement}
f_{r0}=\lambda\frac{d}{dr}\biggl(\frac{\gamma
f^\prime}{\sqrt{\alpha\gamma r^4}}\biggr).
\end{equation}

For a static spherically symmetric gravitational field the line
element has the form
\begin{equation}
\label{eqn.mog.mstg.eom.lineelement}
ds^2=\gamma(r)dt^2-\alpha(r)dr^2-r^2(d\theta^2+\sin^2\theta
d\phi^2),
\end{equation}
and the equations of motion for a test particle are
\begin{eqnarray} 
\nonumber
\frac{d^2r}{d\tau^2}+\frac{\alpha'}{2\alpha}\biggl(\frac{dr}{d\tau}\biggr)^2
-\frac{r}{\alpha}\biggl(\frac{d\theta}{d
\tau}\biggr)^2-r\biggl(\frac{\sin^2\theta}{\alpha}\biggr)\biggl(\frac{d\phi}{d\tau}\biggr)^2
&+&\frac{\gamma'}{2\alpha}\biggl(\frac{dt}{d\tau}\biggr)^2 \\
\label{eqn.mog.mstg.eom.rmotion} + \frac{1}{\alpha}\frac{d}{dr}\biggl(\frac{\lambda\gamma
f'}{\sqrt{\alpha\gamma
r^4}}\biggr)\biggl(\frac{dt}{d\tau}\biggr) &=& 0,\\
\label{eqn.mog.mstg.eom.tequation}
\frac{d^2t}{d\tau^2}+\frac{\gamma'}{\gamma}\biggl(\frac{dt}{d\tau}\biggr)
\biggl(\frac{dr}{d\tau}\biggr)+\frac{1}{\gamma}\frac{d}{dr}\biggl(\frac{\lambda\gamma
f'}{\sqrt{\alpha\gamma
r^4}}\biggr)\biggl(\frac{dr}{d\tau}\biggr)&=&0,\\
\frac{d^2\theta}{d\tau^2}+\frac{2}{r}\biggl(\frac{d\theta}{d\tau}
\biggr)\biggl(\frac{dr}{d\tau}\biggr)-\sin\theta\cos\theta\biggl(\frac{d\phi}{d\tau}\biggr)^2
&=&0,\\
\label{eqn.mog.mstg.eom.phiequation}
\frac{d^2\phi}{d\tau^2}+\frac{2}{r}\biggl(\frac{d\phi}{d\tau}\biggr)\biggl(\frac{dr}{d\tau}\biggr)
+2\cot\theta\biggl(\frac{d\phi}{d\tau}\biggr)\biggl(\frac{d\theta}{d\tau}\biggr)&=&0.
\end{eqnarray}

The motion of a test particle can be shown to lie in the plane, $\theta=\pi/2$,
by an appropriate choice of axes. Integrating \eref{eqn.mog.mstg.eom.phiequation} gives
\begin{equation}
\label{eqn.mog.mstg.eom.angular} r^2\frac{d\phi}{d\tau}=J,
\end{equation}
where $J$ is the conserved orbital angular momentum. Integration
of \eref{eqn.mog.mstg.eom.tequation} gives
\begin{equation}
\label{eqn.mog.mstg.eom.dtequation}
\frac{dt}{d\tau}=-\frac{1}{\gamma}\biggl[\frac{\lambda\gamma
f'}{\sqrt{\alpha\gamma r^4}}+E\biggr],
\end{equation}
where $E>0$ is the conserved orbital energy per unit mass ($E=0$ for the photon). Substituting \eref{eqn.mog.mstg.eom.dtequation} into \eref{eqn.mog.mstg.eom.rmotion}, and using
\eref{eqn.mog.mstg.eom.angular}, we obtain
\begin{equation}
\label{eqn.mog.mstg.eom.reducedrmotion}
\frac{d^2r}{d\tau^2}+\frac{\alpha'}{2\alpha}\biggl(\frac{dr}{d\tau}\biggr)^2
-\frac{J^2}{\alpha
r^3}+\frac{\gamma'}{2\alpha\gamma^2}\biggl(\frac{\lambda\gamma
f'}{\sqrt{\alpha\gamma r^4}}+E\biggr)^2
=\frac{1}{\alpha\gamma}\frac{d}{dr}\biggl(\frac{\lambda\gamma
f'}{\sqrt{\alpha\gamma r^4}}\biggr)\biggl(\frac{\lambda\gamma
f'}{\sqrt{\alpha\gamma r^4}}+E\biggr).
\end{equation}

\subsection{\label{section.mog.mstg.yukawa}MSTG acceleration law}

Approximating the line element of \eref{eqn.mog.mstg.eom.lineelement} by the Schwarzschild
solution:
\begin{equation}
\label{eqn.mog.mstg.eom.Schwarzschild} \alpha(r)\sim \left(1-\frac{2GM}{r}\right)^{-1},\quad
\gamma(r)\sim 1-\frac{2GM}{r},
\end{equation}
and making the approximations  $2GM/r\ll 1,\lambda f'/r^2\ll 1$,
$f/r^2\ll 1$ and the slow motion approximation $dr/dt\ll 1$, \eref{eqn.mog.mstg.eom.reducedrmotion}
becomes
\begin{equation}
\label{eqn.mog.mstg.eom.Newton} \frac{d^2r}{dt^2}-\frac{J_N^2}{r^3}+\frac{GM}{r^2}
=\lambda\frac{d}{dr}\biggl(\frac{f'}{r^2}\biggr),
\end{equation}
where $J_N$ is the Newtonian orbital angular momentum.

Transforming the Kalb-Ramond-Proca field equations of  \eref{eqn.mog.mstg.action.action.krpeom}, to polar coordinates for the components $A_{\theta\phi}=f(r)\sin\theta$,\index{Modified gravity!Kalb-Ramond-Proca}
\begin{equation}\label{eqn.mog.mstg.eom.weakfequation}
\biggl(1-\frac{2GM}{r}\biggr)f''-\frac{2}{r}\biggl(1-\frac{3GM}{r}\biggr)f'
-\biggl(\frac{\mu^2}{r^2}+\frac{8GM}{r}\biggr)f=0,
\end{equation}
which has the solution~\protect\citep{Moffat.JCAP05.2005}
\begin{equation}
\label{eqn.mog.mstg.eom.fweak} f(r)=\frac{1}{3}sG^2M^2\exp(-\mu r)(1+\mu r),
\end{equation}
where $s$ is a dimensionless constant.   The skew field is therefore an excellent candidate for the phantom of dark matter due to the result of \eref{eqn.mog.mstg.eom.fweak}, which leads to {\it gravitationally} strong astrophysical effects, with a geometric originating fifth force, similar to the appearance of a fifth force charge, \(Q_5\), as an integration constant in \eref{eqn.mog.stvg.yukawa.soln}, described in \sref{section.mog.stvg}, for STVG.\index{Modified gravity!Phantom of dark matter}

Substituting \eref{eqn.mog.mstg.eom.fweak} into \eref{eqn.mog.mstg.eom.Newton} gives
\begin{equation}
\label{eqn.mog.mstg.eom.Yukawa} \frac{d^2r}{dt^2}
-\frac{J_N^2}{r^3}=-\frac{GM}{r^2}+\frac{\sigma\exp(-\mu
r)}{r^2}(1+\mu r),
\end{equation}
where $\sigma$ is given by
\begin{equation}
\label{eqn.mog.mstg.eom.sigma} \sigma=\frac{\lambda sG^2M^2\mu^2}{3}.
\end{equation}

It is phenomenologically important for the modified acceleration law to be consistent with the observed Tully-Fisher relation, \(v^4 \propto M\)~\protect\citep{Tully.AAP.1977.54}, so MSTG requires that the constant, \(s\), be of the form
\begin{equation}
\label{eqn.mog.mstg.eom.renormS}
s = g M^{-3/2}, 
\end{equation}
so \eref{eqn.mog.mstg.eom.sigma} becomes:
\begin{equation}
\label{eqn.mog.mstg.eom.sigmammodel}
\sigma=\sqrt{M M_0}, 
\end{equation}
where 
\begin{equation}
\label{eqn.mog.mstg.eom.m0}
{M_0} = \left(\frac{\lambda g G }{3 r_0^2}\right)^2 , 
\end{equation}
is a parameter related to the
strength of the coupling of the Kalb-Ramond-Proca field to matter, and
\begin{equation}\label{eqn.mog.mstg.eom.r0}
r_0 = \frac{1}{\mu} , 
\end{equation}
and the gravitational constant, \(G\), in \eref{eqn.mog.mstg.eom.Yukawa} is taken to be:
\begin{equation}
\label{eqn.mog.mstg.eom.renormG}
G_{\infty}=G_N\biggl(1+\sqrt{\frac{M_0}{M}}\biggr),
\end{equation}
where $G_N$ is Newton's gravitational constant. If the dependence of \(\sigma\) on the source mass distribution is correctly modelled as per \eref{eqn.mog.mstg.eom.sigma}, then the MSTG mass parameter, \(M_0\) , defined by \eref{eqn.mog.mstg.eom.m0}, will be a universal constant.  \(\mu\) denotes the effective mass of the skewon, \(A_{\mu\nu}\), with reciprocal identified as the MSTG length parameter, \(r_0\), defined by \eref{eqn.mog.mstg.eom.r0}.

Substituting \erefss{eqn.mog.mstg.eom.sigmammodel}{eqn.mog.mstg.eom.r0}{eqn.mog.mstg.eom.renormG} into \eref{eqn.mog.mstg.eom.Yukawa}, and neglecting the Newtonian angular momentum, \(J_N\), we obtain the MSTG acceleration law,
\begin{equation}
\label{eqn.mog.mstg.eom.accelerationlaw3} a(r)=-\frac{G_NM}{r^2}
\biggl\{1+\sqrt\frac{M_0}{M}\biggl[1-\exp(-r/r_0)\biggl(1+\frac{r}{r_0}\biggr)
\biggr]\biggr\}.
\end{equation}

We can rewrite \eref{eqn.mog.mstg.eom.accelerationlaw3} in the form
\begin{equation}
\label{eqn.mog.mstg.eom.runG} a(r)=-\frac{G(r)M}{r^2},
\end{equation}
where
\begin{equation}
\label{eqn.mog.mstg.eom.runningG}
G(r)=G_N\biggl\{1+\sqrt\frac{M_0}{M}\biggl[1-\exp(-r/r_0)\biggl(1+\frac{r}{r_0}\biggr)
\biggr]\biggr\}.
\end{equation}
\subsection{\label{section.mog.mstg.mog}Poisson equations }

The experience of a test particle in the MSTG theory, moving in an extended matter distribution, is a combination of the force of gravity due to Einstein's metric gravity theory, and a coupled fifth force due to a Kalb-Ramond-Proca field.  The weak-field,  central potential for a static, spherically symmetric system can be split into two parts:\index{Newton's central potential}
\begin{equation}
\label{eqn.mog.mstg.mog.phipotential} \Phi(r)=\Phi_N(r)+\Phi_Y(r),
\end{equation}
where
\begin{equation}
\label{eqn.mog.mstg.mog.Newtonian} \Phi_N(r)=-\frac{G_{\infty}M}{r},
\end{equation}
and
\begin{equation}
\label{eqn.mog.mstg.mog.Yukawa} \Phi_Y(r)=\frac{G_N \sigma \exp(-\mu r)}{r}
\end{equation}
denote the Newtonian and Yukawa potentials, respectively, where  \(M\) denotes the total constant mass of a point source.  The point source gravitational coupling in \eref{eqn.mog.mstg.mog.Newtonian} is
\begin{equation}
\label{eqn.mog.mstg.mog.renormG} G_\infty  = G_N (1 + \frac{\sigma}{M}),
\end{equation}
where \(\sigma\), defined by \eref{eqn.mog.mstg.eom.sigma}, is dependent on the source mass distribution through a power-law model, derived from the Tully-Fisher relation, leading to the phenomenological parametrizations of \erefs{eqn.mog.mstg.eom.sigmammodel}{eqn.mog.mstg.eom.renormG}.  Since the Schwarzschild solution, according to \eref{eqn.mog.mstg.eom.Schwarzschild}, was used in the derivation of \eref{eqn.mog.mstg.mog.renormG}, \erefs{eqn.mog.mstg.eom.sigmammodel}{eqn.mog.mstg.eom.renormG} may be generalized to static, spherically symmetric matter distributions, using the interior solution of the Schwarzschild metric, and we may set \(M\) to the active mass interior to a sphere of radius, \(r\),
\begin{equation}
\label{eqn.mog.mstg.mog.massr} M=\int d^3{\bf r'}\rho({\bf r}').
\end{equation}
The Poisson equations for \(\Phi_N({\bf r})\) and \(\Phi_Y({\bf r})\) are given by
\begin{equation}
\label{eqn.mog.mstg.mog.NewtonPoiss} \nabla^2\Phi_N({\bf r})=-G_{\infty}\rho({\bf r}),
\end{equation}
and
\begin{equation}
\label{eqn.mog.mstg.mog.YukawaPoiss} (\nabla^2-\mu^2)\Phi_Y({\bf
r})=\frac{\sigma}{M}G_N \rho({\bf r}),
\end{equation}
respectively. For sufficiently weak fields, the
Poisson \erefs{eqn.mog.mstg.mog.NewtonPoiss}{eqn.mog.mstg.mog.YukawaPoiss} are
uncoupled and determine the potentials \(\Phi_N({\bf r})\) and
\(\Phi_Y({\bf r})\) for non-spherically symmetric systems, which can
be solved analytically and numerically. The Green's function for
the Yukawa Poisson equation is given by
\begin{equation}
(\nabla^2-\mu^2)\Delta_Y({\bf r})=-\delta^3({\bf r}).
\end{equation}
The full solutions to the potentials are given by
\begin{equation}
\label{eqn.mog.mstg.mog.fullNewton} \Phi_N({\bf r})=- G_N\int d^3{\bf
r'}\left(1+\frac{\sigma}{M}\right)\frac{\rho({\bf r'})}{4\pi\vert {\bf r}-{\bf r'}\vert},
\end{equation}
and
\begin{equation}
\label{eqn.mog.mstg.mog.fullYukawa} \Phi_Y({\bf r})=G_N\int d^3{\bf
r'}\frac{\sigma}{M}\exp(-\mu \vert{\bf r}-{\bf r'}\vert)\frac{\rho({\bf
r'})}{4\pi \vert {\bf r}-{\bf r'}\vert }.
\end{equation}

The modified acceleration law is the gradient of the potential of \eref{eqn.mog.mstg.mog.phipotential},
\begin{equation} \label{eqn.mog.mstg.mog.moda}
{\bf a}({\bf r})=-{\mathbf\nabla}\Phi=-\bigl({\mathbf\nabla}\Phi_N({\bf
r})+{\mathbf\nabla}\Phi_Y({\bf r})\bigr).
\end{equation}

Combining \erefss{eqn.mog.mstg.mog.fullNewton}{eqn.mog.mstg.mog.fullYukawa}{eqn.mog.mstg.mog.moda},
\begin{equation}\label{eqn.mog.mstg.mog.fullacc}
{\bf a}({\bf r}) = - G_N\int d^3{\bf r}'\frac{({\bf r}-{\bf r'})
\rho({\bf r'})}{4\pi\vert {\bf r}-{\bf
r}'\vert^3}\biggl\{1+\frac{\sigma}{M} \Bigl[1-\exp(-\mu\vert{\bf r}-{\bf
r}'\vert)(1+\mu\vert{\bf r}-{\bf r}'\vert)\Bigr]\biggr\}.
\end{equation}

Dividing the Tully-Fisher relation inspired phenomenological input of \eref{eqn.mog.mstg.eom.sigmammodel} by \(M\), 
\begin{equation}
\label{eqn.mog.mstg.mog.alpha} \frac{\sigma}{M}=\sqrt{\frac{M_0}{M}},
\end{equation}
and substituting \erefs{eqn.mog.mstg.eom.r0}{eqn.mog.mstg.mog.alpha} into \eref{eqn.mog.mstg.mog.fullacc}, we obtain
\begin{equation}
\label{eqn.mog.mstg.mog.fullGaccel} {\bf a}({\bf r})=-\int d^3{\bf
r}'\frac{({\bf r}-{\bf r'}) \rho({\bf r'})}{4\pi\vert {\bf r}-{\bf
r}'\vert^3}G({\bf r}-{\bf r}'),
\end{equation}
where
\begin{equation}\label{eqn.mog.mstg.mog.fullG} G({\bf r}-{\bf
r}') = G_N\biggl\{1+\biggl(\frac{M_0}{\int d^3{\bf r}'\rho({\bf
r}')}\biggr)^{1/2} \Bigl[1-\exp\left(-\frac{\vert{\bf r}-{\bf
r}'\vert}{r_0}\right)\left(1+\frac{\vert{\bf r}-{\bf r}'\vert}{r_0}\right)\Bigr]\biggr\}.
\end{equation}

For a \(\delta\)-function point source,
\begin{equation}
\label{eqn.mog.mstg.mog.delta} \rho({\bf r})=M \delta^3({\bf r}),
\end{equation}
the modified acceleration law of \erefs{eqn.mog.mstg.mog.fullGaccel}{eqn.mog.mstg.mog.fullG} reduces to the point source solution of \erefs{eqn.mog.mstg.eom.runG}{eqn.mog.mstg.eom.runningG}.

For a static, spherically symmetric system, the effective modified acceleration law is:
\begin{equation} \label{eqn.mog.mstg.mog.Gforce}
a(r)=-\frac{G(r)M(r)}{r^2},
\end{equation}
where
\begin{equation}
\label{eqn.mog.mstg.mog.fullGspherical}
G(r)=G_N\biggl\{1+\sqrt{\frac{M_0}{M(r)}}\biggl[1-\exp(-r/r_0)
\biggl(1+\frac{r}{r_0}\biggr)\biggr]\biggr\}.
\end{equation}
We observe that \(G(r)\rightarrow G_N\) as \(r\rightarrow 0\).
\subsection{\label{section.mog.mstg.dynamic}Dynamical mass measurements}\index{Modified gravity!Phantom of dark matter|(}

Comparison of \eref{eqn.mog.mstg.mog.Gforce} with the Newtonian acceleration law:
\begin{equation} \label{eqn.mog.mstg.mog.Newton}
a(r)=-\frac{G_{N}M_{N}(r)}{r^2},
\end{equation}
allows the interpretation of the modified gravity dynamic mass as a scaled version of the Newtonian dynamic mass,
\begin{equation}\label{eqn.mog.mstg.mass} 
M_{\rm MOG}(r)  = \frac{G_{N}M_{N}(r)}{G(r)},
\end{equation}
where the varying gravitation coupling, \(G(r)\), may take the form of \eref{eqn.mog.mstg.mog.fullGspherical}, derived in MSTG.  The MSTG dynamic mass,
\begin{equation}
\label{eqn.mog.mstg.mass.mstg}
M_{\rm MSTG}(r) = M_{N}(r) \biggl\{1+\sqrt{\frac{M_0}{M_{\rm MSTG}(r)}}\biggl[1-\exp(-r/r_0)
\biggl(1+\frac{r}{r_0}\biggr)\biggr]\biggr\}^{-1},
\end{equation}
has the exact analytic solution:
\begin{equation}
\label{eqn.mog.mstg.mass.mstg.soln}
M_{\rm MSTG}(r)=M_{N}(r) + M_{0}\xi(r) - \sqrt{{M_{0}}^2 \xi(r)^2+2 M_{0} M_{N}(r) \xi(r)},
\end{equation}\begin{equation}
\label{eqn.mog.mstg.mass.mstg.xi}
\xi(r)\equiv\frac{1}{2}\left[{1-\exp(-r/r_0)\biggl(1+\frac{r}{r_0}\biggr)}\right]^2,
\end{equation}
which is identified with the total baryonic mass within a separation, \(r\) from the center of the system.

This MSTG acceleration law is applied to galaxy rotation curves in \cref{chapter.galaxy}, in \erefs{eqn.galaxy.dynamics.mog.acceleration}{eqn.galaxy.dynamics.mstg}.  In \sref{section.galaxy.uma} , in order to compute the overall best-fitting mean result, \(M_0\) and \(r_0\) are permitted to vary across the sample of 29 galaxies, as tabulated in \tref{table.galaxy.mstg}.  The galaxy rotation curves, plotted in \fref{figure.galaxy.velocity}, are subsequently {\bf one parameter} best-fits by the stellar mass-to-light ratio, \(\Upsilon\), applying the mean results of \eref{eqn.galaxy.dynamics.mstg.parameters.meanuniversal} universally.  The MSTG acceleration law is applied to clusters of galaxies in \cref{chapter.cluster}, according to \ssref{section.cluster.models.mog}{section.cluster.models.mog.mstg}, in order to compute the scaling of the parameters, \(M_0\) and \(r_0\).  In \sref{section.cluster.models}, the MSTG mass is best-fitted to the X-ray gas mass of a sample of 11 clusters of galaxies, and plotted in \fref{figure.cluster.models.mass} according to the best-fit cluster model parameters tabulated in Panel (c) of \tref{table.cluster.models.bestfit}, for MSTG.   A summary of lessons learned from the application of MSTG to the astrophysics of galaxies and clusters of galaxies is supplied in \sref{section.summary.theory}.\index{Modified gravity!Phantom of dark matter|)}\index{Modified gravity!Metric skew-tensor gravity|)}

\section{\label{section.mog.stvg}Scalar-tensor-vector gravity}\index{Modified gravity!Scalar-tensor-vector gravity|(}

Whereas the metric skew-tensor gravity theory, of \sref{section.mog.mstg}, describes the effective, low energy Yukawa skewon as the Kalb-Ramond-Proca field, a separate solution is to model a gravity theory with a simpler Maxwell-Proca field, such as in the scalar-tensor-vector gravity (STVG) theory, which describes the low energy Yukawa phion, \(\phi_{\mu}\), as a massive spin-1\(^{-}\) vector field, described in \sref{section.mog.stvg.maxproca}.    The action in \sref{section.mog.stvg.action} for the STVG theory, given by \eref{eqn.mog.stvg.action.lagrangian}, includes an Einstein metric background (the gravity sector) to the Maxwell-Proca field in which the gravitational coupling, \(G\), and the phion coupling, \(\omega\), and the phion mass, \(\mu\), are treated as a triplet of scalar fields (scalar-tensor-vector sector).  To address the hypothesis stated in \sref{chapter.introduction.objective.theory}, of fitting galaxy rotation curves and galaxy cluster masses without dominant dark matter, it is sufficient to work in the weak-field spherically symmetric limit of STVG, where the test particle equations of motion, calculated in \sref{section.mog.stvg.eom}, are used to derive the point source acceleration law in \sref{section.mog.stvg.yukawa} and effective Poisson equations are deduced in \sref{section.mog.stvg.mog} for distributions of matter.  The STVG dynamic mass is provided in \sref{section.mog.stvg.dynamic} by \eref{eqn.mog.mstg.mass.stvg} which is nonlinear through \erefs{eqn.mog.stvg.mog.alpha}{eqn.mog.stvg.mog.mu}, and requires a numerical solution unlike the exact analytic solution of \erefs{eqn.mog.mstg.mass.mstg.soln}{eqn.mog.mstg.mass.mstg.xi} for MSTG.  Using the MSTG dynamic mass as the initial guess for the STVG numerical computation led to fast convergence in fewer than ten iterations.  Lessons learned from the application of STVG to the astrophysics of galaxies and clusters of galaxies may be found in \sref{section.summary.theory}.

\subsection{\label{section.mog.stvg.maxproca}Maxwell-Proca field}\index{Modified gravity!Maxwell-Proca|(}

\citet{Nieuwenhuizen.NPB.1973.60} found that the only massive vector fields free of ghosts, tachyons and higher-order poles in the propagator for linearized gravitation are the massive spin-1 Maxwell-Proca fields.

The expectation from the Yukawa skewon theory of \sref{section.mog.mstg.yukawa} is that the gravitational coupling, \(G(r)\) of \eref{eqn.mog.mstg.mog.fullGspherical} and the mass and range parameters, \(M_0\) and \(r_0\), are scale dependent.  Conversely, STVG theory models this astrophysical scale dependence with a renormalized triplet of self-interacting, cosmological, Klein-Gordon scalar fields.

\index{Modified gravity!Scalar-tensor-vector gravity}\citet{Moffat.JCAP03.2006} introduced the scalar-tensor-vector gravity (STVG) theory by including a massive spin-1\(^{-}\) vector phion, which is the Maxwell-Proca field of \sref{section.mog.stvg.action}, self-coupled and coupled to a matter current, to an Einstein-Hilbert action.  Perhaps much simpler than the NGT, and possibly MSTG, the STVG effectively captures the fifth force due to a weak-field emergent Yukawa meson, simulating the predictions of NGT and MSTG, to a first order approximation.  To address the hypothesis stated in \sref{chapter.introduction.objective.theory}, of fitting galaxy rotation curves and galaxy cluster masses without dominant dark matter, it is sufficient to work in the weak-field limit where  the effective, low energy excitation is described by the Yukawa phion theory of \sref{section.mog.stvg.yukawa}.  The cumulative renormalization of the phion mass, \(\mu\), self-coupling, \(\omega\), and the gravitational coupling, \(G\), provide the {\it gravitational} strength.  The central force law, for test particle motion in STVG, is derived in \sref{section.mog.stvg.mog}.
\subsection{\label{section.mog.stvg.action}Action}

The STVG action, with matter present, is based on the Lagrangian density, 
\begin{equation}\label{eqn.mog.stvg.action.lagrangian}
	{\cal L}={\cal L}_{EH}+{\cal L}_{\phi}+{\cal L}_{S}+{\cal L}_M.
\end{equation}
The Einstein-Hilbert Lagrangian density,
\begin{equation}\label{eqn.mog.stvg.action.lagrangian.EH}
	{\cal L}_{EH}=\frac{c^4}{16\pi G}\big(R-2\Lambda\big)\sqrt{-g},
\end{equation}
provides the general relativistic background, where \(\Lambda\) is the cosmological constant.   The Maxwell-Proca spin-1\(^{-}\) vector phion, \(\phi_\mu\), introduces the fifth force by a modification to gravity's action by the inclusion of the Lagrangian density,
\begin{equation}\label{eqn.mog.stvg.action.lagrangian.phion} 
	{\cal L}_{\phi}=-\omega\left[\frac{1}{4}B^{\mu\nu}B_{\mu\nu}-\frac{1}{2}\mu^2\phi_\mu\phi^\mu+V_\phi(\phi)\right]\sqrt{-g},
\end{equation}
where \(\mu\) is the phion mass, \(\omega\) characterizes the coupling strength between the phion and matter,  \(V_\phi\) is the phion self-interaction potential, and the phion field strength tensor is
\begin{equation}\label{eqn.mog.stvg.action.lagrangian.phionstrength} 
B_{\mu\nu}=\partial_\mu\phi_\nu-\partial_\nu\phi_\mu.
\end{equation}\index{Fifth force!Geometric origin}

\citet{Isenberg.AP.1977.107} showed that, when minimally coupled to gravity, both the Maxwell field photon, \(A_{\alpha}\), and the Maxwell-Proca field phion, \(\phi_{\alpha}\), where \(\alpha = (0,a)\), have two constraints:  The primary constraint sets the canonical momentum conjugate to \(A_0\) and \(\phi_0\) to zero, and therefore the longitudinal modes are non-propagating.  A secondary constraint enforces the Gauss law,
\begin{equation}\label{section.mog.stvg.action.gauss} 
G = \nabla_{a}E^{a} = \left\{\begin{array}{lll}
0 & \phantom{xx}&\mbox{photon}\\
\mu^2 \phi_{0} &\phantom{xx}&\mbox{phion},
\end{array}\right.
\end{equation}
where \(E^{a}\) is the canonical momentum conjugate to \(A_a\), or \(\phi_a\), respectively. The Maxwell Hamiltonian has undetermined Lagrange multipliers which generate U(1) gauge invariance, but the Maxwell-Proca Hamiltonian is uniquely determined, since it is U(1) gauge non-invariant.  Therefore, the Maxwell field has \(4-1-1 = 2\) degrees of freedom, whereas the Maxwell-Proca field has \(4-1=3\) degrees of freedom.  Similar arguments apply to the MSTG massive spin-1\(^{+}\) skewon, of \sref{section.mog.mstg.action}, which is a Kalb-Ramond-Proca field.\index{Nonsymmetric gravitation theory!Gauge field theory}

\citet{Moffat.JCAP03.2006} confirmed that there are no pathological singularities in the Maxwell-Proca field coupled to gravity and promotes the three coupling constants of the theory, \(G\), \(\mu\) and \(\omega\), to scalar fields by introducing associated kinetic and potential terms in the Lagrangian density:\index{Modified gravity!Maxwell-Proca|)}
\begin{equation}\label{eqn.mog.stvg.action.lagrangian.scalar}
{\cal L}_{s} = -\frac{c^4}{G}\left[\frac{1}{2}g^{\eta\nu}\left(\frac{\nabla_\eta G\nabla_\nu G}{G^2}+\frac{\nabla_\eta\mu\nabla_\nu\mu}{\mu^2}-\nabla_\eta\omega\nabla_\nu\omega\right)+\frac{V_G(G)}{G^2}+\frac{V_\mu(\mu)}{\mu^2}+V_\omega(\omega)\right]\sqrt{-g},
\end{equation}
where \(\nabla_\eta\) denotes covariant differentiation with respect to the local SO(3,1) invariant, symmetric metric \(g_{\eta\nu}\), while \(V_G\), \(V_\mu\), and \(V_\omega\) are the self-interaction potentials associated with the scalar fields.

The action principle for STVG in the presence of matter,
\begin{equation}\label{eqn.mog.stvg.action.action}
\delta S = \delta \int d^4 x \left({\cal L}_{EH}+{\cal L}_{\phi}+{\cal L}_{S}+{\cal L}_M\right) = 0,
\end{equation}
where \({\cal L}_M\) is the ordinary matter Lagrangian density, with \(S_M=\int d^4 x {\cal L}_M\). 

The total energy-momentum tensor takes the form,
\begin{equation}\label{eqn.mog.stvg.action.energyMomentum}
T_{\mu\nu}=T_{M\mu\nu}+T_{\phi\mu\nu}+T_{S\mu\nu},
\end{equation}
where
\begin{equation}\label{eqn.mog.stvg.action.matterCurrent.S}
\frac{2}{\sqrt{-g}}\frac{\delta S_M}{\delta
g^{\mu\nu}}=-T_{M\mu\nu},\quad\frac{2}{\sqrt{-g}}\frac{\delta
S_\phi}{\delta g^{\mu\nu}} =-T_{\phi\mu\nu},\quad
\frac{2}{\sqrt{-g}}\frac{\delta S_S}{\delta
g^{\mu\nu}}=-T_{S\mu\nu}.
\end{equation}

Variation of the action with respect to \(g^{\mu\nu}\) yields the Einstein field equations in the presence of a massive vector phion:
\begin{equation}\label{eqn.mog.stvg.action.fieldeqns}
G_{\mu\nu} + \Lambda g_{\mu\nu} + Q_{\mu\nu}  = \frac{8\pi G}{c^4} T_{\mu\nu},
\end{equation}
where \(G_{\mu\nu}\) is the Einstein tensor given by \eref{eqn.mog.mstg.action.einsteintensor}, and 
\begin{eqnarray}\nonumber
Q_{\mu \nu} &=& \frac{8\pi G}{c^4} \omega \left\{\left(B_{\mu\kappa}B_\nu{}^\kappa-\frac{1}{4}g_{\mu\nu}B_{\kappa\lambda}B^{\kappa\lambda}\right) + \mu^2\left(\phi_\mu\phi_\nu-\frac{1}{2}g_{\mu\nu}\phi^\kappa\phi_\kappa\right)+g_{\mu\nu}V_\phi(\phi)\right\}\\
\nonumber&-&8\pi\left\{ \left(\frac{\nabla_\alpha G\nabla_\beta G}{G^2}+\frac{\nabla_\alpha\mu\nabla_\beta\mu}{\mu^2}-\nabla_\alpha\omega\nabla_\beta\omega\right)
\left(\delta^\alpha_\mu\delta^\beta_\nu-\frac{1}{2}g^{\alpha\beta}g_{\mu\nu}\right)\right\}\\
\label{eqn.mog.stvg.action.field}&+&8\pi g_{\mu\nu}\left\{\frac{V_G(G)}{G^2}+\frac{V_\mu(\mu)}{\mu^2}+V_\omega(\omega)\right\}.
\end{eqnarray}
A fifth force-matter current arises from extremizing the matter action under variations of the Maxwell-Proca phion field, \(\phi_\mu\):
\begin{equation}\label{eqn.mog.stvg.action.matterCurrent}
J^\mu=-\frac{1}{\sqrt{-g}}\frac{\delta S_M}{\delta\phi_\mu}.
\end{equation}
Variation of the action with respect to \(\phi_\nu\) yields the Maxwell-Proca equations for the massive vector phion:
\begin{equation}\label{eqn.mog.stvg.action.phiEOM}
\nabla_\mu B^{\mu\nu}+\frac{1}{\omega}B^{\mu\nu}\nabla_\mu\omega+\mu^2\phi^\nu-\frac{\partial V_\phi(\phi)}{\partial\phi_\nu}=\frac{1}{\omega}J^\nu.
\end{equation}
Variation of the action with respect to the gravitational coupling, \(G\), the phion coupling, \(\omega\), and the phion mass, \(\mu\) yields the scalar field equations:
\begin{eqnarray}\nonumber
\nabla^\nu\nabla_\nu G-\frac{3}{2}\frac{\nabla^\nu G\nabla_\nu G}{G}+\frac{G}{2}\left(\frac{\nabla^\nu\mu\nabla_\nu\mu}{\mu^2}-\nabla^\nu\omega\nabla_\nu\omega\right)
+\frac{3}{G}V_G(G)&\phantom{=}&\phantom{0}\\
-V'_G(G)+G\left[\frac{V_\mu(\mu)}{\mu^2}+V_\omega(\omega)\right]-\frac{G}{16\pi}(R-2\Lambda)&=&0,\\
\label{eqn.mog.stvg.action.omegaEOM}
\nabla^\nu\nabla_\nu\omega
-\frac{G\mu^2}{2c^4}\phi_\mu\phi^\mu+\frac{G}{4c^4}B^{\mu\nu}B_{\mu\nu}+\frac{G}{c^4}V_\phi(\phi)+V'_\omega(\omega)&=&0,\\
\label{eqn.mog.stvg.action.muEOM}
\nabla^\nu\nabla_\nu\mu-\frac{\nabla^\nu\mu\nabla_\nu\mu}{\mu}
+\frac{G\omega\mu^3}{c^4}\phi_\mu\phi^\mu+\frac{2}{\mu}V_\mu(\mu)-V'_\mu(\mu)&=&0.
\end{eqnarray}
\subsection{\label{section.mog.stvg.eom}Motion under the fifth force}

The equations of motion of a test particle are
\begin{equation}
\label{eqn.mog.stvg.eom.geodesic2} \frac{du^\mu}{d\tau}+ \left\{{\mu\atop
\alpha\beta}\right\}u^\alpha u^\beta
=a_5^{\mu},
\end{equation}
where $\tau$ is the proper time along the path of the particle and
$u^\lambda=dx^\lambda/d\tau$ is the 4-velocity of the particle, and
\begin{equation}
\left\{{\lambda\atop \mu\nu}\right\}={1\over 2}g^{\lambda\rho}
\left(g_{\mu\rho,\nu}+g_{\rho\nu,\mu}-g_{\mu\nu,\rho}\right),
\end{equation}
is the Christoffel connection.  The acceleration, \(a_5^{\mu}\), is due to the fifth force derived from the Euler-Lagrange equations for a test particle, of mass \(m\), and fifth force charge,
\begin{equation}\label{eqn.mog.stvg.eom.q5}
q_5 = \kappa m,
\end{equation}
where \(\kappa\) is a constant, independent of \(m\).  Such a test particle has a point particle action~\protect\citep{Moffat.JCAP03.2006,Moffat.CQG.2009.26},
\begin{equation} \label{eqn.mog.stvg.eom.testparticleaction} 
S_{\rm TP} = -m \int d\tau \sqrt{g_{\alpha\beta}u^\alpha u^\beta} - q_5 \int d\tau \omega \phi_\mu u^\mu.
\end{equation}
Variation of \eref{eqn.mog.stvg.eom.testparticleaction} yields the Euler-Lagrange equations corresponding to \eref{eqn.mog.stvg.eom.geodesic2}, where the velocity-dependent fifth force is given by
\begin{equation}
\label{eqn.mog.stvg.eom.force} f_5^\mu=q_5\left[\omega {B^\mu}_\nu u^{\nu}
+\nabla^\mu
\omega \left( \phi_\alpha u^{\alpha}\right)
-\nabla_\alpha \omega \left( \phi^\mu u^{\alpha}\right)\right].
\end{equation}
Dividing the fifth force of \eref{eqn.mog.stvg.eom.force} by the test particle mass, \(m\), and using \eref{eqn.mog.stvg.eom.q5}, the mass \(m\) cancels, and the acceleration due to the fifth force becomes,
\begin{equation}
\label{eqn.mog.stvg.eom.acceleration} a_5^\mu=\frac{f_5^\mu}{m} =\kappa\omega {B^\mu}_\nu u^{\nu}
+\kappa\nabla^\mu \omega \left( \phi_\alpha u^{\alpha}\right)
-\kappa\nabla_\alpha \omega \left( \phi^\mu u^{\alpha}\right),
\end{equation}
which is independent of the test particle mass, in exact agreement with the weak equivalence principle and the {\it universality of free fall}.  Taking \(\omega\) as constant, 
\begin{equation}
\label{eqn.mog.stvg.eom.a5} a_5^\mu =\kappa \omega {B^\mu}_\nu u^{\nu}.
\end{equation}

For a static spherically symmetric gravitational field the line
element has the form
\begin{equation}
\label{eqn.mog.stvg.eom.lineelement}
ds^2=Bdt^2-Adr^2-r^2(d\theta^2+\sin^2\theta
d\phi^2),
\end{equation}
and the equations of motion for a test particle are
\begin{eqnarray} \nonumber 
\frac{d^2r}{d\tau^2}+\frac{\alpha'}{2\alpha}\biggl(\frac{dr}{d\tau}\biggr)^2
-\frac{r}{\alpha}
\biggl(\frac{d\theta}{d\tau}\biggr)^2-r\biggl(\frac{\sin^2\theta}{\alpha}\biggr)\biggl(\frac{\kappa \omega
d\phi}{d\tau}\biggr)^2
&+&\frac{\gamma'}{2\alpha}\biggl(\frac{dt}{d\tau}\biggr)^2 \\
\label{eqn.mog.stvg.eom.rmotion}
+\kappa \omega\frac{1}{\alpha}\biggl(\frac{d\phi_0}{dr}\biggr)\biggl(\frac{dt}{d\tau}\biggr) &=&0,\\
\label{eqn.mog.stvg.eom.tequation}
\frac{d^2t}{d\tau^2}+\frac{\gamma'}{\gamma}\biggl(\frac{dt}{d\tau}\biggr)
\biggl(\frac{dr}{d\tau}\biggr)
+\kappa \omega\frac{1}{\gamma}\biggl(\frac{d\phi_0}{dr}\biggr)\biggl(\frac{dr}{d\tau}\biggr)&=&0,\\
\label{eqn.mog.stvg.eom.thetaequation}
\frac{d^2\theta}{d\tau^2}+\frac{2}{r}\biggl(\frac{d\theta}{d\tau}
\biggr)\biggl(\frac{dr}{d\tau}\biggr)-\sin\theta\cos\theta\biggl(\frac{d\phi}{d\tau}\biggr)^2
&=&0,\\
\label{eqn.mog.stvg.eom.phiequation}
\frac{d^2\phi}{d\tau^2}+\frac{2}{r}\biggl(\frac{d\phi}{d\tau}\biggr)\biggl(\frac{dr}{d\tau}\biggr)
+2\cot\theta\biggl(\frac{d\phi}{d\tau}\biggr)\biggl(\frac{d\theta}{d\tau}\biggr)&=&0.
\end{eqnarray}

The motion of a test particle can be shown to lie in the plane, $\theta=\pi/2$,
by an appropriate choice of axes. Integrating \eref{eqn.mog.stvg.eom.phiequation} gives
\begin{equation}
\label{eqn.mog.stvg.eom.angular} r^2\frac{d\phi}{d\tau}=J,
\end{equation}
where $J$ is the conserved orbital angular momentum. Integration
of \eref{eqn.mog.stvg.eom.tequation} gives
\begin{equation}
\label{eqn.mog.stvg.eom.dtequation}
\frac{dt}{d\tau}=-\frac{1}{\gamma}\biggl[\kappa \omega \phi_0+E\biggr],
\end{equation}
where $E>0$ is the conserved orbital energy per unit mass ($E=0$ for the photon). Substituting \eref{eqn.mog.stvg.eom.dtequation} into \eref{eqn.mog.stvg.eom.rmotion}, and using
\eref{eqn.mog.stvg.eom.angular}, we obtain
\begin{equation}
\label{eqn.mog.stvg.eom.reducedrmotion}
\frac{d^2r}{d\tau^2}+\frac{\alpha'}{2\alpha}\biggl(\frac{dr}{d\tau}\biggr)^2
-\frac{J^2}{\alpha
r^3}+\frac{\gamma'}{2\alpha\gamma^2}(\kappa \omega\phi_0+E)^2
=\kappa \omega\frac{1}{\alpha\gamma}\biggl(\frac{d\phi_0}{dr}\biggr)(\kappa \omega\phi_0+E).
\end{equation}

\subsection{\label{section.mog.stvg.yukawa}STVG acceleration law}\index{Equivalence principle!Violations}\index{Modified gravity!Phantom of dark matter}\index{Equivalence principle!Universality of free fall}

Approximating the line element of \eref{eqn.mog.stvg.eom.lineelement} by the Schwarzschild
solution:
\begin{equation}
\label{eqn.mog.stvg.eom.Schwarzschild} \alpha(r)\sim \left(1-\frac{2GM}{r}\right)^{-1},\quad
\gamma(r)\sim 1-\frac{2GM}{r},
\end{equation}
and making the approximations  $2GM/r\ll 1,\kappa \omega \phi_0 \ll 1$, and the slow motion approximation $dr/dt\ll 1$, \eref{eqn.mog.stvg.eom.reducedrmotion}
becomes
\begin{equation}
\label{eqn.mog.stvg.eom.Newton} \frac{d^2r}{dt^2}-\frac{J_N^2}{r^3}+\frac{GM}{r^2}
=\kappa \omega\frac{d \phi_0}{dr},
\end{equation}
where $J_N$ is the Newtonian orbital angular momentum.

In the limit of no phion self-interactions, \(V_\phi(\phi) \rightarrow 0\), and with \(\omega\) constant,  \eref{eqn.mog.stvg.action.phiEOM} becomes
\begin{equation}\label{eqn.mog.stvg.yukawa.freephiEOM}
\nabla_\mu B^{\mu\nu}+\mu^2\phi^\nu=\frac{1}{\omega}J^\nu.
\end{equation}
In the weak-field, static spherically symmetric limit with \(J^{\nu}=0\), the only nonpropagating mode, \(\phi_0\), obeys the Maxwell-Proca equation 
\begin{equation}\label{eqn.mog.stvg.yukawa.phi0}
\frac{\partial^2}{\partial r^2}  \phi_0+\frac{2}{r}\frac{\partial}{\partial r}\phi_0-\mu^2\phi_0=0,
\end{equation}
which has the Yukawa solution\index{Fifth force!Yukawa meson}
\begin{equation}\label{eqn.mog.stvg.yukawa.soln}
\phi_0(r)=-Q_5\frac{e^{-\mu r}}{r}.
\end{equation}
The constant \(Q_5\) emerges as a constant of integration, and should be interpreted as an effective Yukawa phion field strength, whereas the mass of the effective Yukawa phion, \(\mu\), should be interpreted as the range of the Yukawa interaction, \(\lambda = 1/\mu\).  Substituting \eref{eqn.mog.stvg.yukawa.soln}  into \eref{eqn.mog.stvg.eom.Newton},
\begin{equation}
\label{eqn.mog.stvg.eom.Yukawa00} \frac{d^2r}{dt^2}
-\frac{J_N^2}{r^3}=-\frac{GM}{r^2}+\frac{\kappa \omega Q_5\exp(-\mu
r)}{r^2}(1+\mu r).
\end{equation}
Since the effective phion field strength is proportional to the source mass, \(M\), with the same constant of proportionality as in \eref{eqn.mog.stvg.eom.q5},~\protect\citep{Moffat.CQG.2009.26}\index{Modified gravity!Maxwell-Proca}
\begin{equation}\label{eqn.mog.stvg.yukawa.q5}
Q_5 = \kappa M,
\end{equation}
we may write \eref{eqn.mog.stvg.eom.Yukawa00} as
\begin{equation}
\label{eqn.mog.stvg.eom.Yukawa0} \frac{d^2r}{dt^2}
-\frac{J_N^2}{r^3}=-\frac{GM}{r^2}+\frac{ \alpha G_N M \exp(-\mu
r)}{r^2}(1+\mu r),
\end{equation}
where 
\begin{equation}
\label{eqn.mog.stvg.eom.couplings} \alpha G_N = \kappa^2 \omega.
\end{equation}
Demanding consistency with the observed Newtonian force law, for small \(r\), when \(\mu r \ll 1\), the difference between \eref{eqn.mog.stvg.eom.Yukawa0} and the Newtonian force law vanishes, 
\begin{equation}
\label{eqn.mog.stvg.eom.Yukawalimit}  \frac{G_N M}{r^2}-\frac{GM}{r^2}+\frac{ \alpha M  G_N}{r^2} = 0,
\end{equation}
and the gravitational constant, G, in \eref{eqn.mog.stvg.eom.Yukawa0} has the solution
\begin{equation}
\label{eqn.mog.stvg.eom.Yukawalimit} G_\infty = G_N(1 + \alpha).
\end{equation}
Substituting \eref{eqn.mog.stvg.eom.Yukawalimit} into \eref{eqn.mog.stvg.eom.Yukawa0} and neglecting the Newtonian angular momentum, \(J_N\) , we obtain the STVG acceleration law, 
\begin{equation}
\label{eqn.mog.stvg.eom.accelerationlaw3}
a(r)=-\frac{G_N M}{r^2}\biggl\{1+\alpha \biggl[1-\frac{\exp(-\mu r)}{r^2}(1+\mu r)\biggr]\biggr\}. 
\end{equation}

Whereas the derivation of the MSTG acceleration law of \eref{eqn.mog.mstg.eom.accelerationlaw3} relied upon the phenomenological input of \erefs{eqn.mog.mstg.eom.renormS}{eqn.mog.mstg.eom.sigmammodel} leading to the MSTG parameters \(M_0\) and \(r_0\), \citet{Moffat.CQG.2009.26} integrated the equations of motion in the weak-field, spherically symmetric limit, obtaining \(\alpha\) and \(\mu\) as functions of the mass \(M\), 
\begin{eqnarray} 
\label{eqn.mog.stvg.yukawa.alpha} \alpha &=& \frac{M}{\left(\sqrt{M} + E\right)^2}\left(\frac{G_{\infty}}{G_{N}}-1\right),\\
\label{eqn.mog.stvg.yukawa.mu} \mu &=& \frac{D}{\sqrt{M}}.
\end{eqnarray}
The parameters \(D\) and \(E\) are universal constants.  We can rewrite \eref{eqn.mog.stvg.eom.accelerationlaw3} in the form
\begin{equation}
\label{eqn.mog.stvg.yukawa.runG} a(r)=-\frac{G(r)M}{r^2},
\end{equation}
where
\begin{equation}
\label{eqn.mog.stvg.yukawa.Geff}
G(r)=G_N \left\{1+\alpha - \alpha e^{-\mu r} {\left(1+\mu r\right)}\right\}.
\end{equation}

\subsection{\label{section.mog.stvg.mog}Poisson equations}

The experience of a test particle in the STVG theory, moving in an extended matter distribution, is a combination of the force of gravity due to Einstein's metric gravity theory, and a fifth force described by a triplet of scalar fields and a Maxwell-Proca field.  The weak-field,  central potential for a static, spherically symmetric system can be split into two parts:\index{Newton's central potential}
\begin{equation}
\label{eqn.mog.stvg.mog.phipotential} \Phi(r)=\Phi_N(r)+\Phi_Y(r),
\end{equation}
where
\begin{equation}
\label{eqn.mog.stvg.mog.Newtonian} \Phi_N(r)=-\frac{G_{\infty}M}{r},
\end{equation}
and
\begin{equation}
\label{eqn.mog.stvg.mog.Yukawa0} \Phi_Y(r)= \frac{ \kappa \omega Q_5 \exp(-\mu r)}{r}
\end{equation}
denote the Newtonian and Yukawa potentials, respectively, where  \(M\) and \(Q_5\) denote the total constant mass and fifth force charge of a point source.  \(G_\infty\) is the gravitational coupling in \eref{eqn.mog.stvg.mog.Newtonian}, and is given by \eref{eqn.mog.stvg.eom.Yukawalimit}, and \(\mu\) denotes the effective mass of the phion, \(\phi^{\mu}\), in STVG.   Since \(Q_5\) is proportional to \(M\) by \eref{eqn.mog.stvg.yukawa.q5} and using \eref{eqn.mog.stvg.eom.couplings}, \eref{eqn.mog.stvg.mog.Yukawa0} becomes
\begin{equation}
\label{eqn.mog.stvg.mog.Yukawa} \Phi_Y(r)= \frac{ \alpha G_N M \exp(-\mu r)}{r}.
\end{equation}
The Poisson equations
for \(\Phi_N({\bf r})\) and \(\Phi_Y({\bf r})\) are given by
\begin{equation}
\label{eqn.mog.stvg.mog.NewtonPoiss} \nabla^2\Phi_N({\bf r})=-G_{\infty}\rho({\bf r}),
\end{equation}
and
\begin{equation}
\label{eqn.mog.stvg.mog.YukawaPoiss} (\nabla^2-\mu^2)\Phi_Y({\bf
r})=  \alpha G_N \rho({\bf r}),
\end{equation}
respectively. For sufficiently weak fields, the
Poisson \erefs{eqn.mog.stvg.mog.NewtonPoiss}{eqn.mog.stvg.mog.YukawaPoiss} are
uncoupled and determine the potentials \(\Phi_N({\bf r})\) and
\(\Phi_Y({\bf r})\) for non-spherically symmetric systems, which can
be solved analytically and numerically. The Green's function for
the Yukawa Poisson equation is given by
\begin{equation}
(\nabla^2-\mu^2)\Delta_Y({\bf r})=-\delta^3({\bf r}).
\end{equation}
The full solutions to the potentials are given by
\begin{equation}
\label{eqn.mog.stvg.mog.fullNewton} \Phi_N({\bf r})=- G_{N}\int d^3{\bf
r'}\frac{(1+\alpha)\rho({\bf r'})}{4\pi\vert {\bf r}-{\bf r'}\vert}
\end{equation}
and
\begin{equation}
\label{eqn.mog.stvg.mog.fullYukawa} \Phi_Y({\bf r})=G_N \int d^3{\bf
r'} \frac{\alpha\rho({\bf
r'})\exp(-\mu \vert{\bf r}-{\bf r'}\vert)}{4\pi\vert {\bf r}-{\bf r'}\vert}.
\end{equation}

The modified acceleration law is the gradient of the potential of \eref{eqn.mog.stvg.mog.phipotential},
\begin{equation} \label{eqn.mog.stvg.mog.moda}
{\bf a}({\bf r})=-{\mathbf\nabla}\Phi=-\bigl({\mathbf\nabla}\Phi_N({\bf
r})+{\mathbf\nabla}\Phi_Y({\bf r})\bigr).
\end{equation}
Combining \erefss{eqn.mog.stvg.mog.fullNewton}{eqn.mog.stvg.mog.fullYukawa}{eqn.mog.stvg.mog.moda},
\begin{equation}\label{eqn.mog.stvg.mog.fullacc}
{\bf a}({\bf r}) = - G_N\int d^3{\bf r}'\frac{({\bf r}-{\bf r'})
\rho({\bf r'})}{4\pi\vert {\bf r}-{\bf
r}'\vert^3}\biggl\{1+\alpha - \alpha \exp(-\mu\vert{\bf r}-{\bf
r}'\vert)(1+\mu\vert{\bf r}-{\bf r}'\vert)\biggr\}.
\end{equation}
Therefore
\begin{equation}
\label{eqn.mog.stvg.mog.fullGaccel} {\bf a}({\bf r})=-\int d^3{\bf
r}'\frac{({\bf r}-{\bf r'}) \rho({\bf r'})}{4\pi\vert {\bf r}-{\bf
r}'\vert^3}G({\bf r}-{\bf r}'),
\end{equation}
where
\begin{equation}\label{eqn.mog.stvg.mog.fullG} G({\bf r}-{\bf
r}') = G_N\biggl\{1+\alpha - \alpha \exp(-\mu\vert{\bf r}-{\bf
r}'\vert)(1+\mu\vert{\bf r}-{\bf r}'\vert)\biggr\}.
\end{equation}

For a \(\delta\)-function point source,
\begin{equation}
\label{eqn.mog.stvg.mog.delta} \rho({\bf r})=M \delta^3({\bf r}),
\end{equation}
the modified acceleration law of \erefs{eqn.mog.stvg.mog.fullGaccel}{eqn.mog.stvg.mog.fullG} reduces to the point source solution of \erefs{eqn.mog.stvg.yukawa.runG}{eqn.mog.stvg.yukawa.Geff}.

For a spherically symmetric system, the total baryonic mass within a separation, \(r\), from the center of the system, is
\begin{equation} \label{eqn.mog.stvg.mog.Gforce}
M(r)=\int_0^r  4\pi {r^\prime}^2 dr^\prime \rho(r^\prime).
\end{equation}
Whereas the MSTG Poisson equations of \erefs{eqn.mog.mstg.mog.NewtonPoiss}{eqn.mog.mstg.mog.YukawaPoiss} relied upon the Tully-Fisher relation inspired phenomenological input of \erefs{eqn.mog.mstg.eom.sigmammodel}{eqn.mog.mstg.eom.renormG} leading to the MSTG parameters \(M_0\) and \(r_0\) in \eref{eqn.mog.mstg.mog.fullGspherical}, \(\alpha\) and \(\mu\) can be obtained, by integrating the equations of motion in the weak-field, spherically symmetric limit~\protect\citep{Moffat.CQG.2009.26}:
\begin{eqnarray} 
\label{eqn.mog.stvg.mog.alpha} \alpha &=& \frac{M(r)}{\left(\sqrt{M(r)} + E\right)^2}\left(\frac{G_{\infty}}{G_{N}}-1\right),\\
\label{eqn.mog.stvg.mog.mu} \mu &=& \frac{D}{\sqrt{M(r)}}.
\end{eqnarray}
For a static, spherically symmetric system, the effective modified acceleration law is:
\begin{equation} \label{eqn.mog.stvg.mog.Gforce}
a(r)=-\frac{G(r)M(r)}{r^2},
\end{equation}
where
\begin{equation}
\label{eqn.mog.stvg.mog.fullGspherical}
G(r)=G_N \left\{1+\alpha - \alpha e^{-\mu r} {\left(1+\mu r\right)}\right\}.
\end{equation}
where \(\alpha\) and \(\mu\) are given by \erefs{eqn.mog.stvg.mog.alpha}{eqn.mog.stvg.mog.mu}, respectively. 

\subsection{\label{section.mog.stvg.dynamic}Dynamical mass measurements}\index{Modified gravity!Phantom of dark matter|(}

Comparison of \eref{eqn.mog.stvg.mog.Gforce} with the Newtonian acceleration law of \eref{eqn.mog.mstg.mog.Newton} allows the interpretation of the modified gravity dynamic mass as a scaled version of the Newtonian dynamic mass,
\begin{equation}\label{eqn.mog.stvg.mass} 
M_{\rm MOG}(r)  = \frac{G_{N}M_{N}(r)}{G(r)},
\end{equation}
where the varying gravitation coupling, \(G(r)\), may take the form of \eref{eqn.mog.stvg.mog.fullGspherical}, derived in STVG.  The STVG dynamic mass,
\begin{equation}
\label{eqn.mog.mstg.mass.stvg}
M_{\rm STVG}(r) = M_{N}(r) \left\{1+\alpha - \alpha e^{-\mu r} {\left(1+\mu r\right)}\right\}^{-1},
\end{equation} 
where \(\alpha\) and \(\mu\) are defined by \erefs{eqn.mog.stvg.mog.alpha}{eqn.mog.stvg.mog.mu}, respectively, is identified with the total baryonic mass within a separation, \(r\) from the center of the system.

This STVG acceleration law is applied to galaxy rotation curves in \cref{chapter.galaxy}, in \erefsss{eqn.galaxy.dynamics.mog.acceleration}{eqn.galaxy.dynamics.stvg}{eqn.galaxy.dynamics.stvg.alpha}{eqn.galaxy.dynamics.stvg.mu}.  In \sref{section.galaxy.uma}, in order to compute the overall best-fitting mean result, \(D,\ E\) and \(G_{\infty}\) are permitted to vary across the sample of 29 galaxies, as tabulated in \tref{table.galaxy.stvg}.  The galaxy rotation curves, plotted in \fref{figure.galaxy.velocity}, are subsequently {\bf one parameter} best-fits by the stellar mass-to-light ratio, \(\Upsilon\), applying the mean results of \eref{eqn.galaxy.dynamics.stvg.parameters.meanuniversal} universally.  The STVG acceleration law is applied to clusters of galaxies in \cref{chapter.cluster}, according to \ssref{section.cluster.models.mog}{section.cluster.models.mog.stvg}, in order to compute the scaling of the asymptotic coupling, \(G_{\infty}\).  In \sref{section.cluster.models}, the STVG mass of \eref{eqn.mog.mstg.mass.stvg} is fitted to the X-ray gas mass of a sample of 11 clusters, and plotted in \fref{figure.cluster.models.mass} according to the best-fit cluster model parameters tabulated in Panel (d) of \tref{table.cluster.models.bestfit}, for STVG.\index{Modified gravity!Phantom of dark matter|)}\index{Modified gravity!Scalar-tensor-vector gravity|)}

\part{\label{part.astroph}Astrophysics}
\chapterquote{To myself I am only a child playing on the beach, while vast oceans of truth lie undiscovered before me.}{Sir Isaac Newton}
\chapter{\label{chapter.galaxy}Galaxy rotation curves}\index{Galaxy rotation, \(v\)|(}

The creation of galaxy rotation curves from astrophysical observations is subject to model dependent assumptions.  The road from photometry, in some observed electromagnetic band, to the mass profile, in some chosen gravity theory, takes its way through the dynamics of the galaxy, with the destination a rotational velocity profile.  Spiral galaxies show a remarkable variation of the distribution and relative abundances of stellar material, distributed in bulges and disks, and the intergalactic medium, distributed in exponentially thin rings with vanishing amounts within galaxy cores.  These are the three visible components --  the sources of photometric data -- that are used to reconstruct the dynamics of the galaxy, and the predicted galaxy rotation curves.

The galaxy mass profiles are determined by a best-fit algorithm, within each gravity theory depending on dynamics in the weak-field, as in \sref{section.galaxy.dynamics}, for a sample from the Ursa Major filament of galaxies, in \sref{section.galaxy.uma}.  Every galaxy studied, from the highest to lowest in surface brightness, from the most giant to the smallest dwarf, require some form of dark matter or some modification of gravity.  Each of the candidates offer robust and distinguishable predictions for the mass luminosity relationship, as in \sref{section.galaxy.uma.masslight}.\index{Stellar mass-to-light ratio, \(\Upsilon\)}  Although Milgrom's modified Newtonian dynamics and Moffat's modified gravity theories are sourced by ordinary baryons, there is evidence that each of these theories lead to measurable, and distinguishable halos of phantom dark matter, as described in \sref{section.galaxy.halos}.  Dark matter distributions are sensitive to baryon distributions because \(\chi^2\)-fitting algorithms recover the kinks and wiggles, repatriating the surface masses of orphan features, described in \sref{section.galaxy.halos.orphans}.  If the best-fit stellar mass-to-light ratio for the model is near unity, \(\Upsilon \sim 1\), then the Tully-Fisher relation, as in \sref{section.galaxy.tullyfisher}, follows from fundamental physics.\index{Galaxy rotation, \(v\)|)}

\addtocontentsheading{lof}{Ursa Major filament of galaxies}

\section{\label{section.galaxy.dynamics}Curve-fitting}

The observational data from galaxy rotation curves is compared to the predictions of cold non-baryonic dark matter (CDM) halos, Milgrom's modified Newtonian dynamics, and Moffat's modified gravity theories, in \sref{section.galaxy.dynamics.dm}, \sref{section.galaxy.dynamics.mond}, and \sref{section.galaxy.dynamics.mog}, respectively.

\subsection{\label{section.galaxy.dynamics.dm}CDM halos}\label{Dark matter!Collisionless}\index{Dark matter!Missing mass problem}

In \cref{chapter.darkmatter}, the halo density power-law function of \eref{eqn.newton.darkmatter.powerlaw}, was shown to have fitting formulae for a power-law index between \(1 \le \gamma(r) \le 3\), for the NFW formulae of \erefs{eqn.newton.darkmatter.nfw} {eqn.newton.darkmatter.nfw.mass}, 
\begin{equation} \label{eqn.galaxy.dynamics.dm.nfw} 
\begin{array}{lll}
\rho(r) & = & \dfrac{\rho_{0} r_{s}^3}{r(r+r_{s})^2},\\
M(r) & = & 4\pi \rho_{0} r_{s}^3 \left\{\ln(r+r_{s})  - \ln(r_{s}) - \frac{r}{r+r_{s}}\right\},\end{array}
\end{equation} 
and between \(0 \le \gamma(r) \le 3\), for the core-modified formulae of \erefs{eqn.newton.darkmatter.coremodified}{eqn.newton.darkmatter.coremodified.mass}, 
\begin{equation} \label{eqn.galaxy.dynamics.dm.coremodified} 
\begin{array}{lll}
\rho(r) & = &  \dfrac{\rho_{0} r_{s}^3}{r^3+r_{s}^3},\\
M(r) & = & \frac{4}{3}\pi\rho_{0} r_{s}^3 \left\{\ln(r^3+r_{s}^3) - \ln(r_{s}^3) \right\}.\end{array}
\end{equation} 
Each profile is self-similar and describes the tremendous variation of galaxy-scale halos, without any further parameters.  In addition, each profile has a simple analytic expression for the integrated mass function, \(M(r)\), relevant for curve-fitting.  The two parameters that must be varied in both \erefs{eqn.galaxy.dynamics.dm.nfw}{eqn.galaxy.dynamics.dm.coremodified}, are \(\rho_{0}\) and \(r_{s}\).  In the core-modified model these can be interpreted as the dark matter central density, and the radius at which the density is one-half the central density, respectively.  Furthermore, the core-modified \(\gamma\rightarrow 0\) behaviour occurs in the baryon dominated galactic core, decreasing the dark matter density where the cusp problem prevents better fits using the NFW profile.  

Best-fits to the mass profiles of the dark matter halos, neglecting the stellar galactic disk \((\Upsilon=0)\) were poor to gross for both the NFW profile and the core-modified model, whereas simultaneously best-fit parameters, \(\rho_0\), \(r_s\), \(\Upsilon\), produced low to very low \(\chi^2\), as shown in \tref{table.galaxy.darkmatter}.  The very low \(\chi^2\) best-fits repatriated many of the orphan features seen within the galaxy rotation curves, as shown in \fref{figure.galaxy.velocity}.  Moreover, the predicted surface mass profile, \(\Sigma(r)\), extends gradually into the galaxy with a much broader center than the predictions of the modified gravity \({\bar \Sigma}\)-maps, as shown in \fref{figure.galaxy.Sigma}. 

\subsubsection{\label{subsection.newton.darkmatter.observations}Observations}\index{Dark matter!Observations|(}

The CDM computations, using HI and K-band photometric surface luminosity data, detailed in \sref{section.galaxy.uma}, with galaxy rotation curves plotted in \fref{figure.galaxy.velocity}, indicate the following: 
\begin{enumerate}
\item The sample may be universally fit with a common NFW profile given by \eref{eqn.galaxy.dynamics.dm.nfw}, where the NFW parameters are varied in order to best-fit the rotation curve -- either with or without baryons.  The fits without baryons lead to gross best-fits of the galaxy rotation curves, with very poor \(\chi^2\).  Including the visible HI (and He) gaseous disk and the available luminous stellar disk with a variable stellar mass-to-light ratio \(\Upsilon\) provides excellent fits to the large galaxies, but suppresses the best-fit \(\Upsilon \ll 1\), particularly in the case of the dwarf galaxies.  The worst of these dwarf galaxies cannot be fitted using the NFW profile with any nonzero value of the stellar mass-to-light ratio.  This confirms the cusp problem due to the singular NFW fitting formula and reinforces the importance of correctly incorporating the baryonic components into the galaxy models.
\item Every galaxy in the UMa sample, from the highest to lowest in surface brightness may be universally fit with a common core-modified profile given by \eref{eqn.galaxy.dynamics.dm.coremodified} -- with no extra parameters beyond those of the NFW parameters -- provided the visible HI (and He) gaseous disk and the available luminous stellar disk are included.  This model provides superior  fits, with the lowest reduced \(\chi^2\) statistic over all of the gravity theories considered, including all of the dwarf galaxies, and yields values of \(\Upsilon \sim 1\) as tabulated in \tref{table.galaxy.darkmatter}.  Moreover, the dark matter to baryon ratio  at the outermost radial point, tabulated in \tref{table.galaxy.mass} with mean values provided by \eref{eqn.galaxy.uma.powerlaw.dmfraction}, is below the upper limit set by \citet{Spergel:ApJS:2007} in the Wilkinson microwave anisotropy probe (WMAP) third year results.  
\item Every galaxy in the UMa sample, from the highest to lowest in surface brightness has a central disk where the dark matter density differs strongly from a single power-law density profile, and \(\gamma(r)\) of \eref{eqn.newton.darkmatter.powerlaw} increases with radii, \(r\), as shown in \fref{figure.galaxy.powerlaw}.  This solution to the dark matter cusp problem is studied in \sref{section.galaxy.uma.powerlaw}.
\item The UMa sample can be fit by Newton's theory alone -- without dark matter -- using the visible HI (and He) gaseous disk and the available luminous stellar disk, within a Newtonian core up to some radius which varies across the galaxy sample.  This maximizes the stellar mass-to-light ratio, \(\Upsilon\), and leads to systematically bad fits beyond the Newtonian core, indicating that the missing mass problem increases with radius.  This point is elaborated in \sref{section.galaxy.halos.core}.  The Newtonian core radii are plotted in \fref{figure.galaxy.masslight}, and the galaxy rotation curves derived from this best Newtonian core model are plotted in \fref{figure.galaxy.velocity}.\index{Newton's central potential!Galaxy core}
\item The total mass and the shape of the dark matter halo varies in all galaxies, independent of the total mass of the visible HI (and He) gaseous and stellar disks.  The dark matter parameters are neither correlated with galactic mass, nor the flat rotation velocity, \(v_{out}\), nor with the extent of the galaxy rotation curve, \(r_{out}\), as listed in \tref{table.galaxy.uma}.  Sub kiloparsec, high-resolution \map{\Sigma} predictions are provided in \fref{figure.galaxy.Sigma}.
 \item Orphan features become traceable to a parent in either the gaseous disk, or the luminous stellar disk, for \(r \lesssim r_{s}\), but become increasingly orphaned for \(r \gg r_{s}\) where the dark matter halo dominates.  This provides the most obvious improvement between the quality of the fits, as compared to those of the NFW profile. 
\end{enumerate}\index{Dark matter!Observations|)}

Conclusions drawn upon identification of the missing mass as CDM is presented in the summary \sref{section.summary.darkmatter}.

\subsection{\label{section.galaxy.dynamics.mond}Milgrom's acceleration law}\index{MOND!acceleration law}\index{MOND!|(}

In \cref{chapter.mog}, Milgrom's acceleration law of \eref{eqn.mog.mond.milgromacc}, 
\begin{equation}
\label{eqn.galaxy.mond.milgromacc} a \mu(x)=a_{\rm N},
\end{equation}
with the interpolating function,
\begin{equation}\label{eqn.galaxy.mond.interpolating}\index{MOND!Interpolating function, \(\mu\)}
\mu(x)=\frac{x}{\sqrt{1+x^2}},
\end{equation}
where
\begin{equation}\label{eqn.galaxy.mond.x}
x \equiv x(r) = \frac{a(r)}{a_{0}},
\end{equation}
was shown in \eref{eqn.mog.mond.milgromlaw} to have the solution,
\begin{equation}
\label{eqn.galaxy.mond.milgromlaw} a(r) = a_0 \sqrt{\frac{1}{2}\left(\frac{a_N(r)}{a_0}\right)^2+\sqrt{\frac{1}{4}\left(\frac{a_N(r)}{a_0}\right)^4 + \left(\frac{a_N(r)}{a_0}\right)^2}},
\end{equation}
written in terms of the Newtonian acceleration of a test particle at a separation, \(r\),
\begin{equation}
\label{eqn.galaxy.mond.newtonianacc} a_N(r) = \frac{G_{N}M(r)}{r^2},
\end{equation}
where \(M(r)\) is the baryonic mass integrated within a sphere of radius, \(r\).  Each of the galaxy rotation curves in \sref{section.galaxy.uma.velocity} are fitted in MOND by substituting the MOND acceleration law of \eref{eqn.galaxy.mond.milgromlaw} into \eref{eqn.galaxy.uma.orbitalv} for the orbital velocity.

Using the interpolating function of \eref{eqn.galaxy.mond.interpolating},  \citet{Sanders.ARAA.2002.40} suggested that using the fits to the rotation curves of \sref{section.galaxy.uma}, using a revised cluster distance of 18.6 Mpc to Ursa Major, from the Cepheid-based re-calibrated Tully-Fisher relation of \citet{Sakai.APJ.2000.529}, would imply that the MOND universal acceleration should be reduced to 
\begin{equation}\label{eqn.galaxy.mond.a0}
a_0 = 1.0 \times 10^{-8}\ \mbox{cm s}^{-2}.
\end{equation}
In \sref{section.galaxy.uma}, \(a_0\) is permitted to vary across the sample of 29 galaxies, in order to compute the MOND universal acceleration parameter, in \tref{table.galaxy.mond}, with the best-fitting results,
\begin{equation}\label{eqn.galaxy.mond.a0.subsample}
a_0 = \left\{\begin{array}{ll}
(1.34\pm0.66) \times 10^{-8}\ \mbox{cm s}^{-2}&\mbox{\tt HSB galaxy subsample},\\
(1.02\pm0.78) \times 10^{-8}\ \mbox{cm s}^{-2}&\mbox{\tt LSB galaxy subsample}.\end{array}\right.
\end{equation}
Because of the gross uncertainty in the mean results, the galaxy rotation curves of \fref{figure.galaxy.velocity} are  {\bf one parameter} best-fits by the stellar mass-to-light ratio, \(\Upsilon\), applying \eref{eqn.galaxy.mond.a0} universally.  

For galaxies of sufficiently high surface brightness, the asymptotic circular velocity,
\begin{equation}\label{eqn.galaxy.mond.vout}
v_{\rm out}^4 = a_{0} G_N M,
\end{equation}
satisfies the empirical Tully-Fisher relation \(L \propto v_{\rm out}^4\) provided one uses a luminosity parameter which is proportional to the observed mass.  This is shown more precisely in \sref{section.galaxy.uma.masslight}, where the stellar mass-to-light ratio, \(\Upsilon\), is the single {\it free parameter}, treated as a constant within a galaxy (recovering the empirical Tully-Fisher relation), but varying from galaxy to galaxy depending on the best-fit, with results tabulated in \tref{table.galaxy.mond}.

\subsubsection{\label{subsection.mog.mond.observations}Observations}\index{MOND!Observations|(}
The MOND computations using HI and K-band photometric surface luminosity data, detailed in \sref{section.galaxy.uma}, with galaxy rotation curves plotted in \fref{figure.galaxy.velocity}, indicate the following: 
\begin{enumerate}
\item The sample may be universally fit with a single MOND interpolating function and MOND acceleration constant given by \erefs{eqn.galaxy.mond.interpolating}{eqn.galaxy.mond.a0} yielding a best-fit stellar mass-to-light ratio \(\Upsilon\) providing excellent to poor fits.\index{MOND!Interpolating function, \(\mu\)}\index{MOND!Interpolating function, \(\mu\)}
\item The sample may be fit with a single  MOND interpolating function but a varying, best-fit MOND acceleration parameter, tabulated in \tref{table.galaxy.mond}, yielding stellar mass-to-light ratios closer to  \(\Upsilon \sim 1\), and correcting those poor fits with the universal MOND acceleration, but providing minor correction to those fits that were already good.  The best-fit MOND acceleration parameter is not correlated with the galactic surface brightness. 
\item Every galaxy from the highest to lowest in surface brightness has a central disk that is dominated by the Newtonian potential, where the MOND interpolating function remains in the Newtonian core, \(\mu \sim 1\), and a MOND regime where \(\mu > 1\) outside of the core.
\item Once within the MOND regime, the dynamics within the galactic disk continue to dominate, rising monotonically with orbital distance, as shown in \fref{figure.galaxy.Gamma} which plot \(\Gamma(r) \equiv 1/\mu\) vs. \(r\).  Unless \(\mu\) is bounded from below, the dynamical mass factor, \(\Gamma(r) \rightarrow \infty\) suggesting a classical instability.  Conversely, since for every galaxy in the sample \(\Gamma(r_{\rm out} < 10\), there is evidence that \(\mu > \mu_{\infty} \sim 0.1\) is bounded by a cosmological lower limit.
\item The best-fit stellar mass-to-light ratio, \(\Upsilon\), is generally too large in the Newtonian core and too small in the MOND regime for those HSB galaxies that show poor fits, however the trend is reversed in those LSB galaxies that MOND does not fit well, as evident in \fref{figure.galaxy.masslight}.  For the lowest surface brightness galaxies, the increased uncertainty in the stellar mass-to-light ratio in the MOND regime lead to dramatic increases in the uncertainty in the total galaxy mass and relatively weak fits in the Newtonian core.
\item Orphan features are traceable to a parent in either the gaseous or luminous stellar disks and are generally independent of choice of either the best-fit or universal acceleration parameter, but become increasingly pronounced toward the outermost radial data point in the velocity rotation curve.
\end{enumerate}\index{MOND!Observations|)}\index{MOND|)}

\subsection{\label{section.galaxy.dynamics.mog}Moffat's modified gravity}\index{Modified gravity!History}

Moffat's modified gravity theory predicts that galaxy rotation curves are explained by the radial acceleration law,
\begin{equation}\label{eqn.galaxy.dynamics.mog.acceleration}
a(r) = - \frac {G(r) M(r)}{r^2},
\end{equation}
where \(G(r)\) is the effective gravitational constant, and varies through the galaxy.  In the cores of each galaxy, where Newtonian gravity dominates the dynamics, \(G(r) \sim G_{N}\), the value of Newton's constant.  However, within a few kiloparsecs away from the core, the repulsive Yukawa forces becomes appreciable, \(G(r) > G_{N}\). For the analysis of galaxy rotation curves, we will consider the effective gravitational constant given in \sref{section.mog.mstg.mog} by \eref{eqn.mog.mstg.mog.fullGspherical}, derived from MSTG~\citep{Moffat.JCAP05.2005}:
\begin{equation} \label{eqn.galaxy.dynamics.mstg}
G(r) = G_N \left\{1+\sqrt{\frac{M_0}{M(r)}}\left[1-\exp(-{r}/{r_0})\left(1+\frac{r}{r_0}\right)\right]\right\}.
\end{equation}
\citet{Brownstein:ApJ:2006} applied the MSTG acceleration law of \erefs{eqn.galaxy.dynamics.mog.acceleration}{eqn.galaxy.dynamics.mstg} to a large sample of LSB and HSB galaxy rotation curves, obtaining satisfactory fits with the
parameters
\begin{equation} \label{eqn.galaxy.dynamics.mstg.parameters}
M_0=96.0\times 10^{10}M_{\solar},\quad r_0=13.92\,{\rm kpc}.
\end{equation}
However, the dwarf LSB and HSB galaxy rotation curves were better fit with smaller values for these parameters.  In \sref{section.galaxy.uma}, \(M_0\) and \(r_0\) are permitted to vary across the sample of 29 galaxies, in order to compute the MSTG mean-universal parameters, in \tref{table.galaxy.mstg}, with the overall best-fitting result,
\begin{equation} \label{eqn.galaxy.dynamics.mstg.parameters.meanuniversal}
M_0=(98.6\pm21.6)\times 10^{10 }M_{\solar},\quad r_0=(16.4\pm6.1)\,{\rm kpc}.
\end{equation}
\fref{figure.galaxy.velocity} are subsequently {\bf one parameter} best-fits by the stellar mass-to-light ratio, \(\Upsilon\), applying the mean results of \eref{eqn.galaxy.dynamics.mstg.parameters.meanuniversal} universally.

\index{Modified gravity!Scalar-tensor-vector gravity}\citet{Moffat.JCAP03.2006} introduced the scalar-tensor-vector gravity theory (STVG) where the weak-field, massive skew symmetric sector of NGT and MSTG are reduced to the simplest representation of the Yukawa meson -- a massive Maxwell-Proca spin-\(1^{-}\) vector field, \(\phi_\mu\) and a triplet of scalar fields, \(G, \mu, \omega\).  The STVG modified acceleration law results from coupling the additional degrees of freedom to the Einstein metric, where Newton's constant and the Yukawa meson's coupling and range are dynamical scalar fields. 

The predictions of STVG mimics those of MSTG and NGT at astrophysical scales, but since the basic excitations of the three theories are qualitatively different, fits to astrophysical phenomena may constrain the phenomenological parameter space.  Moreover, since STVG is a relatively simple gauge theory of gravitation, the static, spherically symmetric solution has been calculated exactly and resembles the Reissner-Nordstr\"om solution, but with the source electromagnetic charge replaced by the source ``fifth force'' charge between fermions and the massive Maxwell-Proca spin-\(1^{-}\) vector field.

A derivation of a new acceleration law in STVG -- from the action principle, but without necessary ad-hoc phenomenological input -- provided a modified acceleration law of the form of \eref{eqn.galaxy.dynamics.mog.acceleration}, where the effective gravitational constant is determined from the modified central force law, given in \sref{section.mog.stvg.mog} by \erefss{eqn.mog.stvg.mog.alpha}{eqn.mog.stvg.mog.mu}{eqn.mog.stvg.mog.fullGspherical}: 
\begin{eqnarray} 
\label{eqn.galaxy.dynamics.stvg} G(r) &=& G_N \left\{1+\alpha - \alpha e^{-\mu r} {\left(1+\mu r\right)}\right\},\\
\label{eqn.galaxy.dynamics.stvg.alpha} \alpha(r) &=& \frac{M(r)}{\left(\sqrt{M(r)} + E\right)^2}\left(\frac{G_{\infty}}{G_{N}}-1\right),\\
\label{eqn.galaxy.dynamics.stvg.mu} \mu(r) &=& \frac{D}{\sqrt{M(r)}},
\end{eqnarray}
obtaining satisfactory fits with universal parameters,~\citep{Moffat.CQG.2009.26}
\begin{equation} \label{eqn.galaxy.dynamics.stvg.parameters}
D = 6.25 M_{\odot}^{1/2}/\mbox{pc},\quad E = 25000 M_{\odot}^{1/2},\quad G_{\infty} = 20 G_{N}.
\end{equation}
In \sref{section.galaxy.uma}, \(D\), \(E\) and \(G_\infty\) are permitted to vary across the sample of 29 galaxies, in order to compute the STVG mean-universal parameters, in \tref{table.galaxy.stvg}, with the overall best-fitting result,
\begin{equation} \label{eqn.galaxy.dynamics.stvg.parameters.meanuniversal}
D = (6.44\pm0.20) \sqrt{M_{\solar}}\mbox{pc}^{-1},\quad E = (28.4\pm7.9) \times 10^3 M_{\odot}^{1/2},\quad G_{\infty} = (24.4\pm18.0) G_{N}.
\end{equation}
The galaxy rotation curves of \fref{figure.galaxy.velocity} are subsequently {\bf one parameter} best-fits by the stellar mass-to-light ratio, \(\Upsilon\), applying the mean results of \eref{eqn.galaxy.dynamics.stvg.parameters.meanuniversal} universally.

\subsubsection{\label{subsection.galaxy.dynamics.mog.observations}Observations}\index{Modified gravity!Observations|(}
The MSTG and STVG computations using HI and K-band photometric surface luminosity data, detailed in \sref{section.galaxy.uma}, with solutions plotted in \fref{figure.galaxy.velocity}, indicate the following:                                               
\begin{enumerate}
\item The sample may be universally fit with the MOG acceleration law given by \erefs{eqn.galaxy.dynamics.mog.acceleration}{eqn.galaxy.dynamics.mstg} in MSTG, or \erefss{eqn.galaxy.dynamics.stvg}{eqn.galaxy.dynamics.stvg.alpha}{eqn.galaxy.dynamics.stvg.mu} in STVG, and either universal MSTG or universal STVG parameters of \erefs{eqn.galaxy.dynamics.mstg.parameters.meanuniversal}{eqn.galaxy.dynamics.stvg.parameters.meanuniversal}, respectively, yielding best-fit stellar mass-to-light ratios, \(\Upsilon\), providing fits and galaxy masses comparable to MOND.
\item The sample may be fit with varying, best-fit MSTG parameters of \eref{eqn.galaxy.dynamics.mstg}, tabulated in \tref{table.galaxy.mstg}, yielding stellar mass-to-light ratios closer to  \(\Upsilon \sim 1\) for the HSB galaxies and farther from unity for the LSB galaxies \(\Upsilon > 1\), and correcting the fits using the mean-universal MSTG parameters. The best-fit MSTG scale parameters are correlated with the galactic surface brightness, with larger values of \(M_{0}\) and smaller values of \(r_{0}\) preferred by HSB galaxies, with the reverse trend in LSB galaxies.
\item The sample may be fit with varying, best-fit STVG parameters of \eref{eqn.galaxy.dynamics.stvg}, tabulated in \tref{table.galaxy.stvg}, yielding the most robust stellar mass-to-light ratios \(\Upsilon \sim 1\) of all the gravity theories tested, and correcting the fits with the mean-universal STVG parameters.  The best-fit STVG parameters are strongly correlated with the galactic surface brightness, with larger values preferred by HSB galaxies, and smaller values preferred by LSB galaxies, with the greatest uncertainty in the STVG cosmological parameter, \(G_{\infty}\).
\item Every galaxy from the highest to lowest in surface brightness has a central disk that is dominated by the Newtonian potential, where the MOG running gravitational coupling remains in the Newtonian core, \(G(r)  \sim G_{N}\), and a MOG regime where \(G(r) > G_{N}\) outside of the core.
\item Once within the MOG regime, the dynamics within the galactic disk continue to dominate, rising monotonically with orbital distance, as shown in \fref{figure.galaxy.Gamma} which plot \(\Gamma(r) \equiv G(r)/G_{N}\) vs. \(r\).  Since \(G(r)\) is bounded from above, the dynamical mass factor, \(\Gamma(r) \rightarrow \Gamma_{\infty}\) confirming the asymptotic stability of MOG.  This is the primary difference between the MOG and MOND predictions, in which the former return to a Keplerian behaviour (with larger than Newton gravitational coupling) whereas MOND favours asymptotically flat galaxy rotation curves until the dynamics are correlated with other systems.
\end{enumerate}\index{Modified gravity!Observations|)}

Lessons learned from modified gravity theory are presented in \sref{section.summary.theory} of the summary chapter of \pref{part.conclusions}.

\section{\label{section.galaxy.uma}Ursa Major filament of galaxies} Surrounding the local group of the Milky Way, the Coma-Sculptor cloud is our home in the Virgo supercluster.  \citet{Tully.AJ.1996.112} identified 79 galaxies of the Ursa Major (UMa) filament, in the first of a trilogy of works, and provided surface brightness measurements in the blue, red and infrared bands.  UMa lies in the plane of the Virgo supercluster at the junction of filamentary structures, beyond the long axis of the filament of the Coma-Sculptor cloud.  

In a sequel, \citet{Tully.APJ.1997.484} explored the differences in surface brightness amongst the 62 galaxies of the {\it complete sample}, and identified two distinct radial configurations of spiral galaxies of varying size, all of which are unevolved and rich in HI gas consistent with observations in Virgo and Fornax.  These are the high surface brightness (HSB) galaxies and the low surface brightness (LSB) galaxies.  Remarkably, even though the configurations differ, both types of HSB and LSB galaxies exhibit a common exponential disk for the central surface brightness profiles.  However, the mass-to-light ratio in LSB galaxies is difficult to explain using Newtonian gravity without dark matter dominated cores, whereas the mass-to-light ratio in HSB galaxies is difficult to explain using Newtonian gravity without baryon dominated cores and extended dark matter halos.  The UMa sample suggests that structure formation avoids the region of parameter space between  LSB and HSB galaxies, possibly due to different angular momentum regimes.  Passing from high  to low specific angular momentum, there is first the transition from LSB to HSB regimes, and at very low specific angular momentum, there is another transition from HSB galaxies which are exponential disk dominated to disk and bulge dominated.  Since the first transition can be modelled by a single parameter -- the mass-to-light ratio of the stellar exponential disk, whereas the second transition requires a second parameter -- the mass-to-light ratio of the bulge, this chapter will focus exclusively on the subsample where the bulge can be neglected; and one parameter fits are possible for both HSB and LSB galaxies.

\citet{Sanders.APJ.1998.503} presented a third paper in the series on the rotation curves of UMa galaxies which focussed on the near-infrared band, because it is relatively free of the effects of dust absorption and less sensitive to recent star formation.  In this work, it was observed that the exercise of fitting dark matter halos to galaxy rotation curve data required at least three free parameters per galaxy (stellar disk mass-to-light ratio, halo core radius and density normalization) and essentially any observed rotation curve can be reproduced.  However, dark matter gained more predictive power when the density law was parametrized by singular \(r^{-\gamma}\) halos, with \(1 \leq \gamma \leq 3\).  However, although these singular halo models such as \citet{Navarro.APJ.1996.462} produced acceptable fits to HSB galaxies, they generally failed for LSB galaxies.  \citet{McGaugh.APJ.1998.499} tested the dark matter hypothesis with LSB galaxies, finding that progressively lower surface brightness galaxies have progressively larger mass deficits, requiring high concentrations of dark matter deep in the galaxy core, rendering the visible components insignificant to the galaxy dynamics and leading to fine-tuning problems.  In comparison, MOND was shown to fit all of the galaxies in the sample with only a single free parameter (disk mass-to-light ratio) although MOND itself has the Milgrom acceleration parameter, \(a_{0}\) and the best-fit mass-to-light ratio also depends on the choice of a universal interpolating function \(\mu(x)\).\index{MOND!Interpolating function, \(\mu\)}  The notion of fitting galaxy rotation curves without dark matter was further explored in \citet{Brownstein:ApJ:2006} utilizing a larger sample (including UMa) where it was confirmed that MOND provided good one parameter fits to the sample's galaxy rotation curves with a universal choice of \(a_{0}\) and \(\mu(x)\).

\subsection{\label{section.galaxy.uma.photometry}Photometry}

According to \citet{Sanders.APJ.1998.503}, the existence of $K$-band surface photometry is a great advantage since the near-infrared
emission, being relatively free of the effects of dust absorption and less sensitive to recent star formation, is a more precise tracer of the mean radial distribution of the dominant stellar population.  The principal advantages of using infrared luminosities is that
stellar mass-to-light ratios are less affected by population differences and extinction corrections are
minimal~\citep{Verheijen.APJ.2001.563}.  

The galaxy rotation curves of \sref{section.galaxy.uma} are divided into high and low surface brightness galaxies , as in \sref{section.galaxy.uma.velocity}.  The {\it component velocities} plotted in \fref{figure.galaxy.velocity} are based on the surface photometric data of the gaseous disk (HI plus He) component and luminous stellar disk component.  The method of generating the rotation curves closely followed \citet{Sanders.APJ.1998.503} and ~\citep{Verheijen.APJ.2001.563}.  The ROTMOD task of that group's Groningen Image Processing System (GIPSY)\footnote{\url{http://www.astro.rug.nl/~gipsy/}} was used to analyse the $HI$ and $K$-band surface photometry data to produce the velocity profiles of the gaseous disk (HI plus He) distribution and luminous stellar disks, accounting for the \citet{Verheijen.AAP.2001.370} revised distance estimate to UMa from $D=15.5\,\mbox{Mpc}$ to $D=18.6\,\mbox{Mpc}$.  

For each galaxy in the UMa sample, the photometric data is best-fit to the galaxy rotation curve data through a non-linear least squares fitting algorithm, which minimizes the weighted sum of squares of deviations between the fit and the data.  The sum of squares of deviations is characterized by the estimated variance of the fit.  The reduced \(\chi^2/\nu\) statistic is computed as the value of \(\chi^2\) per degree of freedom,
\begin{equation} \label{eqn.galaxy.uma.chi2}
\mbox{reduced}-\chi^2 \equiv \chi^2/\nu,\ \mbox{where}\ \nu=N-p,
\end{equation}
where the number of degrees of freedom, \(\nu\), is the difference between the number of data points in the galaxy rotation curve, \(N\), and the number of free parameters, \(p\).

\subsection{\label{section.galaxy.uma.Surfacemass}Surface mass computation}\index{Surface mass, \(\Sigma\)}
The gaseous disk is modelled as an 
infinitely thin, uniform disk and the surface mass density profile is derived numerically by means of a computation allowing for high resolution sampling of the HI gas data.  The UMa sample was resolved at the sub-kiloparsec scale, equivalent to a resolution of
\begin{equation}
\label{eqn.galaxy.uma.varrho}
\varrho = \frac{r_{\rm out}}{100},
\end{equation}
where \(r_{\rm out}\) is the outermost observed radial position, measured in kiloparsecs, of the rotation velocity data, listed for each galaxy in Column (4) of \tref{table.galaxy.uma}; and Column (5) is the observed velocity at the outermost observed radial position.  

The surface brightness computation,
\begin{equation}
\label{eqn.galaxy.uma.rotmodHI}
M_{\rm HI} = 2.36 \times 10^{5} D^2 \int S dv\,[M_{\solar}],
\end{equation}
derives the absolute surface mass density of the HI gas, where $\int S dv$ is the integrated HI flux density in units of Jy km/s as measured from the global HI profile -- taken
from Column (15) of Table 2 of~\citet{Verheijen.APJ.2001.563}, and D is the distance in Mpc. The computation results in a radial surface mass profile, at the resolution, \(\varrho\),  of \eref{eqn.galaxy.uma.varrho}, and a total (integrated) result which is an absolute measurement of the HI disk mass.  Although the computation is free of unspecified parameters, it is systematically affected by changes in distance estimates.

However, since all of the UMa galaxies are located within a filament of the Coma-Sculptor cloud -- and at similar redshift, listed in Column (3) of \tref{table.galaxy.uma} -- luminosity distances are common and the uncertainty in the mass-to-light ratio is greatly reduced.  This improves the certainty in identifying orphan features, as in \sref{section.galaxy.halos.orphans}, which appear from emergent surface mass profiles.  The total (integrated) HI gas masses, determined by the computation, are listed in Column (2) of \tref{table.galaxy.mass}, and the total mass of the HI (and He) gaseous disk is determined by scaling the HI gas mass by the Big bang nucleosynthesis Helium fraction,
\begin{equation}
\label{eqn.galaxy.uma.rotmodGas}
M_{\rm gas} = \frac{4}{3} M_{HI},
\end{equation}
where the \(4/3\) BBN He fraction is enforced across the UMa sample.  This introduces a margin of uncertainty, in \eref{eqn.galaxy.uma.rotmodGas}, which increases radially due to evolutionary changes in the distribution of HI and He, since the formation of the UMa filament of galaxies.  \citet{Hoekstra.MNRAS.2001.323} showed that the BBN scale factor would have to increase by a factor of \(\sim 7\) to fit a sample of 24 spiral galaxies without dark matter, obtaining good fits for most galaxies, but not for those galaxies which show a rapid decline of the HI surface density in the outermost regions.

\addtocontentsheading{lot}{Ursa Major filament of galaxies}
\begin{table}
\caption[Galaxy properties of the sample]{\label{table.galaxy.uma}{\sf Galaxy properties of the Ursa Major sample}}
\begin{center}
\begin{tabular}{c|ccccccc} \multicolumn{8}{c}{} \\ \hline 
{\sc Galaxy} & {\sc Type} & {\sc Redshift} & {\({z_{0}}_{\rm disk}\)} & {\(L_{K}\)} & {\(v_{\rm max}\)} &{\(r_{\rm out}\)} & {\(v_{\rm out}\)} \\ 
&&&\footnotesize(kpc)&\footnotesize(\(10^{10} L_{\odot}\))&\footnotesize(km s\(^{-1}\))&\footnotesize(kpc)&\footnotesize(km s\(^{-1}\)) \\
\footnotesize(1)&\footnotesize(2)&\footnotesize(3)&\footnotesize(4)&\footnotesize(5)&\footnotesize(6)&\footnotesize(7)&\footnotesize(8) \\ \hline\hline
\multicolumn{8}{c}{\fcolorbox{white}{white}{\sf High surface brightness (HSB) galaxies}} \\ \hline
NGC~3726 & SBc & 0.002887 & 0.68 & 6.216 & \( 169_{ -12 }^{+9 }\) & 33.6 & \( 167 \pm 15 \) \\
NGC~3769 & SBb & 0.002459 & 0.356 & 1.678 & \( 126_{ -8 }^{+5 }\) & 38.5 & \( 113 \pm 11 \) \\
NGC~3877 & Sc & 0.002987 & 0.562 & 6.396 & \( 171_{ -6.5 }^{+8 }\) & 11.7 & \( 169 \pm 10 \) \\
NGC~3893 & Sc & 0.003226 & 0.486 & 5.598 & \( 194_{ -8.5 }^{+10 }\) & 21.1 & \( 148_{ -17 }^{+21 }\) \\
NGC~3949 & Sbc & 0.002669 & 0.346 & 2.901 & \( 169_{ -44 }^{+7 }\) & 8.8 & \( 169_{ -44 }^{+7 }\) \\
NGC~3953 & SBbc & 0.00351 & 0.767 & 12.183 & \( 234_{ -7 }^{+10 }\) & 16.2 & \( 215 \pm 10 \) \\
NGC~3972 & Sbc & 0.002843 & 0.389 & 1.124 & \( 134_{ -7 }^{+5 }\) & 9 & \( 134 \pm 5 \) \\
NGC~3992 & SBbc & 0.003496 & 0.832 & 13.482 & \( 272_{ -8.5 }^{+7 }\) & 36 & \( 237_{ -10 }^{+7 }\) \\
NGC~4013 & Sb & 0.002773 & 0.41 & 7.09 & \( 198 \pm 10 \) & 32.2 & \( 170 \pm 10 \) \\
NGC~4051 & SBbc & 0.002336 & 0.54 & 6.856 & \( 170 \pm 7 \) & 12.6 & \( 153 \pm 10 \) \\
NGC~4085 & Sc & 0.002487 & 0.313 & 1.797 & \( 136 \pm 7 \) & 6.4 & \( 136 \pm 7 \) \\
NGC~4088 & SBc & 0.002524 & 0.67 & 8.176 & \( 182 \pm 8.5 \) & 22.1 & \( 174 \pm 8 \) \\
NGC~4100 & Sbc & 0.003584 & 0.508 & 4.909 & \( 195_{ -7 }^{+10 }\) & 23.5 & \( 159_{ -8 }^{+10 }\) \\
NGC~4138 & Sa & 0.002962 & 0.281 & 4.203 & \( 195 \pm 10 \) & 21.7 & \( 150 \pm 21 \) \\
NGC~4157 & Sb & 0.002583 & 0.518 & 9.098 & \( 201 \pm 10 \) & 30.8 & \( 185 \pm 14 \) \\
NGC~4217 & Sb & 0.003426 & 0.583 & 7.442 & \( 191_{-7 }^{+8.5 }\) & 17.3 & \( 178 \pm 12 \) \\
NGC~4389 & SBbc & 0.002396 & 0.292 & 1.782 & \( 110 \pm 8 \) & 5.5 & \( 110 \pm 8 \) \\
UGC~6399 & Sm & 0.00264 & 0.475 & \ldots & \( 88 \pm 5 \) & 8.1 & \( 88 \pm 5 \) \\
UGC~6973 & Sab & 0.002337 & 0.194 & 4.513 & \( 180_{ -10 }^{+5 }\) & 8.1 & \( 180_{ -10 }^{+5 }\) \\ \hline
\multicolumn{8}{c}{\fcolorbox{white}{white}{\sf Low surface brightness (LSB) galaxies}} \\ \hline
NGC~3917 & Scd & 0.003218 & 0.616 & 2.289 & \( 138 \pm 5 \) & 15.3 & \( 137 \pm 8 \) \\
NGC~4010 & SBd & 0.003008 & 0.691 & 1.169 & \( 129_{ -6 }^{+7 }\) & 10.8 & \( 122_{ -6 }^{+5 }\) \\
NGC~4183 & Scd & 0.003102 & 0.637 & 0.924 & \( 115 \pm 8.5 \) & 21.7 & \( 113_{ -10 }^{+13 }\) \\
UGC~6446 & Sd & 0.002149 & 0.356 & \ldots & \( 85 \pm 8 \) & 15.9 & \( 80 \pm 11 \) \\
UGC~6667 & Scd & 0.003246 & 0.583 & 0.173 & \( 86 \pm 5 \) & 8.1 & \( 86 \pm 5 \) \\
UGC~6818 & Sd & 0.002696 & 0.356 & \ldots & \( 74_{ -5 }^{+7 }\) & 7.2 & \( 74_{ -5 }^{+7 }\) \\
UGC~6917 & SBd & 0.003038 & 0.583 & 0.26 & \( 111_{ -7 }^{+5 }\)& 10.8 & \( 111_{ -7 }^{+5 }\) \\
UGC~6923 & Sdm & 0.003556 & 0.259 & 0.237 & \( 81 \pm 5 \) & 5.3 & \( 81 \pm 5 \) \\
UGC~6983 & SBcd & 0.003609 & 0.529 & 0.16 & \( 113 \pm 6 \) & 16.2 & \( 109 \pm 12 \) \\
UGC~7089 & Sdm & 0.002568 & 0.616 & \ldots & \( 79 \pm 7 \) & 9.4 & \( 79 \pm 7 \) \\ \hline \multicolumn{8}{c}{}
\end{tabular} \end{center}
\parbox{6.375in}{\small Notes. --- Relevant galaxy properties of the UMa sample: Column (1) is
the NGC/UGC galaxy number. Column (2) is the galaxy morphological type. Column (3) is the observed redshift from the NASA/IPAC Extragalactic Database. Column (4) is the K-band vertical scale height of the luminous stellar disk, and Column (5) is the K-band luminosity data converted from the 2MASS K-band apparent magnitude via \eref{eqn.galaxy.halos.tullyfisher.luminosity}.  Column (6) is the velocity amplitude (maximum) of the rotation curve.  Column (7) is the outermost observed radial position in the rotation velocity data; and Column (8) is the observed velocity at the outermost observed position.}
\end{table}

The luminous stellar disk was assumed to be described by the Van der
Kruit and Searle law, where the disk density distribution as a function of z (vertical height from the plane of the
disk) is given by
\begin{equation}
\label{eqn.galaxy.uma.rotmodDisk}
\Sigma(z)={\rm sech}^{2}(z/z_{0})/z_{0},
\end{equation}
where $z_{0}$ is the vertical scale height of the luminous stellar disk, and was assumed to be 20\% of the near infrared
exponential disk scale length according to Column (13) of Table 2 of~\citet{Verheijen.APJ.2001.563}, as listed in Column (4) of \tref{table.galaxy.uma}.  The surface brightness computations using \eref{eqn.galaxy.uma.rotmodDisk} return the surface mass density of the stellar disk to within an overall multiplicative factor, \(\Upsilon\), which is strictly set to unity in the computation, as listed in Column (3) of \tref{table.galaxy.mass}.  The idea of a varying the stellar mass-to-light ratio, \(\Upsilon(r)\) throughout a galaxy -- shown in \fref{figure.galaxy.masslight} -- would lead to perfect fits for any gravity theory, but the arbitrariness of such a solution would lead to a fine-tuning problem, and instead a best-fit \(\Upsilon\) is computed by a nonlinear least-squares algorithm -- shown in the same figure -- with results provided for the best-fit NFW and core-modified dark matter profiles  in \tref{table.galaxy.darkmatter}, the best-fit MOND universal acceleration in \tref{table.galaxy.mond}, and the best-fit MSTG and STVG parameters in \tref{table.galaxy.mstg} and \tref{table.galaxy.stvg}, respectively.

\subsection{\label{section.galaxy.uma.velocity}High and low surface brightness galaxies}

The galaxy rotation curves, in \fref{figure.galaxy.velocity}, plot the rotation velocity profiles, 
\begin{equation}
\label{eqn.galaxy.uma.orbitalv}
v(r)=\sqrt{r a(r)},
\end{equation}
in km s\(^{-1}\), vs.\,\(r\) in kpc, where the acceleration law, \(a(r)\), is given by \erefs{eqn.newton.darkmatter.newton}{eqn.galaxy.mond.newtonianacc} for Newton's theory (with and without dark matter), \eref{eqn.galaxy.mond.milgromlaw} for Milgrom's MOND, and \eref{eqn.galaxy.dynamics.mog.acceleration} for Moffat's MOG.

Shown for each galaxy are the mean-universal best-fits according to Moffat's STVG and MSTG theories and Milgrom's MOND; and the best-fitting core-modified dark matter -- and the corresponding core-modified dark matter halo components.  The best-fit Newtonian results (visible baryons only) are plotted for comparison.

\subsubsection{\label{subsection.galaxy.uma.velocity.newtoniancore}Newtonian core}

Each of the gravity theories which fit high and low surface brightness galaxies disagree with Newton's theory without dark matter.  The disagreement is small in the core of each galaxy, but increases with separation from the center.  In each galaxy in the sample, there is a Newtonian core where neither modified gravity nor dark matter is required to fit the galaxy rotation curves.  The computation of the radius of the Newtonian core weights the velocity points inside the core, and discards the velocity points outside the core, and yields a single parameter, the best-fit stellar mass-to-light ratio, \(\Upsilon\), for each galaxy, as detailed in \sref{section.galaxy.uma.masslight}.  For every galaxy in the sample, the best-fit Newtonian core model, plotted in brown dot-dotted lines,  shows the characteristic Keplerian behaviour outside the Newtonian core, which disagrees with the galaxy rotation curves, but the model shows reduced \(\chi^2/\nu\)  comparable to the modified gravity theories within the Newtonian core.\index{Newton's central potential!Galaxy core}

\newcommand{\subvelocity}{\small The rotation velocity profile, \(v(r)\) in km s\(^{-1}\), vs. orbital distance, \(r\) in kpc}
\begin{figure}[ht]\index{Galaxy rotation, \(v\)|(}
\begin{picture}(460,185)(0,0)
\put(0,40){\includegraphics[width=0.48\textwidth]{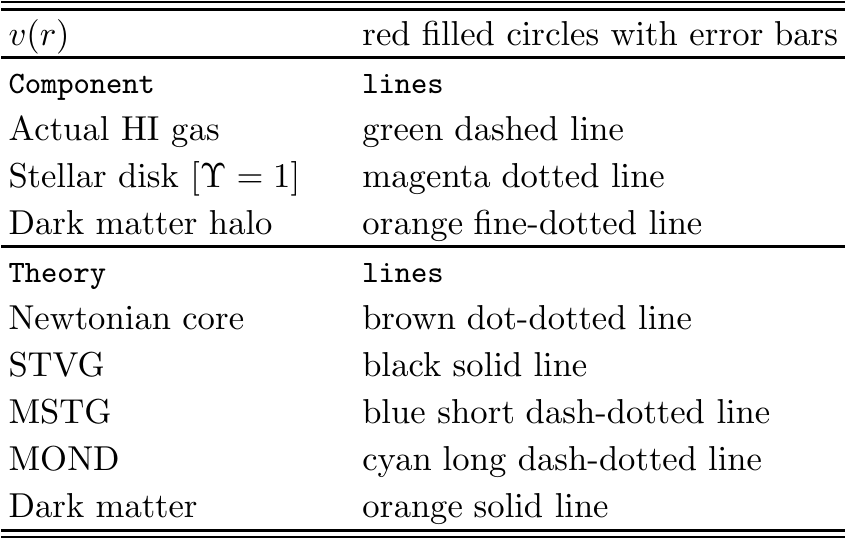}}
\put(225,0){\includegraphics[width=0.5\textwidth]{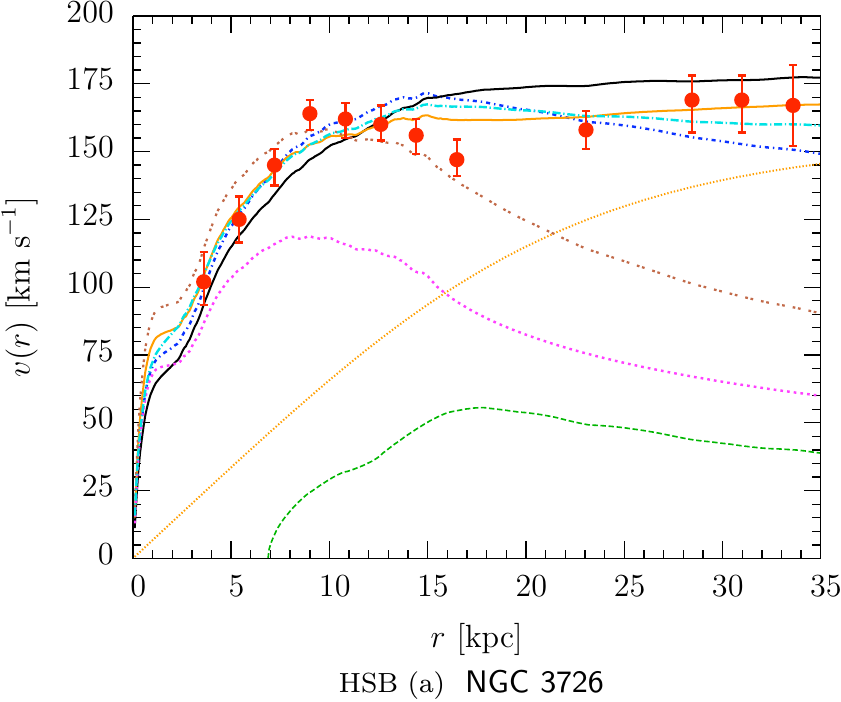}}
\end{picture}
\caption[Galaxy rotation curves]{\label{figure.galaxy.velocity} {UMa --- Rotation curves.}\break\break{\subvelocity} for 19 HSB and 10 LSB galaxies.  The dynamic data consist of the measured orbital velocities. The photometric data sets consist of the actual HI gas component and the stellar disk component, with a normalized stellar mass-to-light ratio, \(\Upsilon=1\). The computed best-fit results by varying the stellar mass-to-light ratio, \(\Upsilon\), are plotted for Moffat's STVG and MSTG theories and Milgrom's MOND theory with mean-universal parameters.  Results are plotted for the best-fit core-modified dark matter theory including visible baryons, and the corresponding dark matter halo component.  The best-fit Newtonian core model (visible baryons only) is plotted for comparison.  {\it The figure is continued.}}
\end{figure}
 
\begin{figure}
\begin{picture}(460,450)(82,190)
\put(30,12){\includegraphics[width=1.28\textwidth]{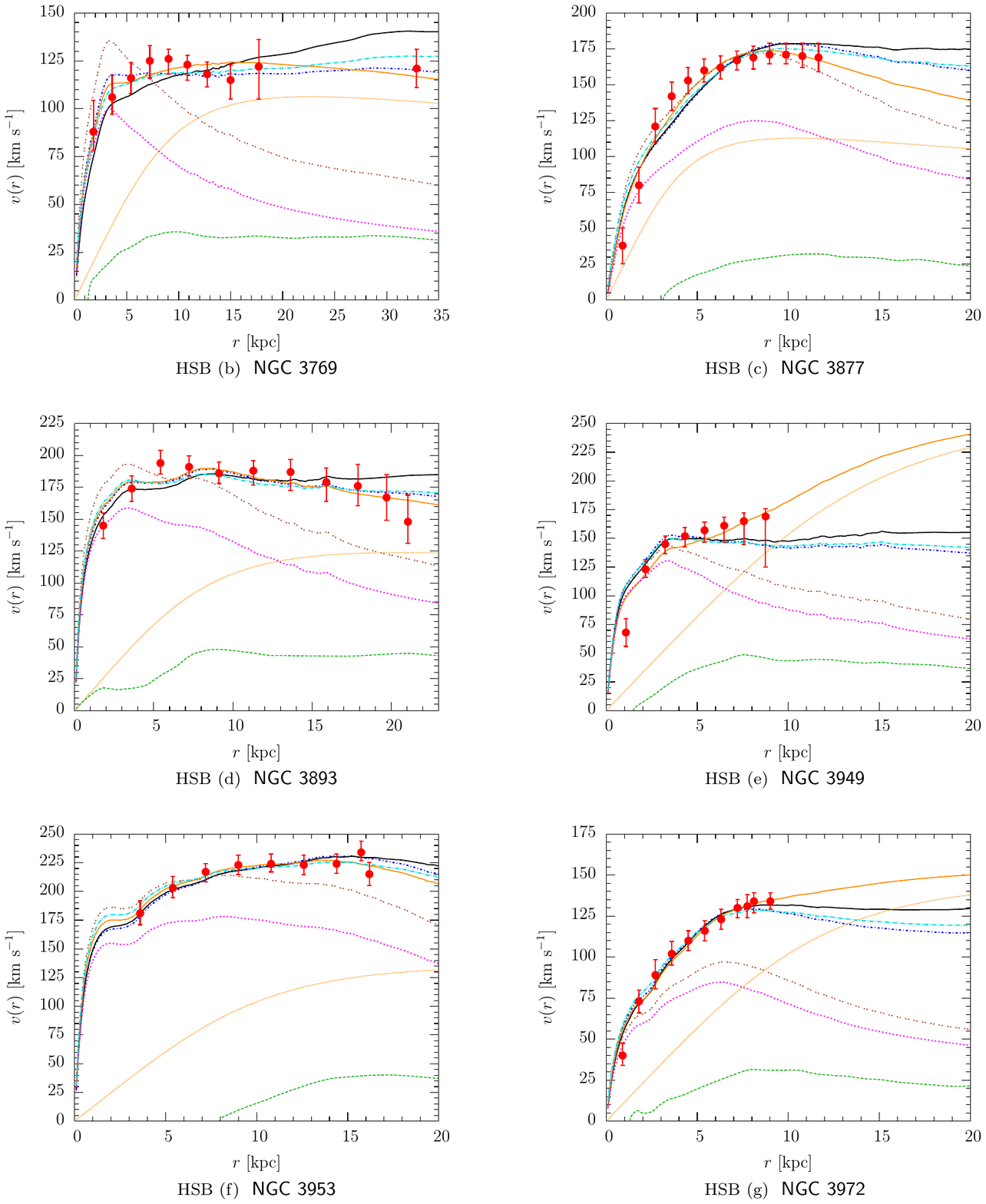}}
\put(82,45){\includegraphics[width=0.98\textwidth]{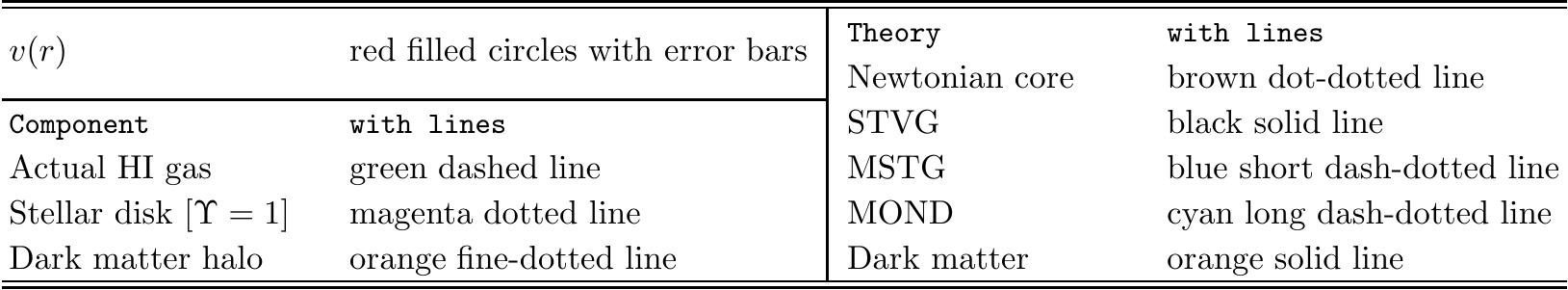}}
\end{picture}
\fcont{figure.galaxy.velocity}{UMa --- Rotation curves.}
{\subvelocity}.
\end{figure}
\begin{figure}
\begin{picture}(460,450)(82,190)
\put(30,12){\includegraphics[width=1.28\textwidth]{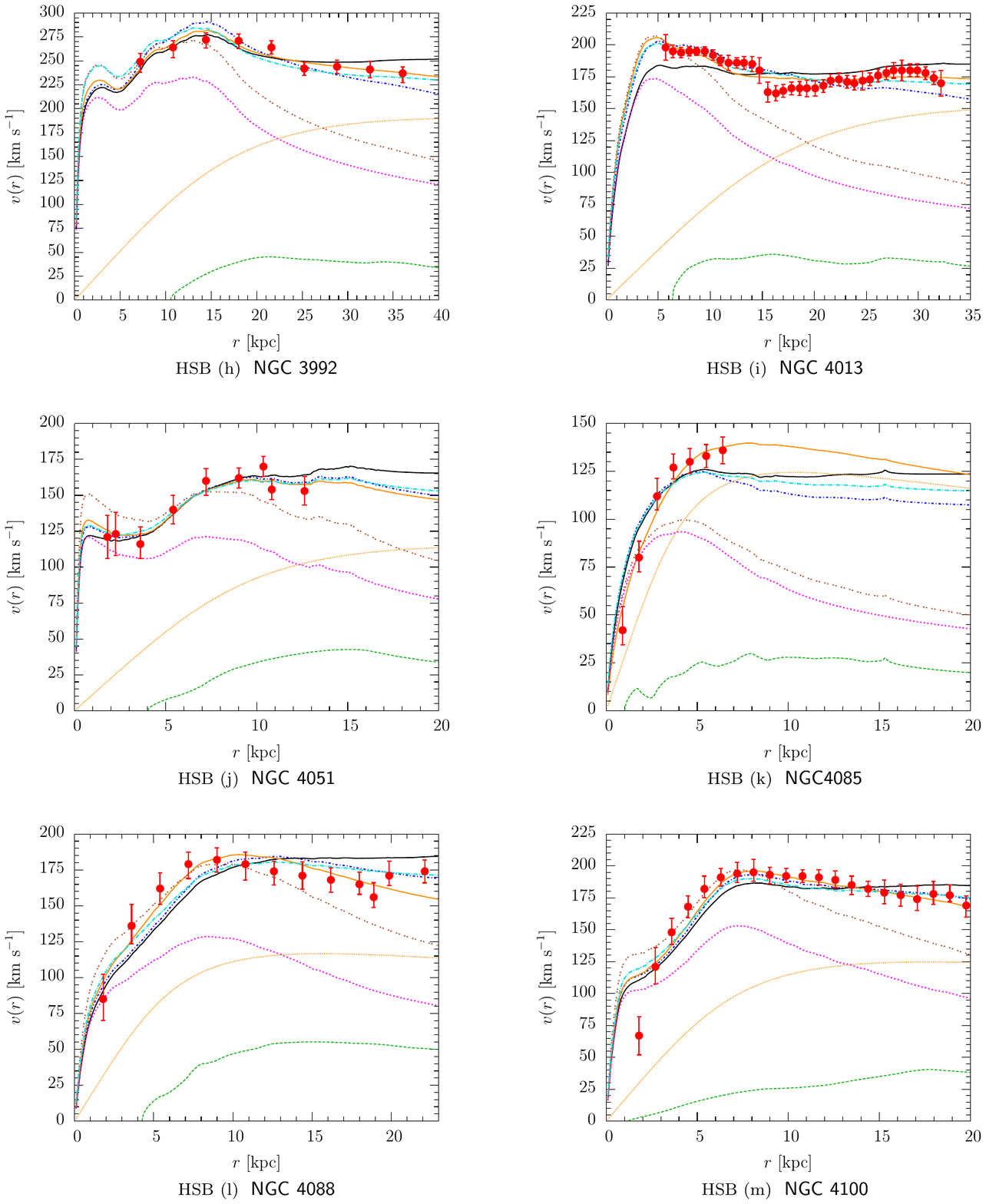}}
\put(82,45){\includegraphics[width=0.98\textwidth]{figure/galaxy_hsb_velocity_legend}}
\end{picture}
\fcont{figure.galaxy.velocity}{UMa --- Rotation curves.}
{\subvelocity}.
\end{figure}
\begin{figure}
\begin{picture}(460,450)(82,190)
\put(30,12){\includegraphics[width=1.28\textwidth]{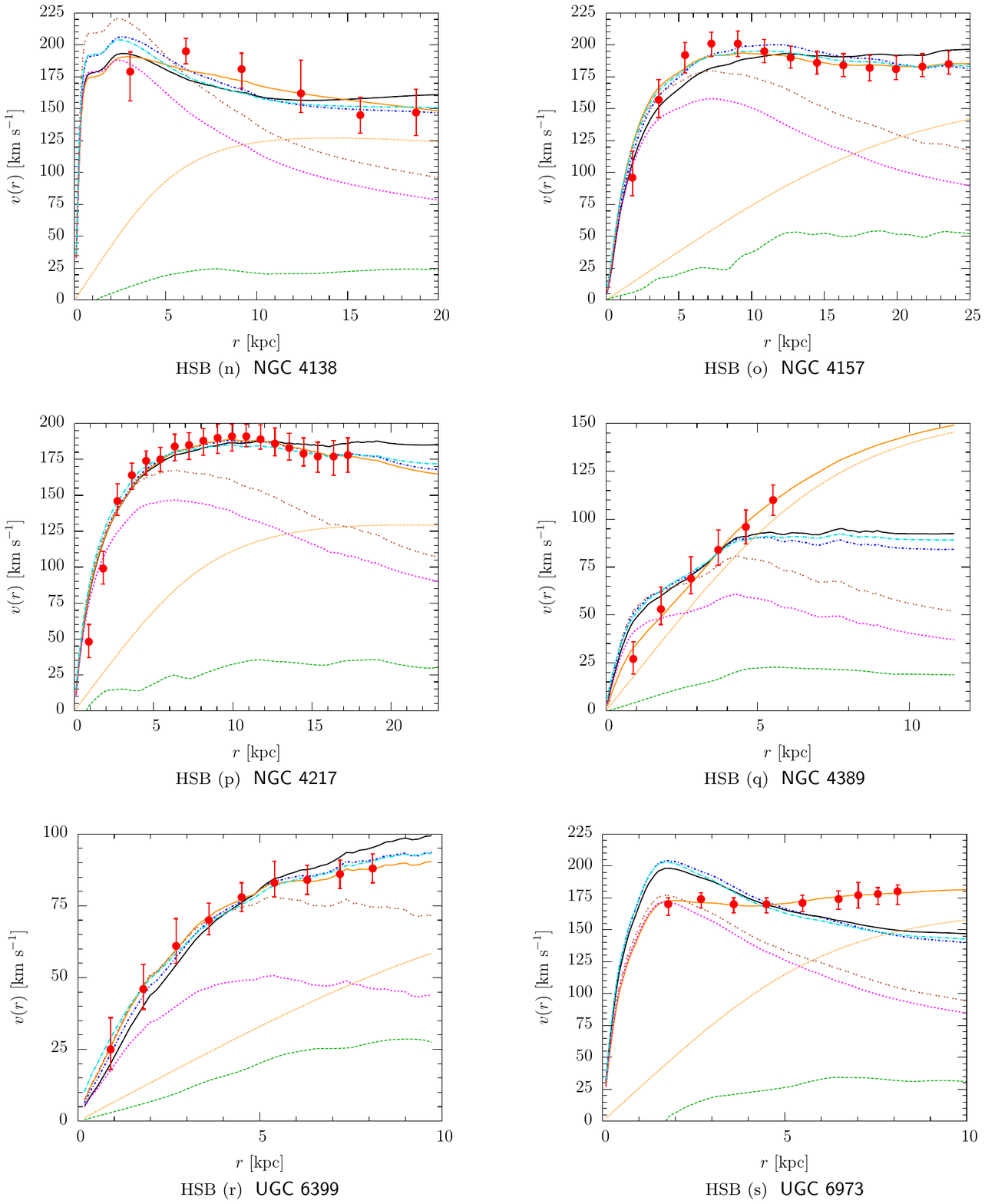}}
\put(82,45){\includegraphics[width=0.98\textwidth]{figure/galaxy_hsb_velocity_legend}}
\end{picture}
\fcont{figure.galaxy.velocity}{UMa --- Rotation curves.}
{\subvelocity}.
\end{figure}\begin{figure}
\begin{picture}(460,450)(82,190)
\put(30,12){\includegraphics[width=1.28\textwidth]{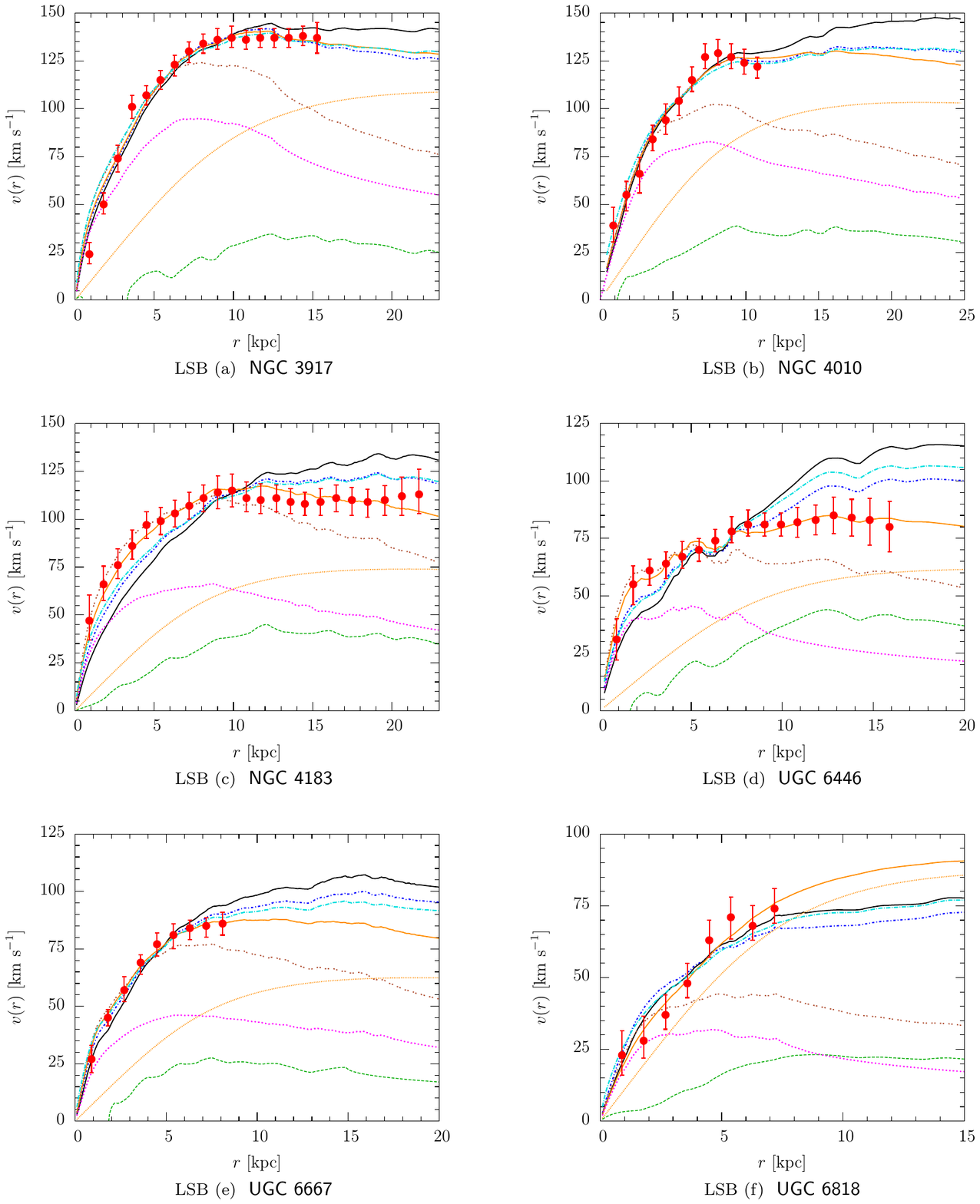}}
\put(82,45){\includegraphics[width=0.98\textwidth]{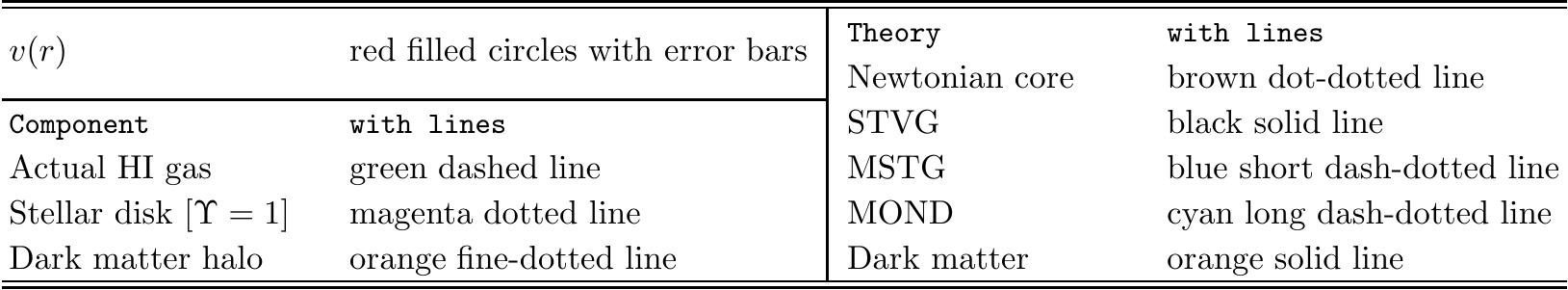}}
\end{picture}
\fcont{figure.galaxy.velocity}{UMa --- Rotation curves.}
{\subvelocity}.
\end{figure}
\begin{figure}
\begin{picture}(460,290)(82,335) 
\put(30,12){\includegraphics[width=1.28\textwidth]{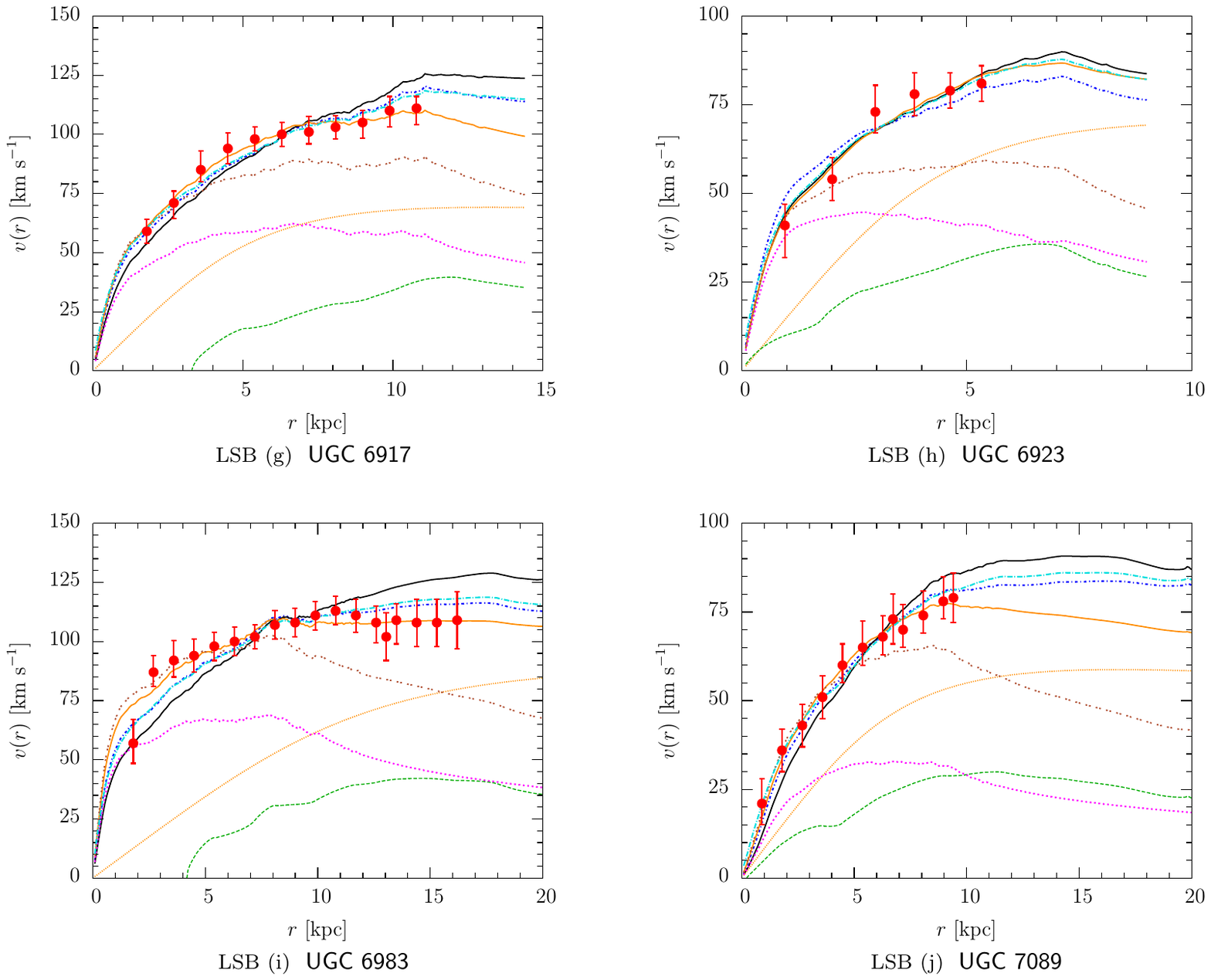}}
\put(82,45){\includegraphics[width=0.98\textwidth]{figure/galaxy_lsb_velocity_legend}}
\end{picture}
\fcont{figure.galaxy.velocity}{UMa --- Rotation curves.\break}
{\subvelocity} for 19 HSB and 10 LSB galaxies.  The dynamic data consist of the measured orbital velocities. The photometric data sets consist of the actual HI gas component and the stellar disk component, with a normalized stellar mass-to-light ratio, \(\Upsilon=1\). The computed best-fit results by varying the stellar mass-to-light ratio, \(\Upsilon\), are plotted for Moffat's STVG and MSTG theories and Milgrom's MOND theory with mean-universal parameters.  Results are plotted for the best-fit core-modified dark matter theory including visible baryons, and the corresponding dark matter halo component.  The best-fit Newtonian core model (visible baryons only) is plotted for comparison.\index{Galaxy rotation, \(v\)|)}
\end{figure}

\subsubsection{\label{subsection.galaxy.uma.velocity.parameters}Best-fit and universal parameters}

Other than Newton's theory without dark matter, each of the gravity theories which fits the sample of high and low surface brightness galaxies contain either one, two or three parameters unique to their respective acceleration laws.  In the case of dark matter, as discussed in \sref{section.galaxy.dynamics.dm}, either the NFW profile of \eref{eqn.galaxy.dynamics.dm.nfw} or the core-modified profile of \eref{eqn.galaxy.dynamics.dm.coremodified} contain two parameters -- the central dark matter density, \(\rho_{0}\), and the scale radius, \(r_{s}\) -- which are taken to vary from galaxy to galaxy.    

Conversely, the hope for the modified gravity theories is that there exist universal parameters which simultaneously fit all of the data for high and low surface brightness galaxies.  To research this possibility, each of the parameters unique to each of the modified gravity theories and the stellar mass-to-light ratio, \(\Upsilon\), were first allowed to vary and a table of best-fit values was constructed.  Then each parameter was averaged across the 

\begin{landscape}
\begin{table}\index{Dark matter!Observations|(}
\caption[Dark matter best-fit NFW and core-modified parameters]{\label{table.galaxy.darkmatter}{\sf Dark matter best-fit NFW and core-modified parameters}}
\begin{center}
\begin{tabular}{c|cccc|cccc} \multicolumn{9}{c}{} \\ \hline 
{\sc Galaxy} & \multicolumn{4}{c|}{\sc NFW} & \multicolumn{4}{c}{\sc core-modified} \\ 
&{\(\rho_{0}\)}&{\(r_{s}\)}&{\(\Upsilon\)}&{\(\chi^2/\nu\)}&{\(\rho_{0}\)}&{\(r_{s}\)}&{\(\Upsilon\)}&{\(\chi^2/\nu\)}\\
&\footnotesize{\((10^{6}\,M_{\odot}/\mbox{kpc}^3)\)}&\footnotesize{(kpc)}&&&\footnotesize{\((10^{6} M_{\odot}/\mbox{kpc}^3)\)}&\footnotesize{(kpc)}&&\footnotesize\\
\footnotesize(1)&\footnotesize(2)&\footnotesize(3)&\footnotesize(4)&\footnotesize(5)&\footnotesize(6)&\footnotesize(7)&\footnotesize(8)&\footnotesize(9)\\ \hline\hline
\multicolumn{9}{c}{\fcolorbox{white}{white}{\sf High surface brightness (HSB) galaxies}} \\ \hline
{NGC~3726} & {\(5.44 \pm 0.05\)} & {\(18.75 \pm 0.09\)} & {\(0.512 \pm 0.013\)} & {\(\phantom{1}89.9\)} & {\(2.52 \pm 0.02\)} & {\(21.25 \pm 0.15\)} & {\(1.344 \pm 0.009\)} & {\(\phantom{1}48.6\)} \\ 
{NGC~3769} & {\(10.7692 \pm 0.1056\)} & {\(9.81 \pm 0.04\)} & {\(0.485 \pm 0.016\)} & {\(\phantom{1}15.2\)} & {\(6.7925 \pm 0.0671\)} & {\(9.06 \pm 0.04\)} & {\(1.080 \pm 0.013\)} & {\(\phantom{1}29.5\)} \\ 
{NGC~3877} & {\(1.2294 \pm 0.0855\)} & {\(36.59 \pm 1.51\)} & {\(1.333 \pm 0.026\)} & {\(150.8\)} & {\(36.4086 \pm 1.1273\)} & {\(4.15 \pm 0.05\)} & {\(1.057 \pm 0.015\)} & {\(\phantom{1}81.6\)} \\ 
{NGC~3893} & {\(19.1700 \pm 0.2243\)} & {\(9.22 \pm 0.05\)} & {\(0.818 \pm 0.010\)} & {\(\phantom{1}97.4\)} & {\(11.0954 \pm 0.1314\)} & {\(8.26 \pm 0.05\)} & {\(1.169 \pm 0.007\)} & {\(107.3\)} \\ 
{NGC~3949} & {\(11.7541 \pm 0.1531\)} & {\(15.81 \pm 0.13\)} & {\(0.295 \pm 0.013\)} & {\(109.1\)} & {\(22.3083 \pm 0.3004\)} & {\(7.73 \pm 0.12\)} & {\(0.837 \pm 0.008\)} & {\(144.5\)} \\ 
{NGC~3953} & {\(19.9878 \pm 0.5249\)} & {\(11.53 \pm 0.13\)} & {\(0.723 \pm 0.017\)} & {\(\phantom{1}30.8\)} & {\(8.5557 \pm 0.2232\)} & {\(10.13 \pm 0.15\)} & {\(1.250 \pm 0.007\)} & {\(\phantom{1}29.1\)} \\ 
{NGC~3972} & {\(3.5376 \pm 0.0919\)} & {\(27.86 \pm 0.47\)} & {\(0.205 \pm 0.039\)} & {\(\phantom{1}24.2\)} & {\(7.4425 \pm 0.2017\)} & {\(11.62 \pm 0.62\)} & {\(1.369 \pm 0.015\)} & {\(\phantom{1}35.3\)} \\ 
{NGC~3992} & {\(12.0503 \pm 0.1071\)} & {\(17.29 \pm 0.07\)} & {\(0.768 \pm 0.008\)} & {\(\phantom{1}30.3\)} & {\(5.8921 \pm 0.0525\)} & {\(17.38 \pm 0.09\)} & {\(1.161 \pm 0.006\)} & {\(\phantom{1}48.1\)} \\ 
{NGC~4013} & {\(6.1682 \pm 0.0305\)} & {\(17.72 \pm 0.04\)} & {\(0.879 \pm 0.006\)} & {\(\phantom{1}53.1\)} & {\(3.4824 \pm 0.0174\)} & {\(18.02 \pm 0.06\)} & {\(1.333 \pm 0.005\)} & {\(\phantom{1}35.7\)} \\ 
{NGC~4051} & {\(6.2598 \pm 0.1504\)} & {\(15.57 \pm 0.21\)} & {\(0.803 \pm 0.016\)} & {\(\phantom{1}51.3\)} & {\(7.1332 \pm 0.1699\)} & {\(9.50 \pm 0.19\)} & {\(1.192 \pm 0.009\)} & {\(\phantom{1}40.6\)} \\ 
{NGC~4085} & {\(6.0469 \pm 0.1323\)} & {\(25.06 \pm 0.39\)} & {\(0.041 \pm 0.030\)} & {\(132.5\)} & {\(44.5464 \pm 0.8612\)} & {\(4.14 \pm 0.05\)} & {\(0.640 \pm 0.015\)} & {\(\phantom{1}56.1\)} \\ 
{NGC~4088} & {\(37.7433 \pm 0.6839\)} & {\(6.83 \pm 0.05\)} & {\(0.490 \pm 0.020\)} & {\(129.4\)} & {\(17.2688 \pm 0.2857\)} & {\(6.24 \pm 0.04\)} & {\(1.182 \pm 0.012\)} & {\(101.3\)} \\ 
{NGC~4100} & {\(27.6264 \pm 0.2878\)} & {\(7.92 \pm 0.03\)} & {\(0.740 \pm 0.010\)} & {\(203.3\)} & {\(15.3878 \pm 0.1440\)} & {\(7.07 \pm 0.03\)} & {\(1.162 \pm 0.006\)} & {\(127.6\)} \\ 
{NGC~4138} & {\(28.4191 \pm 0.3219\)} & {\(7.26 \pm 0.04\)} & {\(0.761 \pm 0.011\)} & {\(136.3\)} & {\(25.3513 \pm 0.2839\)} & {\(5.60 \pm 0.03\)} & {\(0.942 \pm 0.009\)} & {\(116.4\)} \\ 
{NGC~4157} & {\(4.6502 \pm 0.0348\)} & {\(21.47 \pm 0.09\)} & {\(0.876 \pm 0.008\)} & {\(\phantom{1}74.5\)} & {\(3.2823 \pm 0.0723\)} & {\(19.59 \pm 0.37\)} & {\(1.319 \pm 0.006\)} & {\(\phantom{1}67.9\)} \\ 
{NGC~4217} & {\(13.3452 \pm 0.1702\)} & {\(12.43 \pm 0.07\)} & {\(0.689 \pm 0.010\)} & {\(137.4\)} & {\(11.7430 \pm 0.1399\)} & {\(8.39 \pm 0.06\)} & {\(1.141 \pm 0.006\)} & {\(\phantom{1}95.9\)} \\ 
{NGC~4389} & {\(1361.94 \pm 0.00\)} & {\(11.71 \pm 0.16\)} & {\ldots} & {\(212.9\)} & {\(21.71 \pm 0.47\)} & {\(7.33 \pm 0.34\)} & {\(0.440 \pm 0.023\)} & {\(\phantom{1}16.0\)} \\ 
{UGC~6399} & {\(0.5152 \pm 0.0451\)} & {\(44.21 \pm 2.78\)} & {\(1.252 \pm 0.091\)} & {\(\phantom{12}6.5\)} & {\(2.4957 \pm 0.1910\)} & {\(12.07 \pm 2.29\)} & {\(1.949 \pm 0.032\)} & {\(\phantom{12}0.7\)} \\ 
{UGC~6973} & {\(58.92 \pm 0.59\)} & {\(6.10 \pm 0.03\)} & {\(0.516 \pm 0.010\)} & {\(\phantom{1}16.7\)} & {\(37.11 \pm 0.38\)} & {\(5.99 \pm 0.05\)} & {\(0.938 \pm 0.007\)} & {\(\phantom{12}3.3\)} \\ 
\hline
\end{tabular} 
\end{center}
\end{table}
\begin{table}
\tcont{table.galaxy.darkmatter}{\sf Dark matter best-fit NFW and core-modified parameters}
\begin{center}
\begin{tabular}{c|cccc|ccccccc} \multicolumn{9}{c}{} \\ \hline 
{\sc Galaxy} & \multicolumn{4}{c|}{\sc NFW} & \multicolumn{4}{c}{\sc core-modified} \\ 
&{\(\rho_{0}\)}&{\(r_{s}\)}&{\(\Upsilon\)}&{\(\chi^2/\nu\)}&{\(\rho_{0}\)}&{\(r_{s}\)}&{\(\Upsilon\)}&{\(\chi^2/\nu\)}\\
&\footnotesize{\((10^{6}\,M_{\odot}/\mbox{kpc}^3)\)}&\footnotesize{(kpc)}&&&\footnotesize{\((10^{6}\,M_{\odot}/\mbox{kpc}^3)\)}&\footnotesize{(kpc)}&&\footnotesize\\
\footnotesize(1)&\footnotesize(2)&\footnotesize(3)&\footnotesize(4)&\footnotesize(5)&\footnotesize(6)&\footnotesize(7)&\footnotesize(8)&\footnotesize(9)\\ \hline\hline
\multicolumn{9}{c}{\fcolorbox{white}{white}{\sf Low surface brightness (LSB) galaxies}} \\ \hline
{NGC~3917} & {\(6.7085 \pm 0.1519\)} & {\(14.82 \pm 0.15\)} & {\(0.410 \pm 0.025\)} & {\(111.0\)} & {\(5.6671 \pm 0.1017\)} & {\(10.16 \pm 0.12\)} & {\(1.280 \pm 0.011\)} & {\(\phantom{1}41.1\)} \\ 
{NGC~4010} & {\(4.3826 \pm 0.1492\)} & {\(20.37 \pm 0.35\)} & {\(0.059 \pm 0.041\)} & {\(\phantom{1}61.8\)} & {\(6.8776 \pm 0.1972\)} & {\(8.75 \pm 0.20\)} & {\(1.128 \pm 0.015\)} & {\(\phantom{1}30.6\)} \\ 
{NGC~4183} & {\(11.4466 \pm 0.2260\)} & {\(8.07 \pm 0.06\)} & {\(0.565 \pm 0.034\)} & {\(\phantom{12}8.3\)} & {\(3.8300 \pm 0.0772\)} & {\(8.39 \pm 0.08\)} & {\(1.831 \pm 0.018\)} & {\(\phantom{1}18.9\)} \\ 
{UGC~6446} & {\(4.4871 \pm 0.0939\)} & {\(9.00 \pm 0.09\)} & {\(0.790 \pm 0.039\)} & {\(\phantom{12}9.0\)} & {\(2.3983 \pm 0.0505\)} & {\(8.84 \pm 0.13\)} & {\(1.830 \pm 0.026\)} & {\(\phantom{1}10.3\)} \\ 
{UGC~6667} & {\(2.1918 \pm 0.2197\)} & {\(19.60 \pm 0.87\)} & {\(0.443 \pm 0.129\)} & {\(\phantom{1}17.8\)} & {\(3.3321 \pm 0.2427\)} & {\(7.60 \pm 0.55\)} & {\(1.983 \pm 0.038\)} & {\(\phantom{12}6.3\)} \\ 
{UGC~6818} & {\(426.5471 \pm 0.0000\)} & {\(14.80 \pm 0.19\)} & {\ldots} & {\(\phantom{1}88.4\)} & {\(6.7047 \pm 0.1919\)} & {\(7.43 \pm 0.28\)} & {\(0.847 \pm 0.048\)} & {\(\phantom{1}22.6\)} \\ 
{UGC~6917} & {\(8.7271 \pm 0.4399\)} & {\(10.16 \pm 0.20\)} & {\(0.331 \pm 0.069\)} & {\(\phantom{1}11.1\)} & {\(8.4851 \pm 0.3999\)} & {\(5.27 \pm 0.11\)} & {\(1.627 \pm 0.028\)} & {\(\phantom{12}7.5\)} \\ 
{UGC~6923} & {\(3.8520 \pm 0.1547\)} & {\(14.65 \pm 0.40\)} & {\(0.178 \pm 0.067\)} & {\(\phantom{1}15.4\)} & {\(12.5918 \pm 0.5142\)} & {\(4.37 \pm 0.17\)} & {\(1.077 \pm 0.038\)} & {\(\phantom{1}17.1\)} \\ 
{UGC~6983} & {\(2.7906 \pm 0.0471\)} & {\(16.17 \pm 0.15\)} & {\(1.030 \pm 0.023\)} & {\(\phantom{1}23.9\)} & {\(2.7009 \pm 0.0462\)} & {\(11.90 \pm 0.18\)} & {\(1.708 \pm 0.015\)} & {\(\phantom{1}28.4\)} \\ 
{UGC~7089} & {\(0.7878 \pm 0.4587\)} & {\(33.08 \pm 1.32\)} & {\(0.342 \pm 0.170\)} & {\(\phantom{12}6.7\)} & {\(3.8393 \pm 0.2282\)} & {\(6.66 \pm 0.22\)} & {\(2.020 \pm 0.063\)} & {\(\phantom{12}4.4\)} \\ 
\hline \multicolumn{9}{c}{}
\end{tabular} 
\end{center}
\parbox{1.0in}{\phantom{Notes.}}
\parbox{7.5in}{\small Notes. --- Best-fitting dark matter NFW and core-modified parameters, and stellar mass-to-light ratios, \(\Upsilon\), of the UMa sample: Column (1) is
the NGC/UGC galaxy number.  Columns (2) and (3) list the best-fit parameters for the NFW fitting formula of \eref{eqn.newton.darkmatter.nfw}; and Columns (6) and (7) list the corresponding best-fit parameters for the core-modified universal fitting formula of \eref{eqn.newton.darkmatter.coremodified}.  Columns (4) and (8) list the simultaneous best-fitting stellar mass-to-light ratios; and Columns (5) and (9) compare the reduced-\(\chi^2\) statistic of \eref{eqn.galaxy.uma.chi2} in the NFW and core-modified fitting formulae, respectively.}\index{Dark matter!Observations|)}
\end{table}
\end{landscape}
\begin{landscape}
\begin{table}\index{MOND!Observations}
\caption[MOND best-fit and universal acceleration parameter]{\label{table.galaxy.mond}{\sf MOND best-fit and universal acceleration parameter}}
\begin{center}
\begin{tabular}{c|ccc|cc} \multicolumn{6}{c}{} \\ \hline 
{\sc Galaxy} & \multicolumn{3}{c|}{\sc Best-fit} & \multicolumn{2}{c}{\sc Universal} \\ 
&{\(a_{0}\)}&{\(\Upsilon\)}&{\(\chi^2/\nu\)}&{\(\Upsilon\)}&{\(\chi^2/\nu\)} \\
&\footnotesize{(\(10^{-8}\mbox{cm s}^{-2}\))}&&&&\footnotesize \\ 
\footnotesize(1)&\footnotesize(2)&\footnotesize(3)&\footnotesize(4)&\footnotesize(5)&\footnotesize(6) \\ \hline\hline
\multicolumn{6}{c}{\fcolorbox{white}{white}{\sf High surface brightness (HSB) galaxies}} \\ \hline
{NGC~3726} & {\(1.07 \pm 0.03\)} & {\(0.961 \pm 0.022\)} & {\(\phantom{1}87.1\)} & {\(1.020 \pm 0.008\)} & {\(\phantom{1}80.0\)} \\ 
{NGC~3769} & {\(0.76 \pm 0.02\)} & {\(1.073 \pm 0.021\)} & {\(\phantom{1}40.6\)} & {\(0.848 \pm 0.010\)} & {\(\phantom{1}50.9\)} \\ 
{NGC~3877} & {\(0.65 \pm 0.04\)} & {\(1.584 \pm 0.022\)} & {\(149.0\)} & {\(1.375 \pm 0.008\)} & {\(141.8\)} \\ 
{NGC~3893} & {\(1.06 \pm 0.02\)} & {\(1.187 \pm 0.010\)} & {\(171.8\)} & {\(1.215 \pm 0.006\)} & {\(155.6\)} \\ 
{NGC~3949} & {\(1.92 \pm 0.02\)} & {\(0.857 \pm 0.008\)} & {\(204.4\)} & {\(1.148 \pm 0.008\)} & {\(433.0\)} \\ 
{NGC~3953} & {\(1.11 \pm 0.03\)} & {\(1.278 \pm 0.013\)} & {\(\phantom{1}36.5\)} & {\(1.322 \pm 0.005\)} & {\(\phantom{1}33.9\)} \\ 
{NGC~3972} & {\(2.20 \pm 0.03\)} & {\(0.616 \pm 0.008\)} & {\(\phantom{1}27.1\)} & {\(1.221 \pm 0.012\)} & {\(\phantom{1}45.4\)} \\ 
{NGC~3992} & {\(1.34 \pm 0.02\)} & {\(1.166 \pm 0.010\)} & {\(\phantom{1}45.3\)} & {\(1.325 \pm 0.005\)} & {\(\phantom{1}89.2\)} \\ 
{NGC~4013} & {\(1.03 \pm 0.01\)} & {\(1.212 \pm 0.009\)} & {\(\phantom{1}43.1\)} & {\(1.236 \pm 0.004\)} & {\(\phantom{1}42.2\)} \\ 
{NGC~4051} & {\(1.09 \pm 0.03\)} & {\(1.078 \pm 0.018\)} & {\(\phantom{1}44.6\)} & {\(1.122 \pm 0.008\)} & {\(\phantom{1}40.0\)} \\ 
{NGC~4085} & {\(2.45 \pm 0.04\)} & {\(0.685 \pm 0.009\)} & {\(154.8\)} & {\(1.218 \pm 0.012\)} & {\(240.6\)} \\ 
{NGC~4088} & {\(0.63 \pm 0.01\)} & {\(1.556 \pm 0.015\)} & {\(103.6\)} & {\(1.225 \pm 0.008\)} & {\(134.2\)} \\ 
{NGC~4100} & {\(0.82 \pm 0.01\)} & {\(1.387 \pm 0.009\)} & {\(193.8\)} & {\(1.265 \pm 0.005\)} & {\(193.5\)} \\ 
{NGC~4138} & {\(1.35 \pm 0.03\)} & {\(1.011 \pm 0.012\)} & {\(236.2\)} & {\(1.155 \pm 0.008\)} & {\(229.5\)} \\ 
{NGC~4157} & {\(1.07 \pm 0.01\)} & {\(1.173 \pm 0.011\)} & {\(\phantom{1}69.5\)} & {\(1.217 \pm 0.006\)} & {\(\phantom{1}66.6\)} \\ 
{NGC~4217} & {\(1.32 \pm 0.03\)} & {\(1.104 \pm 0.012\)} & {\(122.9\)} & {\(1.253 \pm 0.005\)} & {\(127.1\)} \\ 
{NGC~4389} & {\(2.90 \pm 0.06\)} & {\(0.287 \pm 0.009\)} & {\(129.2\)} & {\(0.924 \pm 0.020\)} & {\(199.0\)} \\ 
{UGC~6399} & {\(0.54 \pm 0.08\)} & {\(1.325 \pm 0.150\)} & {\(\phantom{12}3.9\)} & {\(0.730 \pm 0.017\)} & {\(\phantom{12}5.3\)} \\ 
{UGC~6973} & {\(2.07 \pm 0.02\)} & {\(1.129 \pm 0.007\)} & {\(160.7\)} & {\(1.385 \pm 0.007\)} & {\(594.1\)} \\ 
\hline\hline
{\sc Mean Values}&\(\phantom{0}1.34\pm0.66\)&\(\phantom{0}1.09\pm0.32\)&--&\(\phantom{0}1.17\pm0.18\)&-- \\ \hline
\end{tabular} 
\end{center}
\end{table}
\begin{table}
\tcont{table.galaxy.mond}{\sf MOND best-fit and universal acceleration parameter}
\begin{center}
\begin{tabular}{c|ccc|cc} \multicolumn{6}{c}{} \\ \hline 
{\sc Galaxy} & \multicolumn{3}{c|}{\sc Best-fit} & \multicolumn{2}{c}{\sc Universal} \\ 
&{\(a_{0}\)}&{\(\Upsilon\)}&{\(\chi^2/\nu\)}&{\(\Upsilon\)}&{\(\chi^2/\nu\)} \\
&\footnotesize{(\(10^{-8}\,\mbox{cm s}^{-2}\))}&&&&\footnotesize \\ 
\footnotesize(1)&\footnotesize(2)&\footnotesize(3)&\footnotesize(4)&\footnotesize(5)&\footnotesize(6) \\ \hline\hline
\multicolumn{6}{c}{\fcolorbox{white}{white}{\sf Low surface brightness (LSB) galaxies}} \\ \hline
{NGC~3917} & {\(1.14 \pm 0.06\)} & {\(0.902 \pm 0.042\)} & {\(\phantom{1}59.9\)} & {\(1.008 \pm 0.008\)} & {\(\phantom{1}56.6\)} \\ 
{NGC~4010} & {\(2.29 \pm 0.14\)} & {\(0.334 \pm 0.030\)} & {\(\phantom{1}32.2\)} & {\(0.861 \pm 0.010\)} & {\(\phantom{1}38.6\)} \\ 
{NGC~4183} & {\(0.30 \pm 0.01\)} & {\(2.060 \pm 0.027\)} & {\(\phantom{12}5.8\)} & {\(0.693 \pm 0.009\)} & {\(\phantom{1}95.0\)} \\ 
{UGC~6446} & {\(0.30 \pm 0.01\)} & {\(1.656 \pm 0.042\)} & {\(\phantom{1}12.4\)} & {\(0.399 \pm 0.012\)} & {\(177.4\)} \\ 
{UGC~6667} & {\(0.67 \pm 0.11\)} & {\(1.129 \pm 0.175\)} & {\(\phantom{1}11.1\)} & {\(0.722 \pm 0.018\)} & {\(\phantom{1}10.6\)} \\ 
{UGC~6818} & {\(2.42 \pm 0.10\)} & {\(0.053 \pm 0.010\)} & {\(\phantom{1}18.4\)} & {\(0.441 \pm 0.019\)} & {\(\phantom{1}47.6\)} \\ 
{UGC~6917} & {\(0.52 \pm 0.04\)} & {\(1.669 \pm 0.068\)} & {\(\phantom{1}10.2\)} & {\(1.001 \pm 0.014\)} & {\(\phantom{1}17.7\)} \\ 
{UGC~6923} & {\(1.41 \pm 0.11\)} & {\(0.409 \pm 0.061\)} & {\(\phantom{1}18.7\)} & {\(0.688 \pm 0.024\)} & {\(\phantom{1}17.9\)} \\ 
{UGC~6983} & {\(0.52 \pm 0.01\)} & {\(1.453 \pm 0.030\)} & {\(\phantom{1}25.6\)} & {\(0.796 \pm 0.010\)} & {\(\phantom{1}59.5\)} \\ 
{UGC~7089} & {\(0.62 \pm 0.06\)} & {\(0.905 \pm 0.143\)} & {\(\phantom{12}3.3\)} & {\(0.366 \pm 0.018\)} & {\(\phantom{12}6.0\)} \\ 
\hline\hline
{\sc Mean Values}&\(\phantom{0}1.02\pm0.78\)&\(\phantom{0}1.06\pm0.66\)&--&\(\phantom{0}0.70\pm0.23\)&-- \\ \hline
 \multicolumn{6}{c}{}
\end{tabular} 
\end{center}
\parbox{1.0in}{\phantom{Notes.}}
\parbox{7.5in}{\small Notes. --- Best-fitting and universal MOND acceleration parameters, \(a_0\) and stellar mass-to-light ratios, \(\Upsilon\), of the UMa sample: Column (1) is the NGC/UGC galaxy number.  Columns (2) through (4) are the results of a simultaneous best-fit of the galaxy rotation curve to the photometric data allowing both \(a_0\) and \(\Upsilon\) to be varied; and Columns (5) and (6) are the results of the best-fit allowing only \(\Upsilon\) to vary, but with \(a_0\) universally fixed according to \eref{eqn.galaxy.mond.a0}.   Columns (4) and (6) compare the reduced-\(\chi^2\) statistic of \eref{eqn.galaxy.uma.chi2} in the best-fit and universal cases, respectively.    The MOND acceleration law, used for galaxy rotation curves, is given by \erefs{eqn.mog.mond.milgromacc}{eqn.mog.mond.interpolating}.}
\end{table}
\end{landscape}
\begin{landscape}
\begin{table}\index{Modified gravity!Observations|(}
\caption[MSTG best-fit and mean-universal parameters]{\label{table.galaxy.mstg}{\sf MSTG best-fit and mean-universal parameters}}
\begin{center}
\begin{tabular}{c|cccc|cc} \multicolumn{7}{c}{} \\ \hline 
{\sc Galaxy} & \multicolumn{4}{c|}{\sc Best-fit} & \multicolumn{2}{c}{\sc Mean-universal} \\ 
&{\(M_{0}\)}&{\(r_{0}\)}&{\(\Upsilon\)}&{\(\chi^2/\nu\)}&{\(\Upsilon\)}&{\(\chi^2/\nu\)} \\
&\footnotesize(\(10^{10} M_{\odot}\))&\footnotesize(kpc)&&&\footnotesize \\ 
\footnotesize(1)&\footnotesize(2)&\footnotesize(3)&\footnotesize(4)&\footnotesize(5)&\footnotesize(6)&\footnotesize(7) \\ \hline\hline
\multicolumn{7}{c}{\fcolorbox{white}{white}{\sf High surface brightness (HSB) galaxies}} \\ \hline
{NGC~3726} & {\(152.2 \pm 3.7\)} & {\(17.8 \pm 0.2\)} & {\(0.954 \pm 0.011\)} & {\(131.3\)} & {\(1.028 \pm 0.006\)} & {\(140.0\)} \\ 
{NGC~3769} & {\(\phantom{1}83.5 \pm 2.1\)} & {\(14.2 \pm 0.2\)} & {\(1.004 \pm 0.014\)} & {\(\phantom{1}24.1\)} & {\(1.097 \pm 0.009\)} & {\(\phantom{1}29.3\)} \\ 
{NGC~3877} & {\(\phantom{1}86.1 \pm 10.6\)} & {\(26.0 \pm 0.9\)} & {\(1.577 \pm 0.016\)} & {\(142.7\)} & {\(1.287 \pm 0.006\)} & {\(153.9\)} \\ 
{NGC~3893} & {\(\phantom{1}80.8 \pm 2.3\)} & {\(15.7 \pm 0.2\)} & {\(1.128 \pm 0.009\)} & {\(123.4\)} & {\(1.108 \pm 0.005\)} & {\(114.8\)} \\ 
{NGC~3949} & {\(\phantom{1}88.9 \pm 3.1\)} & {\(\phantom{1}7.0 \pm 0.1\)} & {\(0.577 \pm 0.011\)} & {\(\phantom{1}81.9\)} & {\(1.147 \pm 0.007\)} & {\(429.6\)} \\ 
{NGC~3953} & {\(\phantom{1}94.8 \pm 5.7\)} & {\(20.6 \pm 0.4\)} & {\(1.235 \pm 0.011\)} & {\(\phantom{1}31.1\)} & {\(1.094 \pm 0.004\)} & {\(\phantom{1}44.4\)} \\ 
{NGC~3972} & {\(\phantom{1}93.5 \pm 6.4\)} & {\(12.3 \pm 0.3\)} & {\(1.131 \pm 0.025\)} & {\(\phantom{1}30.0\)} & {\(1.411 \pm 0.010\)} & {\(\phantom{1}39.1\)} \\ 
{NGC~3992} & {\(163.7 \pm 4.2\)} & {\(15.6 \pm 0.3\)} & {\(0.861 \pm 0.009\)} & {\(\phantom{1}81.2\)} & {\(1.052 \pm 0.004\)} & {\(126.5\)} \\ 
{NGC~4013} & {\(146.2 \pm 2.1\)} & {\(15.9 \pm 0.1\)} & {\(0.943 \pm 0.006\)} & {\(\phantom{1}62.9\)} & {\(1.128 \pm 0.003\)} & {\(\phantom{1}84.1\)} \\ 
{NGC~4051} & {\(\phantom{1}93.5 \pm 5.2\)} & {\(16.6 \pm 0.3\)} & {\(1.112 \pm 0.013\)} & {\(\phantom{1}37.2\)} & {\(1.095 \pm 0.006\)} & {\(\phantom{1}32.8\)} \\ 
{NGC~4085} & {\(\phantom{1}64.1 \pm 4.6\)} & {\(\phantom{1}4.9 \pm 0.2\)} & {\(0.451 \pm 0.021\)} & {\(\phantom{1}49.6\)} & {\(1.325 \pm 0.010\)} & {\(245.1\)} \\ 
{NGC~4088} & {\(103.8 \pm 4.1\)} & {\(23.8 \pm 0.3\)} & {\(1.454 \pm 0.012\)} & {\(112.7\)} & {\(1.119 \pm 0.006\)} & {\(169.8\)} \\ 
{NGC~4100} & {\(\phantom{1}83.4 \pm 2.0\)} & {\(18.1 \pm 0.2\)} & {\(1.271 \pm 0.007\)} & {\(154.3\)} & {\(1.142 \pm 0.004\)} & {\(157.0\)} \\ 
{NGC~4138} & {\(\phantom{1}70.5 \pm 2.2\)} & {\(11.2 \pm 0.2\)} & {\(0.926 \pm 0.010\)} & {\(160.0\)} & {\(1.123 \pm 0.007\)} & {\(208.7\)} \\ 
{NGC~4157} & {\(115.3 \pm 2.2\)} & {\(16.4 \pm 0.2\)} & {\(1.025 \pm 0.008\)} & {\(\phantom{1}76.6\)} & {\(1.080 \pm 0.004\)} & {\(\phantom{1}76.1\)} \\ 
{NGC~4217} & {\(\phantom{1}85.6 \pm 2.5\)} & {\(14.5 \pm 0.2\)} & {\(1.074 \pm 0.008\)} & {\(\phantom{1}90.7\)} & {\(1.128 \pm 0.004\)} & {\(\phantom{1}89.8\)} \\ 
{NGC~4389} & {\(\phantom{1}97.1 \pm 7.6\)} & {\(\phantom{1}4.5 \pm 0.2\)} & {\(0.149 \pm 0.017\)} & {\(\phantom{1}11.8\)} & {\(1.427 \pm 0.019\)} & {\(219.9\)} \\ 
{UGC~6399} & {\(\phantom{1}99.1 \pm 17.1\)} & {\(21.7 \pm 1.1\)} & {\(1.819 \pm 0.054\)} & {\(\phantom{12}0.8\)} & {\(1.480 \pm 0.020\)} & {\(\phantom{12}5.8\)} \\ 
{UGC~6973} & {\(106.4 \pm 3.1\)} & {\(\phantom{1}6.7 \pm 0.1\)} & {\(0.701 \pm 0.009\)} & {\(\phantom{1}16.0\)} & {\(1.355 \pm 0.006\)} & {\(618.2\)} \\ 
\hline\hline
{\sc Mean Values}&\(100.3 \pm 26.8\)&\(14.9\pm6.1\)&\(1.02 \pm 0.39\)&--&\(1.19\pm0.14\)&-- \\ \hline
\end{tabular} 
\end{center}
\end{table}
\begin{table}
\tcont{table.galaxy.mstg}{\sf MSTG best-fit and mean-universal parameters}
\begin{center}
\begin{tabular}{c|cccc|cc} \multicolumn{7}{c}{} \\ \hline 
{\sc Galaxy} & \multicolumn{4}{c|}{\sc Best-fit} & \multicolumn{2}{c}{\sc Mean-universal} \\ 
&{\(M_{0}\)}&{\(r_{0}\)}&{\(\Upsilon\)}&{\(\chi^2/\nu\)}&{\(\Upsilon\)}&{\(\chi^2/\nu\)} \\
&\footnotesize(\(10^{10} M_{\odot}\))&\footnotesize(kpc)&&&\footnotesize \\ 
\footnotesize(1)&\footnotesize(2)&\footnotesize(3)&\footnotesize(4)&\footnotesize(5)&\footnotesize(6)&\footnotesize(7) \\ \hline\hline
\multicolumn{7}{c}{\fcolorbox{white}{white}{\sf Low surface brightness (LSB) galaxies}} \\ \hline
{NGC~3917} & {\(\phantom{1}95.5 \pm 4.3\)} & {\(16.1 \pm 0.3\)} & {\(1.144 \pm 0.016\)} & {\(\phantom{1}40.1\)} & {\(1.156 \pm 0.007\)} & {\(\phantom{1}37.7\)} \\ 
{NGC~4010} & {\(\phantom{1}94.5 \pm 6.8\)} & {\(14.8 \pm 0.4\)} & {\(1.026 \pm 0.023\)} & {\(\phantom{1}30.3\)} & {\(1.115 \pm 0.009\)} & {\(\phantom{1}28.8\)} \\ 
{NGC~4183} & {\(104.4 \pm 4.7\)} & {\(29.9 \pm 0.4\)} & {\(1.998 \pm 0.020\)} & {\(\phantom{12}9.8\)} & {\(1.104 \pm 0.009\)} & {\(119.6\)} \\ 
{UGC~6446} & {\(\phantom{1}95.3 \pm 4.4\)} & {\(24.9 \pm 0.4\)} & {\(1.846 \pm 0.029\)} & {\(\phantom{1}15.1\)} & {\(1.135 \pm 0.017\)} & {\(\phantom{1}93.6\)} \\ 
{UGC~6667} & {\(\phantom{1}93.6 \pm 15.5\)} & {\(20.4 \pm 1.0\)} & {\(1.880 \pm 0.061\)} & {\(\phantom{12}5.9\)} & {\(1.560 \pm 0.022\)} & {\(\phantom{12}8.4\)} \\ 
{UGC~6818} & {\(\phantom{1}95.0 \pm 8.1\)} & {\(\phantom{1}9.0 \pm 0.3\)} & {\(0.594 \pm 0.048\)} & {\(\phantom{1}24.2\)} & {\(1.704 \pm 0.034\)} & {\(\phantom{1}62.0\)} \\ 
{UGC~6917} & {\(\phantom{1}96.1 \pm 10.7\)} & {\(23.4 \pm 0.8\)} & {\(1.909 \pm 0.034\)} & {\(\phantom{1}10.3\)} & {\(1.476 \pm 0.013\)} & {\(\phantom{1}25.6\)} \\ 
{UGC~6923} & {\(\phantom{1}93.0 \pm 9.7\)} & {\(11.1 \pm 0.3\)} & {\(1.012 \pm 0.047\)} & {\(\phantom{1}22.6\)} & {\(1.472 \pm 0.027\)} & {\(\phantom{1}38.5\)} \\ 
{UGC~6983} & {\(\phantom{1}97.4 \pm 3.8\)} & {\(20.9 \pm 0.3\)} & {\(1.541 \pm 0.019\)} & {\(\phantom{1}26.6\)} & {\(1.202 \pm 0.010\)} & {\(\phantom{1}49.3\)} \\ 
{UGC~7089} & {\(100.3 \pm 14.1\)} & {\(22.2 \pm 0.9\)} & {\(2.344 \pm 0.090\)} & {\(\phantom{12}3.9\)} & {\(1.635 \pm 0.029\)} & {\(\phantom{12}9.6\)} \\ 
\hline\hline
{\sc Mean Values}&\(95.5 \pm 3.4\)&\(19.2\pm6.5\)&\(1.53 \pm 0.56\)&--&\(1.36\pm0.24\)&-- \\ \hline
\hline \multicolumn{7}{c}
{}
\end{tabular} 
\end{center}
\parbox{1.0in}{\phantom{Notes.}}
\parbox{7.5in}{\small Notes. --- Best-fitting and universal MSTG parameters, \(M_{0},\,r_{0}\), and stellar mass-to-light ratios, \(\Upsilon\), of the UMa sample: Column (1) is the NGC/UGC galaxy number.  Columns (2) through (5) are the results of a simultaneous best-fit of the galaxy rotation curve to the photometric data allowing both \(M_{0},\,r_{0}\) and \(\Upsilon\) to be varied; and Columns (6) and (7) are the results of the best-fit allowing only \(\Upsilon\) to vary, but with \(M_{0},\,r_{0}\) universally fixed according to \eref{eqn.galaxy.dynamics.mstg.parameters}.   Columns (5) and (7) compare the reduced-\(\chi^2\) statistic of \eref{eqn.galaxy.uma.chi2} in the best-fit and universal cases, respectively.  The MSTG acceleration law is given by \erefs{eqn.galaxy.dynamics.mog.acceleration}{eqn.galaxy.dynamics.mstg}.}
\end{table}
\end{landscape}
\begin{landscape}
\begin{table}
\caption[STVG best-fit and mean-universal parameters]{\label{table.galaxy.stvg}{\sf STVG best-fit and mean-universal parameters}}
\begin{center}
\begin{tabular}{c|ccccc|cc} \multicolumn{8}{c}{} \\ \hline 
{\sc Galaxy} & \multicolumn{5}{c|}{\sc Best-fit} & \multicolumn{2}{c}{\sc Mean-universal} \\ 
&{\(D\)}&{\(E\)}&{\(G_{\infty}\)}&{\(\Upsilon\)}&{\(\chi^2/\nu\)}&{\(\Upsilon\)}&{\(\chi^2/\nu\)} \\
&\footnotesize(\(\sqrt{M_{\solar}}\)pc\(^{-1}\))&\footnotesize(\(10^{3}\sqrt{M_{\solar}}\))&\footnotesize(\(G_{N}\))&&&\footnotesize \\
\footnotesize(1)&\footnotesize(2)&\footnotesize(3)&\footnotesize(4)&\footnotesize(5)&\footnotesize(6)&\footnotesize(7)&\footnotesize(8) \\ \hline\hline
\multicolumn{8}{c}{\fcolorbox{white}{white}{\sf High surface brightness (HSB) galaxies}} \\ \hline
{NGC~3726} & {\(6.21 \pm 0.05\)} & {\(19.1 \pm 1.5\)} & {\(17.5 \pm 0.2\)} & {\(1.020 \pm 0.015\)} & {\(\phantom{1}67.6\)} & {\(0.739 \pm 0.007\)} & {\(158.9\)} \\ 
{NGC~3769} & {\(7.22 \pm 0.08\)} & {\(37.3 \pm 1.2\)} & {\(15.0 \pm 0.2\)} & {\(1.081 \pm 0.017\)} & {\(\phantom{1}47.7\)} & {\(0.669 \pm 0.009\)} & {\(205.0\)} \\ 
{NGC~3877} & {\(6.47 \pm 0.18\)} & {\(29.6 \pm 13.7\)} & {\(13.8 \pm 0.7\)} & {\(1.453 \pm 0.048\)} & {\(134.8\)} & {\(1.177 \pm 0.007\)} & {\(143.4\)} \\ 
{NGC~3893} & {\(6.51 \pm 0.05\)} & {\(34.3 \pm 1.9\)} & {\(18.0 \pm 0.2\)} & {\(1.176 \pm 0.009\)} & {\(167.6\)} & {\(0.991 \pm 0.006\)} & {\(226.9\)} \\ 
{NGC~3949} & {\(6.42 \pm 0.05\)} & {\(28.1 \pm 1.3\)} & {\(56.4 \pm 0.8\)} & {\(0.607 \pm 0.015\)} & {\(\phantom{1}98.6\)} & {\(1.022 \pm 0.008\)} & {\(332.0\)} \\ 
{NGC~3953} & {\(6.39 \pm 0.10\)} & {\(34.4 \pm 7.0\)} & {\(19.5 \pm 0.5\)} & {\(1.210 \pm 0.017\)} & {\(\phantom{1}33.6\)} & {\(1.088 \pm 0.005\)} & {\(\phantom{1}43.6\)} \\ 
{NGC~3972} & {\(6.39 \pm 0.12\)} & {\(7.8 \pm 4.0\)} & {\(26.4 \pm 0.8\)} & {\(0.790 \pm 0.095\)} & {\(\phantom{1}21.4\)} & {\(1.152 \pm 0.011\)} & {\(\phantom{1}24.6\)} \\ 
{NGC~3992} & {\(6.40 \pm 0.04\)} & {\(28.7 \pm 2.2\)} & {\(21.3 \pm 0.2\)} & {\(1.104 \pm 0.008\)} & {\(\phantom{1}60.5\)} & {\(1.009 \pm 0.004\)} & {\(\phantom{1}83.1\)} \\ 
{NGC~4013} & {\(6.33 \pm 0.02\)} & {\(25.8 \pm 0.8\)} & {\(18.5 \pm 0.1\)} & {\(1.142 \pm 0.006\)} & {\(\phantom{1}32.3\)} & {\(0.873 \pm 0.003\)} & {\(\phantom{1}99.7\)} \\ 
{NGC~4051} & {\(6.41 \pm 0.09\)} & {\(34.9 \pm 3.6\)} & {\(20.5 \pm 0.5\)} & {\(1.088 \pm 0.018\)} & {\(\phantom{1}41.9\)} & {\(0.967 \pm 0.007\)} & {\(\phantom{1}46.9\)} \\ 
{NGC~4085} & {\(6.42 \pm 0.09\)} & {\(27.1 \pm 1.8\)} & {\(67.9 \pm 1.6\)} & {\(0.546 \pm 0.034\)} & {\(\phantom{1}66.0\)} & {\(1.151 \pm 0.011\)} & {\(189.7\)} \\ 
{NGC~4088} & {\(6.33 \pm 0.07\)} & {\(23.4 \pm 3.0\)} & {\(13.0 \pm 0.2\)} & {\(1.417 \pm 0.016\)} & {\(\phantom{1}99.4\)} & {\(0.941 \pm 0.007\)} & {\(278.0\)} \\ 
{NGC~4100} & {\(6.50 \pm 0.04\)} & {\(43.1 \pm 1.8\)} & {\(15.8 \pm 0.2\)} & {\(1.343 \pm 0.008\)} & {\(176.9\)} & {\(1.000 \pm 0.004\)} & {\(288.2\)} \\ 
{NGC~4138} & {\(6.59 \pm 0.06\)} & {\(34.0 \pm 1.4\)} & {\(20.9 \pm 0.3\)} & {\(1.058 \pm 0.010\)} & {\(267.4\)} & {\(0.959 \pm 0.007\)} & {\(251.2\)} \\ 
{NGC~4157} & {\(6.37 \pm 0.03\)} & {\(30.2 \pm 1.3\)} & {\(18.7 \pm 0.1\)} & {\(1.132 \pm 0.008\)} & {\(\phantom{1}55.8\)} & {\(0.897 \pm 0.005\)} & {\(141.3\)} \\ 
{NGC~4217} & {\(6.45 \pm 0.05\)} & {\(41.8 \pm 2.1\)} & {\(23.2 \pm 0.3\)} & {\(1.119 \pm 0.010\)} & {\(104.3\)} & {\(1.036 \pm 0.004\)} & {\(103.7\)} \\ 
{NGC~4389} & {\(6.36 \pm 0.14\)} & {\(24.7 \pm 1.5\)} & {\(82.0 \pm 2.6\)} & {\(0.310 \pm 0.045\)} & {\(\phantom{1}15.1\)} & {\(1.123 \pm 0.020\)} & {\(149.7\)} \\ 
{UGC~6399} & {\(6.34 \pm 0.32\)} & {\(20.4 \pm 7.8\)} & {\(10.6 \pm 0.8\)} & {\(1.649 \pm 0.135\)} & {\(\phantom{12}0.8\)} & {\(1.037 \pm 0.021\)} & {\(\phantom{1}27.3\)} \\ 
{UGC~6973} & {\(6.36 \pm 0.04\)} & {\(26.9 \pm 0.9\)} & {\(64.4 \pm 0.7\)} & {\(0.788 \pm 0.010\)} & {\(\phantom{12}2.1\)} & {\(1.252 \pm 0.006\)} & {\(458.6\)} \\ 
\hline\hline
{\sc Mean Values}&\(6.48 \pm 0.25\)&\(29.0 \pm 8.3\)&\(28.5 \pm 21.6\)&\(1.054 \pm 0.328\)&--&\(1.004 \pm 0.145\)&-- \\ \hline
\end{tabular} 
\end{center}
\end{table}
\begin{table}
\tcont{table.galaxy.stvg}{\sf STVG best-fit and mean-universal parameters}
\begin{center}
\begin{tabular}{c|ccccc|cc} \multicolumn{8}{c}{} \\ \hline 
{\sc Galaxy} & \multicolumn{5}{c|}{\sc Best-fit} & \multicolumn{2}{c}{\sc Mean-universal} \\ 
&{\(D\)}&{\(E\)}&{\(G_{\infty}\)}&{\(\Upsilon\)}&{\(\chi^2/\nu\)}&{\(\Upsilon\)}&{\(\chi^2/\nu\)} \\
&\footnotesize(\(\sqrt{M_{\solar}}\)pc\(^{-1}\))&\footnotesize(\(10^{3}\sqrt{M_{\solar}}\))&\footnotesize(\(G_{N}\))&&&\footnotesize \\
\footnotesize(1)&\footnotesize(2)&\footnotesize(3)&\footnotesize(4)&\footnotesize(5)&\footnotesize(6)&\footnotesize(7)&\footnotesize(8) \\ \hline\hline
\multicolumn{8}{c}{\fcolorbox{white}{white}{\sf Low surface brightness (LSB) galaxies}} \\ \hline
{NGC~3917} & {\(6.40 \pm 0.08\)} & {\(29.1 \pm 2.5\)} & {\(19.4 \pm 0.4\)} & {\(1.086 \pm 0.027\)} & {\(\phantom{1}35.3\)} & {\(0.883 \pm 0.007\)} & {\(\phantom{1}43.6\)} \\ 
{NGC~4010} & {\(6.41 \pm 0.12\)} & {\(44.9 \pm 4.2\)} & {\(22.5 \pm 0.7\)} & {\(1.059 \pm 0.039\)} & {\(\phantom{1}34.2\)} & {\(0.840 \pm 0.010\)} & {\(\phantom{1}34.3\)} \\ 
{NGC~4183} & {\(6.24 \pm 0.09\)} & {\(16.4 \pm 2.3\)} & {\(7.9 \pm 0.1\)} & {\(1.838 \pm 0.031\)} & {\(\phantom{12}9.1\)} & {\(0.618 \pm 0.009\)} & {\(310.4\)} \\ 
{UGC~6446} & {\(6.40 \pm 0.10\)} & {\(25.3 \pm 1.5\)} & {\(8.9 \pm 0.2\)} & {\(1.694 \pm 0.037\)} & {\(\phantom{1}12.8\)} & {\(0.654 \pm 0.016\)} & {\(319.2\)} \\ 
{UGC~6667} & {\(6.44 \pm 0.29\)} & {\(22.3 \pm 7.3\)} & {\(11.6 \pm 0.8\)} & {\(1.675 \pm 0.149\)} & {\(\phantom{12}4.4\)} & {\(1.071 \pm 0.023\)} & {\(\phantom{1}25.2\)} \\ 
{UGC~6818} & {\(6.37 \pm 0.18\)} & {\(27.1 \pm 1.7\)} & {\(36.2 \pm 1.4\)} & {\(0.757 \pm 0.080\)} & {\(\phantom{1}25.7\)} & {\(1.170 \pm 0.034\)} & {\(\phantom{1}33.1\)} \\ 
{UGC~6917} & {\(6.40 \pm 0.20\)} & {\(24.6 \pm 6.3\)} & {\(10.9 \pm 0.5\)} & {\(1.754 \pm 0.075\)} & {\(\phantom{12}9.1\)} & {\(1.063 \pm 0.015\)} & {\(\phantom{1}60.1\)} \\ 
{UGC~6923} & {\(6.37 \pm 0.18\)} & {\(23.8 \pm 2.1\)} & {\(26.6 \pm 1.2\)} & {\(0.929 \pm 0.064\)} & {\(\phantom{1}20.7\)} & {\(1.078 \pm 0.028\)} & {\(\phantom{1}17.9\)} \\ 
{UGC~6983} & {\(6.38 \pm 0.08\)} & {\(26.4 \pm 1.7\)} & {\(12.3 \pm 0.2\)} & {\(1.454 \pm 0.025\)} & {\(\phantom{1}25.6\)} & {\(0.777 \pm 0.010\)} & {\(162.2\)} \\ 
{UGC~7089} & {\(6.34 \pm 0.27\)} & {\(32.0 \pm 5.4\)} & {\(10.5 \pm 0.6\)} & {\(2.304 \pm 0.154\)} & {\(\phantom{12}4.7\)} & {\(0.937 \pm 0.028\)} & {\(\phantom{1}36.9\)} \\ 
\hline\hline
{\sc Mean Values}&\(6.38 \pm 0.05\)&\(27.2 \pm 7.5\)&\(16.7 \pm 9.3\)&\(1.455 \pm 0.485\)&--&\(0.909 \pm 0.189\)&-- \\ \hline
\hline \multicolumn{8}{c}
{}
\end{tabular} 
\end{center}
\parbox{1.0in}{\phantom{Notes.}}
\parbox{7.5in}{\small Notes. --- Best-fitting and universal STVG parameters, \(D,\,E,\,G_{\infty}\), and stellar mass-to-light ratios, \(\Upsilon\), of the UMa sample: Column (1) is the NGC/UGC galaxy number.  Columns (2) through (6) are the results of a simultaneous best-fit of the galaxy rotation curve to the photometric data allowing  \(D,\,E,\,G_{\infty}\) and \(\Upsilon\) to be varied; and Columns (7) and (8) are the results of the best-fit allowing only \(\Upsilon\) to vary, but with \(D,\,E,\,G_{\infty}\) universally fixed according to \eref{eqn.galaxy.dynamics.stvg.parameters}.   Columns (6) and (8) compare the reduced-\(\chi^2\) statistic of \eref{eqn.galaxy.uma.chi2} in the best-fit and universal cases, respectively.    The STVG acceleration law is given by \erefsss{eqn.galaxy.dynamics.mog.acceleration}{eqn.galaxy.dynamics.stvg}{eqn.galaxy.dynamics.stvg.alpha}{eqn.galaxy.dynamics.stvg.mu}}\index{Modified gravity!Observations|)}
\end{table}
\end{landscape}

\noindent subsamples of HSB and LSB galaxies, and across the complete sample, and mean-universal parameters were chosen.  Next the galaxy rotation curves were refitted and replotted with one free parameter, the stellar mass-to-light ratio, \(\Upsilon\), using the tabulated universal parameters. Therefore all of the dark matter fits have three free parameters, whereas all of the modified gravity theories have only one free parameter, and should be compared with this in mind.

Overall, the core-modified dark matter model shows the lowest reduced \(\chi^2/\nu\) statistic, but the model is less predictive than MOND, MSTG, or STVG due to the variation across the sample in the \(\rho_0\) and \(r_s\) parameters in the dark matter fitting formulae. The only theory that fails to produce \(\chi^2\) best-fits for some dwarf galaxies was the NFW model for NGC 4389 and UGC 6818, as shown in \tref{table.galaxy.darkmatter}.

When MOND as in \tref{table.galaxy.mond}, MSTG as in \tref{table.galaxy.mstg}, and STVG as in \tref{table.galaxy.stvg}, use the best-fit parameters instead of the mean-universal parameters, the reduced \(\chi^2/\nu\) statistic of \eref{eqn.galaxy.uma.chi2} decreases, but there is no case where using mean-universal parameters leads to disagreement with the photometry.  It is reasonable to conclude that MOND, MSTG and STVG provide acceptable fits to galaxy rotation curves with universal parameters and variable mass-to-light ratios, as shown in \fref{figure.galaxy.velocity}.

\newcommand{\subSigma}{\small The surface mass density, \(\Sigma(r)\) in \(M_{\solar}/\mbox{pc}^2\), vs. orbital distance, \(r\) in kpc}
\begin{figure}[ht]\index{Surface mass, \(\Sigma\)|(}
\begin{picture}(460,185)(0,0)
\put(0,40){\includegraphics[width=0.48\textwidth]{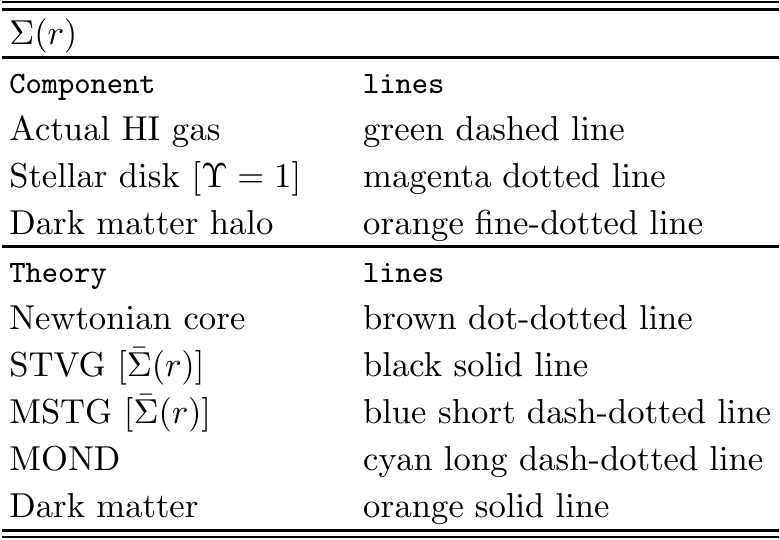}}
\put(225,0){\includegraphics[width=0.5\textwidth]{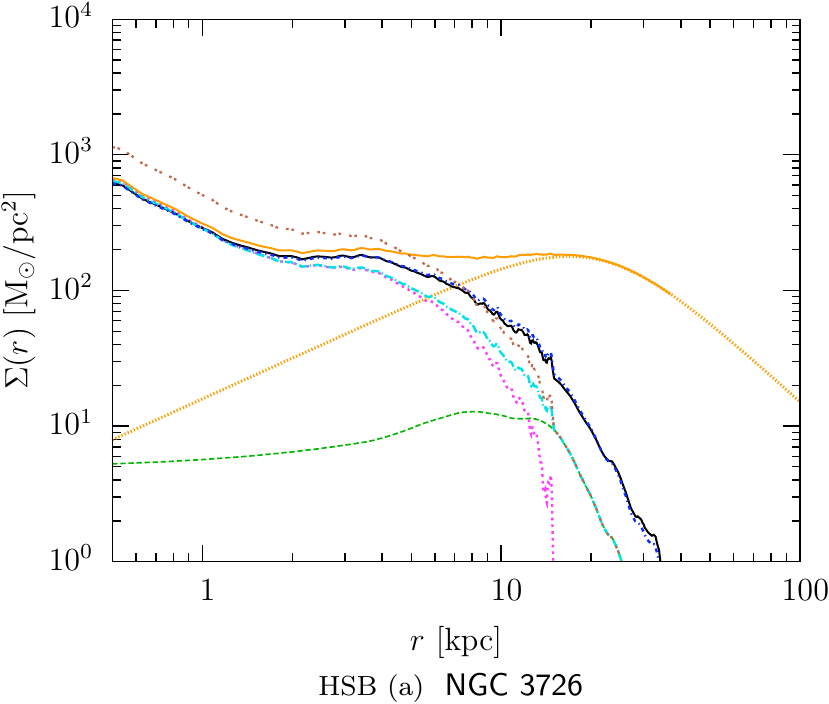}}
\end{picture}
\caption[Surface mass densities]{\label{figure.galaxy.Sigma} {{\sf\small UMa --- Surface mass densities.}}\break\break{\subSigma} for 19 HSB and 10 LSB galaxies.  The photometric data sets consist of the actual HI gas component and the stellar disk component, with a normalized stellar mass-to-light ratio, \(\Upsilon=1\). The computed best-fit results by varying the stellar mass-to-light ratio, \(\Upsilon\), are plotted for Moffat's STVG and MSTG theories and Milgrom's MOND theory with mean-universal parameters.  Results are plotted for the best-fit core-modified dark matter theory including visible baryons, and the corresponding dark matter halo component.  The best-fit Newtonian core model (visible baryons only) is plotted for comparison.  {\it The figure is continued.}}
\end{figure}

\begin{figure}
\begin{picture}(460,450)(82,190)
\put(30,12){\includegraphics[width=1.28\textwidth]{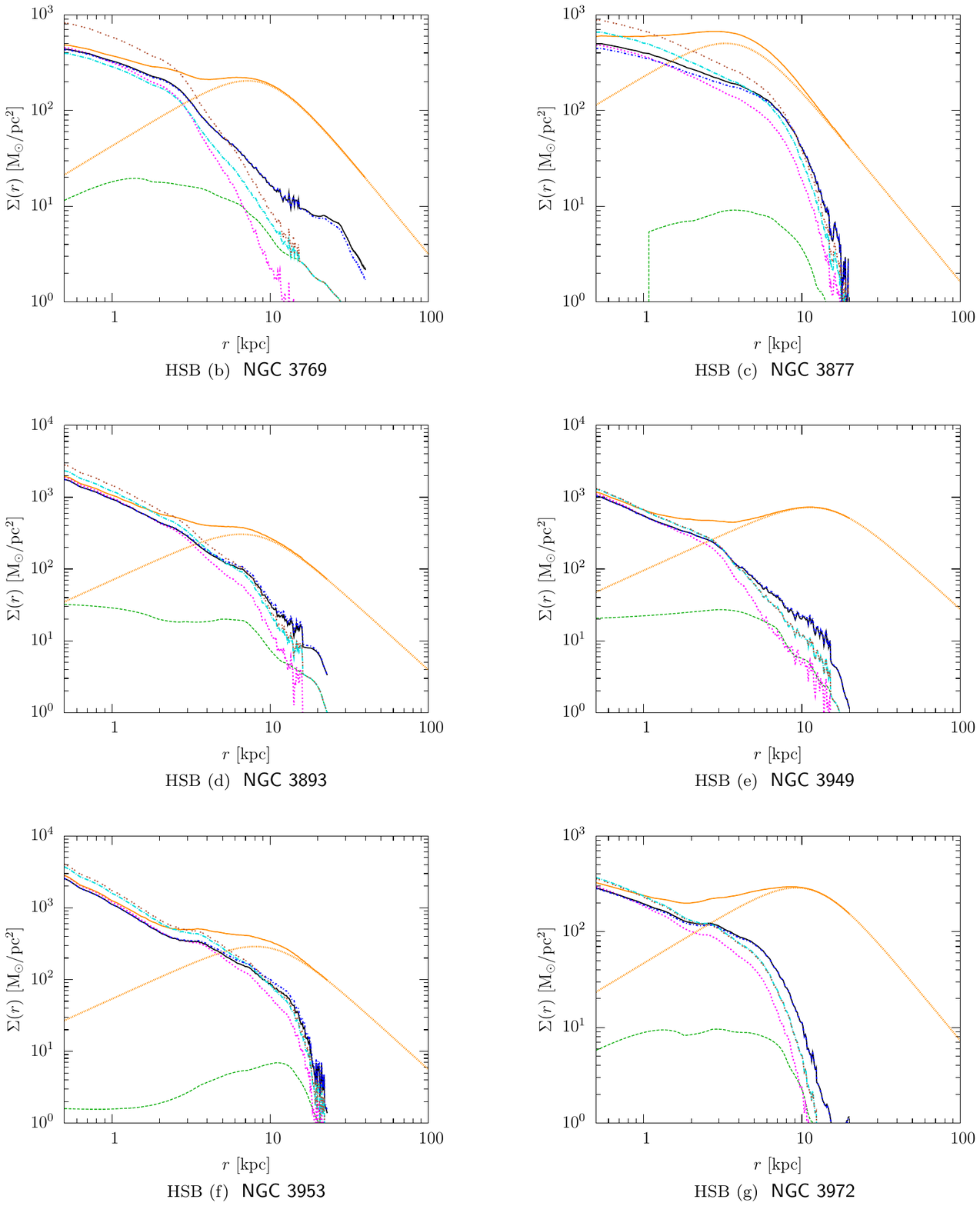}}
\put(82,45){\includegraphics[width=0.98\textwidth]{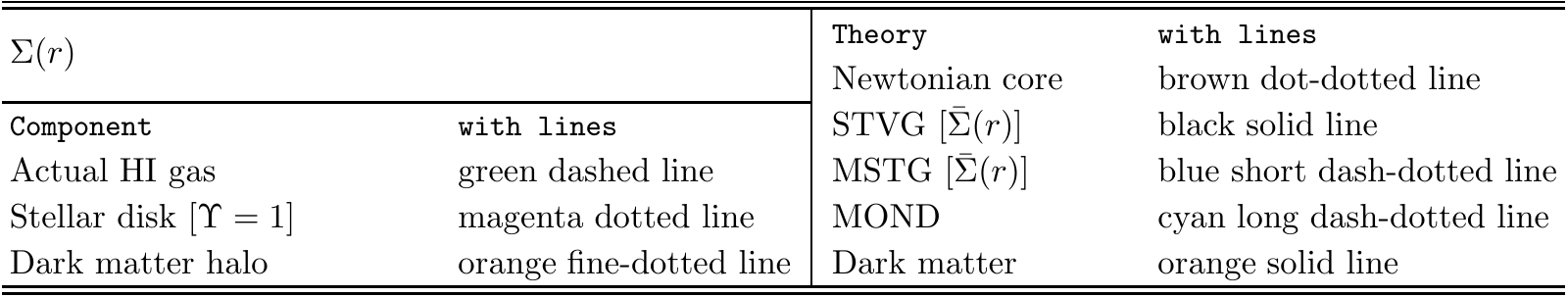}}
\end{picture}
\fcont{figure.galaxy.Sigma}{\sf\small UMa --- Surface mass densities.}
{\subSigma}.
\end{figure}
\begin{figure}
\begin{picture}(460,450)(82,190)
\put(30,12){\includegraphics[width=1.28\textwidth]{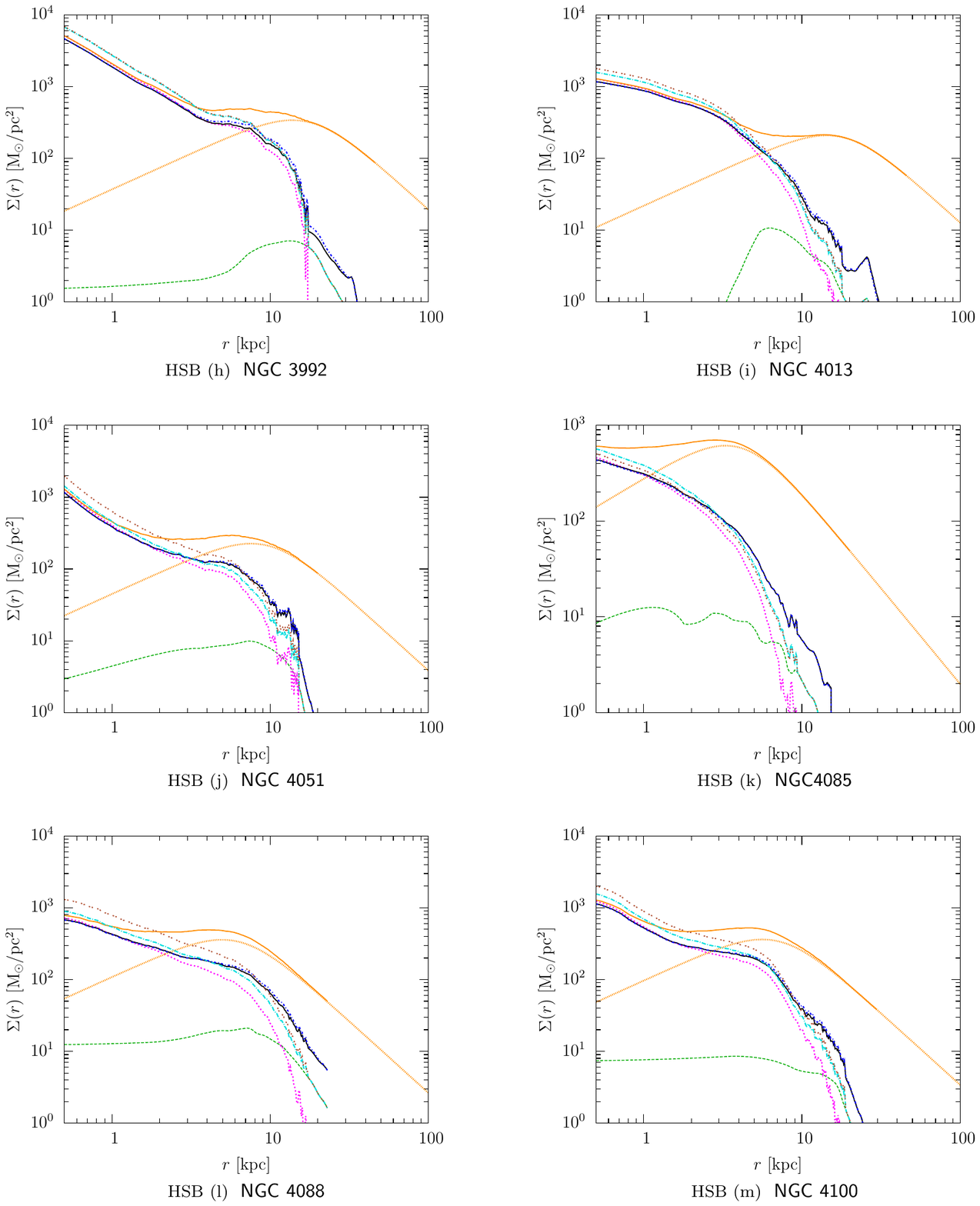}}
\put(82,45){\includegraphics[width=0.98\textwidth]{figure/galaxy_hsb_Sigma_legend}}
\end{picture}
\fcont{figure.galaxy.Sigma}{\sf\small UMa --- Surface mass densities.}
{\subSigma}.
\end{figure}
\begin{figure}
\begin{picture}(460,450)(82,190)
\put(30,12){\includegraphics[width=1.28\textwidth]{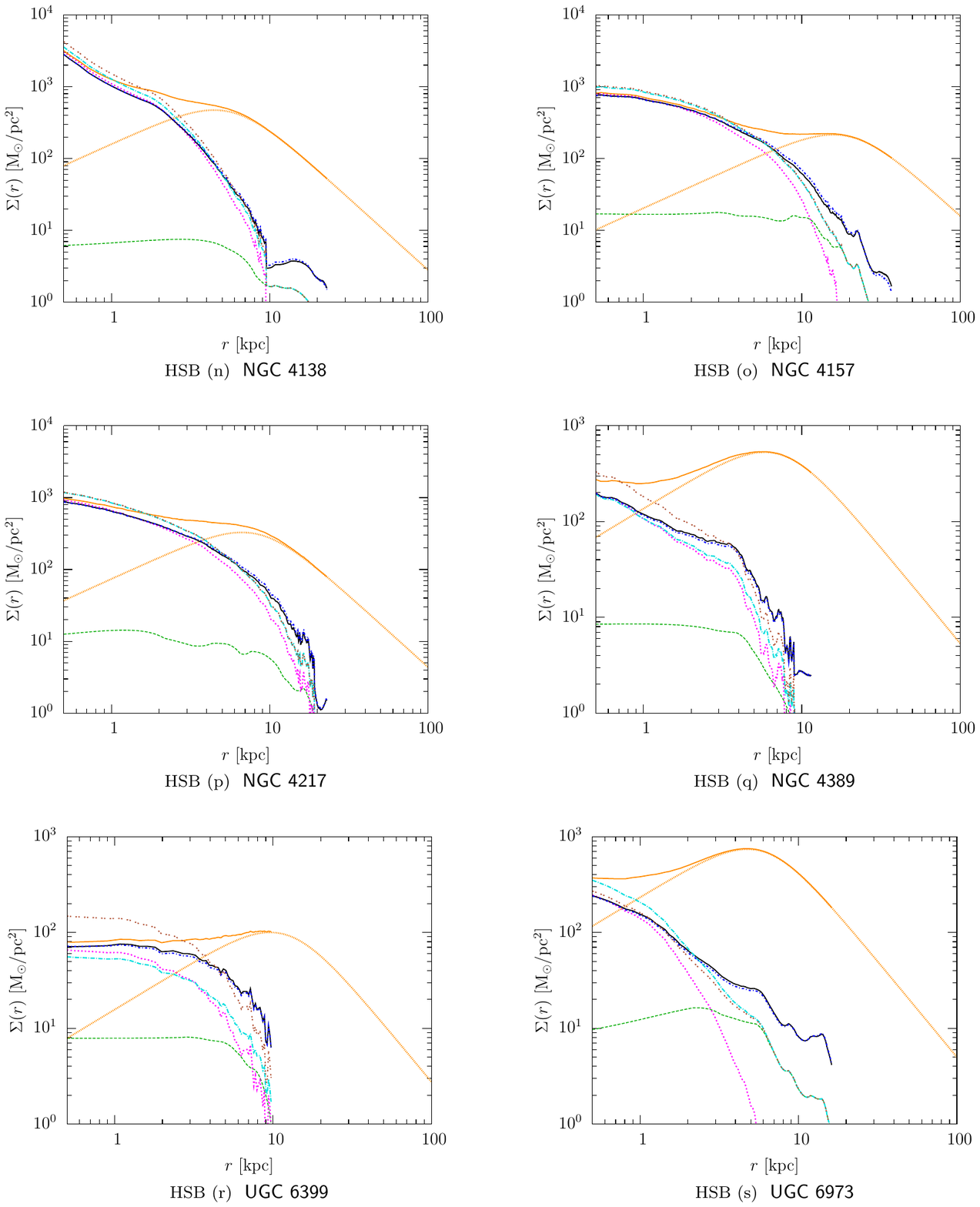}}
\put(82,45){\includegraphics[width=0.98\textwidth]{figure/galaxy_hsb_Sigma_legend}}
\end{picture}
\fcont{figure.galaxy.Sigma}{\sf\small UMa --- Surface mass densities.}
{\subSigma}.
\end{figure}
\begin{figure}\index{Surface mass, \(\Sigma\)|)}
\begin{picture}(460,450)(82,190)
\put(30,12){\includegraphics[width=1.28\textwidth]{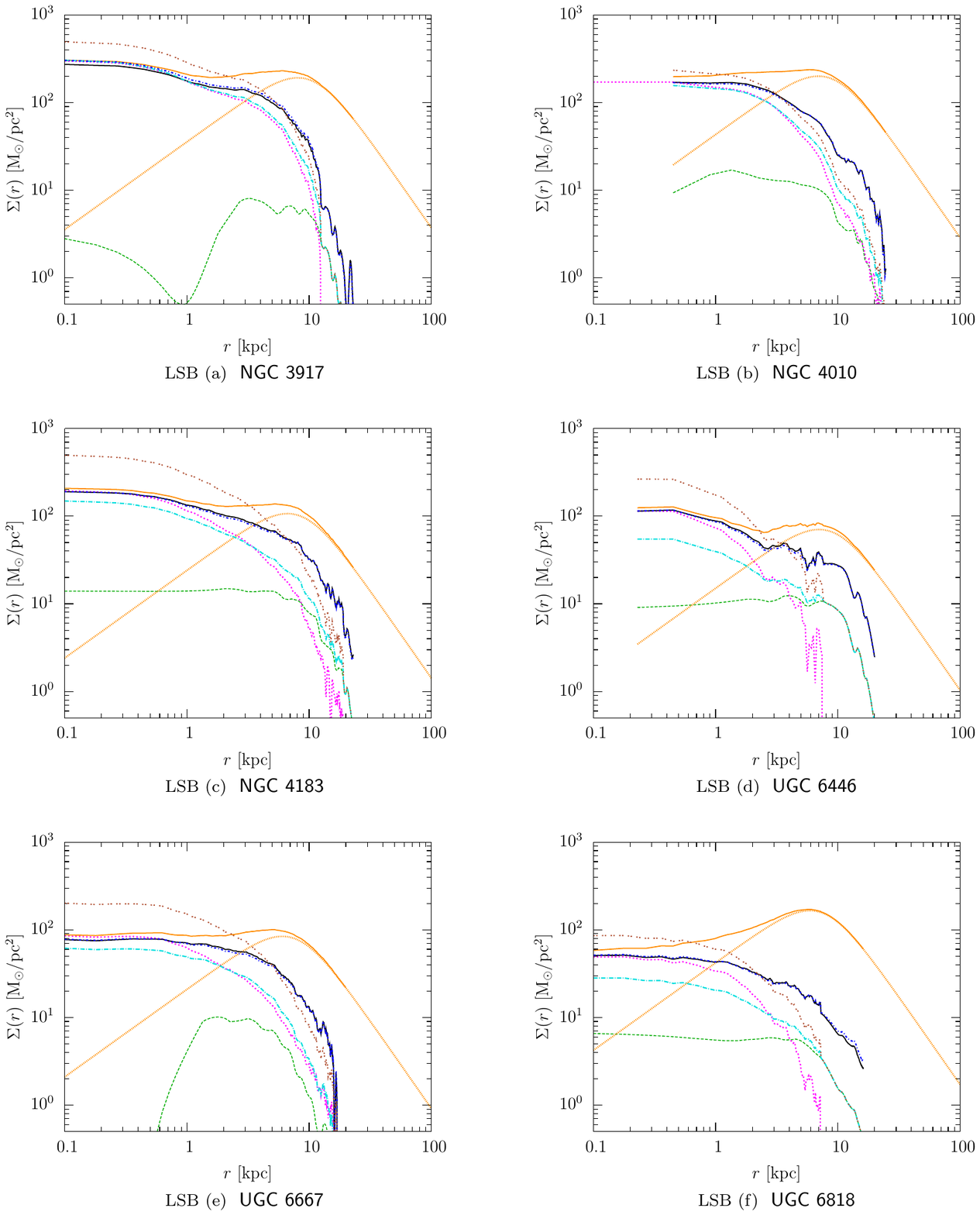}}
\put(82,45){\includegraphics[width=0.98\textwidth]{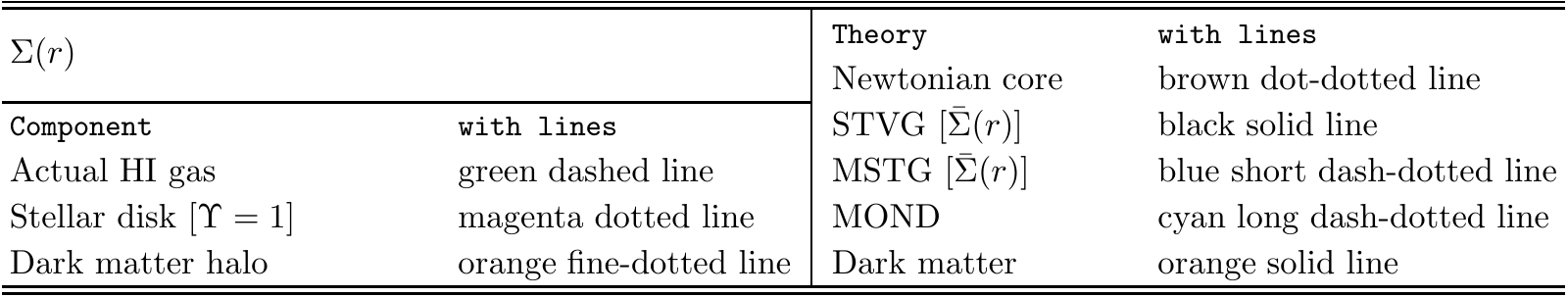}}
\end{picture}
\fcont{figure.galaxy.Sigma}{\sf\small UMa --- Surface mass densities.}
{\subSigma}.
\end{figure}
\begin{figure}
\begin{picture}(460,292)(82,325) 
\put(30,12){\includegraphics[width=1.28\textwidth]{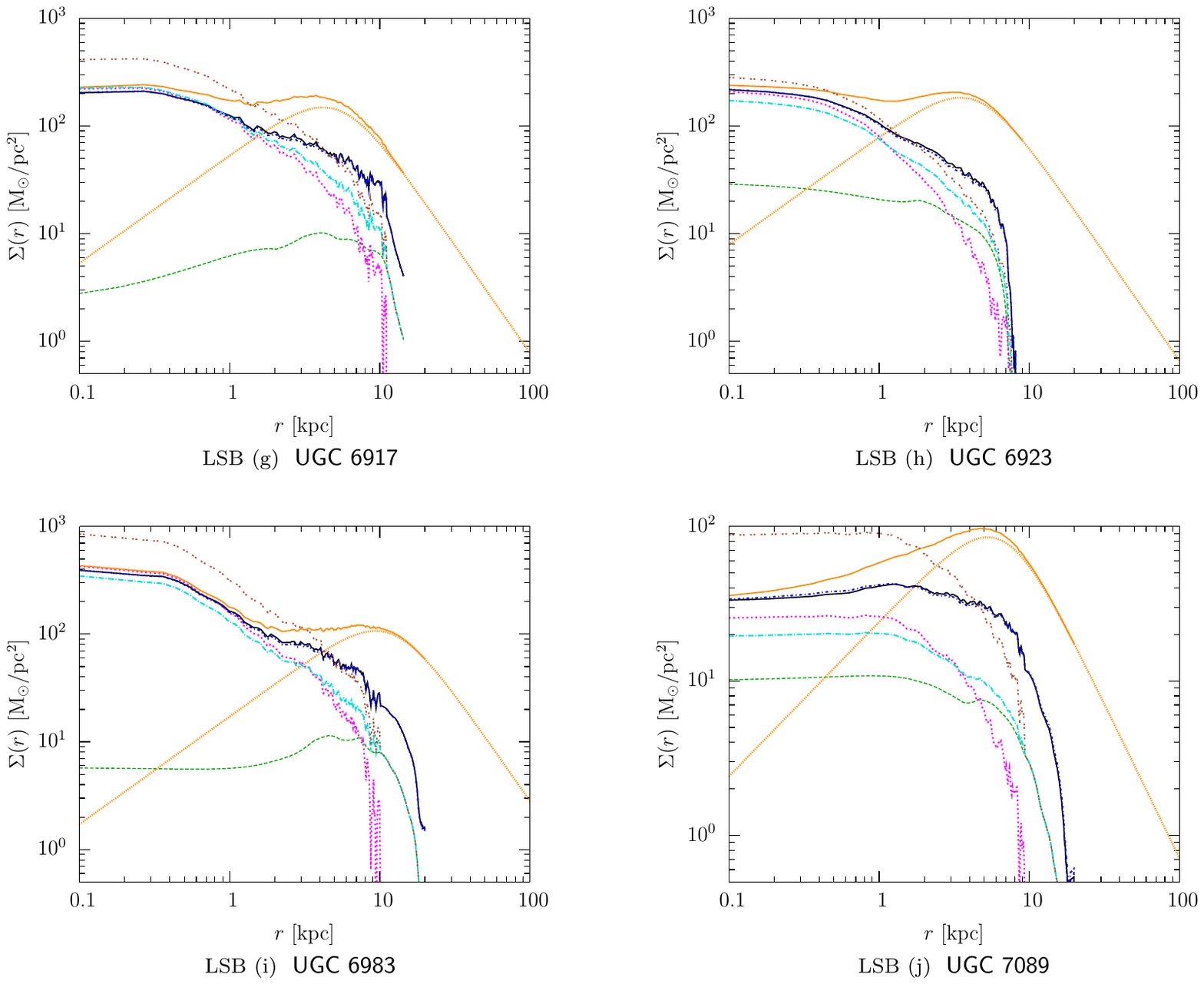}}
\put(82,45){\includegraphics[width=0.98\textwidth]{figure/galaxy_lsb_Sigma_legend}}
\end{picture}
\fcont{figure.galaxy.Sigma}{\sf\small UMa --- Surface mass densities.\break}
{\subSigma} for 19 HSB and 10 LSB galaxies.  The photometric data sets consist of the actual HI gas component and the stellar disk component, with a normalized stellar mass-to-light ratio, \(\Upsilon=1\). The computed best-fit results by varying the stellar mass-to-light ratio, \(\Upsilon\), are plotted for Moffat's STVG and MSTG theories and Milgrom's MOND theory with mean-universal parameters.  Results are plotted for the best-fit core-modified dark matter theory including visible baryons, and the corresponding dark matter halo component.  The best-fit Newtonian core model (visible baryons only) is plotted for comparison.
\end{figure}

\subsection{\label{section.galaxy.uma.Sigma}Surface mass density maps}

The importance of being able to determine the distribution of matter in astronomical objects is that it allows  predictions for ongoing and future experiments, such as galaxy-galaxy lensing, which measures the line-of-sight surface mass density,\index{Gravitational lensing!\map{\Sigma}|(}
\begin{equation}
\label{eqn.galaxy.uma.Sigmarho}
\Sigma(r)\equiv\int \rho(r) dz,
\end{equation}
through the convergence,
\begin{equation}
\label{eqn.galaxy.uma.kappa}
\kappa(r) = \frac{\Sigma(r)}{\Sigma_{c}},
\end{equation}
where 
\begin{equation}
\label{eqn.galaxy.uma.SigmaC.Newton}
\Sigma_{c} = \frac{c^{2}}{4\pi G_{N}} \frac{D_{\rm s}}{D_{\rm l}D_{\rm ls}}
\end{equation}
is the Newtonian critical surface mass density (with vanishing shear), $D_{\rm s}$  is the angular distance to the source, background galaxy, $D_{\rm l}$ is the angular distance to the lens, foreground galaxy.~\protect\citep[Chapter 4]{Peacock:2003}.

\index{Gravitational lensing!MOG|(}
The \map{\Sigma}s plotted in \fref{figure.galaxy.Sigma} provide high resolution sub-kiloparsec predictions, whereas the current state of the art in galaxy-galaxy lensing yield only course grained observations, with resolutions of \(\lesssim\) 10 kpc/pixel.  The MOG predictions for future high resolution \map{\kappa}s must account for the modified acceleration law of \eref{eqn.galaxy.dynamics.mog.acceleration}~\protect\citet{Brownstein.MNRAS.2007.382}:

\begin{equation}\index{Gravitational lensing!\map{\kappa}}
\label{eqn.galaxy.uma.scaledSurfaceMassDensity.MOG}
\kappa(r) =  \int \frac{4\pi G(r)}{c^{2}} \frac{D_{\rm l}D_{\rm ls}}{D_{\rm s}} \rho(r) dz  \equiv \frac{\bar \Sigma(r)}{\Sigma_{c}},
\end{equation}
where
\begin{equation}
\label{eqn.galaxy.uma.Sigma.MOG} {\bar \Sigma}(r) = \int \frac{G(r)}{G_N}\rho(r) dz,
\end{equation}
is the weighted surface mass density, and \(\Sigma_{c}\) is the usual Newtonian critical surface mass density \eref{eqn.galaxy.uma.SigmaC.Newton}.  \citet{Moffat.ArXiv:0805.4774} simplified \erefs{eqn.galaxy.uma.scaledSurfaceMassDensity.MOG}{eqn.galaxy.uma.Sigma.MOG}, in STVG, in the case that the lens may be treated as a {\it point source}, but not for extended mass distributions relevant for galaxy-galaxy lensing.
\index{Gravitational lensing!MOG|)}

The surface mass density due to the visible component is,
\begin{equation} \label{eqn.galaxy.uma.baryonSigma}
\Sigma_{\rm baryon}(r) = \Sigma_{\rm gas}(r) + \Upsilon \Sigma_{\rm disk}(r),
\end{equation}
and therefore the \map{\Sigma} computed for each galaxy depends on the best-fitting stellar mass-to-light ratio, \(\Upsilon\), determined separately for each gravity theory:
\begin{equation}\label{eqn.galaxy.uma.Sigma}
\Sigma(r) = \left\{ \begin{array}{ll} \Sigma_{\rm baryon}(r) & \mbox{\tt Newtonian core, MOND}\\ {\bar \Sigma}_{\rm baryon}(r) & \mbox{\tt MSTG, STVG}\\ \Sigma_{\rm baryon}(r) +\Sigma_{\rm halo}(r) &\mbox{\tt Dark matter}.\end{array}\right.
\end{equation}\index{Gravitational lensing!\map{\Sigma}|)}

\subsection{\label{section.galaxy.uma.mass}Radial mass profiles}

\newcommand{\submass}{\small The radial mass profile, \(M(r)\) in \(M_{\solar}\), vs. orbital distance, \(r\) in kpc}
\begin{figure}[ht]\index{Dark matter!Mass profile|(}\index{Modified gravity!Mass profile|(}\index{MOND!Mass profile|(}
\begin{picture}(460,185)(0,0)
\put(0,40){\includegraphics[width=0.48\textwidth]{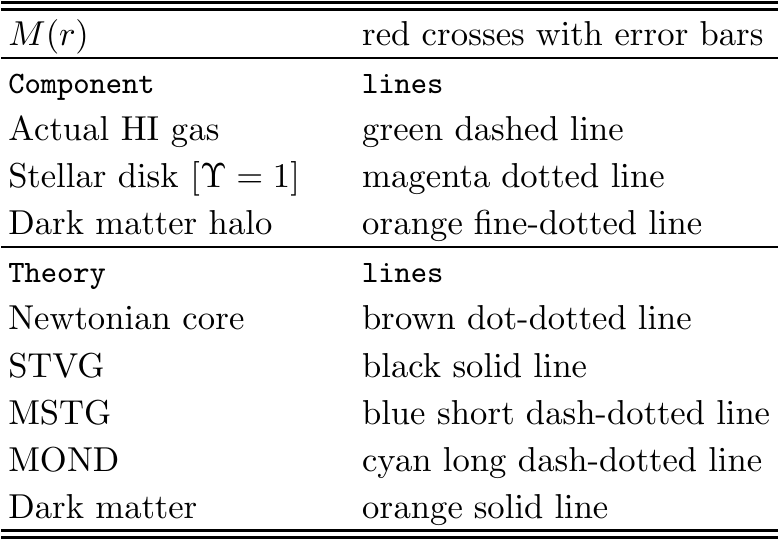}}
\put(225,0){\includegraphics[width=0.5\textwidth]{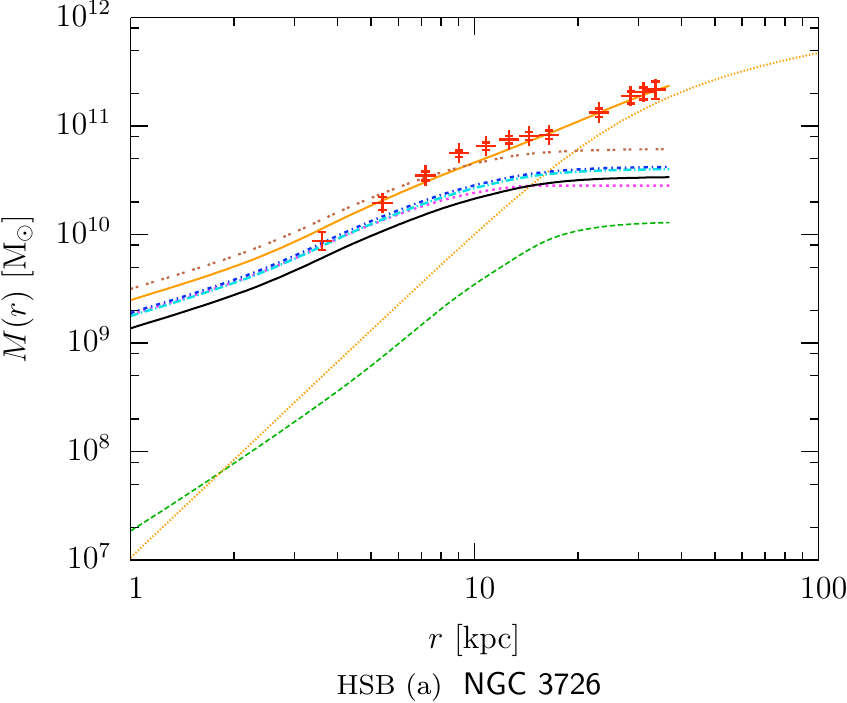}}
\end{picture}
\caption[Mass profiles]{\label{figure.galaxy.mass} {{\sf\small UMa --- Mass profiles.}}\break\break{\submass} for 19 HSB and 10 LSB galaxies.  The dynamic data consist of the Newtonian dynamic mass due to the measured orbital velocities. The photometric data sets consist of the actual HI gas component and the stellar disk component, with a normalized stellar mass-to-light ratio, \(\Upsilon=1\). The computed best-fit results by varying the stellar mass-to-light ratio, \(\Upsilon\), are plotted for Moffat's STVG and MSTG theories and Milgrom's MOND theory with mean-universal parameters.  Results are plotted for the best-fit core-modified dark matter theory including visible baryons, and the corresponding dark matter halo component.  The best-fit Newtonian core model (visible baryons only) is plotted for comparison.  {\it The figure is continued.}}
\end{figure}

\begin{figure}
\begin{picture}(460,450)(82,190)
\put(30,12){\includegraphics[width=1.28\textwidth]{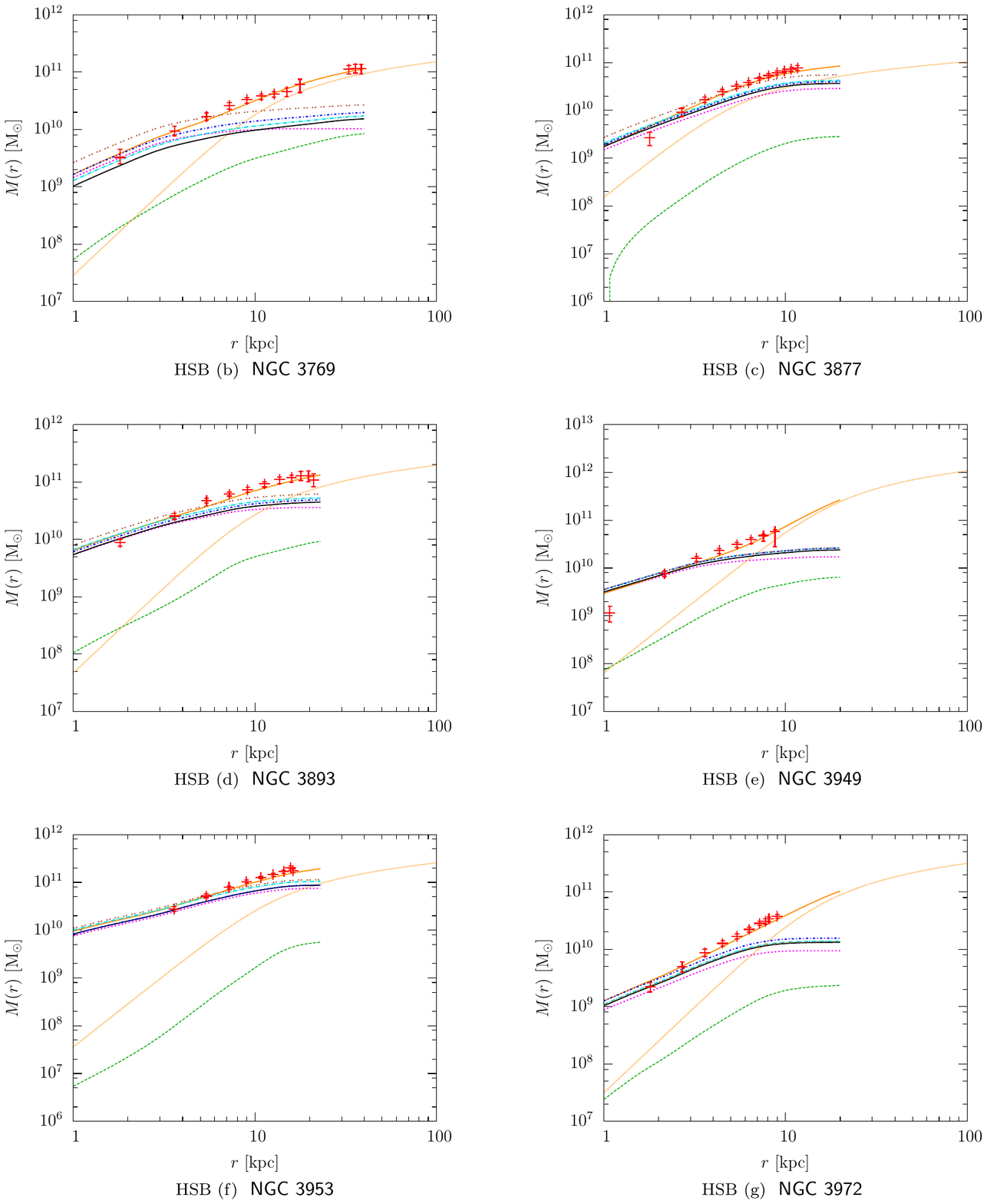}}
\put(82,45){\includegraphics[width=0.98\textwidth]{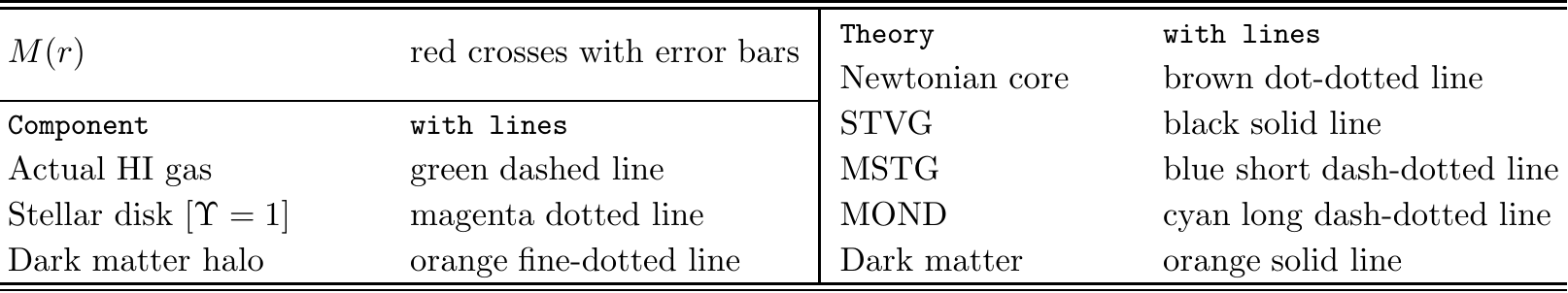}}
\end{picture}
\fcont{figure.galaxy.mass}{\sf\small UMa --- Mass profiles.}
{\submass}.
\end{figure}
\begin{figure}
\begin{picture}(460,450)(82,190)
\put(30,12){\includegraphics[width=1.28\textwidth]{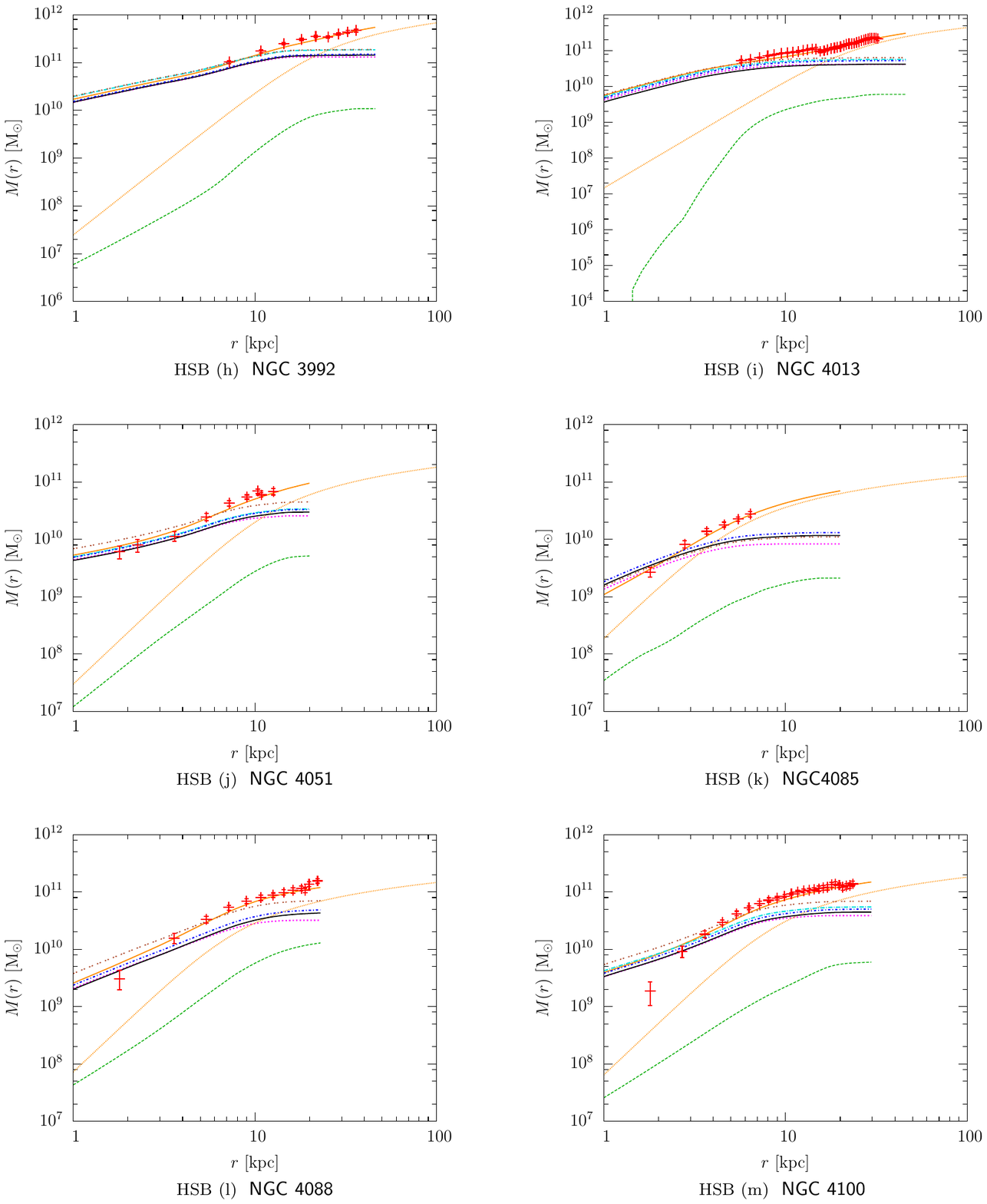}}
\put(82,45){\includegraphics[width=0.98\textwidth]{figure/galaxy_hsb_mass_legend}}
\end{picture}
\fcont{figure.galaxy.mass}{\sf\small UMa --- Mass profiles.}
{\submass}.
\end{figure}
\begin{figure}
\begin{picture}(460,450)(82,190)
\put(30,12){\includegraphics[width=1.28\textwidth]{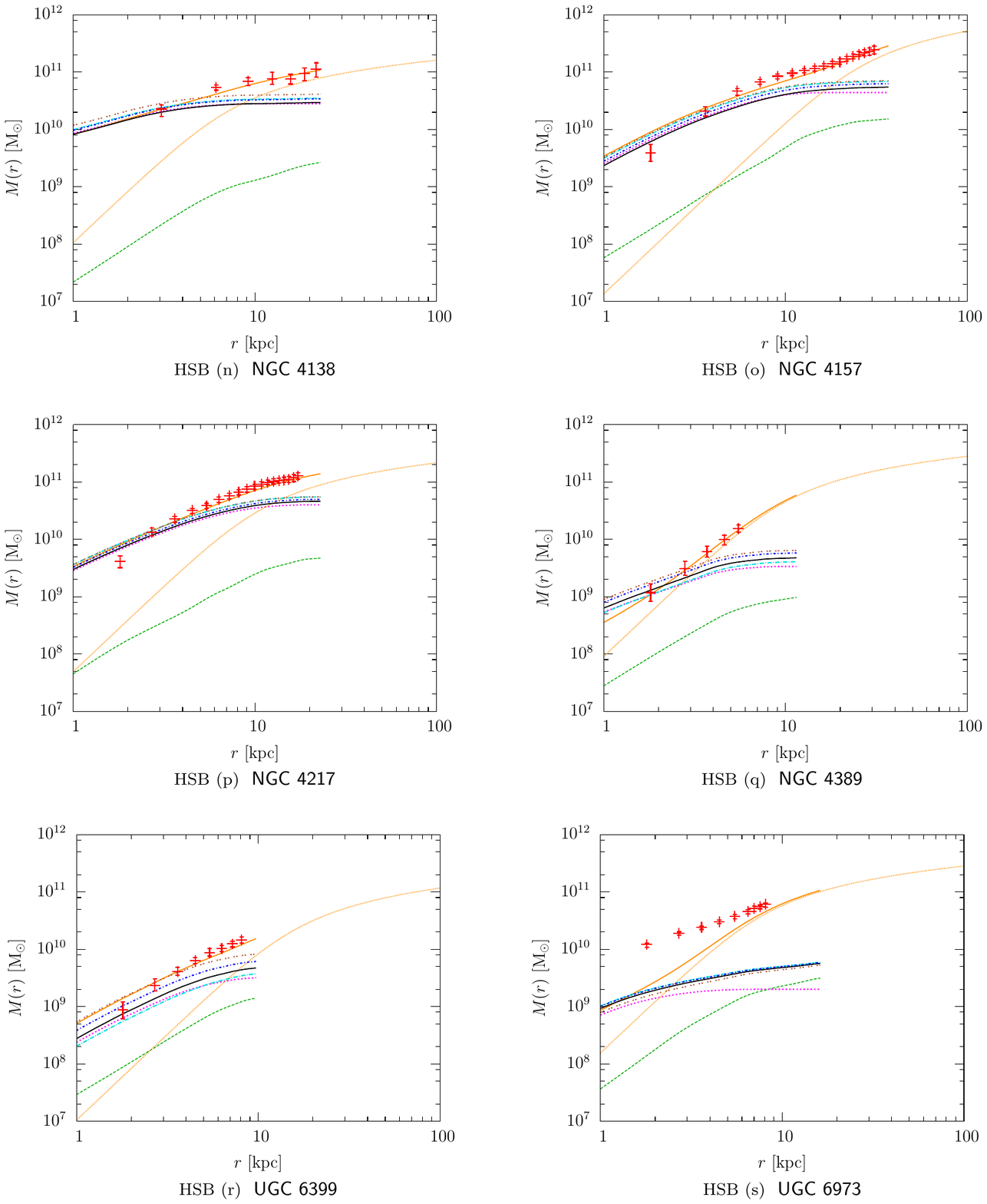}}
\put(82,45){\includegraphics[width=0.98\textwidth]{figure/galaxy_hsb_mass_legend}}
\end{picture}
\fcont{figure.galaxy.mass}{\sf\small UMa --- Mass profiles.}
{\submass}.
\end{figure}

\begin{figure}
\begin{picture}(460,450)(82,190)
\put(30,12){\includegraphics[width=1.28\textwidth]{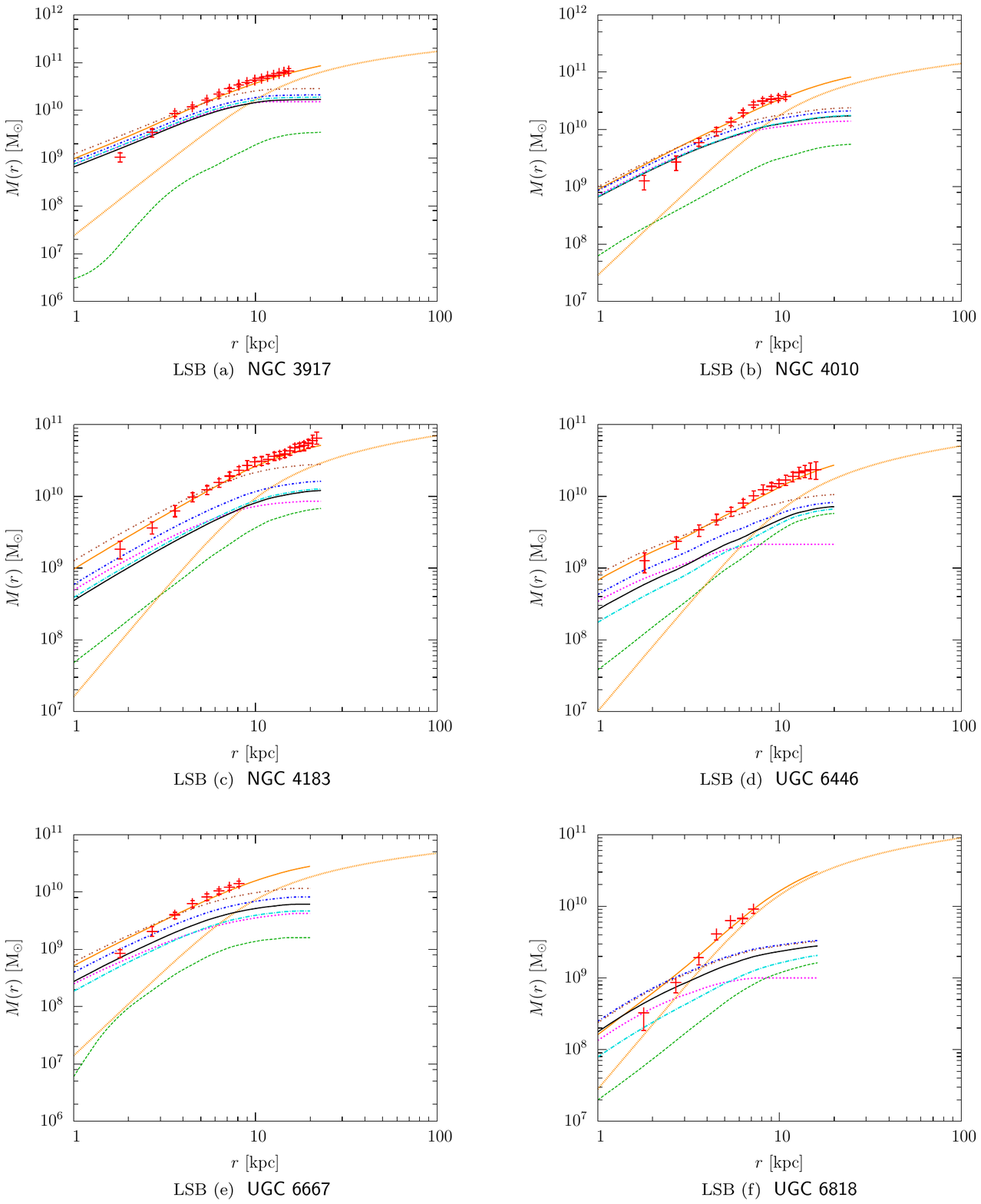}}
\put(82,45){\includegraphics[width=0.98\textwidth]{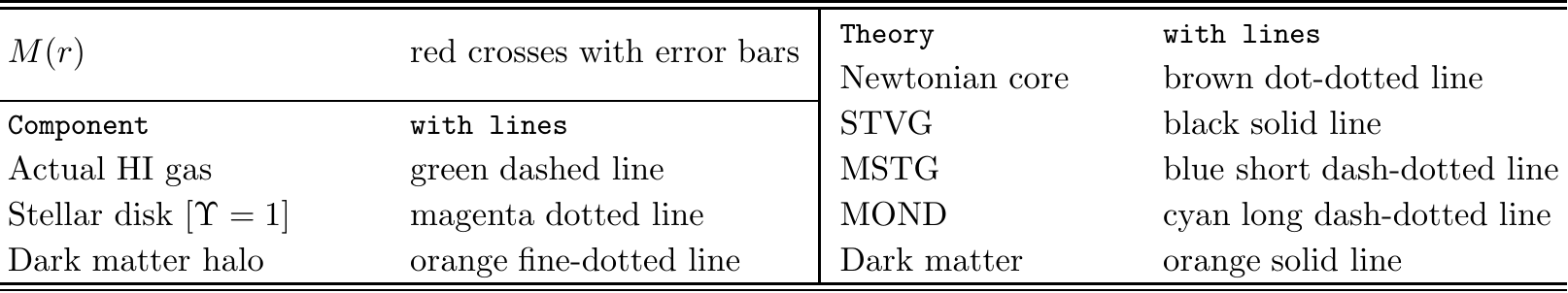}}
\end{picture}
\fcont{figure.galaxy.mass}{\sf\small UMa --- Mass profiles.}
{\submass}.
\end{figure}
\begin{figure}
\begin{picture}(460,290)(82,335) 
\put(30,12){\includegraphics[width=1.28\textwidth]{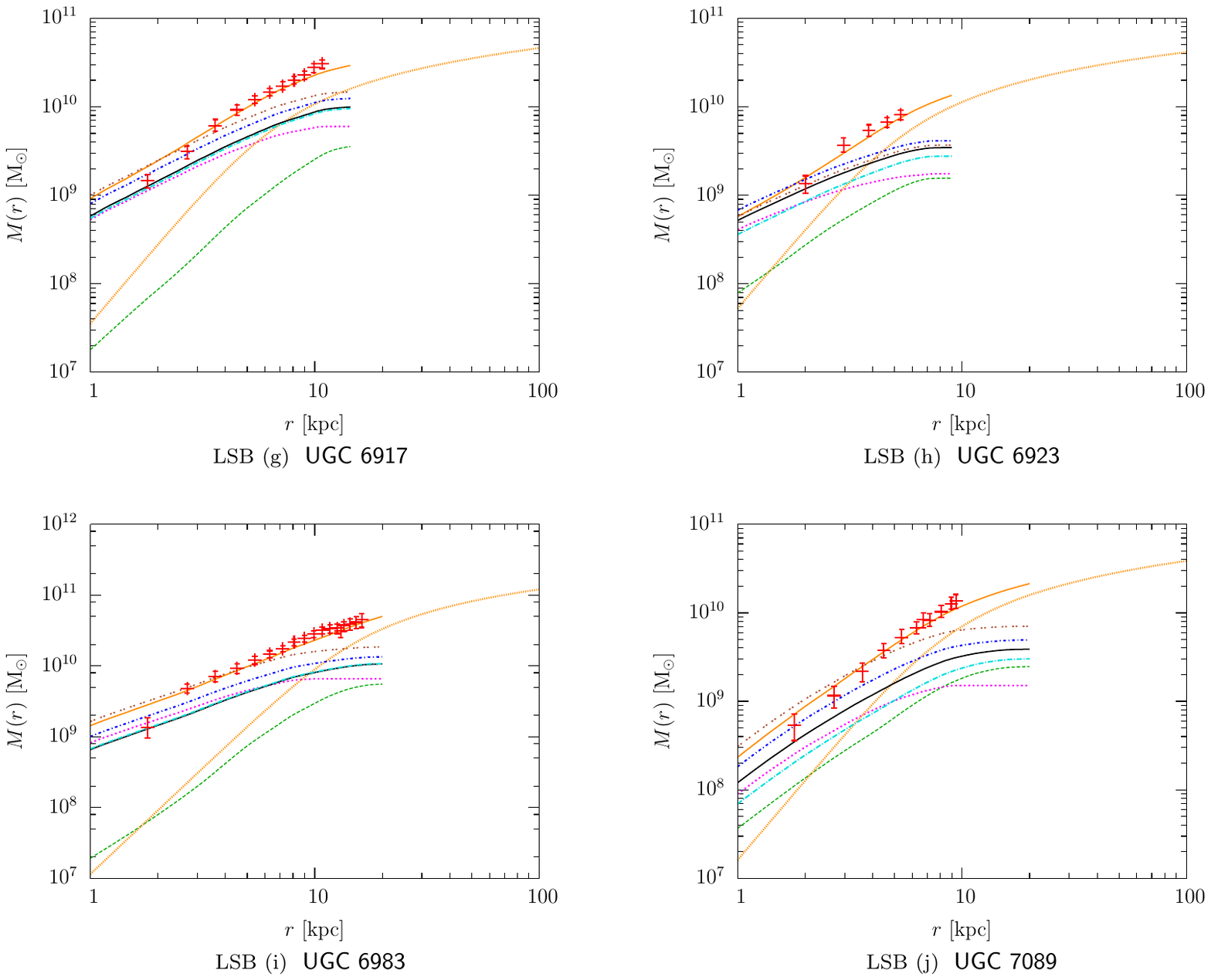}}
\put(82,45){\includegraphics[width=0.98\textwidth]{figure/galaxy_lsb_mass_legend}}
\end{picture}
\fcont{figure.galaxy.mass}{\sf\small UMa --- Mass profiles.\break}
{\submass} for 19 HSB and 10 LSB galaxies.  The dynamic data consist of the Newtonian dynamic mass due to the measured orbital velocities. The photometric data sets consist of the actual HI gas component and the stellar disk component, with a normalized stellar mass-to-light ratio, \(\Upsilon=1\). The computed best-fit results by varying the stellar mass-to-light ratio, \(\Upsilon\), are plotted for Moffat's STVG and MSTG theories and Milgrom's MOND theory with mean-universal parameters.  Results are plotted for the best-fit core-modified dark matter theory including visible baryons, and the corresponding dark matter halo component.  The best-fit Newtonian core model (visible baryons only) is plotted for comparison.
\index{MOND!Mass profile|)}\index{Modified gravity!Mass profile|)}\index{Dark matter!Mass profile|)}
\end{figure}

The missing mass problem is best visualized by solving \eref{eqn.galaxy.uma.orbitalv} for the Newtonian dynamic mass of \eref{eqn.galaxy.mond.newtonianacc}, 
\begin{equation}
\label{eqn.galaxy.uma.mass.dynamic}
M_N(r) = \frac{r \big(v(r)\big)^2}{G_N} , 
\end{equation}
where the velocity points are dynamic variables determined from the galaxy rotation curves of \fref{figure.galaxy.velocity}.  

The visible component mass profiles, plotted in \fref{figure.galaxy.mass}, are based on the surface photometric data of the gaseous disk (HI plus He) component and luminous stellar disk component,
\begin{eqnarray} \label{eqn.galaxy.uma.mass.gas}
M_{\rm gas}(r) &=& \int_0^r 2\pi r^{\prime} \Sigma_{\rm gas}(r^{\prime}) d{r^{\prime}},\\ 
\label{eqn.galaxy.uma.mass.disk}M_{\rm disk}(r) &=& \int_0^r 2\pi r^{\prime} \Sigma_{\rm disk}(r^{\prime}) d{r^{\prime}}.
\end{eqnarray}
The integrated mass profile due to the visible component is therefore,
\begin{equation} \label{eqn.galaxy.uma.mass.baryon}
M_{\rm baryon}(r) = M_{\rm gas}(r) + \Upsilon M_{\rm disk}(r),
\end{equation}
which depends on the best-fitting stellar mass-to-light ratio, \(\Upsilon\), for each galaxy, determined separately for each gravity theory.  
\begin{equation}\label{eqn.galaxy.uma.mass}
M(r) = \left\{ \begin{array}{ll} M_{\rm baryon}(r) & \mbox{\tt Newtonian core, MOND, MSTG, STVG}\\ M_{\rm baryon}(r) + M_{\rm halo}(r) &\mbox{\tt Dark matter},\end{array}\right.
\end{equation}
where the dark matter halo may be computed according to either the NFW formula of \eref{eqn.galaxy.dynamics.dm.nfw}, or alternatively the core-modified formula of \eref{eqn.galaxy.dynamics.dm.coremodified}.  

The total mass of each galaxy due to the visible components is
\begin{equation} \label{eqn.galaxy.uma.masstolight}
M_{\rm baryon} = M_{\rm gas} + \Upsilon M_{\rm disk},
\end{equation}
and therefore the total mass of each galaxy depends on the best-fitting stellar mass-to-light ratio, \(\Upsilon\), determined separately for each gravity theory:
\begin{equation}\label{eqn.galaxy.uma.masstotal}
M_{\rm total} = \left\{ \begin{array}{ll} M_{\rm baryon} & \mbox{\tt Newtonian core, MOND, MSTG, STVG}\\ M_{\rm baryon} + M_{\rm halo} &\mbox{\tt Dark matter},\end{array}\right.
\end{equation}
where the dark matter halo mass may be computed according to \eref{eqn.galaxy.dynamics.dm.nfw} according to the NFW fitting formula, or alternatively according to the core-modified fitting formula of \eref{eqn.galaxy.dynamics.dm.coremodified}.  The final results for the total galaxy masses, according to the best-fitting stellar mass-to-light ratio, \(\Upsilon\), for each gravity theory are provided in \tref{table.galaxy.mass}.

Every galaxy studied, from the highest to lowest in surface brightness, from the most giant to the smallest dwarf, exhibit Newtonian dynamic masses far in excess of the mass profiles due to the visible components, outside the Newtonian core.  The situation within the Newtonian core depends on the particular gravity theory being applied.  Milgrom's theory provides a region inside the MOND regime where accelerations are larger than \(a_0\), where Moffat's MOG theories provide a region inside the MOG regime where the gravitational coupling \(G(r) \sim G_{N}\).  The core-modified dark matter halo is spherical, and does not dominate the visible disks until a critical radius is reached.  In all cases, there is a transition region just outside the Newtonian core where either some form of dark matter is required, or some modification of gravity provides sufficient violations of the strong equivalence principle.

\begin{landscape}
\begin{table}\index{Dark matter!Halo masses}\index{MOND!Mass profile}\index{Modified gravity!Mass profile}\index{Newton's central potential!Galaxy core masses}
\caption[Galaxy masses of the sample]{\label{table.galaxy.mass}{\sf Galaxy masses of the Ursa Major sample}}
\begin{center}
\begin{tabular}{c|cc|c|c|c|c|ccc} \multicolumn{10}{c}{} \\ \hline 
{\sc Galaxy} &\multicolumn{2}{c}{\sc Photometry}& {\sc Newton} & {\sc STVG} & {\sc MSTG} & {\sc MOND} & \multicolumn{3}{c}{\sc Dark matter}  \\ 
&{\(M_{\rm HI}\)}&{\(M_{\rm disk}\)}&{\(M_{\rm baryon}\)}&{\(M_{\rm baryon}\)}&{\(M_{\rm baryon}\)}&{\(M_{\rm baryon}\)}&{\(M_{\rm baryon}\)}&{\(M_{\rm halo}\)}&{\(\frac{M_{\rm halo}}{M_{\rm baryon}}\)} \\
&\footnotesize(\(10^{10} M_{\odot}\))&\footnotesize(\(10^{10} M_{\odot}\))&\footnotesize(\(10^{10} M_{\odot}\))&\footnotesize(\(10^{10} M_{\odot}\))&\footnotesize(\(10^{10} M_{\odot}\))&\footnotesize(\(10^{10} M_{\odot}\))&\footnotesize(\(10^{10} M_{\odot}\))&\footnotesize(\(10^{10} M_{\odot}\))&\\
\footnotesize(1)&\footnotesize(2)&\footnotesize(3)&\footnotesize(4)&\footnotesize(5)&\footnotesize(6)&\footnotesize(7)&\footnotesize(8)&\footnotesize(9)&\footnotesize(10) \\ \hline\hline
\multicolumn{10}{c}{\fcolorbox{white}{white}{\sf High surface brightness (HSB) galaxies}} \\ \hline
NGC~3726 & 0.978 & 2.827 & 6.149 & 3.392 & 4.211 & 4.021 & 5.104 & 16.21 & 3.176 \\
NGC~3769 & 0.678 & 1.030 & 2.743 & 1.593 & 2.034 & 1.777 & 2.016 & 9.206 & 4.566 \\
NGC~3877 & 0.212 & 2.878 & 5.551 & 3.671 & 3.987 & 4.241 & 3.325 & 3.445 & 1.036 \\
NGC~3893 & 0.761 & 3.58 & 6.261 & 4.564 & 4.983 & 5.364 & 5.199 & 7.519 & 1.446 \\
NGC~3949 & 0.488 & 1.711 & 2.645 & 2.399 & 2.613 & 2.614 & 2.242 & 3.893 & 1.736 \\
NGC~3953 & 0.428 & 7.474 & 11.37 & 8.702 & 8.744 & 10.45 & 9.914 & 6.061 & 0.611 \\
NGC~3972 & 0.181 & 0.943 & 1.362 & 1.327 & 1.572 & 1.393 & 1.533 & 1.866 & 1.218 \\
NGC~3992 & 0.812 & 13.116 & 18.76 & 14.32 & 14.88 & 18.46 & 16.31 & 29.68 & 1.819 \\
NGC~4013 & 0.452 & 4.127 & 6.373 & 4.207 & 5.258 & 5.702 & 6.104 & 16.25 & 2.662 \\
NGC~4051 & 0.388 & 2.563 & 4.490 & 2.997 & 3.323 & 3.394 & 3.572 & 3.086 & 0.864 \\
NGC~4085 & 0.159 & 0.831 & 1.108 & 1.168 & 1.313 & 1.224 & 0.744 & 2.049 & 2.755 \\
NGC~4088 & 1.120 & 3.192 & 7.234 & 4.497 & 5.064 & 5.404 & 5.267 & 6.697 & 1.272 \\
NGC~4100 & 0.453 & 3.856 & 6.875 & 4.460 & 5.007 & 5.480 & 5.084 & 8.258 & 1.624 \\
NGC~4138 & 0.209 & 2.805 & 4.129 & 2.967 & 3.429 & 3.519 & 2.922 & 7.623 & 2.608 \\
NGC~4157 & 1.169 & 4.427 & 7.144 & 5.531 & 6.341 & 6.945 & 7.398 & 16.57 & 2.240 \\
NGC~4217 & 0.368 & 3.999 & 5.554 & 4.633 & 5.003 & 5.501 & 5.055 & 6.617 & 1.309 \\
NGC~4389 & 0.083 & 0.337 & 0.656 & 0.489 & 0.592 & 0.422 & 0.259 & 1.262 & 4.873 \\
UGC~6399 & 0.114 & 0.320 & 0.837 & 0.484 & 0.626 & 0.385 & 0.776 & 0.486 & 0.626 \\
UGC~6973 & 0.249 & 0.100 & 0.438 & 0.457 & 0.467 & 0.470 & 0.426 & 4.157 & 9.763 \\
\hline
\end{tabular} 
\end{center}
\end{table}
\begin{table}
\tcont{table.galaxy.mass}{\sf Galaxy masses of the Ursa Major sample}
\begin{center}
\begin{tabular}{c|cc|c|c|c|c|ccc} \multicolumn{10}{c}{} \\ \hline 
{\sc Galaxy} &\multicolumn{2}{c}{\sc Photometry}& {\sc Newton} & {\sc STVG} & {\sc MSTG} & {\sc MOND} & \multicolumn{3}{c}{\sc Dark matter}  \\ 
&{\(M_{\rm HI}\)}&{\(M_{\rm disk}\)}&{\(M_{\rm baryon}\)}&{\(M_{\rm baryon}\)}&{\(M_{\rm baryon}\)}&{\(M_{\rm baryon}\)}&{\(M_{\rm baryon}\)}&{\(M_{\rm halo}\)}&{\(\frac{M_{\rm halo}}{M_{\rm baryon}}\)} \\
&\footnotesize(\(10^{10} M_{\odot}\))&\footnotesize(\(10^{10} M_{\odot}\))&\footnotesize(\(10^{10} M_{\odot}\))&\footnotesize(\(10^{10} M_{\odot}\))&\footnotesize(\(10^{10} M_{\odot}\))&\footnotesize(\(10^{10} M_{\odot}\))&\footnotesize(\(10^{10} M_{\odot}\))&\footnotesize(\(10^{10} M_{\odot}\))&\\
\footnotesize(1)&\footnotesize(2)&\footnotesize(3)&\footnotesize(4)&\footnotesize(5)&\footnotesize(6)&\footnotesize(7)&\footnotesize(8)&\footnotesize(9)&\footnotesize(10) \\ \hline\hline
\multicolumn{10}{c}{\fcolorbox{white}{white}{\sf Low surface brightness (LSB) galaxies}} \\ \hline
NGC~3917 & 0.271 & 1.525 & 2.865 & 1.708 & 2.124 & 1.899 & 2.313 & 3.699 & 1.599 \\
NGC~4010 & 0.416 & 1.400 & 2.395 & 1.731 & 2.116 & 1.760 & 2.133 & 2.042 & 0.957 \\
NGC~4183 & 0.532 & 0.855 & 2.820 & 1.238 & 1.653 & 1.302 & 2.275 & 2.757 & 1.212 \\
UGC~6446 & 0.442 & 0.215 & 1.070 & 0.730 & 0.833 & 0.675 & 0.983 & 1.331 & 1.354 \\
UGC~6667 & 0.120 & 0.420 & 1.152 & 0.610 & 0.815 & 0.463 & 0.993 & 0.486 & 0.489 \\
UGC~6818 & 0.151 & 0.100 & 0.363 & 0.318 & 0.372 & 0.245 & 0.286 & 0.745 & 2.606 \\
UGC~6917 & 0.285 & 0.600 & 1.493 & 1.018 & 1.266 & 0.981 & 1.356 & 1.178 & 0.869 \\
UGC~6923 & 0.116 & 0.175 & 0.367 & 0.343 & 0.412 & 0.275 & 0.343 & 0.451 & 1.315 \\
UGC~6983 & 0.419 & 0.658 & 1.859 & 1.070 & 1.350 & 1.083 & 1.680 & 2.400 & 1.428 \\
UGC~7089 & 0.185 & 0.151 & 0.706 & 0.388 & 0.493 & 0.302 & 0.552 & 0.636 & 1.153 \\
\hline \multicolumn{10}{c}{}
\end{tabular} 
\end{center}
\parbox{1.0in}{\phantom{Notes.}}
\parbox{7.5in}{\small Notes. --- Galaxy masses and dark matter fractions of the UMa sample: Column (1) is
the NGC/UGC galaxy number.  Column (2) is the total computed mass of the infinitely thin gaseous HI disk which determines the total HI (plus He) gas mass via \eref{eqn.galaxy.uma.rotmodGas}. Column (3) is the total computed mass of the K-band stellar disk \((\Upsilon=1)\).  Columns (4) through (8) are the total baryon mass of each galaxy, via \eref{eqn.galaxy.uma.masstolight}, respective of each gravity theory.  Column (9) is the total dark matter integrated to the outermost radial point in the rotation curve; and Column (10) is the corresponding dark matter to baryon mass fraction to the outermost radial point.}

\end{table}
\end{landscape}
\subsection{\label{section.galaxy.uma.Gamma}Dynamic mass factor}

Since the Newtonian dynamic mass greatly exceeds the baryonic mass outside the Newtonian core, each gravity theory must make up the difference in order to fit the data.  For dark matter, the difference is the halo component,
\begin{equation}\label{eqn.galaxy.uma.Gamma.darkmatter.mass}
M_{N}(r) = M_{\rm baryon} + M_{\rm halo}.
\end{equation}

\noindent The dynamic mass factor is defined as the Newtonian dynamic mass per unit baryonic mass
\begin{equation}\label{eqn.galaxy.uma.Gamma.darkmatter}
\Gamma_{\rm dark~matter}(r) \equiv \frac{M_{N}(r)}{M_{\rm baryon}(r)} = 1 + \frac{M_{\rm halo}(r)}{M_{\rm baryon}(r)}.
\end{equation}
For Milgrom's MOND, the difference is due to the reciprocal factor of the smaller than unity MOND interpolating function,\index{MOND!Interpolating function, \(\mu\)}
\begin{equation}\label{eqn.galaxy.uma.Gamma.mond.mass}
M_{N}(r) = \frac{M_{\rm baryon}(r)}{\mu(r)},
\end{equation}
and the dynamic mass factor is defined as
\begin{equation}\label{eqn.galaxy.uma.Gamma.mond}
\Gamma_{\rm MOND}(r) \equiv \frac{M_{N}(r)}{M_{\rm baryon}(r)} = \frac{1}{\mu(r)}.
\end{equation}
For Moffat's MOG, the difference is due to the multiplicative factor of the larger than Newton gravitational coupling,
\begin{equation}\label{eqn.galaxy.uma.Gamma.mog.mass}
M_{N}(r) = \frac{G(r) M_{\rm baryon}(r)}{G_N},
\end{equation}
and the dynamic mass factor is defined as
\begin{equation}\label{eqn.galaxy.uma.Gamma.mog}
\Gamma_{\rm MOG}(r) \equiv \frac{M_{N}(r)}{M_{\rm baryon}(r)} = \frac{G(r)}{G_N}.
\end{equation}

\newcommand{\subGamma}{\small The Dynamical mass factor, \(\Gamma(r)\), vs. orbital distance, \(r\) in kpc}
\begin{figure}[t]\index{Dynamic mass factor, \(\Gamma\)|(}
\begin{picture}(460,185)(0,0)
\put(0,40){\includegraphics[width=0.48\textwidth]{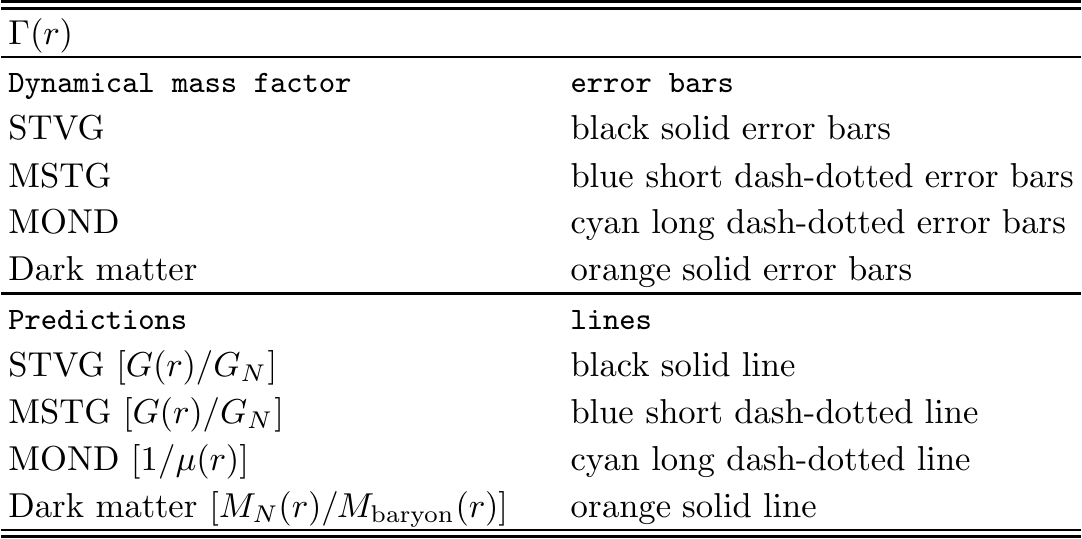}}
\put(225,0){\includegraphics[width=0.5\textwidth]{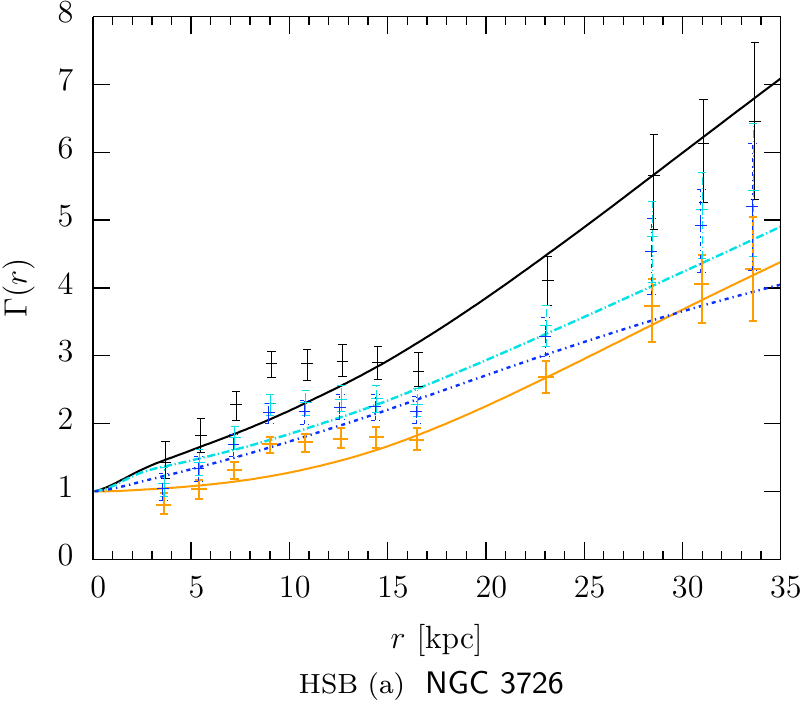}}
\end{picture}
\caption[Dynamical mass factors]{\label{figure.galaxy.Gamma} {{\sf\small UMa --- Dynamical mass factors.}}\break\break{\subGamma} for 19 HSB and 10 LSB galaxies.  The dynamic data consist of the Newtonian dynamic mass due to the measured orbital velocities per unit baryonic mass per gravity theory, shown with error bars. The computed best-fit results by varying the stellar mass-to-light ratio, \(\Upsilon\), are plotted for Moffat's STVG and MSTG theories and Milgrom's MOND theory with mean-universal parameters.  Results are plotted for the best-fit core-modified dark matter theory including visible baryons. {\it The figure is continued.}}
\end{figure}

\begin{figure}
\begin{picture}(460,450)(82,190)
\put(30,12){\includegraphics[width=1.28\textwidth]{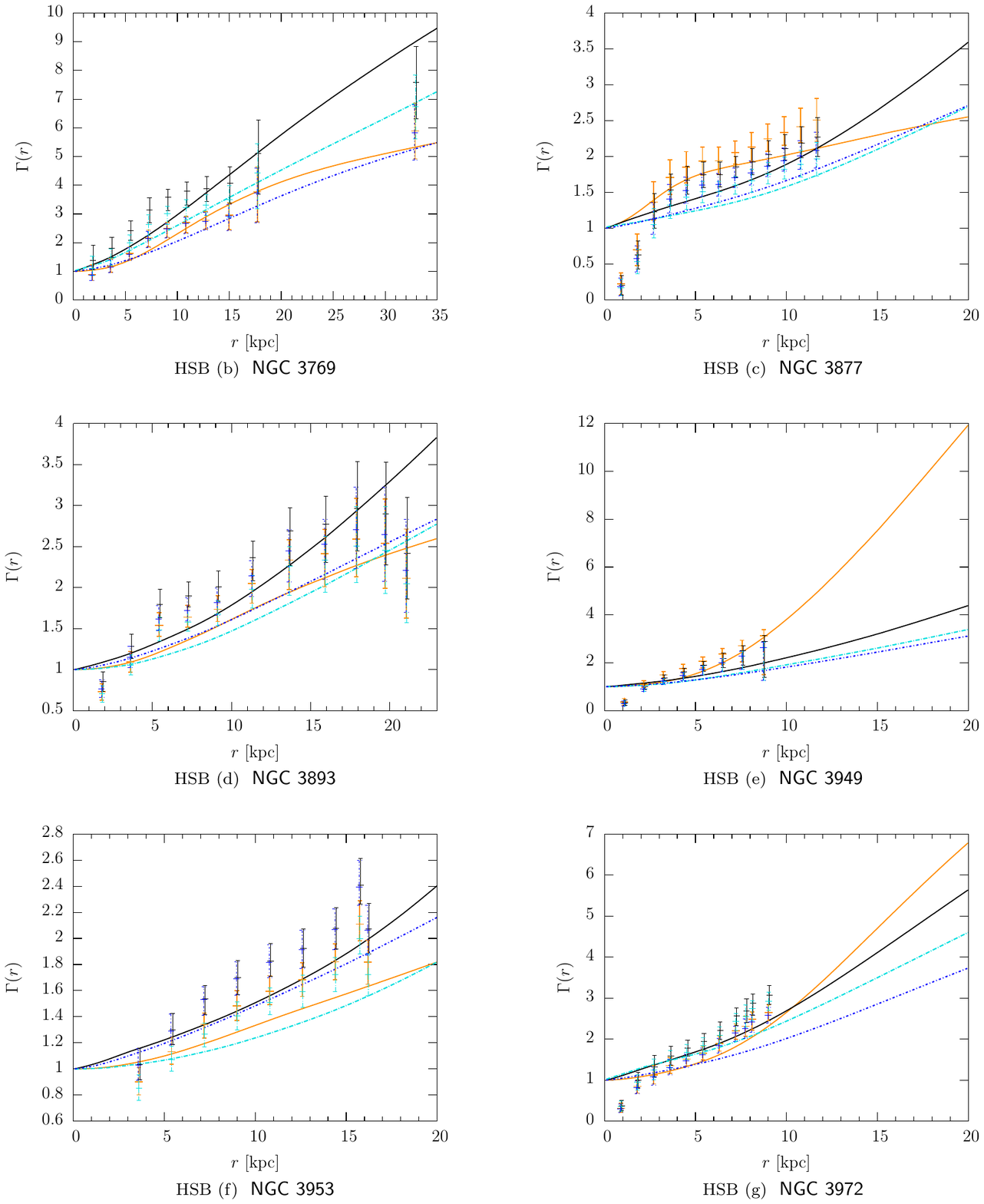}}
\put(82,45){\includegraphics[width=0.98\textwidth]{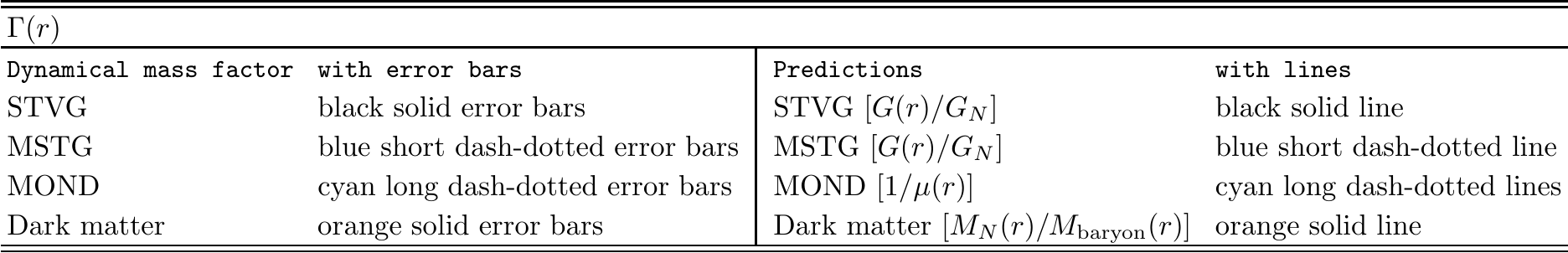}}
\end{picture}
\fcont{figure.galaxy.Gamma}{\sf\small UMa --- Dynamical mass factors.}
{\subGamma}.
\end{figure}
\begin{figure}
\begin{picture}(460,450)(82,190)
\put(30,12){\includegraphics[width=1.28\textwidth]{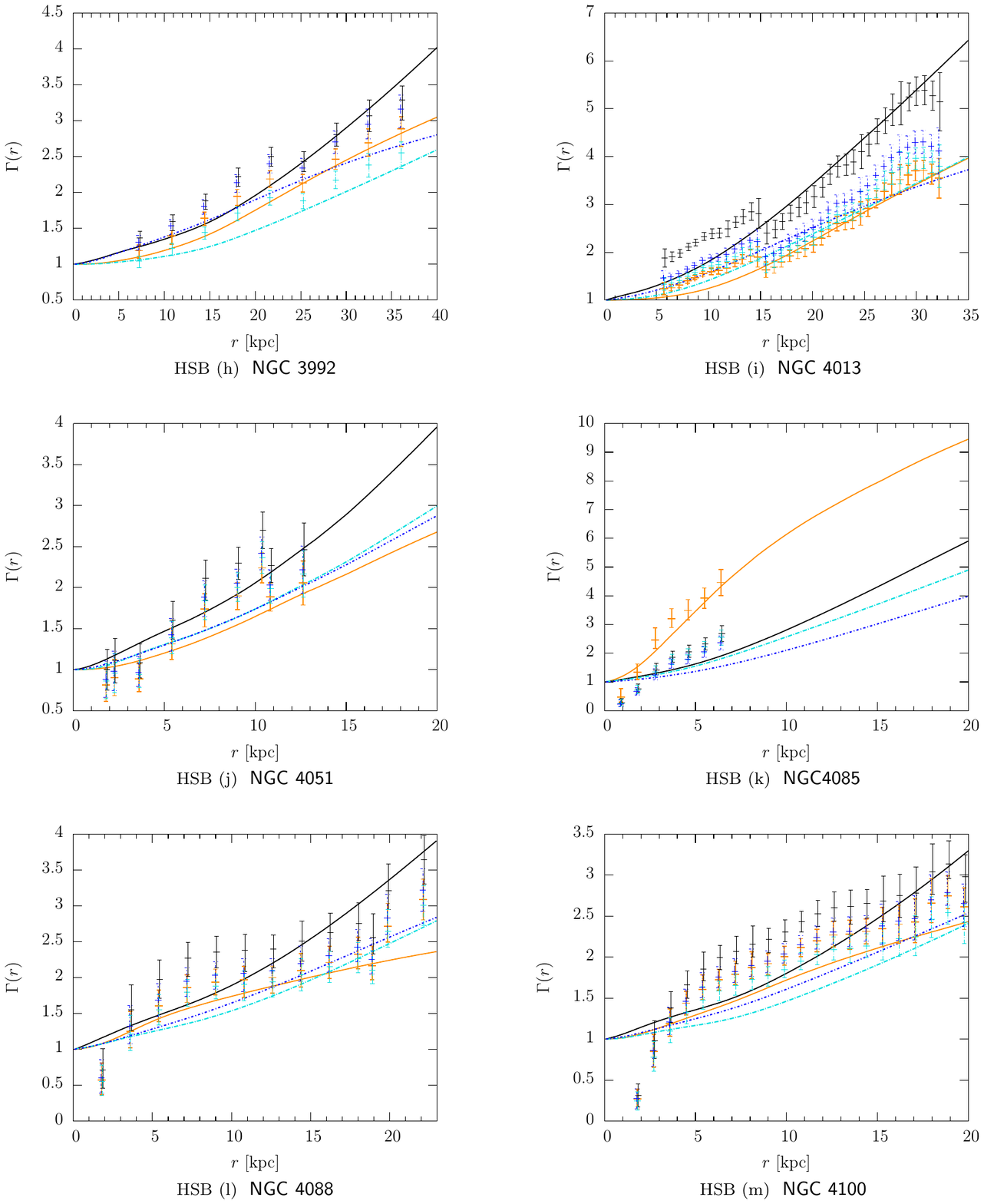}}
\put(82,45){\includegraphics[width=0.98\textwidth]{figure/galaxy_hsb_Gamma_legend}}
\end{picture}
\fcont{figure.galaxy.Gamma}{\sf\small UMa --- Dynamical mass factors.}
{\subGamma}.
\end{figure}
\begin{figure}
\begin{picture}(460,450)(82,190)
\put(30,12){\includegraphics[width=1.28\textwidth]{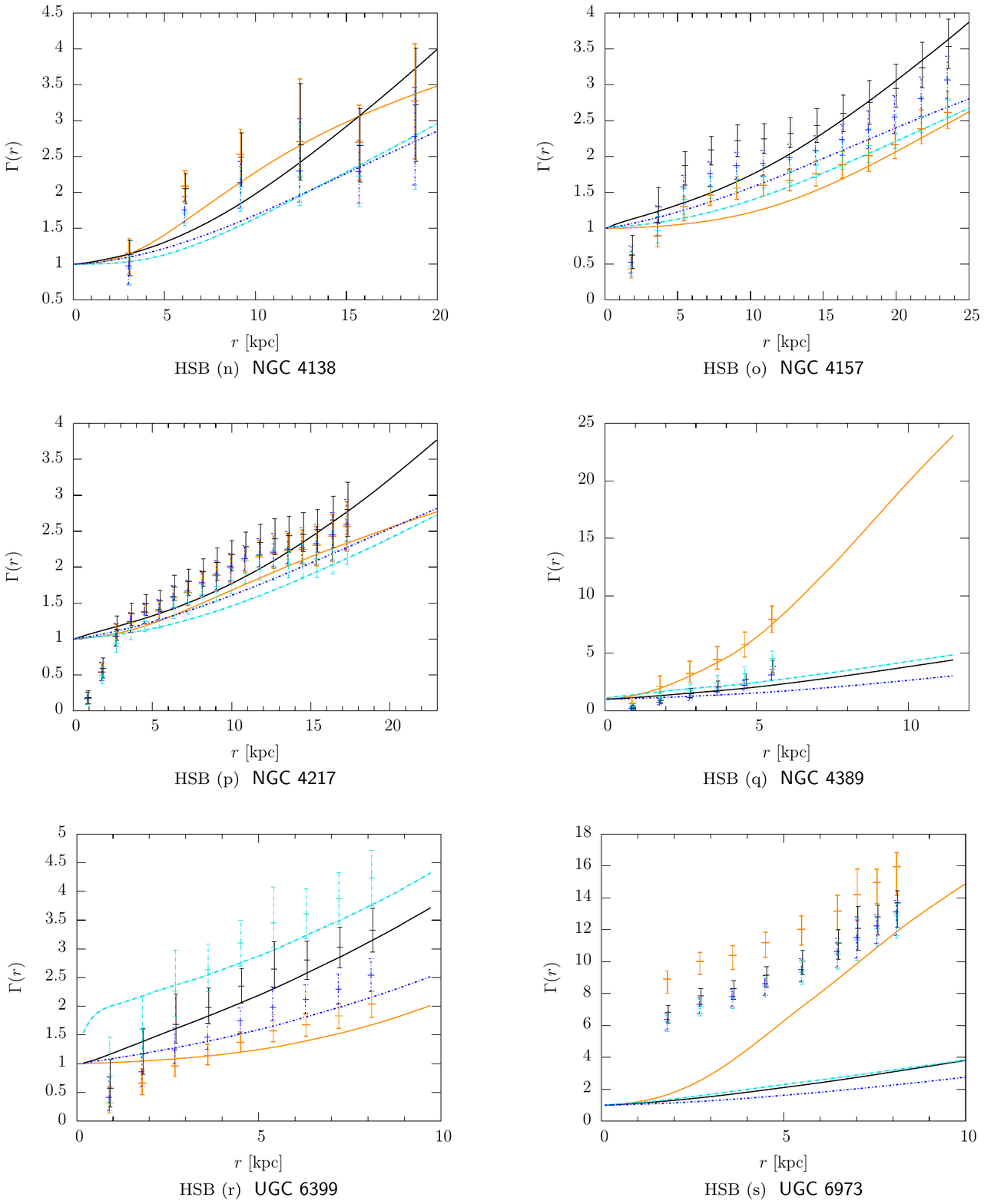}}
\put(82,45){\includegraphics[width=0.98\textwidth]{figure/galaxy_hsb_Gamma_legend}}
\end{picture}
\fcont{figure.galaxy.Gamma}{\sf\small UMa --- Dynamical mass factors.}
{\subGamma}.\index{Dynamic mass factor, \(\Gamma\)|)}
\end{figure}
\begin{figure}\index{Dynamic mass factor, \(\Gamma\)|(}
\begin{picture}(460,450)(82,190)
\put(30,12){\includegraphics[width=1.28\textwidth]{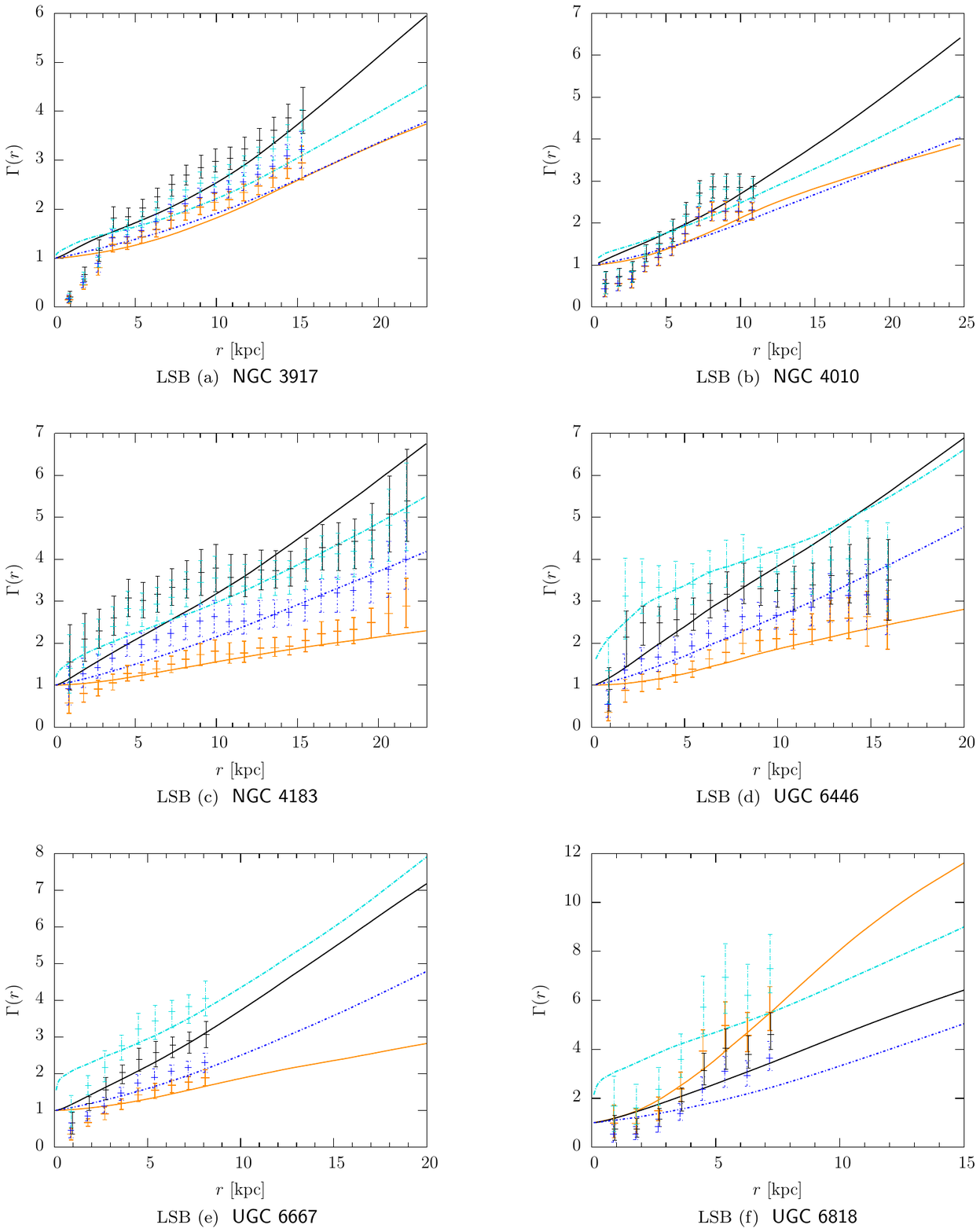}}
\put(82,45){\includegraphics[width=0.98\textwidth]{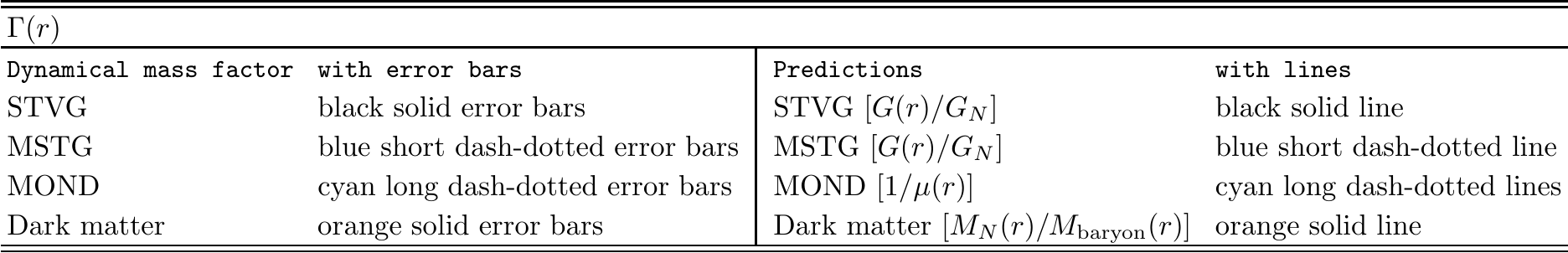}}
\end{picture}
\fcont{figure.galaxy.Gamma}{\sf\small UMa --- Dynamical mass factors.}
{\subGamma}.
\end{figure}
\begin{figure}
\begin{picture}(460,327)(82,295) 
\put(30,12){\includegraphics[width=1.28\textwidth]{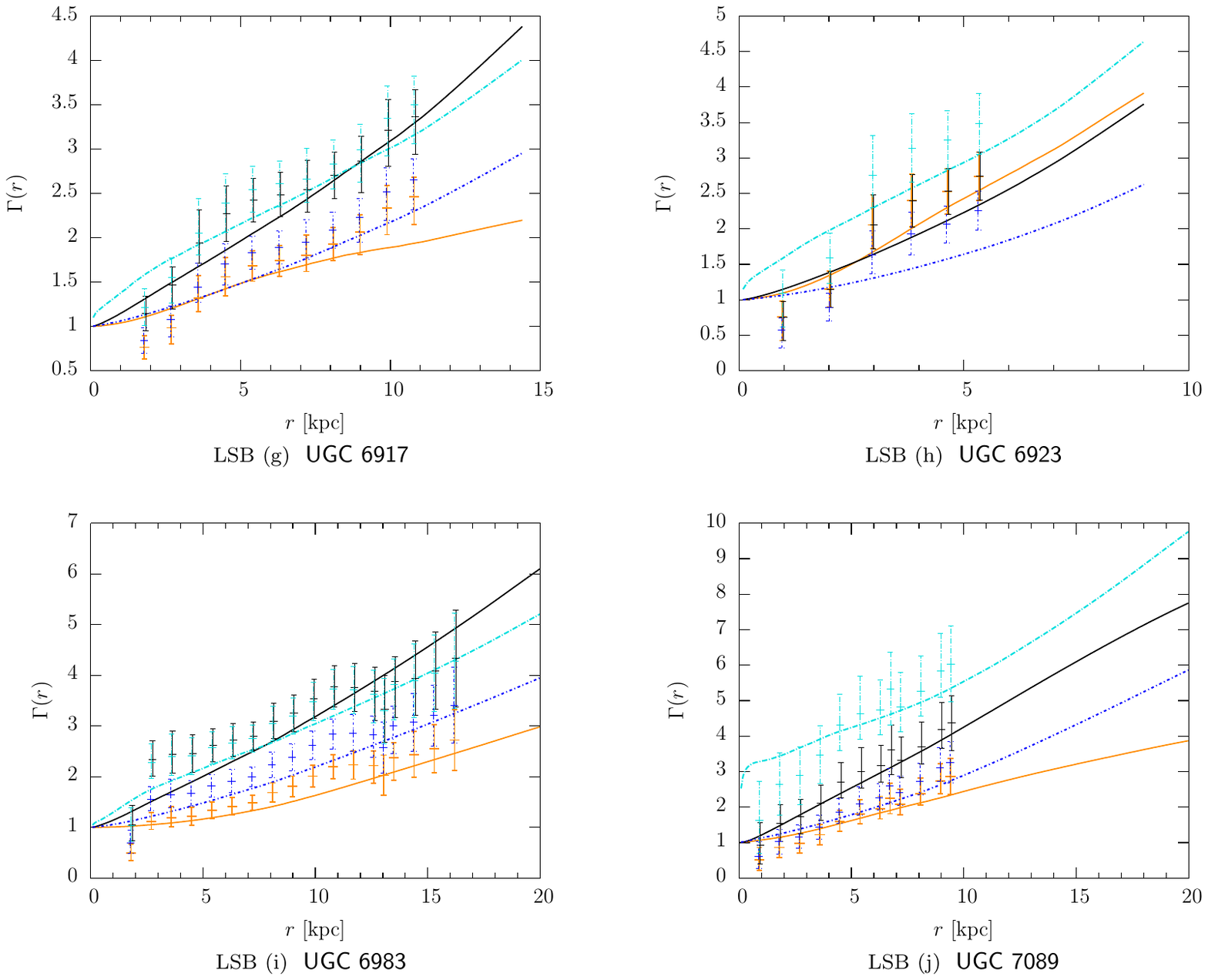}}
\put(82,45){\includegraphics[width=0.98\textwidth]{figure/galaxy_lsb_Gamma_legend}}
\end{picture}
\fcont{figure.galaxy.Gamma}{\sf\small UMa --- Dynamical mass factors.\break}
{\subGamma} for 19 HSB and 10 LSB galaxies.  The dynamic data consist of the Newtonian dynamic mass due to the measured orbital velocities per unit baryonic mass per gravity theory, shown with error bars. The computed best-fit results by varying the stellar mass-to-light ratio, \(\Upsilon\), are plotted for Moffat's STVG and MSTG theories and Milgrom's MOND theory with mean-universal parameters.  Results are plotted for the best-fit core-modified dark matter theory including visible baryons.\index{Dynamic mass factor, \(\Gamma\)|)}
\end{figure}

The dynamic mass factors, plotted in \fref{figure.galaxy.Gamma} for each galaxy in the UMa sample, are shown to be a monotonically rising (near linear) functions, with similar properties:
\begin{equation}\label{eqn.galaxy.uma.Gamma.properties}\begin{array}{lcll}
\Gamma(r)  & \approx & 1 & \mbox{\tt within Newtonian core},\\
\Gamma(r)  & \gg & 1 & \mbox{\tt beyond Newtonian core},\\
\Gamma(r)  & \lesssim & 10 & \mbox{\tt within galaxy},\end{array}
\end{equation}
and is the measure of the missing mass factor.  Each theory may be judged by how well the Newtonian dynamic mass due to the measured orbital velocities per unit baryonic mass per gravity theory, shown with error bars, corresponds to the predictions of \erefss{eqn.galaxy.uma.Gamma.darkmatter}{eqn.galaxy.uma.Gamma.mond}{eqn.galaxy.uma.Gamma.mog}.  The dynamic mass factor provides a unifying picture for dark matter and phantom dark matter and can be phenomenologically applied to constrain the choice of the MOND interpolating function -- without ad hoc choices -- and the form of Moffat's varying gravitational coupling.\index{MOND!Interpolating function, \(\mu\)}\index{Modified gravity!Phantom of dark matter}\index{Newton's constant!Renormalized}

\subsection{\label{section.galaxy.uma.powerlaw}Core-modified dark matter halos}\index{Dark matter!Core-modified|(}

The simple observation that galaxy rotation curves are approximately flat at large radii, where the orbital velocity of \eref{eqn.galaxy.uma.orbitalv} is constant, leads to the conclusion of \eref{eqn.newton.darkmatter.divergent} that the Newtonian dynamic mass of \eref{eqn.galaxy.uma.mass.dynamic} grows linearly with radius, and therefore, since we are not neglecting baryons,
\begin{equation}\label{eqn.galaxy.uma.powerlaw.rho}
\rho(r) \equiv \rho_{\rm baryon}(r) + \rho_{\rm halo}(r) \propto r^{-2},
\end{equation}
is valid where the galaxy rotation curves are approximately flat.  However, the radial distribution of spherically averaged dark matter halos is unlike either of the baryonic components which accumulate in exponentially thin HI (and He) gaseous disks or luminous stellar disks of \eref{eqn.galaxy.uma.rotmodDisk}, and
\begin{eqnarray}
\label{eqn.galaxy.uma.powerlaw.galacticplane.on} &\rho_{\rm baryon}(r) \gg \rho_{\rm halo}(r)\ &\mbox{\tt on the galactic plane},\\
\label{eqn.galaxy.uma.powerlaw.galacticplane.off} &\rho_{\rm baryon}(r) \ll \rho_{\rm halo}(r)\ &\mbox{\tt off the galactic plane},
\end{eqnarray}

\newcommand{\subpowerlaw}{\small The logarithm slope profile, \(\gamma(r)\), vs. orbital distance, \(r\) in kpc}
\begin{figure}[ht]\index{Dark matter!Power law, \(\gamma\)|(}
\begin{picture}(460,185)(0,0)
\put(0,40){\includegraphics[width=0.48\textwidth]{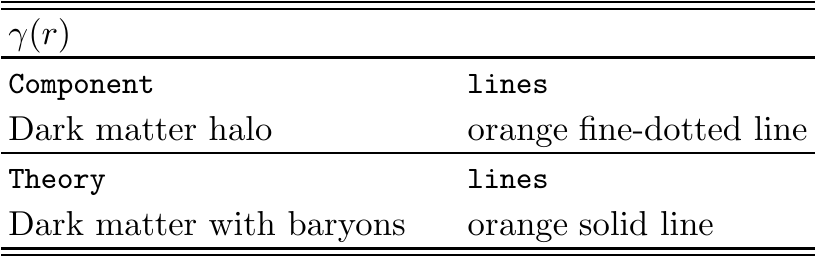}}
\put(225,0){\includegraphics[width=0.5\textwidth]{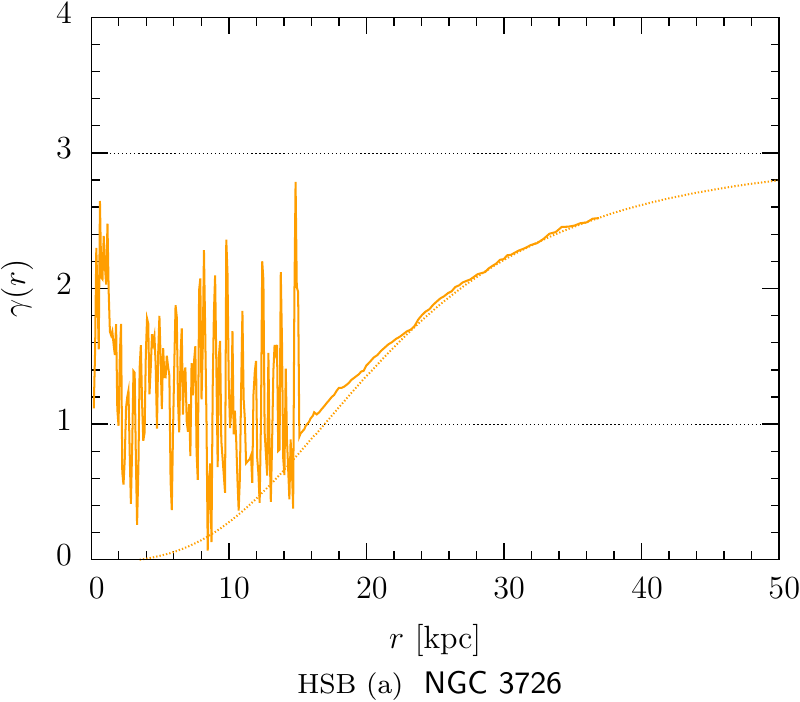}}
\end{picture}
\caption[Dark matter power-law]{\label{figure.galaxy.powerlaw} {{\sf\small UMa --- Dark matter power-law logarithm slopes.}}\break\break{\subpowerlaw} for 19 HSB and 10 LSB galaxies. The logarithm slope, \(\gamma(r)\) of the dark matter power-law, vs. orbital distance, \(r\) in kpc.  The computed best-fit results are plotted for core-modified dark matter theory including visible baryons -- and the corresponding dark matter halo component.  {\it The figure is continued.}}
\end{figure}

\begin{figure}
\begin{picture}(460,450)(82,190)
\put(30,12){\includegraphics[width=1.28\textwidth]{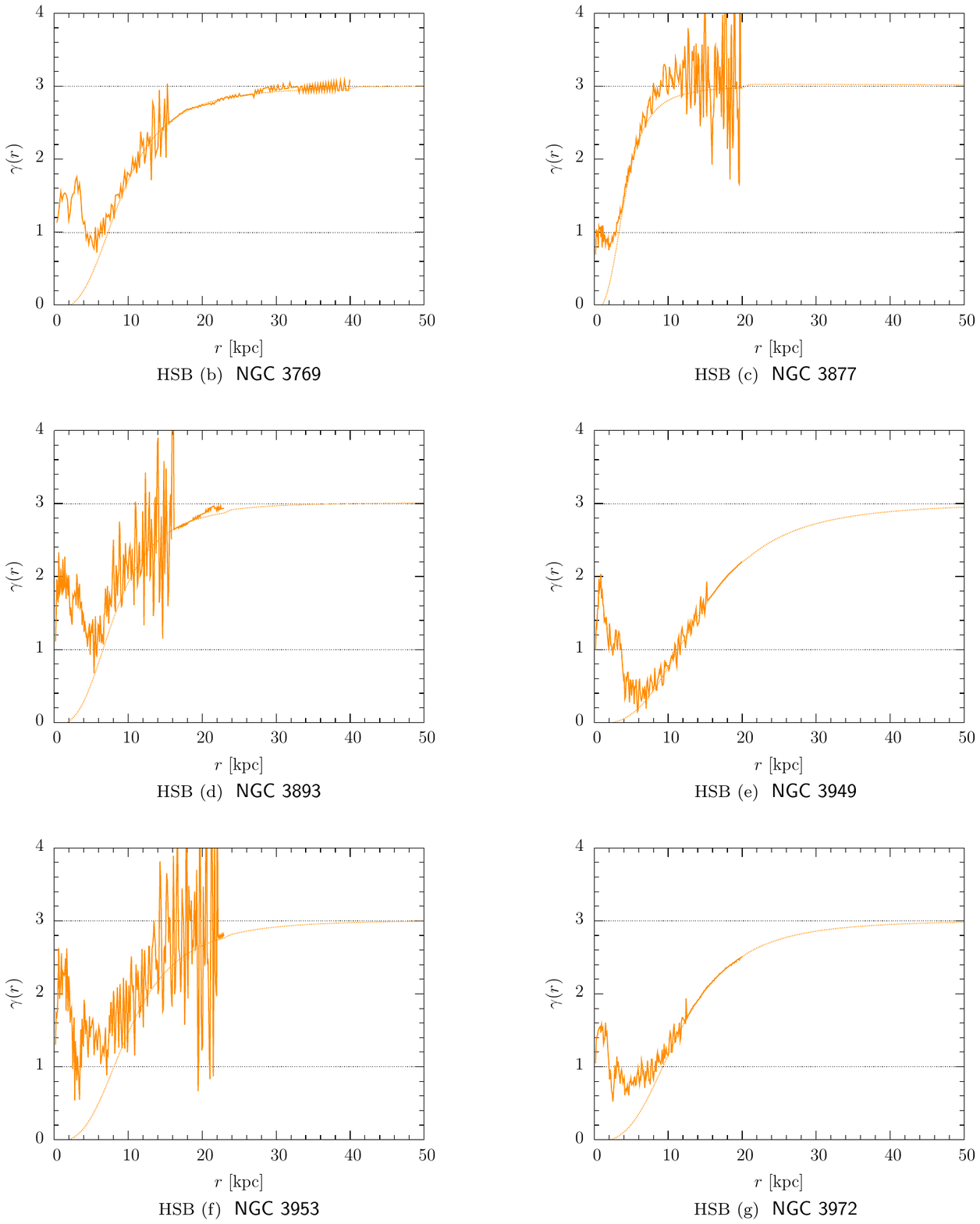}}
\put(82,45){\includegraphics[width=0.98\textwidth]{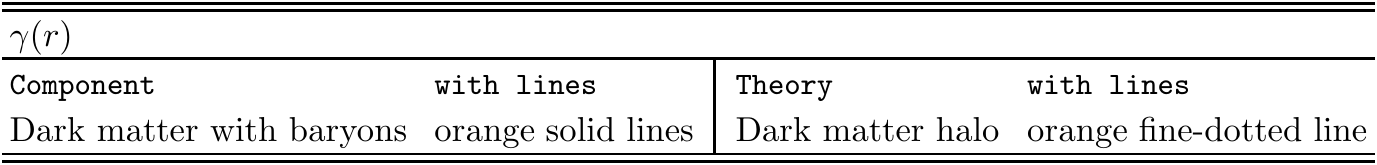}}
\end{picture}
\fcont{figure.galaxy.powerlaw}{\sf\small UMa --- Dark matter power-law logarithm slopes.}
{\subpowerlaw}.
\end{figure}
\begin{figure}
\begin{picture}(460,450)(82,190)
\put(30,12){\includegraphics[width=1.28\textwidth]{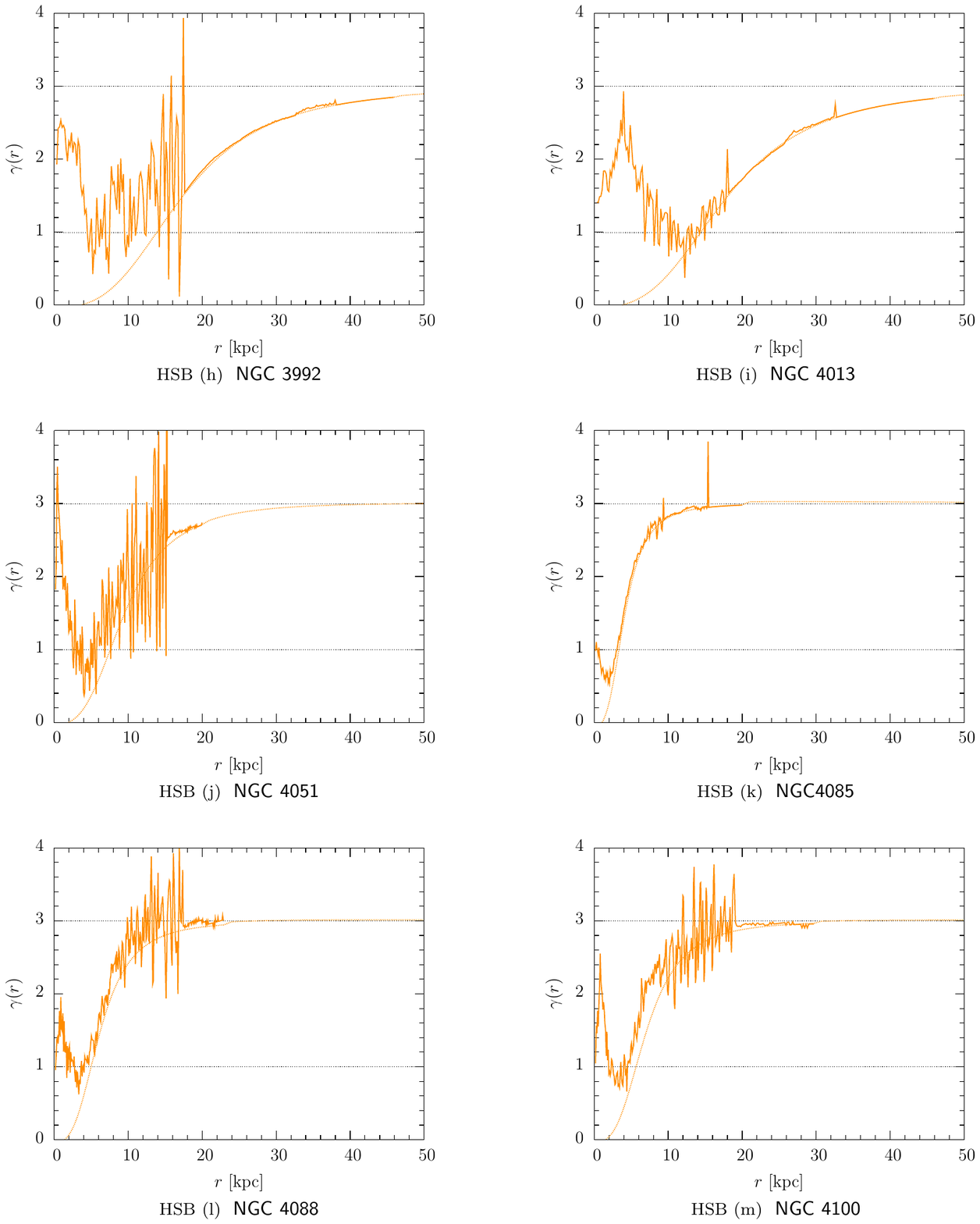}}
\put(82,45){\includegraphics[width=0.98\textwidth]{figure/galaxy_hsb_powerlaw_legend}}
\end{picture}
\fcont{figure.galaxy.powerlaw}{\sf\small UMa --- Dark matter power-law logarithm slopes.}
{\subpowerlaw}.
\end{figure}
\begin{figure}
\begin{picture}(460,450)(82,190)
\put(30,12){\includegraphics[width=1.28\textwidth]{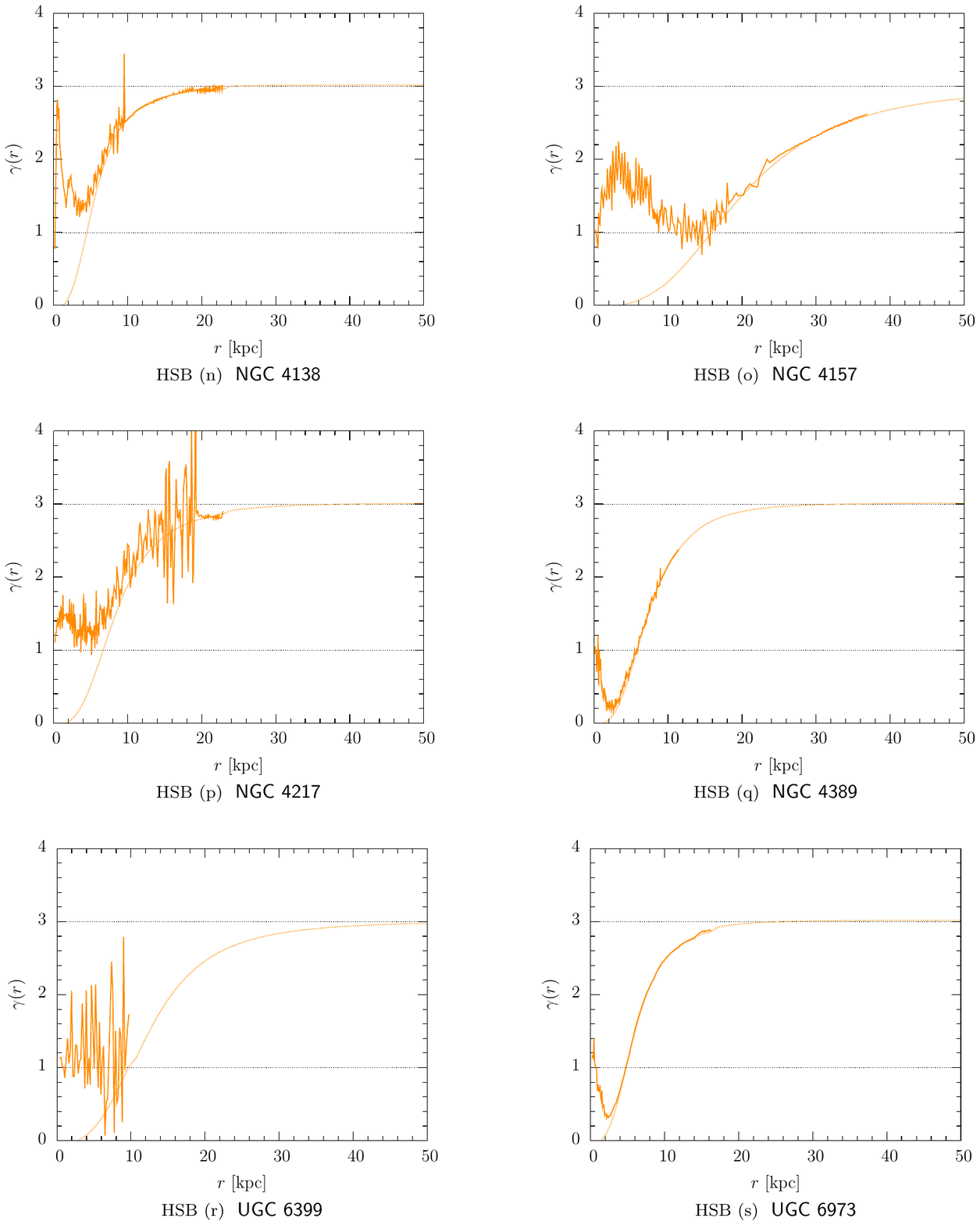}}
\put(82,45){\includegraphics[width=0.98\textwidth]{figure/galaxy_hsb_powerlaw_legend}}
\end{picture}
\fcont{figure.galaxy.powerlaw}{\sf\small UMa --- Dark matter power-law logarithm slopes.}
{\subpowerlaw}.
\end{figure}

\begin{figure}
\begin{picture}(460,450)(82,190)
\put(30,12){\includegraphics[width=1.28\textwidth]{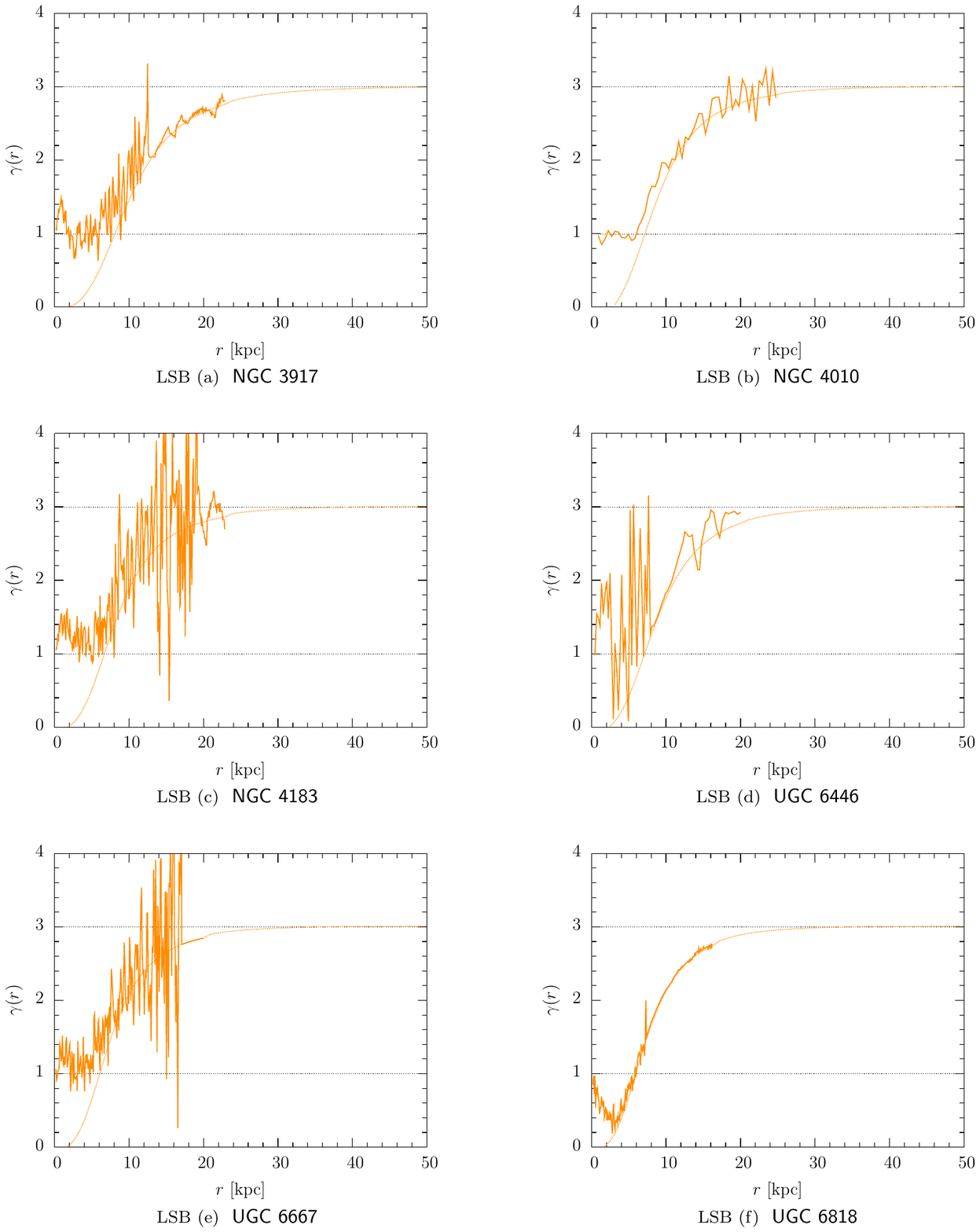}}
\put(85,45){\includegraphics[width=\textwidth]{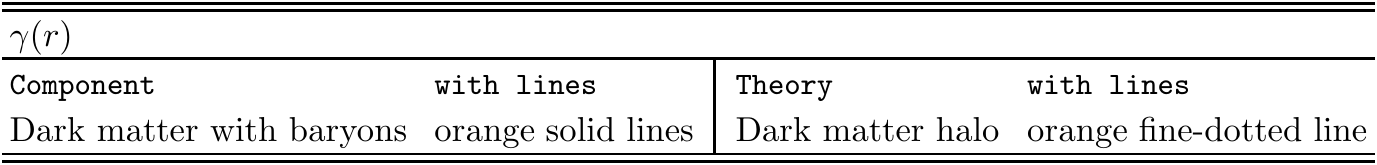}}
\end{picture}
\fcont{figure.galaxy.powerlaw}{\sf\small UMa --- Dark matter power-law logarithm slopes.}
{\subpowerlaw}.
\end{figure}
\begin{figure}
\begin{picture}(460,350)(82,269) 
\put(30,12){\includegraphics[width=1.28\textwidth]{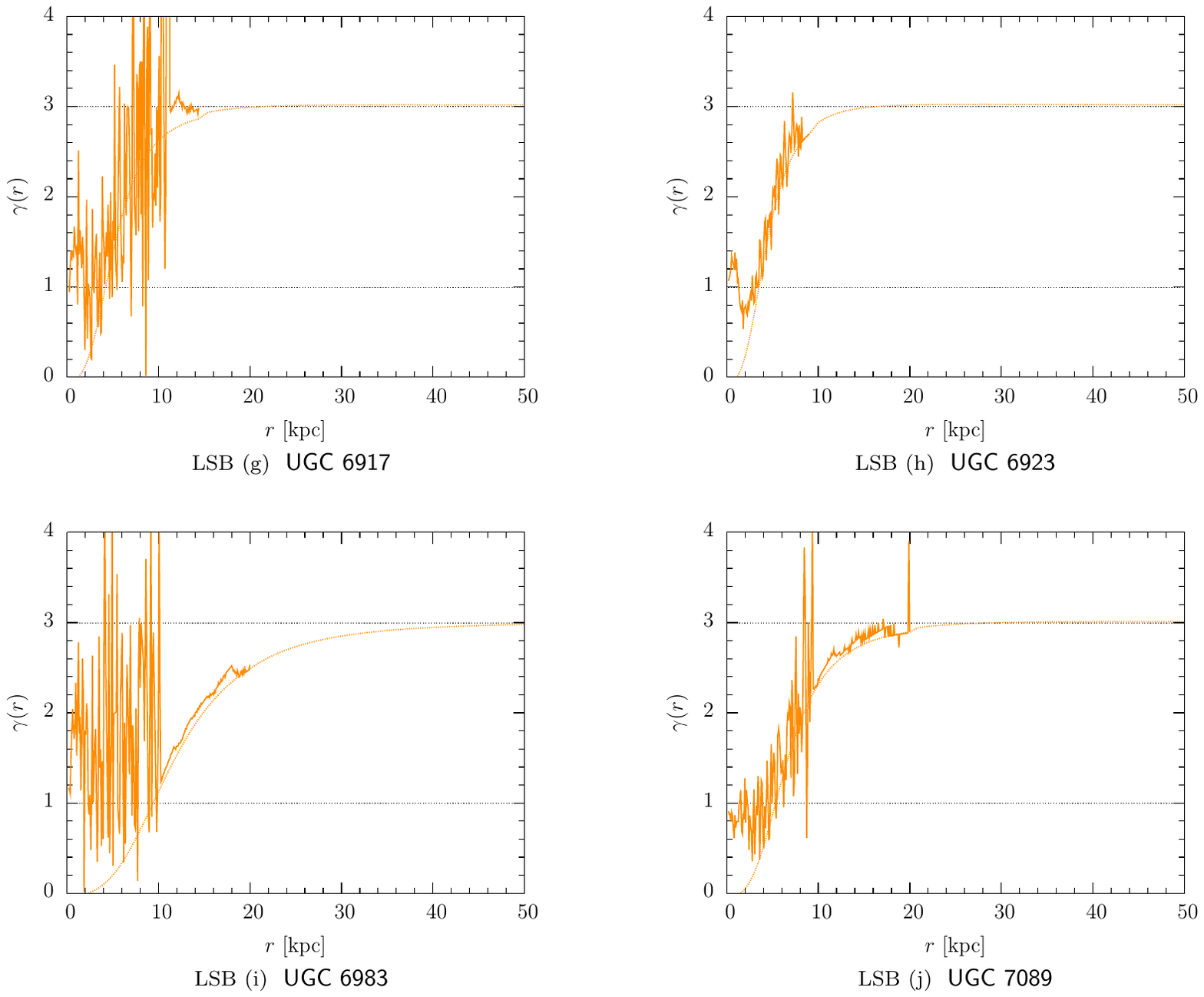}}
\put(82,45){\includegraphics[width=0.98\textwidth]{figure/galaxy_lsb_powerlaw_legend}}
\end{picture}
\fcont{figure.galaxy.powerlaw}{\sf\small UMa --- Dark matter power-law logarithm slopes.\break}
{\subpowerlaw} for 19 HSB and 10 LSB galaxies. The logarithm slope, \(\gamma(r)\) of the dark matter power-law, vs. orbital distance, \(r\) in kpc.  The computed best-fit results are plotted for core-modified dark matter theory including visible baryons -- and the corresponding dark matter halo component.\index{Dark matter!Power law, \(\gamma\)|)}
\end{figure}

Therefore in order to generalize the power-law index of \eref{eqn.newton.darkmatter.coremodified.gamma} to include baryons, it is convenient to derive the relation using the spherically integrated power-law density of \eref{eqn.newton.darkmatter.powerlaw},
\begin{equation}\label{eqn.galaxy.uma.powerlaw.totalmass}
M_{N}(r) \propto \int_0^r \frac{ {r^{\prime}}^2}{{r^{\prime}}^{\gamma({r^{\prime}})}} d{r^{\prime}},
\end{equation}
and 
\begin{equation}\label{eqn.galaxy.uma.powerlaw.dtotalmass}
d M_{N}(r) \propto  {r}^{2-\gamma(r)} dr.
\end{equation}
Thus the spherically averaged power-law index may be defined in terms of the logarithm slope,
\begin{equation}\label{eqn.galaxy.uma.powerlaw.gamma}
\gamma(r) = 2 - \frac{d \ln M_{N}(r)}{d \ln r}.
\end{equation}
The power-law indices for the best-fit core-modified dark matter halo, given by \eref{eqn.newton.darkmatter.coremodified.gamma}, and the Newtonian dynamic mass including baryons, according to \eref{eqn.galaxy.uma.powerlaw.gamma}, are plotted in \fref{figure.galaxy.powerlaw}.

Since the virial radius of the halo naturally extends beyond the outermost radial point in the galaxy rotation curve, \(r_{\rm out}\), the dark matter to baryon fraction can grow without bound until the cosmological limit is reached.  Within each galaxy in the sample, the dark matter to baryon fraction is tabulated to the outermost radial point in the galaxy rotation curve -- in Column (8) of \tref{table.galaxy.mass} -- with mean values:
\begin{equation}\label{eqn.galaxy.uma.powerlaw.dmfraction}
\frac{M_{\rm halo}(r_{\rm out})}{M_{\rm baryon}(r_{\rm out})} = \left\{ \begin{array}{ll} 2.4 \pm 2.1 & \mbox{\tt HSB galaxies}\\1.3 \pm 0.5 & \mbox{\tt LSB galaxies}\\2.0 \pm 1.8 & \mbox{\tt full sample}.\end{array}\right.
\end{equation}
which are consistently below the upper limit set by \citet{Spergel:ApJS:2007} in the Wilkinson microwave anisotropy probe (WMAP) third year results.\index{Dark matter!Core-modified|)}

\subsubsection{\label{subsection.galaxy.uma.powerlaw.cuspproblem}Solution to the dark matter cusp problem}\index{Dark matter!Cusp problem|(}

The conflict between the cuspy dark matter halos predicted by N-body simulations and the constant density cores preferred by dwarf and low surface brightness galaxies may be resolved by a universal core-modified fitting formula  with a constant density core, while including the visible baryons which are dominant in the galaxy core.

As the plot of the dark matter power-law proves in \fref{figure.galaxy.powerlaw}, at large distances from the center of each galaxy in the sample, the density profile of the dark matter halo is well described by a steep power-law, with power-law index \(\gamma \rightarrow 3\), whereas at distances toward the center of the galaxy an increasingly shallow power-law is observed.  For distances less than the dark matter halo core radius, \(r < r_s\),the total density profile including baryons shows a universal \(\gamma \rightarrow 1\) power-law index, and the density profile of the dark matter component alone approaches a rarified, constant density core.

A comparison of \tref{table.galaxy.darkmatter}, show a statistically significant reduction of the \(\chi^2/\nu\) test in \(\sim 90\%\) of the  galaxies results from using the core-modified profile of \eref{eqn.galaxy.dynamics.dm.coremodified} instead of the NFW profile of \eref{eqn.galaxy.dynamics.dm.nfw}. Moreover, in those galaxies that the {\it singular} NFW profile fits well, the best-fit stellar mass-to-light ratio, \(\Upsilon \ll 1\), which has prompted the dark matter community of physicists to disregard the baryonic component in their simulations.   Most strikingly, in one HSB and one LSB galaxy, there does not exist a best-fit NFW profile with any nonzero stellar mass-to-light ratio, forcing \(\Upsilon \equiv 0\) for these galaxies. Alternatively, the core-modified profile prefers values for the  best-fit stellar mass-to-light ratio, \(\Upsilon \sim 1\), which are physically acceptable for every galaxy in the sample.

This core-modified dark matter galaxy model produces excellent fits to the galaxy rotation curves of \fref{figure.galaxy.velocity}, and enables predictions of detailed surface mass density maps, as shown in \fref{figure.galaxy.Sigma}, and demonstrates excellent fits to the mass profiles of \fref{figure.galaxy.mass}, with dark matter to baryon fractions consistent with cosmologically observed values. \index{Dark matter!Cusp problem|)}

\subsection{\label{section.galaxy.uma.masslight}The mass luminosity relationship}

Throughout this work, the stellar mass-to-light ratio, \(\Upsilon\), is treated as a free parameter, with results near unity considered reasonable.  Each gravity theory which attempts to fit the galaxy rotation curve to the integrated surface mass densities of the HI (and He) gas and stellar disk components will select a {\it best-fit} stellar mass-to-light ratio, for each galaxy.  

The behaviour of the mass-to-light ratio within each galaxy and the change in the behaviour from galaxy to galaxy are important concerns of a good fit.  For the sample of galaxies considered in \sref{section.galaxy.uma}, the basic computation is that of the surface mass computation of \sref{section.galaxy.uma.Surfacemass} of the individually detected components: the exponentially thin gaseous (HI and He) disk of \erefs{eqn.galaxy.uma.rotmodHI}{eqn.galaxy.uma.rotmodGas} and the luminous stellar disk of \eref{eqn.galaxy.uma.rotmodDisk}, with the bulge neglected for the reasons stipulated at the start of this section in regards to \citet{Tully.APJ.1997.484}.

In order to calculate the total mass of each galaxy from the photometry, within the context of each gravity theory, the mass luminosity relationship is based on the best-fitting stellar mass-to-light ratio, \(\Upsilon\), according to \eref{eqn.galaxy.uma.masstolight}.  The HI (and He) gaseous component is the only computation that is independent of assumptions on the mass-to-light ratio and is determined by big bang nucleosynthesis according to \eref{eqn.galaxy.uma.rotmodGas}.

\newcommand{\submasslight}{\small The stellar mass-to-light ratio, \(\Upsilon(r)\), vs. orbital distance, \(r\) in kpc}
\begin{figure}\index{Stellar mass-to-light ratio, \(\Upsilon\)|(}
\begin{picture}(460,185)(0,0)
\put(0,40){\includegraphics[width=0.48\textwidth]{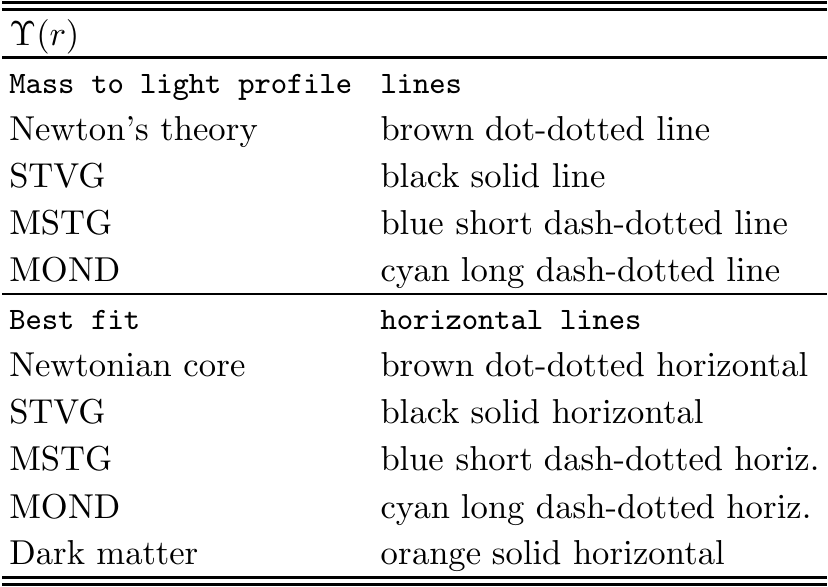}}
\put(225,0){\includegraphics[width=0.5\textwidth]{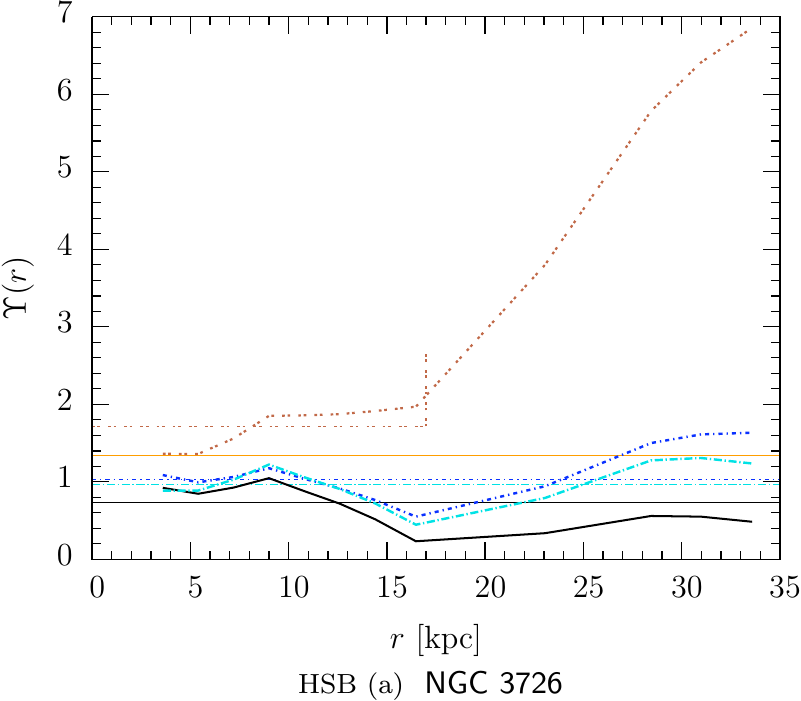}}
\end{picture}
\caption[Stellar mass-to-light ratios]{\label{figure.galaxy.masslight} {{\sf\small UMa --- Stellar mass-to-light ratios.}}\break\break{\submasslight} for 19 HSB and 10 LSB galaxies.  The stellar mass-to-light ratio, \(\Upsilon\), required to fit the galaxy rotation curve at each data point without dark matter is plotted for Moffat's STVG and MSTG theories, Milgrom's MOND theory, and Newton's theory.  Within each theory, the best-fit value of the stellar mass-to-light ratio, \(\Upsilon\), is shown with a horizontal line, including the best-fit core-modified dark matter; and the best-fit Newtonian core model -- the extent of the core shown with a vertical line.  {\it The figure is continued.}}
\end{figure}
\begin{figure}
\begin{picture}(460,450)(82,190)
\put(30,12){\includegraphics[width=1.28\textwidth]{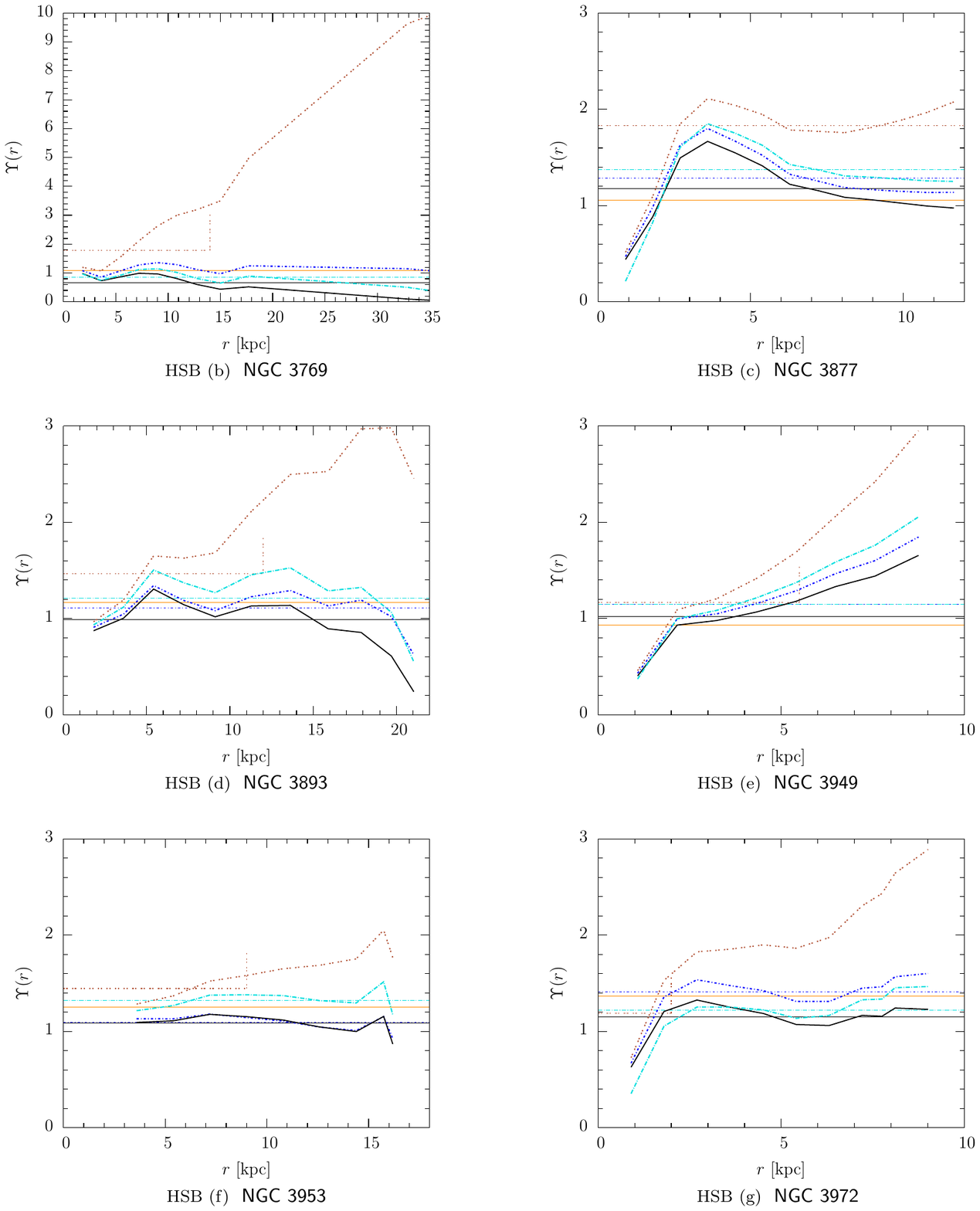}}
\put(80,36){\includegraphics[width=\textwidth]{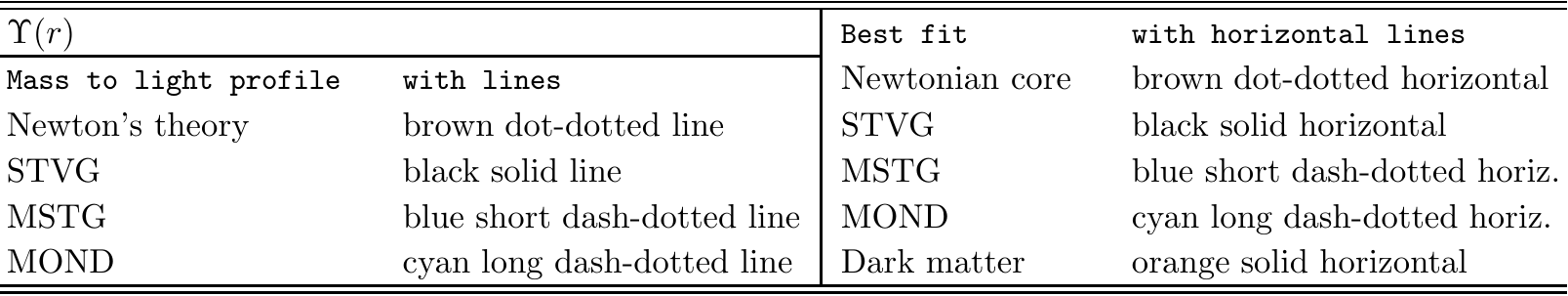}}
\end{picture}
\fcont{figure.galaxy.masslight}{\sf\small UMa --- Stellar mass-to-light ratios.}
{\submasslight}.
\end{figure}
\begin{figure}
\begin{picture}(460,450)(82,190)
\put(30,12){\includegraphics[width=1.28\textwidth]{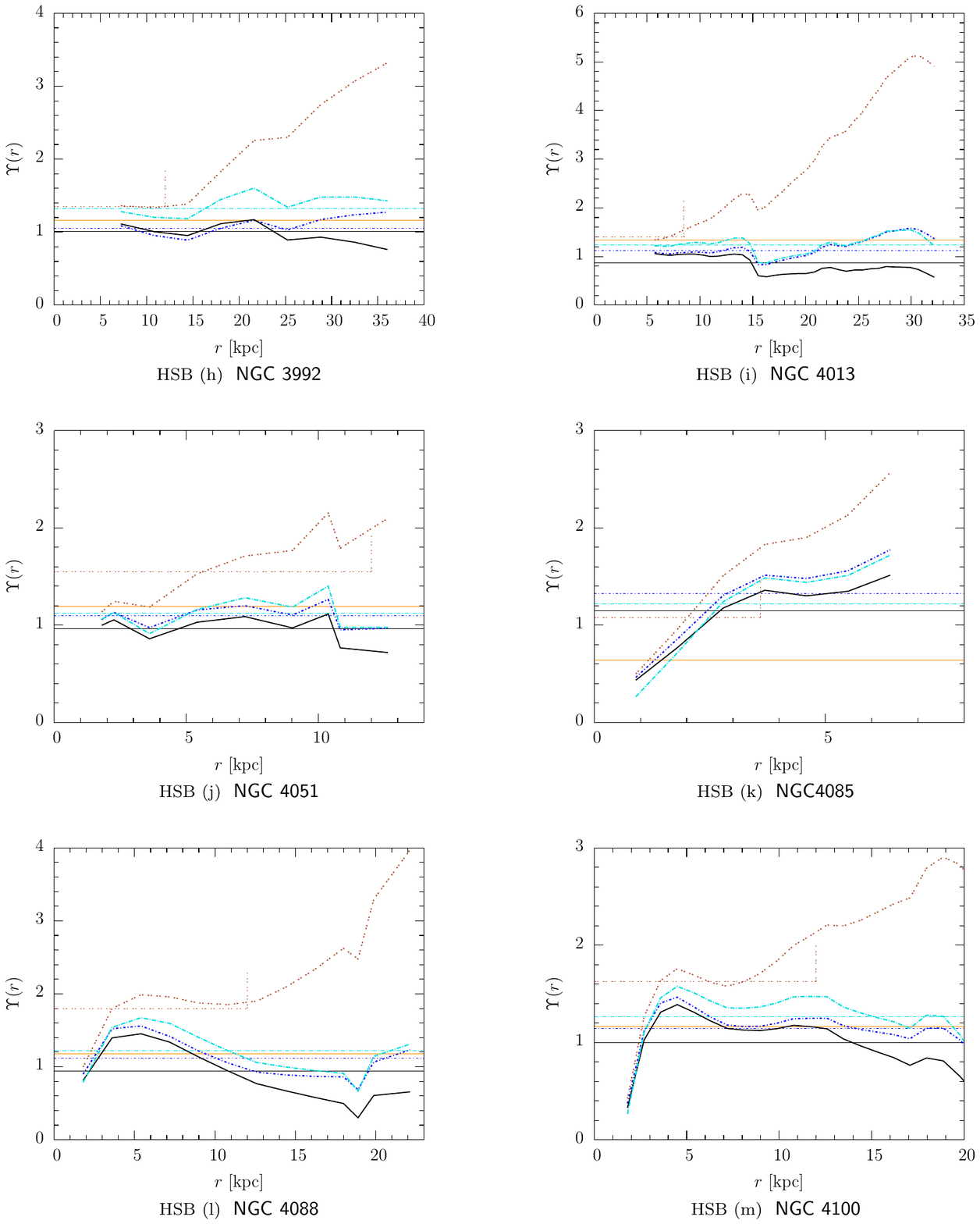}}
\put(80,36){\includegraphics[width=\textwidth]{figure/galaxy_hsb_masslight_legend}}
\end{picture}
\fcont{figure.galaxy.masslight}{\sf\small UMa --- Stellar mass-to-light ratios.}
{\submasslight}.\end{figure}
\begin{figure}
\begin{picture}(460,450)(82,190)
\put(30,12){\includegraphics[width=1.28\textwidth]{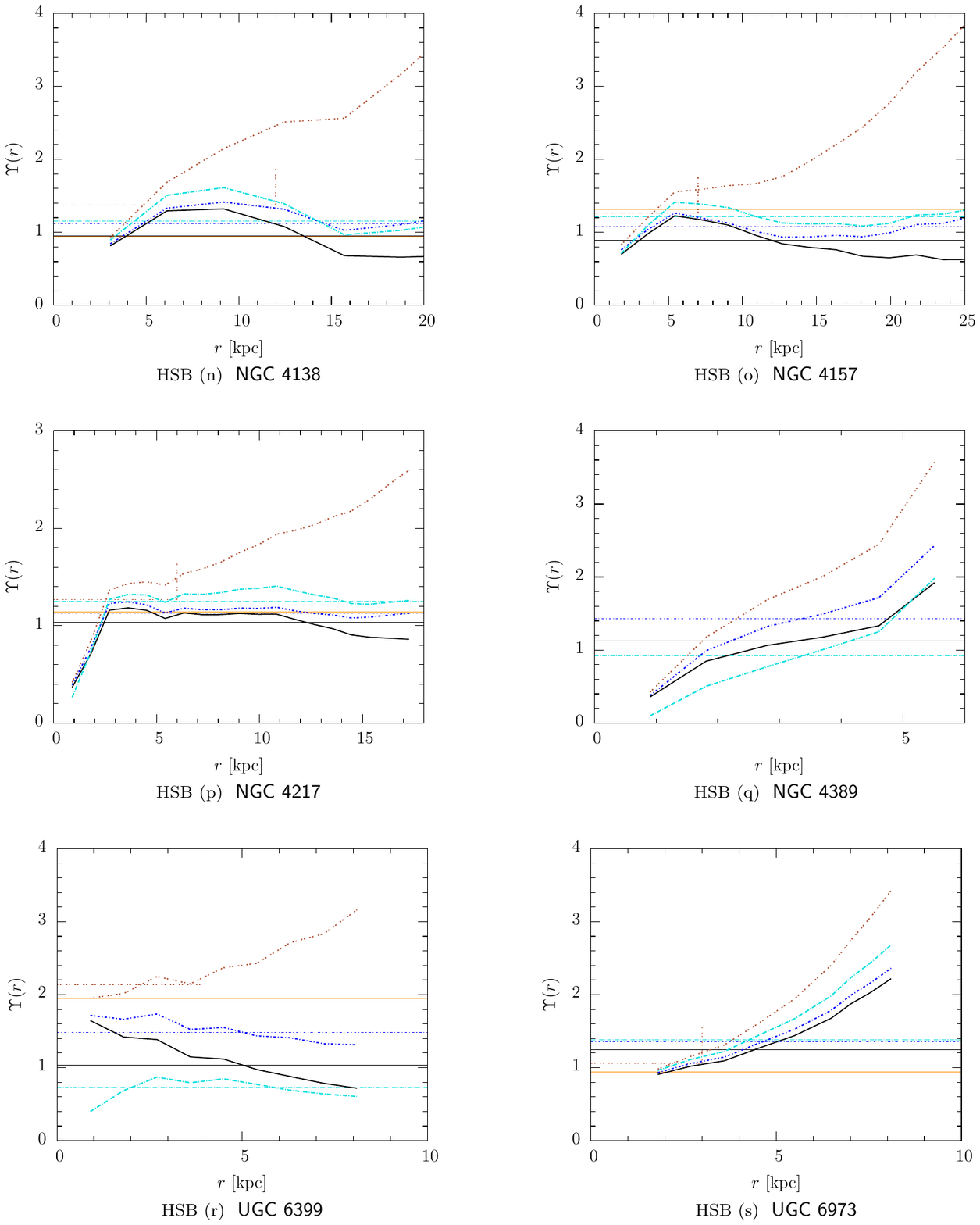}}
\put(80,36){\includegraphics[width=\textwidth]{figure/galaxy_hsb_masslight_legend}}
\end{picture}
\fcont{figure.galaxy.masslight}{\sf\small UMa --- Stellar mass-to-light ratios.}
{\submasslight}.
\end{figure}\begin{figure}
\begin{picture}(460,450)(82,190)
\put(30,12){\includegraphics[width=1.28\textwidth]{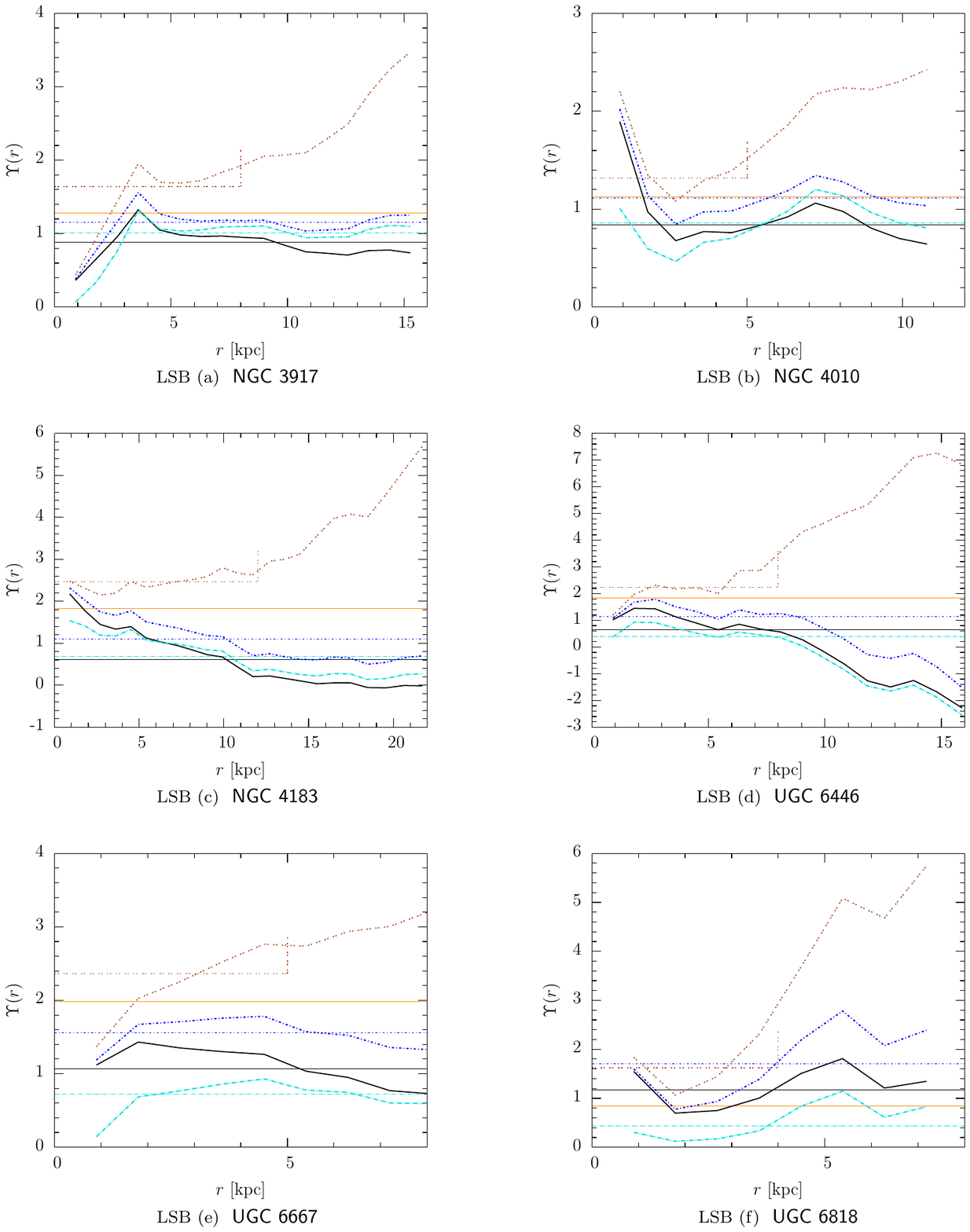}}
\put(82,45){\includegraphics[width=0.98\textwidth]{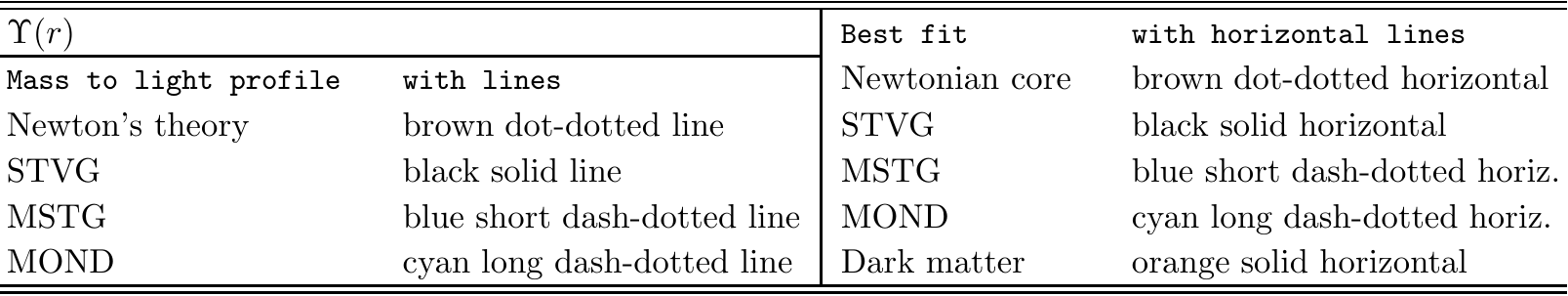}}
\end{picture}
\fcont{figure.galaxy.masslight}{\sf\small UMa --- Stellar mass-to-light ratios.}
{\submasslight}.
\end{figure}
\begin{figure}
\begin{picture}(460,308)(82,311) 
\put(30,12){\includegraphics[width=1.28\textwidth]{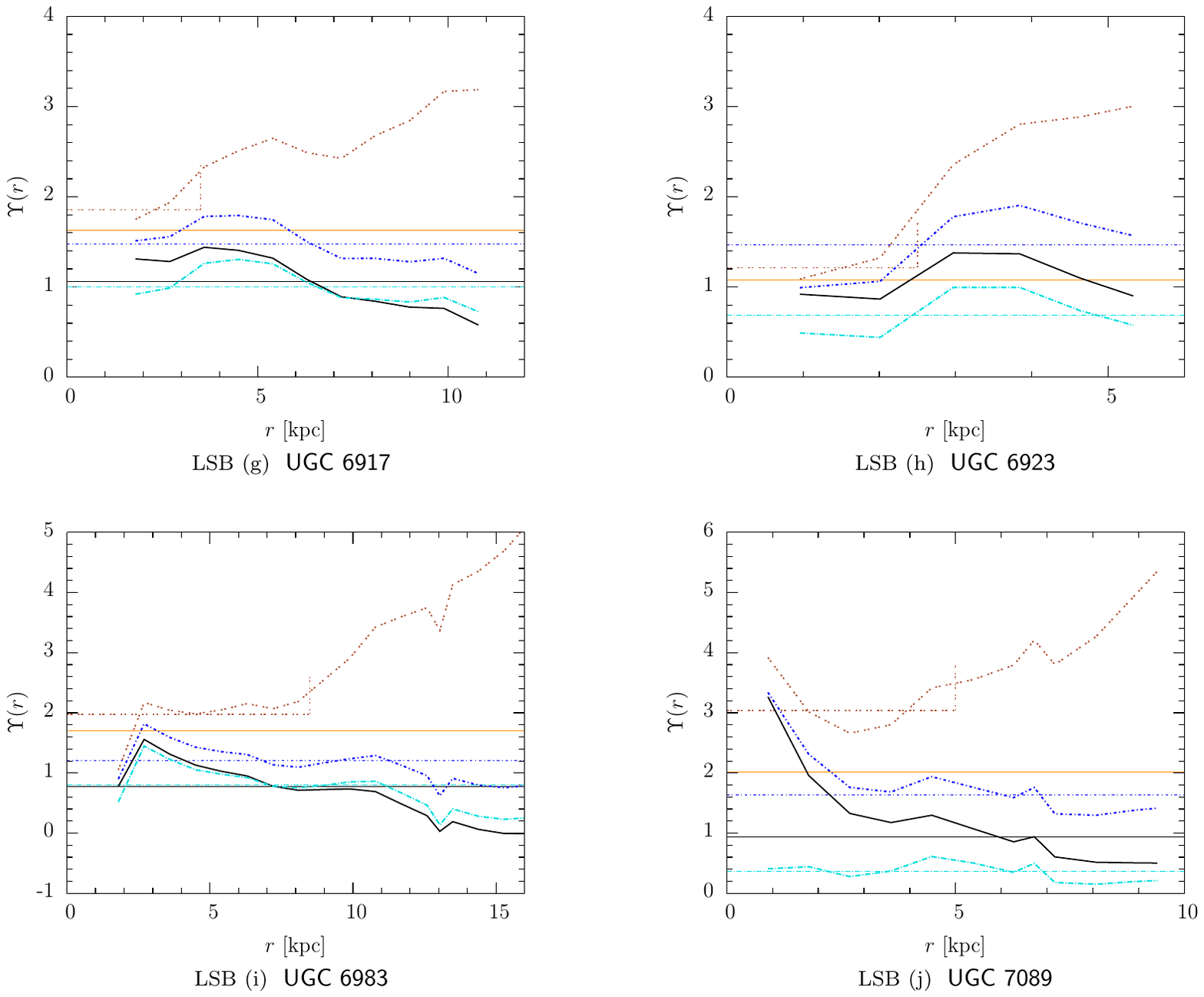}}
\put(80,45){\includegraphics[width=\textwidth]{figure/galaxy_lsb_masslight_legend}}
\end{picture}
\fcont{figure.galaxy.masslight}{\sf\small UMa --- Stellar mass-to-light ratios.\break}
{\submasslight} for 19 HSB and 10 LSB galaxies.  The stellar mass-to-light ratio, \(\Upsilon\), required to fit the galaxy rotation curve at each data point without dark matter is plotted for Moffat's STVG and MSTG theories, Milgrom's MOND theory, and Newton's theory.  Within each theory, the best-fit value of the stellar mass-to-light ratio, \(\Upsilon\), is shown with a horizontal line, including the best-fit core-modified dark matter; and the best-fit Newtonian core model -- the extent of the core shown with a vertical line.\index{Stellar mass-to-light ratio, \(\Upsilon\)|)}
\end{figure}

The best-fit stellar mass-to-light ratio, \(\Upsilon\), and the computed total galaxy mass is listed for each galaxy in \tref{table.galaxy.mass}.  It is clear that the best-fit mass-to-light ratio varies from galaxy to galaxy; and none of the galaxy rotation curves in the UMa sample can be fit by a universal-mean stellar mass-to-light ratio.  The possibility that the stellar mass-to-light ratio changes within a galaxy, \(\Upsilon = \Upsilon(r)\), may be explored by calculating \(\Upsilon(r)\) independently at every point in the galaxy rotation curve, for each gravity theory.  

\fref{figure.galaxy.masslight} plots the stellar mass-to-light ratio, \(\Upsilon(r)\) vs.\,\(r\), showing that the variation in \(\Upsilon(r)\) in every galaxy exceeds the variation in the best-fit stellar mass-to-light ratio, \(\Upsilon\), across the sample of galaxies.

\section{\label{section.galaxy.halos}Halos of phantom dark matter}

Whether galaxy rotation curves are best described by cold non-baryonic dark matter (CDM) halos, as in \sref{section.galaxy.dynamics.dm}, or Milgrom's modified Newtonian dynamics, as in \sref{section.mog.mond}, or Moffat's modified gravity, as in \sref{section.galaxy.dynamics.mog}, or other sensible theories, there are certain regularities in galactic structure which may have theoretical underpinning.  Each of the galaxies studied in \citet{Brownstein:ApJ:2006} and the Ursa Major filament of galaxies, in \sref{section.galaxy.uma}, exhibit a core region where Newton's theory provides acceptable fits and the galaxy may be modelled by a Newtonian core model, as in \sref{section.galaxy.halos.core}.  The observation that mass follows light and the appearance of orphan features beyond the Newtonian core, as in \sref{section.galaxy.halos.orphans}, confirm that the baryons are dynamically important.  This result is natural in MOND and MOG theories which are sourced by baryons alone, and also supports the alternative model of core-modified dark matter, as described in \sref{section.galaxy.halos.coremodified}, which fits the galaxy rotation curves of \sref{section.galaxy.uma.velocity}, including all of the dwarfs, with physically reasonable stellar mass-to-light ratios of \(\Upsilon \approx 1\).  This is difficult to achieve using the cuspy NFW profile for some LSB and dwarf galaxies which prefer \(\Upsilon \sim 0\).  Comparison of the dynamic mass distribution inferred from galaxy rotation curves and the visible baryon distribution derived from each gravity theory enable a fundamental explanation to the Tully-Fisher relation, as in \sref{section.galaxy.tullyfisher}.

\subsection{\label{section.galaxy.halos.core}Newtonian cores}\index{Newton's central potential!Galaxy core}

The radius of the Newtonian core is easily measured by plotting the stellar mass-to-light ratio, as in \fref{figure.galaxy.masslight}, which shows the variation in the Newtonian \(\Upsilon(r)\) in brown dot-dotted lines.  The profile is flat \(\Upsilon \sim 1\) in the Newtonian core, and then rises rapidly outside the Newtonian core, as shown on the figure for each galaxy.  Thus the best-fitting stellar mass-to-light ratio for Newton's theory without dark matter may be computed by weighting the region inside the core radius, for each galaxy, as shown by the horizontal brown-dot-dotted lines in the figure.  As a result, the best-fit Newtonian core model predicts values of \(\Upsilon\) larger than the other theories because there is less gravity due to the visible baryons, without dark matter.  

\subsection{\label{section.galaxy.halos.orphans}Orphan features}\index{Modified gravity!Phantom of dark matter}

\citet{Kent.AJ.1986.91} presented a sample of 37 Sb and Sc galaxies with photometric data, discovering that the component mass distributions admit decompositions into baryon and dark matter components, but could not simultaneously constrain the dark matter distribution and the stellar mass-to-light ratio for the baryons.  At one extreme, the stellar mass-to-light ratio was set to the maximum value permitted by the rotation curves, and a modest halo component produced good fits, but most galaxies were also well fit by models at the other extreme, with much more massive dark matter halos and correspondingly reduced stellar mass-to-light ratios.

\citet{Kent.AJ.1987.93} presented a sample of 16 spiral galaxies with photometric data and extended HI gas, and provided least-squares fits to the rotation curves, finding that a halo component is needed in each galaxy, but is tightly coupled to the stellar mass-to-light ratio.  Although \citet{Kent.AJ.1986.91,Kent.AJ.1987.93} assumed the constant density core dark matter distribution of \eref{eqn.newton.darkmatter.isothermal}, the uncertainty in the stellar mass-to-light ratio is a result of the uncertainty in the dark matter distribution.

However, this fine-tuning problem, which is known as the {\it disk-halo conspiracy}, is resolved by a correlation between the shape of the rotation curve and the shape of the baryonic luminosity measurements, first observed by \citet{Burstein.APJ.1982.253}, which suggests the presence of some features in the rotation curves at the transition from the baryon dominated core to the dominant dark matter halo.  \citet{Salucci.MNRAS.1989.237} showed that the fractional amount of mass from the luminous disk is an increasing function of the luminosity, and argued that the shape of the rotation curve near the edge of the optical disk should vary systematically with luminosity, leading to distinct features in galaxy rotation curves.

\citet{Hoekstra.MNRAS.2001.323} applied a mass model in which the dark matter surface density is a scaled version of the observed HI surface density to a sample of 24 spiral galaxies, obtaining good fits for most galaxies, but not for those galaxies which show a rapid decline of the HI surface density in the outermost regions.
 
\citet{Noordermeer.thesis.2006} provided a systematic study of HI rotation curves in spiral galaxies, finding that galaxy rotation curves have distinct features that may be traced back to the luminous components in the form of bumps and wiggles, and that the declines in the rotation curves at intermediate to large radii are rarely featureless.

\subsection{\label{section.galaxy.halos.coremodified}Core-modified dark matter}\index{Dark matter!Core-modified|(}

It is important to notice that the \map{\Sigma} components of \erefs{eqn.galaxy.uma.baryonSigma}{eqn.galaxy.uma.Sigma} combine to produce maps with features that can be traced back to the luminous stellar disk component for \(r<r_{s}\) and to the gaseous disk component for \(r \approx r_{s}\), whereas the dark matter halo dominates for \(r>r_{s}\), as shown by the surface mass density maps, plotted in \fref{figure.galaxy.Sigma}.  The core-modified dark matter surface mass density distribution including the visible baryons is remarkably flattened in the galaxy core, as compared to the best-fit Newtonian core, in the absence of dark matter, as in \sref{section.galaxy.halos.core}.  Moreover, for every galaxy in the sample, the central surface mass density,
\begin{equation}
\label{eqn.galaxy.Sigma0}
\Sigma_0 \equiv \Sigma(0),
\end{equation}
is determined by the baryonic component alone, where the dark matter halo is rarified as a result of the core-modified model of \eref{eqn.galaxy.dynamics.dm.coremodified}.  This is precisely the reverse situation for the singular halo models of \ssref{section.galaxy.dynamics.dm}{subsection.newton.darkmatter.nfw} which dominate the Newtonian dynamic mass throughout the galaxy, including the core leading to artificially small stellar mass-to-light ratios, \(\Upsilon \ll 1\) as shown by the best-fit NFW and core-modified parameters, listed in \tref{table.galaxy.darkmatter}.  

Therefore the reasoning that dominant dark matter erases the orphan features visible in galaxy rotation curves and the derived \map{\Sigma}s applies only to the NFW profile, and not to the core-modified profile.  The excellent fits to the galaxy rotation curves, in \fref{figure.galaxy.velocity}, confirm that the baryonic components are dynamically important, but the Newtonian force law of \eref{eqn.galaxy.mond.newtonianacc} without dark matter fits none of the galaxy rotation curves outside the Newtonian core.  A comparison of \tref{table.galaxy.darkmatter}, show a statistically significant reduction of the \(\chi^2/\nu\) test in \(\sim 90\%\) of the  galaxies results from using the core-modified profile of \eref{eqn.galaxy.dynamics.dm.coremodified} instead of the NFW profile of \eref{eqn.galaxy.dynamics.dm.nfw} because the orphan features in the dynamic data are correctly repatriated with the baryonic surface density maps.\index{Dark matter!Core-modified|)}

\subsection{\label{section.galaxy.tullyfisher}The Tully-Fisher relation}\index{Galaxy rotation, \(v\)!Tully-Fisher relation|(}

The observational \citet{Tully.AAP.1977.54} relation is an empirical relation between the measured total luminosity of a galaxy in a particular band (proportional to the stellar mass) and the amplitude of the  rotation curve (the maximum velocity) of the form:
\begin{equation}
\label{eqn.galaxy.halos.tullyfisher.relation} L \propto v_{\rm max}^{a} {\rm \ where\ } a \approx 3 - 4,
\end{equation}
where the total luminosity
\begin{equation}
\label{eqn.galaxy.halos.tullyfisher.luminosity} L = 4\pi \Phi d^2,
\end{equation}
may be inferred by measuring the isotropic flux, \(\Phi\), and knowing the distance, \(d\), between earth and the galaxy. Since all of the galaxies in the sample of \sref{section.galaxy.uma} are at a common distance from the Milky Way, and because of the improvements identified in  \sref{section.galaxy.uma.photometry} from using the available near-infrared \(K\)-band, the large astronomical uncertainties are mitigated, leaving an ideal laboratory to study the relationship between the luminosity of a galaxy, and theoretical predictions from \sref{section.galaxy.dynamics}, for each gravity theory, independent of the galactic mass distribution. 
 
\citet{Tully.APJ.2000.533} showed that, although the exponent in \eref{eqn.galaxy.halos.tullyfisher.relation} depends on the wavelength of the measured luminosity, and increases systematically from B to K bands, there appears to be convergence in the near-infrared where
\begin{equation}
\label{eqn.galaxy.halos.tullyfisher.exponent} a = 3.4 \pm 0.1.
\end{equation}

\citet{McGaugh.APJL.2000.533,McGaugh.APJ.2005.632} studied a large sample of galaxies, with stellar masses ranging over five decades, and observed a change in slope in the Tully-Fisher relation which disappears when using the total baryonic mass including both stellar and gaseous components, instead of just the luminous stellar mass, and concluded that the Tully-Fisher relation is fundamentally a relation between the total baryonic mass and the rotational velocity.  Using a combination of \(K\)-band photometry and high resolution rotation curves, \citet{Noordermeer.MNRAS.2007.381} discovered a second change in slope at the high mass end of the Tully-Fisher relation which disappeared when using, in combination, the total baryonic mass and the asymptotic, outermost velocity instead of the velocity amplitude.

Considering the asymptotic behaviour of the galaxy rotation curves of \fref{figure.galaxy.velocity}, most high-resolution galaxy rotation curves are either slowly rising or slowly declining at large radii.  \citet{Verheijen.APJ.2001.563} considered an alternate definition of the ``flat rotation velocity'', categorizing galaxies according to three kinds of behaviour depending on the shape of the rotation curve.

\citet{Avila-Reese.AJ.2008.136} explored the variation in the Tully-Fisher relation using a large sample of 76 high and low surface brightness galaxies, and obtained \(a=3.40\) for the ordinary Tully-Fisher relation (where the stellar luminosity is taken proportional to the stellar mass), confirming \eref{eqn.galaxy.halos.tullyfisher.exponent}.  However, the value of the exponent in \eref{eqn.galaxy.halos.tullyfisher.relation} may be as shallow as \(a=3.00\) for the baryonic Tully-Fisher relation, \(a=2.77\) for the actual \(B\)-band, and \(a=3.67\) for the actual K-band, based on their sample.

\begin{SCfigure}[0.9][h]
\includegraphics[width=0.5\textwidth]{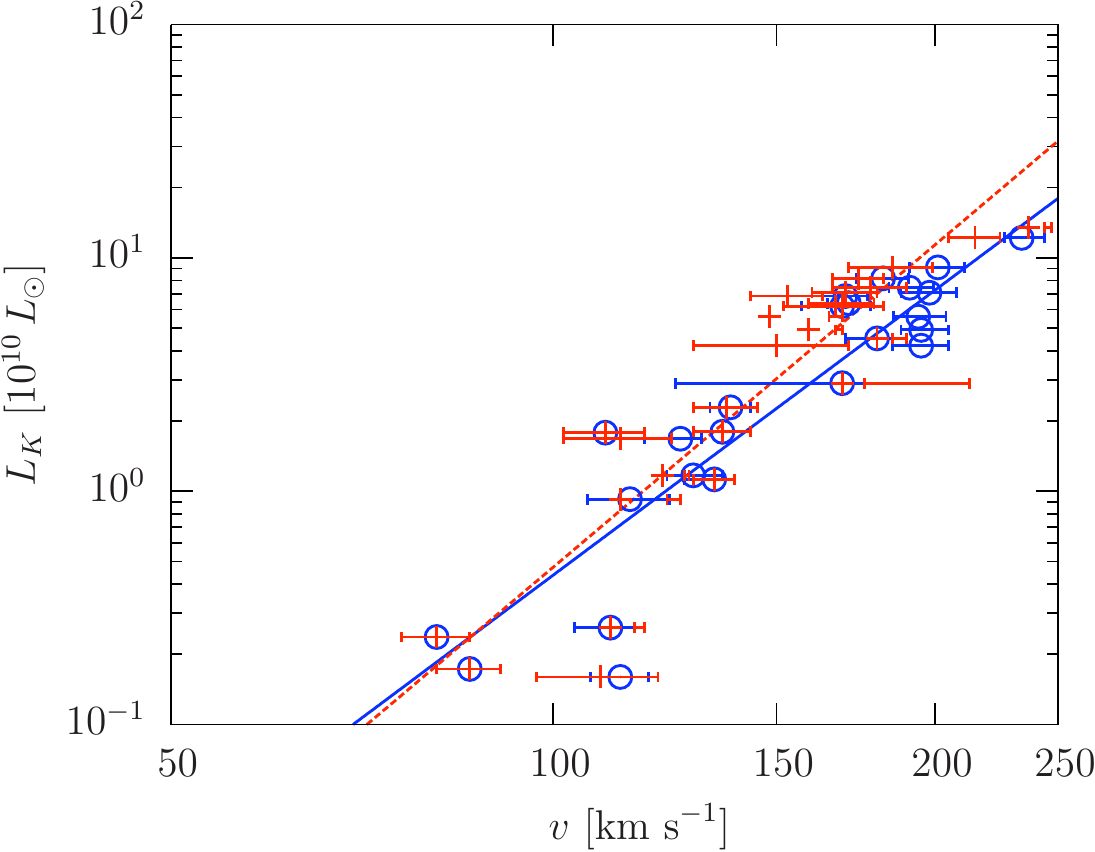}
\caption[Empirical K-band Tully-Fisher relation]{\label{figure.galaxy.halos.tfr.kband}{\sf\small UMa --- \break Empirical K-band Tully-Fisher relation.}\break\break The K-band luminosity, \(L_{K}\) of \eref{eqn.galaxy.halos.tullyfisher.luminosity}, in \(10^{10}\,L_{\solar}\), vs.\,velocity, \(v\) in km s\(^{-1}\) for 19 HSB and 10 LSB galaxies.  The velocity is identified according to \eref{eqn.galaxy.halos.tullyfisher.maxout} as either \(v=v_{\rm max}\), with blue circles with error bars, or as \(v=v_{\rm out}\), with red crosses with error bars, with best-fits shown with  solid blue and red dashed lines, respectively, with parameters listed in the top row of \tref{table.galaxy.halos.tfr}.}
\end{SCfigure}

In order to calculate the total \(K\)-band luminosity,
apparent \(K\)-band magnitudes from the 2MASS survey were used.  Given an apparent \(K\)-band magnitude it is possible to calculate the \(K\)-band luminosity as
\begin{equation} \label{eqn.galaxy.halos.tullyfisher.luminosity}
\log_{10}(L_{K}) = 1.364 - \frac{2}{5} K_{T} + \log_{10}(1+z) + 2 \log_{10} d,
\end{equation}
where \(L_{K}\) is the \(K\)-band luminosity in units of \(10^{10} L_{\solar}\), \(K_{T}\) is the \(K\)-band apparent magnitude and z
is the
redshift of the galaxy (determined from the NASA/IPAC Extragalactic Database), and \(d=18.6\) Mpc is the distance to the galaxy in the Ursa Major filament.  The \(\log_{10}(1+z)\) term is a first order \(K\)-correction.  

The empirical K-band Tully-Fisher relation is plotted in \SCfref{figure.galaxy.halos.tfr.kband}{Empirical K-band Tully-Fisher relation}, with the ordinary relation in blue, including the best-fit power-law, 
\begin{equation} \label{eqn.galaxy.halos.tullyfisher.index.kband}
L_{K} \propto {v_{\rm max}}^{4.1 \pm 0.4},
\end{equation}
of the form of \eref{eqn.galaxy.halos.tullyfisher.relation}.  To consider the effect of identifying the velocity in the Tully-Fisher relation with the asymptotic velocity, \(v_{\rm out}\), instead of the maximum velocity, \(v_{\rm max}\), the asymptotic K-band Tully-Fisher relation is plotted in red, including the best-fit power-law with results listed in \tref{table.galaxy.halos.tfr}.

\begin{table}[h]
\caption[Tully-Fisher relation]{\label{table.galaxy.halos.tfr}{\sf Tully-Fisher relation}}
\begin{center} \begin{tabular}{cc|cc|cc} \\ \hline
&& \multicolumn{2}{c|}{\(v_{\rm max}\)} & \multicolumn{2}{c}{\(v_{\rm out}\)} \\
\multicolumn{2}{c|}{\sc Relation Type} & a & b & a & b \\
\multicolumn{2}{c|}{\footnotesize(1)} & \footnotesize(2) & \footnotesize(3) & \footnotesize(4) & \footnotesize(5) \\ \hline
{\sc Empirical} & K-band & \( 4.1 \pm 0.4 \) & \( -8.5 \pm 0.8 \) & \( 4.6 \pm 0.4 \) & \( -9.5 \pm 0.9 \) \\ \hline \hline
\multirow{4}{*}{\sc Stellar mass} & STVG & \( 3.2 \pm 0.3 \) & \( -6.9 \pm 0.7 \) & \( 3.5 \pm 0.4 \) & \( -7.4 \pm 0.9 \) \\
& MSTG & \( 2.9 \pm 0.3 \) & \( -6.0 \pm 0.7 \) & \( 3.1 \pm 0.4 \) & \( -6.4 \pm 0.9 \) \\
& MOND & \( 4.1 \pm 0.4 \) & \( -8.7 \pm 0.8 \) & \( 4.4 \pm 0.5 \) & \( -9.3 \pm 1.0 \) \\
& Dark Matter & \( 3.0 \pm 0.4 \) & \( -6.2 \pm 1.0 \) & \( 3.1 \pm 0.5 \) & \( -6.5 \pm 1.2 \) \\ \hline
\multirow{4}{*}{\sc Baryonic mass} & STVG & \( 2.6 \pm 0.2 \) & \( -5.4 \pm 0.5 \) & \( 2.8 \pm 0.3 \) & \( -5.7 \pm 0.7 \) \\ 
& MSTG & \( 2.5 \pm 0.2 \) & \( -5.0 \pm 0.5 \) & \( 2.6 \pm 0.3 \) & \( -5.3 \pm 0.7 \) \\
& MOND & \( 3.0 \pm 0.3 \) & \( -6.3 \pm 0.5 \) & \( 3.3 \pm 0.3 \) & \( -6.7 \pm 0.7 \) \\
& Dark Matter & \( 2.5 \pm 0.3 \) & \( -5.1 \pm 0.7 \) & \( 2.7 \pm 0.4 \) & \( -5.4 \pm 0.9 \) \\ \hline
\sc Total mass & Dark Matter & \( 2.9 \pm 0.2 \) & \( -5.5 \pm 0.5 \) & \( 3.1 \pm 0.3 \) & \( -5.9 \pm 0.7 \) \\
\hline \multicolumn{6}{c}{}
\end{tabular} \end{center}
\parbox{6.375in}{\small Notes. --- The empirical and theoretical Tully-Fisher relation:  Column (1) lists the relation type, where the empirical Tully-Fisher relation is the measured K-band luminosity, \(L_{K}\) vs. velocity, plotted in \SCfref{figure.galaxy.halos.tfr.kband}{Empirical K-band Tully-Fisher relation}; and the theoretical Tully-Fisher relation identifies the total luminosity with either the stellar disk mass, in the ordinary case, or the baryonic mass including the stellar disk and HI (plus He) gas mass, each per gravity theory: Moffat's STVG and MSTG theories and Milgrom's MOND theory and the core-modified dark matter theory.  The total mass relation identifies the combined masses of the stellar disk, HI (and He) gaseous disks and dark matter halo as the source of galactic dynamics. Columns (2) and (3) list the power-law index and proportionality constant of \eref{eqn.galaxy.halos.tullyfisher.ab}, respectively, for the ordinary Tully-Fisher relation with \(v=v_{\rm max}\).  Columns (4) and (5) list the power-law index and proportionality constant of \eref{eqn.galaxy.halos.tullyfisher.ab}, respectively, for the asymptotic Tully-Fisher relation with \(v=v_{\rm out}\).}
\end{table}

The empirical Tully-Fisher relation involves the total luminosity in a particular band, such as the K-band, which is proportional to the stellar disk mass through the stellar mass-to-light ratio,
\begin{equation}
\label{eqn.galaxy.halos.tullyfisher.mean} L = \frac{M}{\Upsilon},
\end{equation}
\begin{equation}
\label{eqn.galaxy.halos.tullyfisher.actual}
\log(M) = a \log(v) + b - \log\left({\Upsilon}\right).
\end{equation}

Thus, the effect of \(\Upsilon\ne 1\) is to shift the \(\log(M)\)-intercept; but does not affect the slope.  Theoretical predictions may be may quantified by either computing the appropriate \(\Upsilon\) values which depend on the particular band
of the luminosity measurements, or by considering the respective curve fits to the actual Tully-Fisher relation:
\begin{equation}\label{eqn.galaxy.halos.tullyfisher.ab}
\log(M) = a \log(v) + b.
\end{equation}

\begin{SCfigure}[0.9][h]
\includegraphics[width=0.5\textwidth]{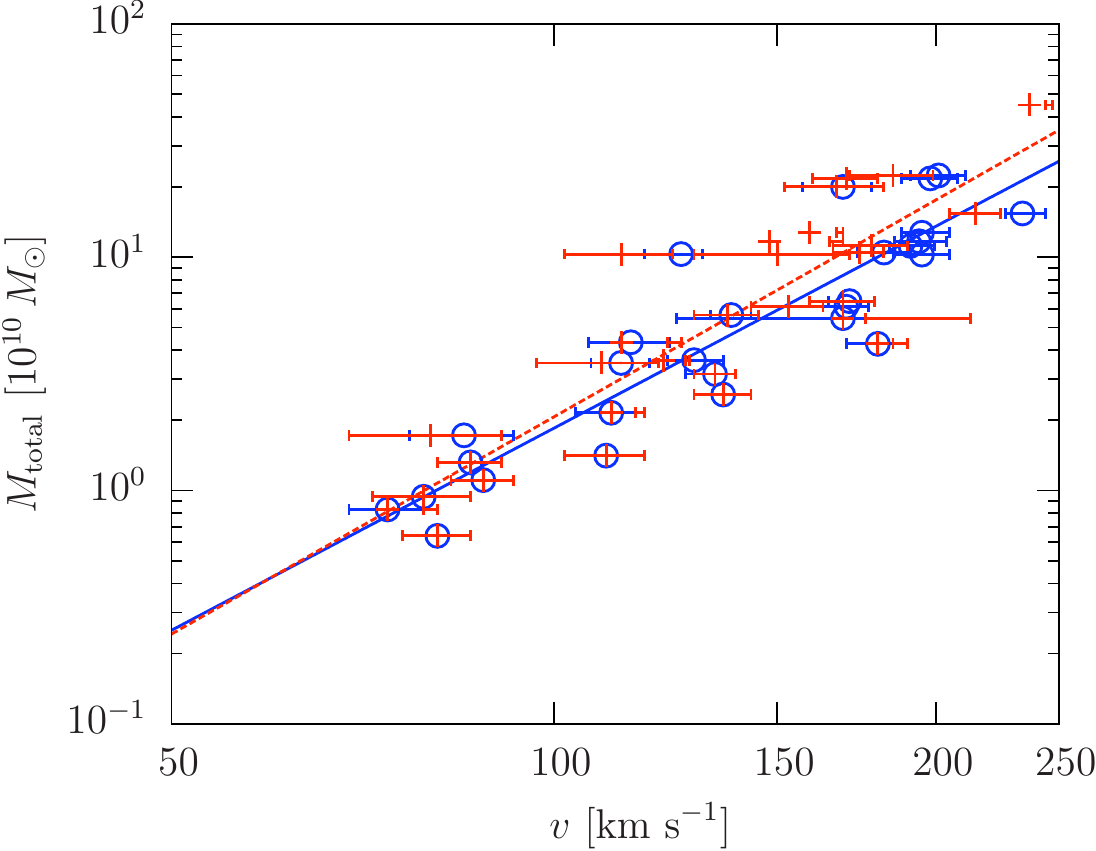}
\caption[Total mass with dark matter vs.\,velocity]{\label{figure.galaxy.halos.tfr.darkmatter} {\sf\small UMa --- Total mass with dark matter vs.\,velocity relation}\break\break The total mass, \(M_{\rm total}=M_{\rm baryon}+M_{\rm halo}\), in \(10^{10}\,M_{\solar}\), vs.\,velocity, \(v\) in km s\(^{-1}\) for 19 HSB and 10 LSB galaxies, with \(M_{\rm halo}\) of \eref{eqn.galaxy.dynamics.dm.coremodified}.  The velocity is identified according to \eref{eqn.galaxy.halos.tullyfisher.maxout} as either \(v=v_{\rm max}\), with blue circles, or as \(v=v_{\rm out}\), with red crosses, with best-fits shown with solid blue and red dashed lines, respectively, with parameters listed in the bottom row of \tref{table.galaxy.halos.tfr}.}
\end{SCfigure}

As regards dark matter, \SCfref{figure.galaxy.halos.tfr.darkmatter}{Dark matter total mass vs.\,velocity} plots the total mass including the luminous baryonic components and the dark matter halo, \(M_{\rm total}=M_{\rm baryon} + M_{\rm halo}\), according to the best-fit core-modified dark matter theory, including the best-fit power-law, with results listed in the bottom row of \tref{table.galaxy.halos.tfr}, finding a minimum of scatter in the best-fit power-law, 
\begin{equation} \label{eqn.galaxy.halos.tullyfisher.index.darkmatter}
M_{\rm total} \propto \left\{\begin{array}{l}{v_{\rm max}}^{2.9 \pm 0.2},\\{v_{\rm out}}^{3.1 \pm 0.3}.\end{array}\right.
\end{equation}
The result is significant because it provides an empirical relation to determine the total mass of a galaxy, from a few simple dynamical velocity measurements.

\newcommand{\subdiskbary}{\small The mass, \(M\), in \(10^{10}\,M_{\solar}\), vs.\,velocity, \(v\) in km s\(^{-1}\)}
\begin{figure}
\begin{picture}(460,320)(82,315) 
\put(30,12){\includegraphics[width=1.28\textwidth]{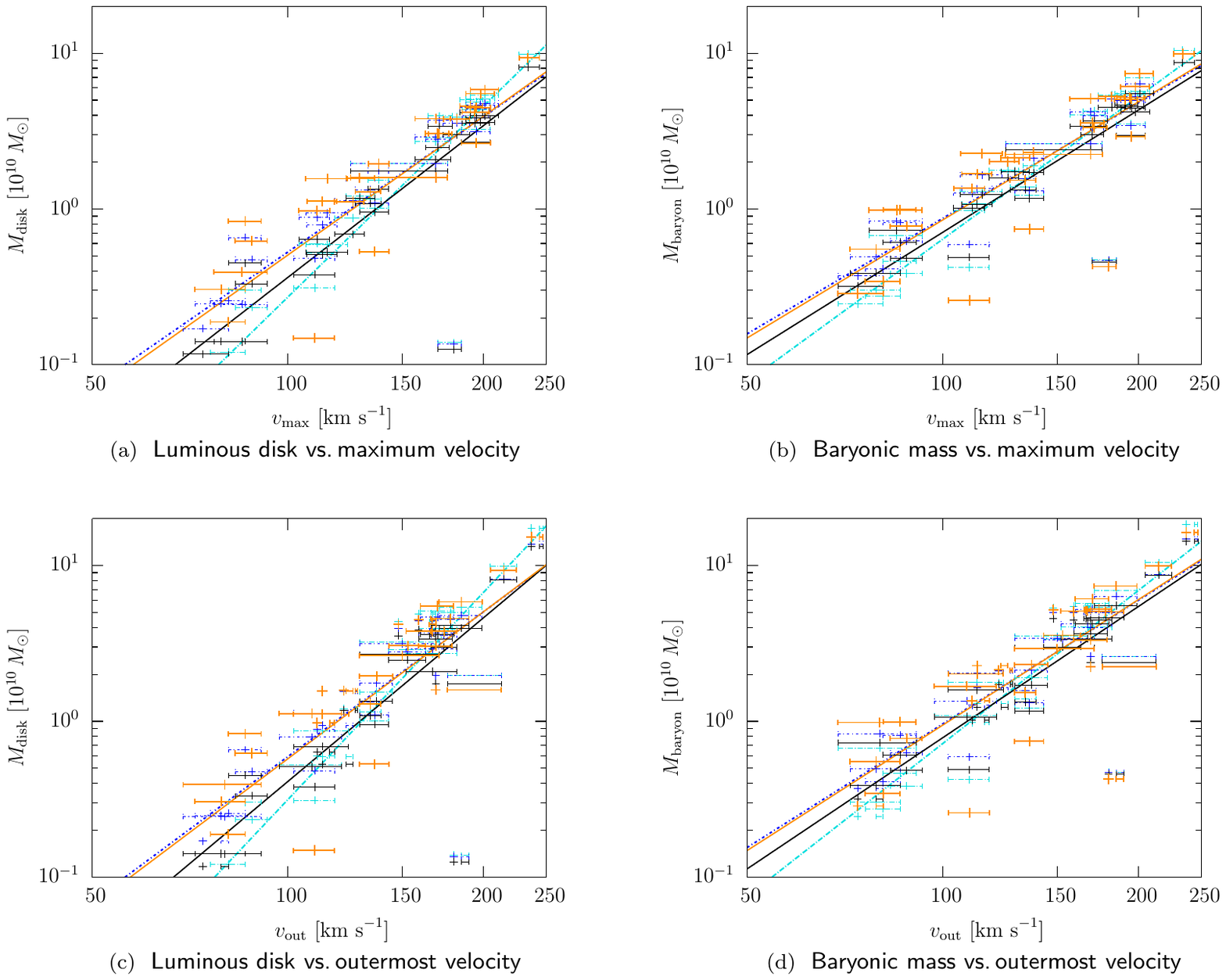}}
\put(82,45){\includegraphics[width=0.98\textwidth]{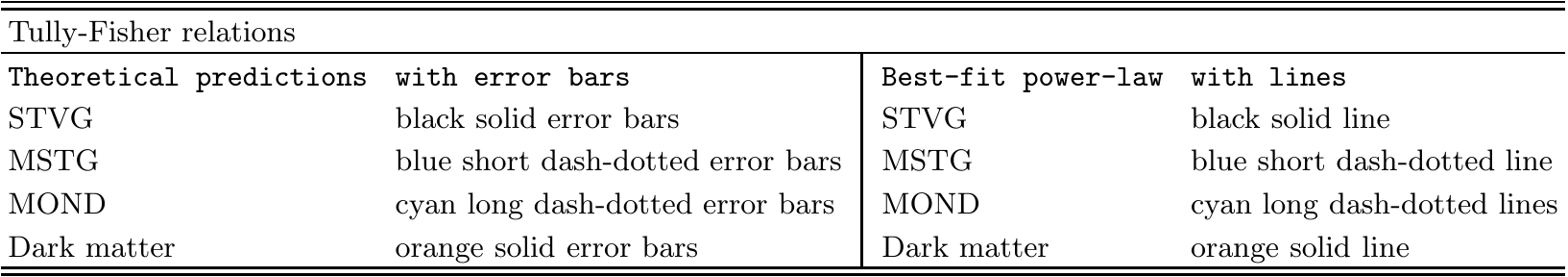}}
\end{picture}
\caption[Theoretical Tully-Fisher relations]{\label{figure.galaxy.halos.tfr.diskbary} {{\sf\small UMa --- Theoretical Tully-Fisher relations.}}\break\break{\subdiskbary} for 19 HSB and 10 LSB galaxies.   The mass is identified according to \tref{table.galaxy.mass} as either \(M=M_{\rm disk}\) in the left panels, or as \(M=M_{\rm baryon}\), in the right panels, for Moffat's STVG and MSTG theories and Milgrom's MOND theory and the core-modified dark matter theory.   The velocity is identified according to \eref{eqn.galaxy.halos.tullyfisher.maxout} as either \(v=v_{\rm max}\), in the top panels, or as \(v=v_{\rm out}\), in the bottom panels, with best-fits shown per gravity theory, with parameters tabulated in \tref{table.galaxy.halos.tfr}.}
\end{figure}

In the ordinary case, without dark matter, \(M\) is identified with \(M_{\rm disk}\).  
However, the theoretical underpinning of the Tully-Fisher relation suggests that the empirical relation is an approximation to the baryonic Tully-Fisher relation which identifies \(M\) with \(M_{\rm baryon}\), instead, because in the absence of non-baryonic dark matter, galactic dynamics are sourced by the total baryonic mass, and not just the luminous stellar disk.  \fref{figure.galaxy.halos.tfr.diskbary} plots four distinct Tully-Fisher relations, where the luminosity is taken proportional to the stellar mass in the left panels and the baryonic mass in the right panels, and where the velocity is taken as the velocity amplitude (maximum) in the top panels, and the asymptotic velocity at the position \(r_{\rm out}\) in the bottom panels:

\begin{equation}\label{eqn.galaxy.halos.tullyfisher.barydisk}
M = \left\{ \begin{array}{ll}
M_{\rm disk} & \mbox{Ordinary Tully-Fisher relation},\\
M_{\rm baryon} & \mbox{Baryonic Tully-Fisher relation},\end{array}\right.
\end{equation}
and for each, two depending on whether
\begin{equation}\label{eqn.galaxy.halos.tullyfisher.maxout}
v = \left\{ \begin{array}{ll}
v_{\rm max} & \quad\mbox{Ordinary Tully-Fisher relation},\\
v_{\rm out} & \quad\mbox{Asymptotic Tully-Fisher relation}.\end{array}\right.
\end{equation}

The scatter in the theoretical Tully-Fisher relations is minimized in the case of the baryonic Tully-Fisher relation across all gravity theories, except dark matter, implying that the empirical Tully-Fisher relation -- which involves only the luminous disk -- is an approximate law.  

\index{Galaxy rotation, \(v\)!Tully-Fisher relation|)}

\chapterquote{A table, a chair, a bowl of fruit and a violin; what else does a man need to be happy?}{Albert Einstein}
\chapter{\label{chapter.cluster}Clusters of galaxies}

\citet{Smail.MNRAS.1995.273} argued that as clusters of galaxies are the largest bound structures known in the universe, their mass-to-light ratios and baryonic fractions should approach that for the cosmos as a whole.  Whereas X-ray luminosity measurements typically give temperature distributions an order of magnitude larger than observed from fits to observed isothermal gas spheres, there is a remarkable variation in the size and shapes of the X-ray distributions, to foil the search for a universal description of the phantom of dark matter, within the modified gravitation theory of \cref{chapter.mog}.

\citet{Brownstein:MNRAS:2006} applied the modified acceleration law of metric skew-tensor gravity (MSTG), as described in \sref{section.mog.mstg}, obtained from the Yukawa skewon theory of \sref{section.mog.mstg.yukawa} in which Einstein gravity is coupled to a Kalb-Ramond Proca field, as in \sref{section.mog.mstg.action}.  Utilizing X-ray observations to fit the gas mass profile and temperature profile of the hot intracluster medium (ICM) with King \(\beta\)-models, the predicted X-ray surface brightnesses of the sample of 106 X-ray clusters were consistent without introducing a non-baryonic dark matter component.  The sub-kiloparsec X-ray surface brightness distributions, as functions of radial distance, are well matched across the sample, including the correct shape, by the \(\beta\)-model fits to the X-ray surface brightness distributions arising from the modified acceleration law.

Prompted by the observed ring-like feature of the weak-lensing map of the galaxy cluster CL~0024+1654, \citet{Milgrom.APJ.2008.678} argued that despite any underlying feature in the baryon distribution, the ring may be observed as the image of the MOND transition region.   The possibility that these emergent features appear as phantom dark matter in the strong and weak lensing mass reconstruction of \citet{Jee.APJ.2007.661} indicates the degree to which MOG theories violate the strong equivalence principle in order to describe clusters of galaxies in the absence of dark matter.  The same phenomenon applied to the {\bc} produces the observed phantom dark matter in the strong and weak lensing mass reconstruction of \citet{Clowe.APJL.2006.648,Bradac.APJ.2006.652,Clowe.NPBPS.2007.173}, in the form of spatially dislocated peaks.  

The physics of X-ray clusters is derived in \sref{section.cluster.xraymass}.  Dark matter distributions are computed and compared to actual gas mass measurements for each of the clusters of galaxies, with best-fit cluster models presented in \sref{section.cluster.models}.  Direct evidence from the {\bc} gravitational lensing experiment is presented in \sref{section.cluster.bullet} which supports the necessity of dominant dark matter, or the modified gravity alternative.
\section{\label{section.cluster.xraymass}X-ray clusters}

The creation of X-ray mass profiles from astrophysical observations, as described in \sref{section.cluster.xraymass.astroph} for clusters of galaxies, is subject to model dependent assumptions based on the isotropic isothermal model, as in \sref{section.cluster.xraymass.isothermal}.  The road from measuring radial, X-ray temperature profiles, to surface mass density maps, as in \sref{section.cluster.xraymass.Sigma}, in some chosen gravity theory, takes its way through the dynamics of the isothermal sphere and dynamical mass computations, as described in \sref{section.cluster.xraymass.Gamma}, which result in best-fit cluster models, presented in \sref{section.cluster.models}.

\subsection{\label{section.cluster.xraymass.astroph}Astrophysical observations}

Clusters of galaxies have been known to require some form of energy density that makes its presence felt only by its gravitational effects since \citet{Zwicky:1933} analysed the velocity dispersion for the Coma cluster.  The more than 1000 galaxies spherically distributed within the Coma cluster comprise a small fraction (10\%) of the baryonic mass, the larger fraction consisting of a diffuse cloud of 100 million degree X-ray emitting plasma -- the intracluster medium (ICM).  The ICM itself comprises only a small fraction (10\%) of the Newtonian dynamic mass as determined from X-ray luminosity measurements.  

Much closer to the Milky Way, the Virgo cluster forms the heart of the Local Supercluster, and has a galaxy population as rich as Coma distributed in three groups.  Messier 49 -- an elliptical / lenticular galaxy -- is the brightest member of the Virgo cluster and is the center of one of the subdominant groups.  The ICM surrounding Messier 49 is a diffuse cloud of 10 million degree X-ray emitting plasma, but only accounts for a tiny fraction (1\%) of the Newtonian dynamic mass as determined from X-ray luminosity measurements.  The Fornax cluster is much smaller than Virgo, but at a similar distance from the Milky Way.  The 15 million degree ICM surrounding the Fornax core -- which is in the preliminary preheating stage of an imminent merger as determined from peculiar velocity measurements along a filament -- comprises a similar fraction (3\%) of the Newtonian dynamic mass.  

Abell 400 is an ongoing cluster-cluster merger, with multiple subclusters around a central main cluster containing the Dumbbell galaxy, which is the result of a galaxy-galaxy merger and is the topic of ongoing X-ray and radiowave analysis due to a pair of suspected supermassive black holes, bound and moving together.  The 30 million degree X-ray emitting plasma accounts for 10\% of the Newtonian dynamic mass. The Hydra-Centaurus supercluster contains two distinct X-ray clusters, each with 100 member galaxies near their respective centers, and have 50 and 40 million degree X-ray emitting ICM plasmas, respectively, which account for 10\% of the Newtonian dynamic masses.  At the center of the Great Attractor 65 Mpc distant, the Norma cluster is half the size of the Coma cluster, but larger than Centaurus, Hydra-A, Fornax and Messier 49 combined.  It is in the process of swallowing a galaxy which shows a comet-like tail nearly twice as long as the galaxy itself, consisting of a 70 million degree X-ray plasma, and accounting for 10\% of the Newtonian dynamic mass.  

Perseus is the brightest X-ray cluster in the sky and is nearly the size of the Coma cluster, but is not as rich in galaxies.  The 80 million degree X-ray plasma accounts for 20\% of the Newtonian dynamic mass.  Chandra has measured concentric ripples in the X-ray surface mass density surrounding a strong source of gravitation inside an X-ray cavity -- a candidate for a \(10^{8}\,M_{\solar}\) black hole.  Abell 2255 is only slightly larger than Perseus, and slightly less than the size of Coma, but the X-ray peak is offset from the brightest cluster galaxy, which has a large peculiar velocity, (1200 km s\(^{-1}\)), indicating an ongoing cluster merger.  The 80 million degree plasma accounts for 8\% of the Newtonian dynamic mass.  The giant Abell 2142 is one-and-a-half times larger than the Coma cluster, and is in the later stages of a cluster-cluster merger showing bow-shock waves.  The 110 million degree plasma accounts for 15\% of the Newtonian dynamic mass.  

The {\bc} is a merger between a giant main cluster with thousands of galaxies and a supersonic subcluster, aligned in the plane perpendicular to the line-of-sight.  The 170 million degree main cluster ICM accounts for 10\% of the Newtonian dynamic mass, and provides strong and weak gravitational lensing observations which show structure offset from the X-ray surface density map.
  
The sample selection includes the {\bc}, Abell 2142, Coma, Abell 2255, Perseus, Norma, Hydra-A, Centaurus, Abell 400, Fornax, and Messier 49, with cluster properties listed in \tref{table.cluster.sample} -- ordered from the hottest X-ray emitting to the coolest of the clusters.  The Newtonian dynamic masses and the ICM gas masses, for each cluster, are plotted in \fref{figure.cluster.models.mass}, and the ratio of the Newtonian dynamic masses to the ICM gas masses are plotted as dynamic mass factors in \fref{figure.cluster.models.Gamma} -- each as a function of radial position and compared to the theoretical predictions of core-modified dark matter halos, as in \sref{section.cluster.models.darkmatter}, Milgrom's MOND as in \sref{section.cluster.models.mond}, and Moffat's MOG as in \sref{section.cluster.models.mog}. 
 
The study of the {\bc}, in \sref{section.cluster.bullet}, is a detailed analysis of the X-ray gas surface density map in relation to the Newtonian dynamic mass inferred from the strong and weak gravitational lensing map.  The {\it missing mass} in MOG is explained by galactic surface mass density maps, presented in \fref{figure.cluster.bullet.galaxy}.  The {\it missing mass} in terms of dark matter is presented in \fref{figure.cluster.bullet.distribution}.

\subsection{\label{section.cluster.xraymass.isothermal}Isotropic isothermal model}

Recent observations from the {\sc XMM-Newton} satellite suggest that the intracluster medium (ICM) is very nearly isothermal
inside the region defined by the X-ray emission with temperatures ranging from \(\approx\) 1--15 keV (or \(10^{7}\) -- \(2 \times
10^{8}\) K) for different clusters~\protect\citep{Arnaud.AAP.2001.365}.  The combination of the observed density profile, \(n_{e}(r)\), and the
temperature profile, \(T(r)\), obtained from X-ray observations
of the galaxy cluster leads to a pressure profile, \(P(r)\), which directly leads to a mass profile, \(M(r)\), by assuming the
gas is in nearly hydrostatic equilibrium with the gravitational potential of the galaxy cluster. Within a
few core radii, the distribution of gas within a galaxy cluster may be fit by a King ``\(\beta\)-model''.

\addtocontentsheading{lot}{X-ray clusters of galaxies}
\begin{landscape}
\begin{table}
\caption[X-ray cluster properties of the sample]{\label{table.cluster.sample}{\sf X-ray cluster properties of the sample}}
\begin{center}
\begin{tabular}{c|cccc|ccc} \multicolumn{8}{c}{} \\ \hline  
{\sc Cluster}   & {\(T\)} & {\(\rho_{0}\)} & {\(\beta\)} & {\(r_{c}\)} & {\(r_{\rm out}\)}  & {\(M_{\rm gas}\)} & {\(M_{N}\)} \\ 
&\footnotesize(keV)&\footnotesize(\(10^{-25}\,\mbox{g/cm}^3\))&\footnotesize(kpc)&\footnotesize(kpc)&\footnotesize(\(10^{14}\,M_{\solar}\))&\footnotesize(\(10^{14}\,M_{\solar}\)) \\
\phantom{XX}\footnotesize(1)&\footnotesize(2)&\footnotesize(3)&\footnotesize(4)&\footnotesize(5)&\footnotesize(6)&\footnotesize(7)&\footnotesize(8)\\ \hline\hline
{\bul}& $14.80^{+1.70}_{-1.20}$&0.226&$0.803^{+0.013}_{-0.013}$&$278.0^{+6.8}_{-6.8}$&$2623_{-97}^{+97}$&$3.89_{-0.24}^{+0.24}$&$33.82_{-2.85}^{+3.96}$\\
Abell 2142&$9.70^{+1.50}_{-1.10}$&0.268&$0.591^{+0.006}_{-0.006}$&$108.5^{+6.2}_{-7.4}$&$2537_{-192}^{+167}$&$2.39_{-0.30}^{+0.26}$&$15.93_{-1.82}^{+2.47}$\\
Coma&$8.38^{+0.34}_{-0.34}$&0.061&$0.654^{+0.019}_{-0.021}$&$242.3^{+18.6}_{-20.1}$&$1954_{-202}^{+201}$&$1.13_{-0.19}^{+0.18}$&$11.57_{-0.70}^{+0.67}$\\
Abell 2255&$6.87^{+0.20}_{-0.20}$&0.032&$0.797^{+0.033}_{-0.030}$&$417.6^{+30.3}_{-32.6}$&$1730_{-174}^{+160}$&$0.85_{-0.13}^{+0.12}$&$9.82_{-0.60}^{+0.65}$\\
Perseus &$6.79^{+0.12}_{-0.12}$&0.632&$0.540^{+0.006}_{-0.004}$&$45.1^{+2.4}_{-2.9}$&$2414_{-189}^{+145}$&$1.88_{-0.22}^{+0.18}$&$9.70_{-0.20}^{+0.23}$\\
Norma &$6.02^{+0.08}_{-0.08}$&0.037&$0.555^{+0.056}_{-0.044}$&$210.6^{+40.4}_{-36.5}$&$1830_{-515}^{+474}$&$0.78_{-0.29}^{+0.28}$&$6.62_{-0.75}^{+0.95}$\\
Hydra-A&$4.30^{+0.40}_{-0.40}$&0.634&$0.573^{+0.003}_{-0.003}$&$35.2^{+1.6}_{-2.1}$&$1502_{-95}^{+76}$&$0.49_{-0.05}^{+0.04}$&$4.06_{-0.38}^{+0.38}$\\
Centaurus &$3.68^{+0.06}_{-0.06}$&0.286&$0.495^{+0.011}_{-0.010}$&$26.1^{+3.7}_{-3.2}$&$1175_{-174}^{+189}$&$0.20_{-0.04}^{+0.05}$&$2.35_{-0.08}^{+0.08}$\\
Abell 400 &$2.31^{+0.14}_{-0.14}$&0.039&$0.534^{+0.013}_{-0.014}$&$108.5^{+8.8}_{-7.8}$&$1062_{-108}^{+97}$&$0.149_{-0.02}^{+0.02}$&$1.42_{-0.10}^{+0.10}$\\
Fornax &$1.20^{+0.04}_{-0.04}$&0.018&$0.804^{+0.098}_{-0.084}$&$122.5^{+13.0}_{-12.6}$&$387_{-74}^{+67}$&$0.009_{-.003}^{+.004}$&$0.373_{-0.06}^{+0.07}$\\
Messier~49 &$0.95^{+0.02}_{-0.01}$&0.259&$0.592^{+0.007}_{-0.007}$&$7.7^{+0.8}_{-0.8}$&$177_{-20}^{+19}$&$0.001_{-.000}^{+.000}$&$0.109_{-.002}^{+.003}$\\
\hline \multicolumn{8}{c}{}
\end{tabular} \end{center}
\parbox{1.0in}{\phantom{Notes.}}
\parbox{7.5in}{\small Notes. ---  Relevant X-ray cluster properties of the sample: Column (1) is
the name of the cluster.  Column (2) is the X-ray temperature.  Columns (3), (4) and (5) are the best-fitting King \(\beta\)-model parameters of \eref{eqn.cluster.xraymass.isothermal.betaRhoModel}, consisting of the X-ray central density, the \(\beta\) parameter, and the core radius, respectively.  Column (6) is the computed radial position of \eref{eqn.cluster.xraymass.isothermal.rout.0} at which the density drops to $\approx 10^{-28}\,\mbox{g/cm}^{3}$, or 250 times the mean cosmological density of baryons.  Column (7) is the total computed mass of the ICM gas inside a sphere of radius \(r_{\rm out}\) according to \eref{eqn.cluster.xraymass.isothermal.Mgas.0}, and Column (8) is the total Newtonian dynamic mass of \eref{eqn.cluster.xraymass.newtonsMass}, determined from X-ray temperature measurements, integrated to \(r_{\rm out}\).}
\end{table}
\end{landscape}

The observed surface brightness of the X-ray cluster can be fit to a radial distribution profile~\protect\citep{Chandrasekhar:1960,King:AJ:1966}:
\begin{equation}
\label{eqn.cluster.xraymass.isothermal.betaModel} I(r)= I_{0}\left[  1+\left(\frac{r}{r_{c}}\right)^{2}\right]^{-3 \beta + 1/2},
\end{equation}
resulting in best-fit parameters, \(\beta\) and \(r_{c}\). A deprojection of the \(\beta\)-model of \eref{eqn.cluster.xraymass.isothermal.betaModel} assuming a nearly isothermal gas sphere then results in a physical gas density
distribution~\protect\citep{Cavaliere:AAP:1976}:
\begin{equation}
\label{eqn.cluster.xraymass.isothermal.betaRhoModel} \rho(r)= \rho_{0}\left[  1+\left(\frac{r}{r_{c}}\right)^{2}\right]^{-3 \beta /2},
\end{equation}
where \(\rho(r)\) is the ICM mass density profile, and $\rho_{0}$ denotes the central density.  The mass profile associated with this density is given by 
\begin{equation}
\label{eqn.cluster.xraymass.isothermal.massProfile}
M(r)  = 4\pi\int_{0}^{r} \rho({r^{\prime}}) {r^{\prime}}^{2} d{r^{\prime}},
\end{equation}
where $M(r)$ is the total mass contained within a sphere of radius \(r\).  Galaxy clusters are observed to have  luminous distributions with finite spatial extent.  This allows an approximate determination of the total mass of the galaxy cluster by first solving \eref{eqn.cluster.xraymass.isothermal.betaRhoModel} for the position, $r_{\rm out}$, at which the density, $\rho(r_{\rm out})$, drops to $\approx 10^{-28}\,\mbox{g/cm}^{3}$, or 250 times the mean cosmological density of baryons:
\begin{equation}
\label{eqn.cluster.xraymass.isothermal.rout.0}
r_{\rm out} = r_{c} \left[\left(
\frac{\rho_{0}}{10^{-28}\,\mbox{g/cm}^{3}}\right)^{2/3\beta}-1\right]^{1/2}.
\end{equation}
Then, the total mass of the ICM gas may be taken as $M_{\rm gas} \approx M(r_{\rm out})$:
\begin{equation}
\label{eqn.cluster.xraymass.isothermal.Mgas.0}
M_{\rm gas}  = 4\pi\int_{0}^{r_{\rm out}}\rho_{0} \left[  1+\left(\frac{{r^{\prime}}}{r_{c}}\right)^{2}\right]^{-3 \beta /2} {r^{\prime}}^{2}
d{r^{\prime}}.
\end{equation}

Provided the number density, \(n\),  traces the actual mass, we may
assume that \(n(r) \propto \rho(r)\), which according to \citet{Reiprich:2001,Reiprich:2002} is explicitly
\begin{equation}
\label{eqn.cluster.xraymass.isothermal.numberMass} \rho_{\rm gas} \approx 1.17 n_{e} m_{p},
\end{equation}
and rewrite \eref{eqn.cluster.xraymass.isothermal.betaRhoModel} 
\begin{equation}
\label{eqn.cluster.xraymass.isothermal.betaNumberModel} n_{e}(r)= n_{0}\left[  1+\left(\frac{r}{r_{c}}\right)^{2}\right]^{-3 \beta /2}.
\end{equation}
For a spherical system in hydrostatic equilibrium, the structure equation can be derived from the collisionless Boltzmann
equation
\begin{equation}
\label{eqn.cluster.xraymass.isothermal.CBE} \frac{d}{dr}(\rho(r) \sigma_{r}^{2}) + \frac{2\rho(r)}{r}\left(\sigma_{r}^{2} - \sigma_{\theta,\phi}^{2}\right) =
-\rho(r) \frac{d\Phi(r)}{dr},
\end{equation}
where \(\Phi(r)\) is the gravitational potential for a point source, \(\sigma_{r}\) and \(\sigma_{\theta,\phi}\) are mass-weighted
velocity dispersions in the radial (\(r\)) and tangential (\(\theta, \phi\)) directions, respectively.  For an isotropic
system,
\begin{equation}
\label{eqn.cluster.xraymass.isothermal.isotropicSystem}
\sigma_{r} = \sigma_{\theta,\phi}.
\end{equation}
The pressure profile, \(P(r)\), can be related to these quantities by
\begin{equation}
\label{eqn.cluster.xraymass.isothermal.pressureProfile}
P(r) = \sigma_{r}^{2} \rho(r).
\end{equation}
Combining \erefss{eqn.cluster.xraymass.isothermal.CBE}{eqn.cluster.xraymass.isothermal.isotropicSystem}{eqn.cluster.xraymass.isothermal.pressureProfile}, the result for the isotropic
sphere is
\begin{equation}
\label{eqn.cluster.xraymass.isothermal.isotropicSphere}
\frac {dP(r)}{dr} = -\rho(r) \frac{d\Phi(r)}{dr}.
\end{equation}
For a gas sphere with temperature profile, \(T(r)\), the velocity dispersion becomes
\begin{equation}
\label{eqn.cluster.xraymass.isothermal.velocityDispersion}
\sigma_{r}^{2} = \frac{kT(r)}{\mu_{A} m_{p}},
\end{equation}
where \(k\) is Boltzmann's constant, \(\mu_{A} \approx 0.609\) is the mean atomic weight and \(m_{p}\) is the proton mass.  We
may now substitute \erefs{eqn.cluster.xraymass.isothermal.pressureProfile}{eqn.cluster.xraymass.isothermal.velocityDispersion} into \eref{eqn.cluster.xraymass.isothermal.isotropicSphere} to obtain
\begin{equation}
\label{eqn.cluster.xraymass.isothermal.isotropicSphere2}
\frac {d}{dr}\left(\frac{kT(r)}{\mu_{A} m_{p}} \rho(r)\right) = -\rho(r) \frac{d\Phi(r)}{dr}.
\end{equation}
Performing the differentiation on the left hand side of \eref{eqn.cluster.xraymass.isothermal.isotropicSphere}, we may solve for the
gravitational acceleration:
\begin{eqnarray}
\nonumber
a(r) & \equiv  & - \frac{d\Phi(r)}{dr} \\
\label{eqn.cluster.xraymass.isothermal.accelerationProfile}
& = & \frac{kT(r)}{\mu_{A} m_{p} r} \left[ \frac{d \ln(\rho(r))}{d \ln(r)} + \frac{d \ln(T(r))}{d \ln(r)}\right].
\end{eqnarray}
For the isothermal isotropic gas sphere, the temperature derivative on the right-hand side of \eref{eqn.cluster.xraymass.isothermal.accelerationProfile} vanishes and the remaining derivative can be evaluated using the \(\beta\)-model of \eref{eqn.cluster.xraymass.isothermal.betaRhoModel}:
\begin{equation}
\label{eqn.cluster.xraymass.isothermal.isothermalAccelerationProfile}
a(r) = -\frac{3\beta kT}{\mu_{A} m_{p}} \left(\frac{r}{r^{2}+r_{c}^{2}}\right).
\end{equation}

\subsection{\label{section.cluster.xraymass.Sigma}Surface mass density map}\index{Surface mass, \(\Sigma\)|(}

To make contact with the
experimental data, we must calculate the surface mass density by integrating \(\rho(r)\) of \eref{eqn.cluster.xraymass.isothermal.betaRhoModel} along the
line-of-sight:
\begin{equation}
\label{eqn.cluster.xraymass.Sigma.surfaceMassDensity.1}
\Sigma(x,y)  = \int_{-z_{\rm out}}^{z_{\rm out}} \rho(x,y,z) dz,
\end{equation} where
\begin{equation}
\label{eqn.cluster.xraymass.Sigma.zout}
z_{\rm out} = \sqrt{r_{\rm out}^{2} - x^{2} - y^{2}}.
\end{equation}
Substituting \eref{eqn.cluster.xraymass.isothermal.betaRhoModel} into \eref{eqn.cluster.xraymass.Sigma.surfaceMassDensity.1}, we obtain
\begin{equation}
\label{eqn.cluster.xraymass.Sigma.surfaceMassDensity.2}
\Sigma(x,y)  = \rho_{0} \int_{-z_{\rm out}}^{z_{\rm out}} \left[  1+\frac{{x^{2}+y^{2}+z^{2}}}{r_{c}^{2}}\right]^{-3
\beta /2} dz.
\end{equation}
This integral becomes tractable by making a substitution of variables:
\begin{equation}
\label{eqn.cluster.xraymass.Sigma.substitutedVariable}
u^{2} = 1+\frac{x^{2}+y^{2}}{r_{c}^{2}},
\end{equation}
so that
\begin{eqnarray} \nonumber
\Sigma(x,y)  &=& \rho_{0} \int_{-z_{\rm out}}^{z_{\rm out}} \left[ u^{2} + \left(\frac{z}{r_{c}}\right)^{2}\right]^{-3
\beta /2} dz\\
\nonumber &=& \frac{\rho_{0}}{u^{3\beta}}\int_{-z_{\rm out}}^{z_{\rm out}} \left[1 + \left(\frac{z}{u
r_{c}}\right)^{2}\right]^{-3 \beta /2} dz\\
\label{eqn.cluster.xraymass.Sigma.surfaceMassDensity.3} &=& 2 \frac{\rho_{0}}{u^{3\beta}} z_{\rm out}
F\left(\left[\frac{1}{2},\frac{3}{2}\beta\right],\left[\frac{3}{2}\right],-\left(\frac{z_{\rm out}}{u
r_{c}}\right)^{2}\right),
\end{eqnarray}
where we have made use of the hypergeometric function, \(F([a,b],[c],z)\).  Substituting \eref{eqn.cluster.xraymass.Sigma.substitutedVariable} into
\eref{eqn.cluster.xraymass.Sigma.surfaceMassDensity.3} gives
\begin{equation}
\label{eqn.cluster.xraymass.Sigma.surfaceMassDensity.4}
\Sigma(x,y)  =  2 \rho_{0} z_{\rm out}\left(1+\frac{x^{2}+y^{2}}{r_{c}^{2}}\right)^{-3\beta/2} 
F\left(\left[\frac{1}{2},\frac{3}{2}\beta\right],\left[\frac{3}{2}\right],-\frac{z_{\rm
out}^{2}}{x^{2}+y^{2}+r_{c}^{2}}\right).
\end{equation}
We next define
\begin{equation}
\label{eqn.cluster.xraymass.Sigma.surfaceMassDensity0.1}
\Sigma_{0} \equiv \Sigma(0,0)  =  2 \rho_{0} z_{\rm out} 
F\left(\left[\frac{1}{2},\frac{3}{2}\beta\right],\left[\frac{3}{2}\right],-\left(\frac{z_{\rm
out}}{r_{c}}\right)^{2}\right),
\end{equation}
which we substitute into \eref{eqn.cluster.xraymass.Sigma.surfaceMassDensity.4}, yielding
\begin{equation}
\label{eqn.cluster.xraymass.Sigma.surfaceMassDensity.5}
\Sigma(x,y)  = \Sigma_{0} \left(1+\frac{x^{2}+y^{2}}{r_{c}^{2}}\right)^{-3\beta/2} 
\frac{F\left(\left[\frac{1}{2},\frac{3}{2}\beta\right],\left[\frac{3}{2}\right],-\frac{z_{\rm
out}^{2}}{x^{2}+y^{2}+r_{c}^{2}}\right)}{F\left(\left[\frac{1}{2},\frac{3}{2}\beta\right],\left[\frac{3}{2}\right],-\frac{z_{\rm
out}^{2}}{r_{c}^{2}}\right)}.
\end{equation}
In the limit \(z_{\rm out} \gg r_{c}\), the Hypergeometric functions simplify to \(\Gamma\) functions, and
\erefs{eqn.cluster.xraymass.Sigma.surfaceMassDensity0.1}{eqn.cluster.xraymass.Sigma.surfaceMassDensity.5} result in the simple, approximate solutions:
\begin{equation}
\label{eqn.cluster.xraymass.Sigma.surfaceMassDensity0}
\Sigma_{0} =  \sqrt{\pi} \rho_{0} r_{c}  \frac{\Gamma\left(\frac{3\beta-1}{2}\right)}{\Gamma\left(\frac{3}{2}\beta\right)}
\end{equation}
and
\begin{equation}
\label{eqn.cluster.xraymass.Sigma.surfaceMassDensity}
\Sigma(x,y)  = \Sigma_{0} \left(1+\frac{x^{2}+y^{2}}{r_{c}^{2}}\right)^{-(3\beta-1)/2 },
\end{equation}
which we may, in principle, fit to the \map{\Sigma} data to determine the King \(\beta\)-model parameters, \(\beta\), \(r_{c}\) and \(\rho_{0}\).\index{Surface mass, \(\Sigma\)|)}

\subsection{\label{section.cluster.xraymass.Gamma}Dynamical mass}

The Newtonian dynamical mass can be obtained as a function of radial position by equating the gravitational acceleration of \eref{eqn.cluster.xraymass.isothermal.accelerationProfile} --  derived in \sref{section.cluster.xraymass.isothermal} for the isotropic isothermal model -- with Newton's acceleration law:
\begin{equation}
\label{eqn.cluster.xraymass.newtonsLaw}
\frac{kT(r)}{\mu_{A} m_{p} r} \left[ \frac{d \ln(\rho(r))}{d \ln(r)} + \frac{d \ln(T(r))}{d \ln(r)}\right] =  \frac{G_{N} M_N(r)}{r^{2}},
\end{equation}
with the solution,
\begin{equation}
\label{eqn.cluster.xraymass.newtonsMass}
M_{N}(r) = - \frac{r}{G_{N}}\frac{kT}{\mu_{A} m_{p}} \left[ \frac{d \ln(\rho(r))}{d \ln(r)} + \frac{d \ln(T(r))}{d
\ln(r)}\right],
\end{equation}
and the isothermal $\beta$-model result of \eref{eqn.cluster.xraymass.isothermal.accelerationProfile} can be
rewritten as
\begin{equation}
\label{eqn.cluster.xraymass.isothermalNewtonsMass}
M_{N}(r) = \frac{3\beta kT}{\mu_{A} m_{p}G_{N}} \left(\frac{r^{3}}{r^{2}+r_{c}^{2}}\right).
\end{equation}

\section{\label{section.cluster.models}Best-fit cluster models}

The study of X-ray clusters, according to \sref{section.cluster.xraymass}, provides valuable information on their mass profiles and insight into their formation and evolution.  It is no longer a matter of fitting the total masses of these systems, but a powerful means to model the spatial distribution of each component, as in the case of the {\bc} presented in \sref{section.cluster.bullet}, which may further constrain cosmological models.  Although X-ray luminosity measurements typically give temperature distributions an order of magnitude larger than observed from fits to observed isothermal gas spheres, this does not guarantee that the missing mass has the form predicted by \(\Lambda\)-CDM cosmological models, particularly because of the remarkable variation in the shapes and scales of the X-ray distributions.

\begin{table}\index{Modified gravity!Observations}\index{MOND!Observations}\index{Dark matter!Observations}
\caption[Best-fit cluster model parameters]{\label{table.cluster.models.bestfit}{\sf Best-fit cluster model parameters}}
\begin{picture}(460,500)(110,320)
\put(0,0){\scalebox{1.1}{\includegraphics{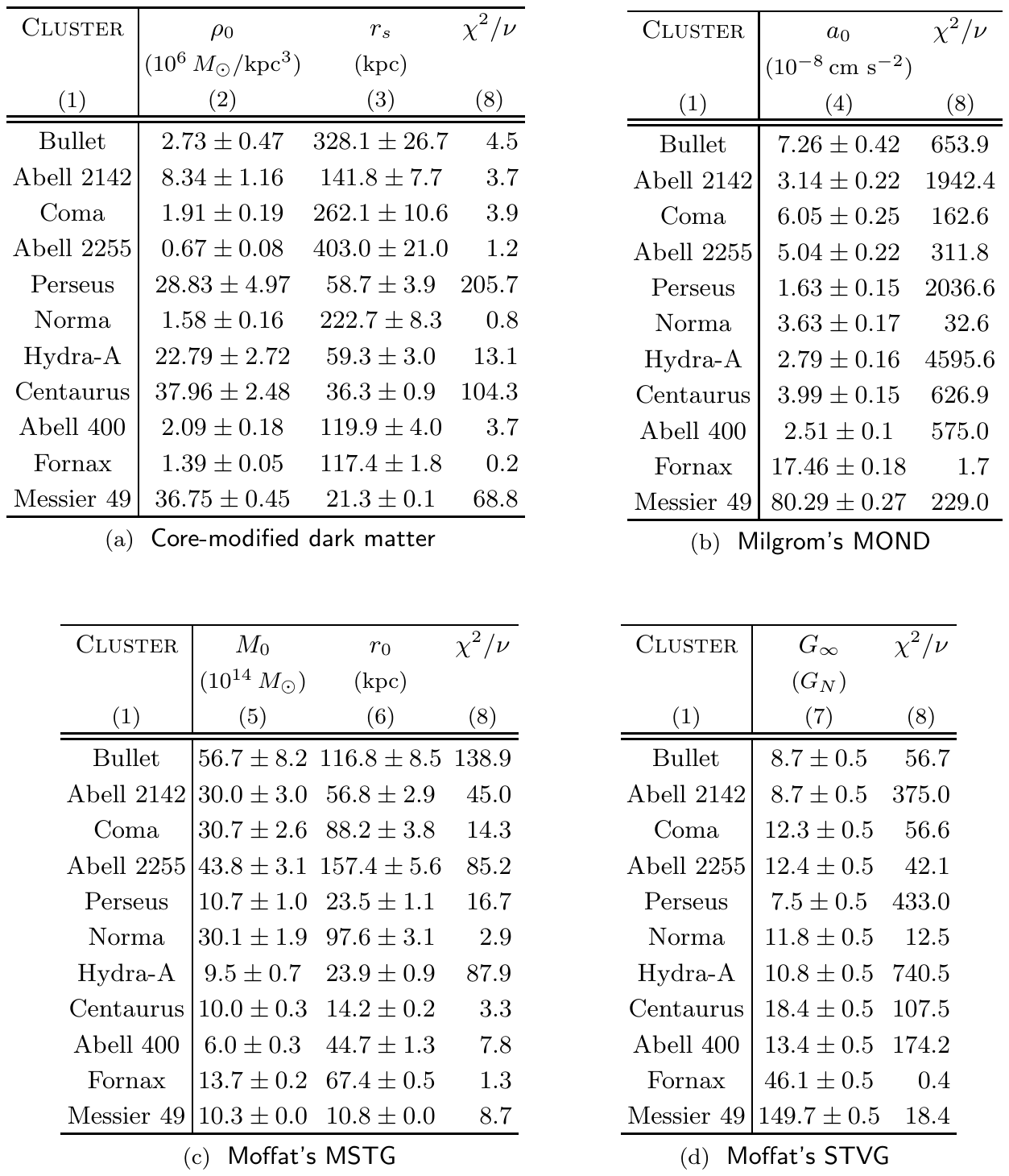}}}
\end{picture}
\parbox{6.375in}{\small Notes. --- Best-fitting parameters of the X-ray cluster sample for (a) Core-modified dark matter, (b) Milgrom's MOND, (c) Moffat's MSTG and (d) Moffat's STVG: Column (1) is the cluster name.  Columns (2) and (3) list the best-fit parameters for the core-modified universal fitting formula of \eref{eqn.newton.darkmatter.coremodified}.  Column (4) list the best-fit MOND acceleration parameter.  Columns (5) and (6) list the best-fit MSTG mass and range parameters, respectively.  Column (7) list the best-fit STVG asymptotic gravitationally coupling.  Column (8) list the reduced-\(\chi^2\) statistic of \eref{eqn.galaxy.uma.chi2} per gravity theory.}
\end{table}

The core-modified dark matter halo, described in \sref{section.cluster.models.darkmatter}, provides an alternative which does not suffer from the cusp problem of the singular NFW fitting formula and does fit high and low surface brightness and dwarf galaxies to low \(\chi^2\).  Alternatively, Milgrom's MOND as in \sref{section.cluster.models.mond} and Moffat's MOG as in \sref{section.cluster.models.mog} are to be explored, and the means by which each of the candidate theories provides best-fit cluster models are compared in \sref{section.cluster.models.mass}.  The common and unusual features of the fits, across the sample of clusters of galaxies, are presented in \ssref{section.cluster.models.mass}{subsection.cluster.models.mass.observations} -- with statistics provided in \tref{table.cluster.models.bestfit}.

\subsection{\label{section.cluster.models.darkmatter}Core-modified dark matter halos}\index{Dark matter!Core-modified|(}

The simplicity of the dark matter paradigm allows predictions which justifiably ignore the X-ray surface mass.  However, it has been known since \citet{Bahcall.ARAA.1977.15,Hoffman.APJ.1985.297} that the halo density profile of virialized clusters of galaxies cannot be fitted by a single power law,
\begin{equation} \label{eqn.cluster.xraymass.dm.powerlaw} 
\rho(r) \propto r^{-\gamma},
\end{equation}
but \(\gamma\) seems to increase with \(r\).  The NFW fitting formula of \eref{eqn.newton.darkmatter.nfw} bridges this behaviour of \eref{eqn.cluster.xraymass.dm.powerlaw} with a singular, cuspy core, \(\gamma \rightarrow 1\) at small \(r\), and \(\gamma\rightarrow 3\) at large r.  

This point has led to controversy over the cuspy shape of the singular NFW profile, which appears as a robust prediction of N-body simulations without baryons, but is not actually observed (with low \(\chi^2\)) in X-ray cluster data.  Conversely, \citet{Ettori.AAP.2002.391} found 2 X-ray clusters out of a sample of 22 clusters that could not be \(\chi^2\) fitted to the NFW profile at all.  Further Chandra studies of the cores of clusters, such as \citet{Ettori.MNRAS.2002.331}, indicated that the NFW profile is only reliable outside the cluster core, as is the case in certain low surface brightness and dwarf galaxies.  \citet{Sand.APJ.2004.604} studied the dark matter distribution in the central region of 6 clusters of galaxies by combining constraints from gravitational lensing and the  stellar velocity dispersion profile of the brightest central galaxy, confirming that the core behaviour is statistically inconsistent with a singular NFW profile, and that the inclusion of baryonic matter affects the dark matter distribution not accounted for in conventional CDM simulations.

In a gravitational lensing study of two X-ray clusters of galaxies, \citet{Smail.MNRAS.1995.273} found that the Newtonian dynamic mass is distributed similarly to the visible baryonic mass with the same core radius, but while it is more concentrated at the center, it is less cuspy than CDM predictions.  \citet{Tyson.APJL.1998.498,Shapiro.APJL.2000.542} argued that the singular density profiles based on NFW fitting formula are in apparent conflict with the observed mass distributions inside dark matter dominated halos on two extremes of the halo mass function -- dwarf galaxies and clusters of galaxies; each of which is better described by a less cuspy or constant density core.    

The core-modified fitting formula of \eref{eqn.newton.darkmatter.coremodified} bridges the behaviour of \eref{eqn.cluster.xraymass.dm.powerlaw} with a constant-density core, \(\gamma\rightarrow 0\) at small \(r\) but \(\gamma\rightarrow 3\) at large r, which fits the high and low surface brightness galaxies in the Ursa Major sample of \sref{section.galaxy.uma} including all of the dwarf galaxies.   The dark matter power-law profile, plotted in \fref{figure.galaxy.powerlaw}, confirms that the variation in the exponent of \eref{eqn.cluster.xraymass.dm.powerlaw} agrees with CDM predictions, provided the visible baryonic components are not neglected.  It is therefore important to test the core-modified dark matter fitting formula of \eref{eqn.newton.darkmatter.coremodified} at the scale of X-ray clusters.

\citet{Arieli.NA.2003.8} compared the best-fits to a sample of 24 X-ray clusters of galaxies, and concluded that a core-modified dark matter profile of the form of \eref{eqn.newton.darkmatter.coremodified} is statistically more consistent with ROSAT observational results than either the NFW profile of \eref{eqn.newton.darkmatter.nfw} or a family of simple polytropic fitting formulae.

Whereas, attempts to fit cluster mass distributions to NFW profiles lead to large uncertainties due to a parameter degeneracy between the central density parameter, \(\rho_{0}\), and the scale radius, \(r_{s}\), which prevented the computation of the best-fit \(\rho_{0}\) and \(r_{s}\) from converging, regardless of the  \(\chi^2\).  Without numerical convergence, the NFW results either over-predict the density at the core or under-predict the total mass.  However, the core-modified fitting formula of \eref{eqn.newton.darkmatter.coremodified} provides excellent fits with low \(\chi^2\) to the mass profiles, plotted in \fref{figure.cluster.models.mass}, and a reasonable explanation of the variation in the dynamic mass factors, plotted in \fref{figure.cluster.models.Gamma}, providing one solution to the missing mass problem, presented in \sref{section.cluster.models.mass}.\index{Dark matter!Core-modified|)}

\subsection{\label{section.cluster.models.mond}Milgrom's MOND without dark matter}\index{MOND|(}

\citet{Brownstein:MNRAS:2006} predicted convergent MOND X-ray surface brightness profiles which did not match any observed distributions of a sample~\protect\citet{Reiprich:2001,Reiprich:2002} of 106 X-ray clusters.  Without treating the MOND acceleration, \(a_{0}\), as a free parameter as opposed to a universal constant, or considering improved but as yet undiscovered MOND interpolating functions, MOND cannot account for the observed X-ray luminosities without the addition of an unseen component to explain away the missing mass.  \citet{Sanders.MNRAS.2003.342,Sanders:2007MNRAS.380..331S} considered adding a neutrino halo, modelled as a nonluminous constant density rigid sphere, discussed in \sref{section.cluster.bullet.neutrino}.  

Conversely, \citet{The.AJ.1988.95} were able to decrease the MOND discrepancy between the X-ray observationally determined gas mass and the X-ray surface brightness of the Coma cluster by increasing the MOND acceleration by a factor of four greater than \eref{eqn.galaxy.mond.a0}.  However, \citet{Aguirre.APJ.2001.561} presented evidence from the central 200 kpc of three clusters which inflates the discrepancy in the MOND acceleration to a factor of \(\sim\) 10.  More recently, \citet{Pointecouteau:MNRAS:2005} used X-ray data from the {\sc XMM-Newton} satellite for eight clusters of varying temperature and masses to place constraints on the general use of MOND phenomenology.

Furthermore, every galaxy rotation curve that produced a weak fitting MOND one-parameter best-fit by a variable stellar mass-to-light ratio, \(\Upsilon\), plotted in \fref{figure.galaxy.velocity} for the Ursa Major sample of \sref{section.galaxy.uma}, shows dramatic improvement and reduction in the reduced \(\chi^2/\nu\) statistic using a two-parameter best-fit including a variable MOND acceleration parameter.  The tabulation of \(a_0\) in Column (2) of \tref{table.galaxy.mond}  provides no statistical support that \(a_{0}\) is a universal constant due to gross uncertainties in the mean results of \eref{eqn.galaxy.mond.a0.subsample}.

A varying choice of the MOND interpolating function, including those of \citet{Bekenstein:PRD:2004} and \citet{Famaey.MNRAS.2005.363}, does not lead to significant improved behaviour since \(a(r) < a_{0}\) or \(x<1\) at all radii within clusters of galaxies.\index{MOND!Interpolating function, \(\mu\)}

Therefore the alternatives for MOND are either add two additional parameters (or scaling relations) per system to include a dark matter component, or to determine if sensible fits are possible without dark matter using a one-parameter best-fit by a variable acceleration parameter.  The absence of a universal acceleration parameter violates the notion that MOND is a fundamental theory, but the notion of a scale dependent acceleration parameter may be a dynamic, more natural effect of a covariant, but Lorentz-violating theory with a preferred frame, as in \sref{section.mog.mond.aether} and is not inconsistent with Bekenstein's TEVES action, as in \sref{section.mog.mond.aether.bekenstein}.\index{MOND|)}

\subsection{\label{section.cluster.models.mog}Moffat's MOG with running couplings}

In the absence of non-baryonic dark matter, the modified gravity dynamical mass may be obtained as a function of radial position by substituting the MOG acceleration law of \eref{eqn.mog.mstg.mog.Gforce} -- with a varying gravitational coupling, \(G(r)\) -- so that the result for the isothermal $\beta$-model of  \eref{eqn.cluster.xraymass.isothermalNewtonsMass} becomes
\begin{equation}
\label{eqn.cluster.models.mog.mass}
M_{\rm MOG}(r) = \frac{3\beta kT}{\mu_{A} m_{p}G(r)} \left(\frac{r^{3}}{r^{2}+r_{c}^{2}}\right).
\end{equation}

\citet{Brownstein:MNRAS:2006} predicted X-ray surface brightness profiles from X-ray luminosity observations  consistent with the observed X-ray gas distributions of a sample of 106 X-ray clusters~\protect\citep{Reiprich:2001,Reiprich:2002} using the modified acceleration law based upon metric skew-tensor gravity, as in \sref{section.mog.mstg}.

\subsubsection{\label{section.cluster.models.mog.mstg}Metric skew-tensor gravity}\index{Modified gravity!Metric skew-tensor gravity|(}

The MSTG dynamic mass is obtained by substituting \(G(r)\) of \eref{eqn.mog.mstg.mog.fullGspherical} into  \eref{eqn.cluster.models.mog.mass} and may be written explicitly as a function of the Newtonian dynamic mass of \eref{eqn.cluster.xraymass.isothermalNewtonsMass} and two parameters, \(M_{0}\) and \(r_{0}\):
\begin{equation}\label{eqn.cluster.models.mog.mstg.soln}
M_{\rm MSTG}(r)=M_{N}(r) + M_{0}\xi(r) - \sqrt{{M_{0}}^2 \xi(r)^2+2 M_{0} M_{N}(r) \xi(r)},
\end{equation}\begin{equation}
\label{eqn.cluster.models.mog.mstg.xi}
\xi(r)\equiv\frac{1}{2}\left[{1-\exp(-r/r_0)\biggl(1+\frac{r}{r_0}\biggr)}\right]^2,
\end{equation}
which are MSTG mass and range parameters, respectively.  However, it is not possible to fit any of the clusters of galaxies with the MSTG mass and range parameters of \eref{eqn.galaxy.dynamics.mstg.parameters.meanuniversal}, which were applied universally to high and low surface brightness galaxies including all of the dwarfs, in the Ursa Major sample of \sref{section.galaxy.uma}, with galaxy rotation curves plotted in \fref{figure.galaxy.velocity}.  Whereas every weak fitting MSTG one-parameter best-fit by a variable stellar mass-to-light ratio, \(\Upsilon\),  shows dramatic improvement and reduction in the reduced \(\chi^2/\nu\) statistic using a three-parameter best-fit including variable MSTG mass and range parameters, the tabulation of \(M_0\)  and \(r_0\) in Column (2) and (3) of \tref{table.galaxy.mstg} provides no statistical support that the MSTG parameters are universal constants, but does provide very strong statistical support that the MSTG parameters are scale dependent.

\citet{Brownstein:MNRAS:2006} used an empirically determined power-law scale relation to set the MSTG mass scale  parameter,
\begin{equation} \label{eqn.cluster.models.mog.prescription.M0}
M_{0} = (60.4 \pm 4.1) \times 10^{14} M_{\solar} \left(\frac{M_{\rm gas}}{10^{14} M_{\solar}}\right)^{0.39 \pm
0.10},
\end{equation}
where \(M_{\rm gas}\), given by \eref{eqn.cluster.xraymass.isothermal.Mgas.0}, is the mass of the ICM integrated to the distance at which the density drops to $\approx 10^{-28}\,\mbox{g/cm}^{3}$, or 250 times the mean cosmological density.

In order to better determine the scale dependence of the parameters, it is reasonable to treat the MSTG mass and range parameters as variable and to perform two-parameter best-fits to the X-ray gas masses of the sample of 11 clusters of galaxies, using \eref{eqn.cluster.models.mog.prescription.M0} as initial value only. The mass profiles are plotted in \fref{figure.cluster.models.mass} according to the best-fit cluster model parameters tabulated in Panel (c) of \tref{table.cluster.models.bestfit}, for MSTG.\index{Modified gravity!Metric skew-tensor gravity|)}

\subsubsection{\label{section.cluster.models.mog.stvg}Scalar-tensor-vector gravity}\index{Modified gravity!Scalar-tensor-vector gravity|(}

\citet{Moffat.CQG.2009.26} investigated a fundamental parameter-free solution to the running couplings using the modified acceleration law based upon scalar-tensor-vector gravity, as in \sref{section.mog.stvg}.  The STVG dynamic mass of \eref{eqn.mog.mstg.mass.stvg} may be written as a function of the Newtonian dynamic mass of \eref{eqn.cluster.xraymass.isothermalNewtonsMass} and two functions \(\alpha(r)\) and \(\mu(r)\) which are derived from an action principle, with the equations of motion given by \erefs{eqn.mog.stvg.mog.alpha}{eqn.mog.stvg.mog.mu}, respectively, in terms of three constants of integration, \(D\), \(E\), and \(G_{\infty}\).  

However, it is not possible to fit any of the clusters of galaxies with the values of \eref{eqn.galaxy.dynamics.stvg.parameters.meanuniversal}, which were applied universally to high and low surface brightness galaxies including all of the dwarfs, in the Ursa Major sample of \sref{section.galaxy.uma}, with galaxy rotation curves plotted in \fref{figure.galaxy.velocity}.  Whereas every weak fitting STVG one-parameter best-fit by a variable stellar mass-to-light ratio, \(\Upsilon\),  show dramatic improvement and reduction in the reduced \(\chi^2/\nu\) statistic using a four-parameter best-fit including variable parameters, the tabulation of \(D\), \(E\), and \(G_{\infty}\) in Columns (2), (3) and (4) of \tref{table.galaxy.stvg} provides no statistical support that the STVG integration constants are universal.

For values of \(D\) sufficiently large and values of \(E\) sufficiently small, the STVG gravitational coupling of \eref{eqn.mog.stvg.yukawa.Geff} simplifies to its asymptotic form,
\begin{equation} \label{eqn.cluster.models.stvg.asymptotic} 
G(r) = G_{\infty},
\end{equation}
and is independent of \(r\).  Substituting this form of the gravitational coupling into \eref{eqn.cluster.models.mog.mass}, we obtain the STVG dynamic mass for clusters of galaxies:
\begin{equation}
\label{eqn.cluster.models.mog.mass.stvg}
M_{\rm STVG}(r) = \frac{3\beta kT}{\mu_{A} m_{p}G_{\infty}} \left(\frac{r^{3}}{r^{2}+r_{c}^{2}}\right).
\end{equation}
Therefore, in order to determine the scale dependence of the STVG asymptotic coupling, it is reasonable to treat \(G_\infty\) as variable and to perform one-parameter best-fits to the X-ray gas masses of the sample of 11 clusters of galaxies,  plotted in \fref{figure.cluster.models.mass} according to the best-fit cluster models parameters tabulated in Panel (d) of \tref{table.cluster.models.bestfit}, STVG.\index{Modified gravity!Scalar-tensor-vector gravity|)}

\subsection{\label{section.cluster.models.mass}The missing mass problem}\index{Dark matter!Missing mass problem|(}

\addtocontentsheading{lof}{X-ray clusters of galaxies}
\newcommand{\subclustermass}{\small The radial mass profile, \(M(r)\) in \(M_{\solar}\), vs. orbital distance, \(r\) in kpc}
\begin{figure}[ht]\index{Dark matter!Mass profile|(}\index{Modified gravity!Mass profile|(}\index{MOND!Mass profile|(}
\begin{picture}(460,185)(0,0)
\put(0,40){\includegraphics[width=0.48\textwidth]{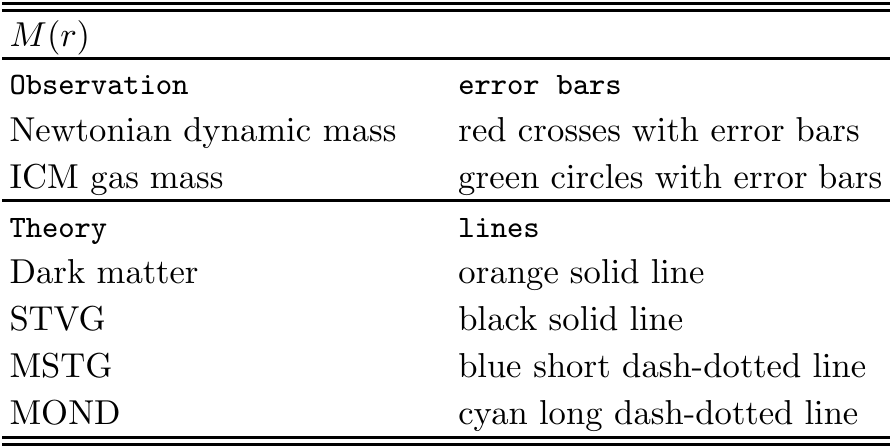}}
\put(225,0){\includegraphics[width=0.5\textwidth]{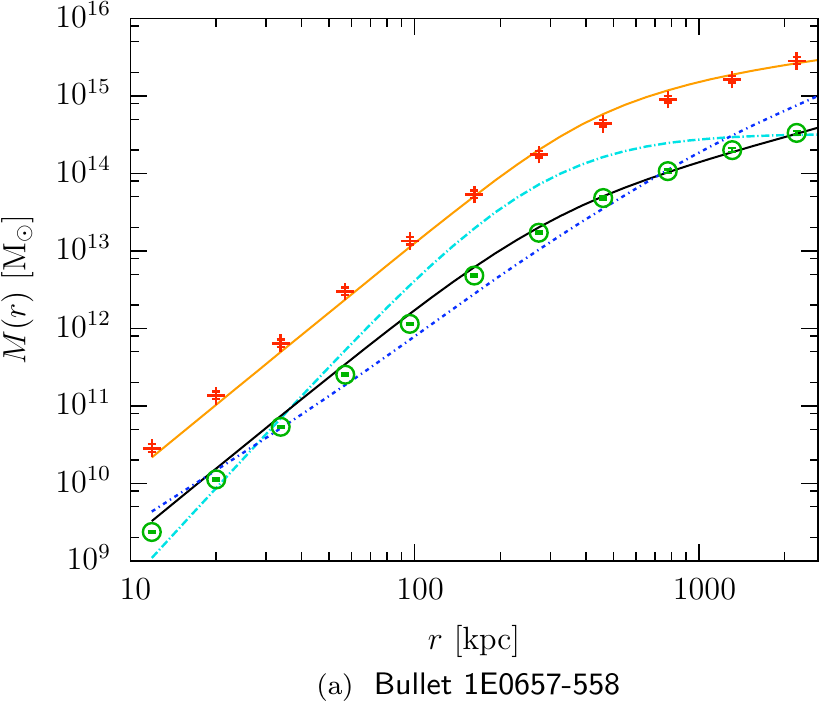}}
\end{picture}
\caption[Mass profiles]{\label{figure.cluster.models.mass} {{\sf\small X-ray clusters of galaxies -- Mass profiles.}}\break\break{\subclustermass} for a sample of  X-ray clusters.  The dynamic data consist of the Newtonian dynamic mass of \eref{eqn.cluster.xraymass.isothermalNewtonsMass}, due to the measured isothermal temperature.  The observed ICM gas masses are derived from \erefs{eqn.cluster.xraymass.isothermal.betaRhoModel}{eqn.cluster.xraymass.isothermal.massProfile} using the best-fit King \(\beta\)-model parameters listed in \tref{table.cluster.sample}.  The computed best-fitted results are plotted for Moffat's STVG and MSTG theories and Milgrom's MOND theory with variable parameters.  Results are plotted for the best-fit core-modified dark matter theory including the X-ray gas mass component.  The reduced-\(\chi^2\) statistic is included in \tref{table.cluster.models.bestfit}.  {\it The figure is continued.}}
\end{figure}

\begin{figure}
\begin{picture}(460,450)(82,190)
\put(30,12){\includegraphics[width=1.28\textwidth]{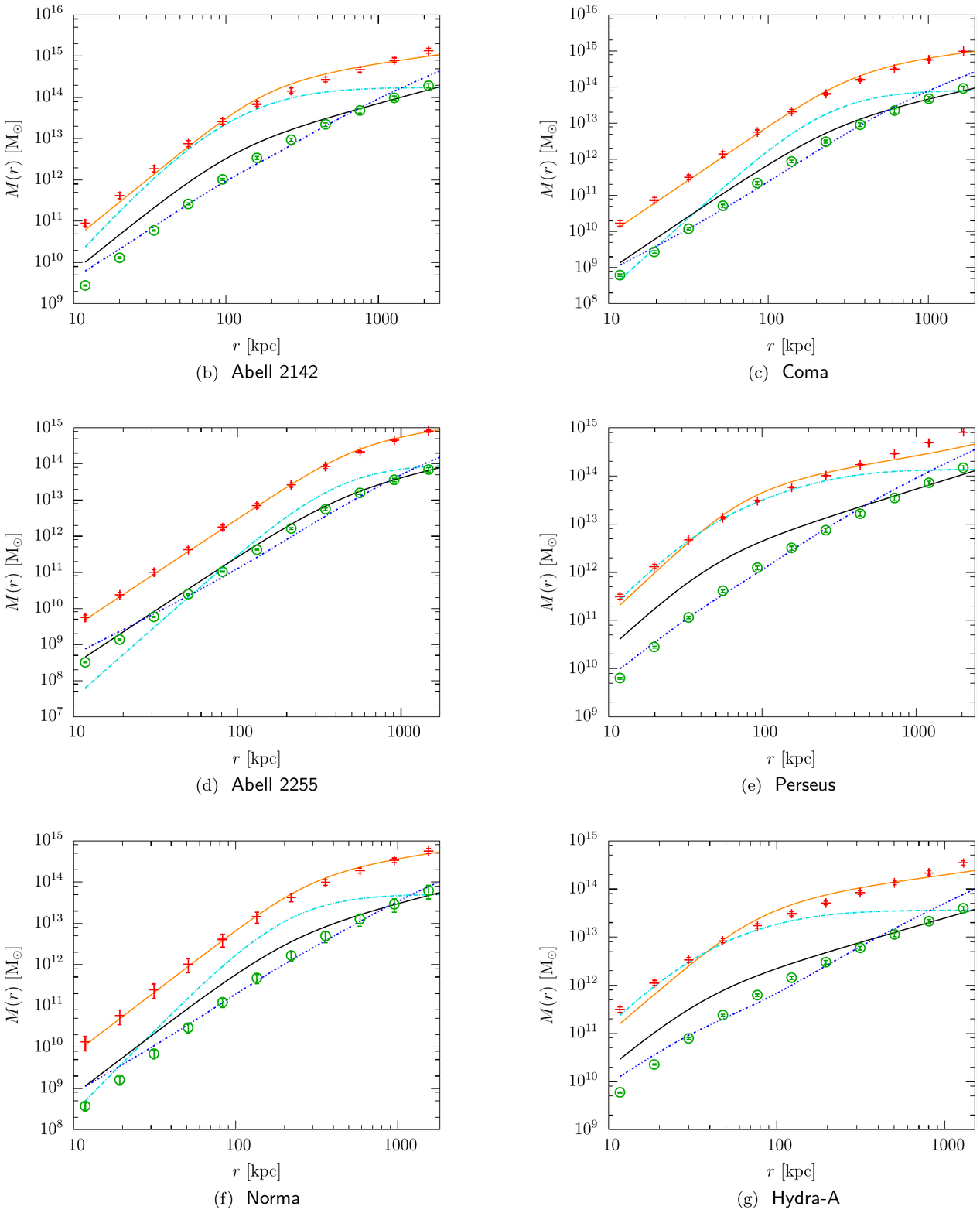}}
\put(82,45){\includegraphics[width=0.98\textwidth]{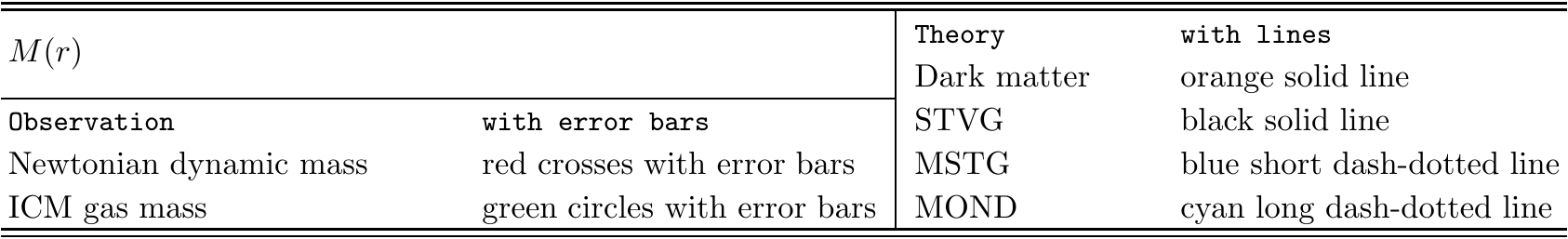}}
\end{picture}
\fcont{figure.cluster.models.mass}{\sf\small X-ray clusters of galaxies -- Mass profiles.}
{\subclustermass}.
\end{figure}
\begin{figure}
\begin{picture}(460,290)(82,335) 
\put(30,12){\includegraphics[width=1.28\textwidth]{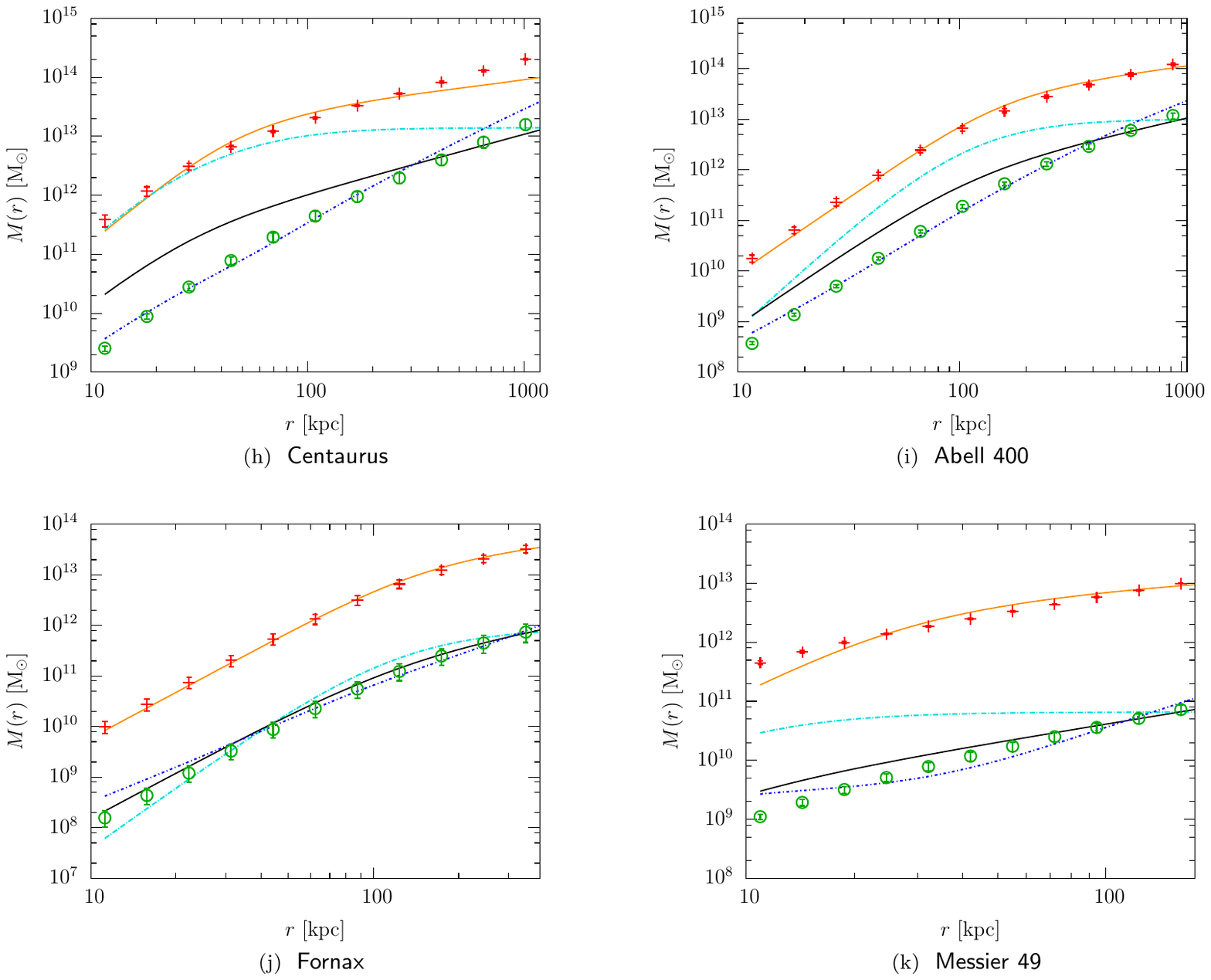}}
\put(82,45){\includegraphics[width=0.98\textwidth]{figure/cluster_models_mass_legend}}
\end{picture}
\fcont{figure.cluster.models.mass}{\sf\small X-ray clusters of galaxies -- Mass profiles.\break}{\subclustermass} for a sample of  X-ray clusters.  The dynamic data consist of the Newtonian dynamic mass of \eref{eqn.cluster.xraymass.isothermalNewtonsMass}, due to the measured isothermal temperature.  The observed ICM gas masses are derived from \erefs{eqn.cluster.xraymass.isothermal.betaRhoModel}{eqn.cluster.xraymass.isothermal.massProfile} using the best-fit King \(\beta\)-model parameters listed in \tref{table.cluster.sample}.  The computed best-fitted results are plotted for Moffat's STVG and MSTG theories and Milgrom's MOND theory with variable parameters.  Results are plotted for the best-fit core-modified dark matter theory including the X-ray gas mass component.  The reduced-\(\chi^2\) statistic is included in \tref{table.cluster.models.bestfit}.\index{MOND!Mass profile|)}\index{Modified gravity!Mass profile|)}\index{Dark matter!Mass profile|)}
\end{figure}

The Newtonian dynamical mass of \eref{eqn.cluster.xraymass.isothermalNewtonsMass} is a derived relation between the density profile for the X-ray gas component, according to the isotropic isothermal model of \sref{section.cluster.xraymass.isothermal}, and the measured isothermal temperature, \(T\).  The ICM gas mass is a spherical integral of the King \(\beta\)-model of \erefs{eqn.cluster.xraymass.isothermal.betaRhoModel}{eqn.cluster.xraymass.isothermal.massProfile}.  \tref{table.cluster.sample} includes the total ICM gas mass and total Newtonian dynamical mass within the position, $r_{\rm out}$, at which the density, $\rho(r_{\rm out})$, drops to $\approx 10^{-28}\,\mbox{g/cm}^{3}$, or 250 times the mean cosmological density of baryons. The total fraction of ICM gas mass is between 1\% and 20\% of the total Newtonian dynamic mass, and is typically 10\%, as demonstrated in \sref{section.cluster.xraymass.astroph}.  Therefore, according to Newtonian dynamics, between 80\% to 99\% of the mass needed to explain the isothermal profiles is missing.

However, whereas the solution that there is just enough dark matter to fill the total difference is consistent with the NFW fitting formula of \citet{Navarro.APJ.1996.462,Navarro.APJ.1997.490}, the cusped profile does not correctly fit the shape of the dynamic mass profile.  \citet{Arieli.NA.2003.8} suggested that there is a clear need to explore modifying the NFW profile, which has been adopted in  hydrodynamic N-body simulations of the structure and evolution of \(\Lambda\)-CDM halos, or finding an alternative which provides a reasonable fit to the X-ray cluster masses.  

Similarly, it is not enough for any gravity theory to solve the missing mass problem, in the absence of dark matter, without providing a reasonable fit to the observed X-ray gas mass distribution for each cluster.  The mass profiles of \fref{figure.cluster.models.mass} plot the Newtonian dynamical mass and observed ICM gas mass profiles, including the best-fits resulting from Moffat's STVG and MSTG theories, Milgrom's MOND theory with variable parameters.  Results are plotted for the best-fit core-modified dark matter theory including the X-ray gas mass component.   The reduced-\(\chi^2\) statistic is included in \tref{table.cluster.models.bestfit}, and reveals that the dark matter solutions of \sref{section.cluster.models.darkmatter} and the modified gravity solutions of \sref{section.cluster.models.mog} are reasonable, although the MOND solution without dark matter of \sref{section.cluster.models.mond} is wrong -- and a variable MOND acceleration parameter only allows a correct fit to the total cluster mass.  This has prompted \citet{Sanders.MNRAS.2003.342,Sanders:2007MNRAS.380..331S} to consider the possibility of 2 eV neutrino halos as providing the missing 80\% to 99\% of cluster dark matter, but \citet{Angus.MNRAS.2008.387} showed that MOND-neutrino-baryon models will not provide reasonable fits to the X-ray gas mass profile, particularly in the inner 100 to 150 kiloparsecs of the cluster.  This neutrino halo hypothesis is explored in \sref{section.cluster.bullet.neutrino} as part of the analysis of the strong and weak lensing map of the {\bc}, presented in \sref{section.cluster.bullet}.

\citet{Biviano.AAP.2006.452} derived mass profiles of the different luminous and dark components of 59 X-ray clusters of galaxies and confirmed that the baryonic components are relevant to mass models of clusters of galaxies both near the center because of the substantial contribution from the central dominant galaxy and in the outer regions, because of the increasing mass fraction of the ICM gas -- and the corresponding decreasing dynamic mass factor.  Therefore the missing mass problem is most serious in the core of galaxy clusters, in complete opposition to the situation in the galaxy rotation curves of \cref{chapter.galaxy} where the dynamical mass factors of \fref{figure.galaxy.Gamma} show a maximum at the outermost observed radial position, for each gravity theory.

The dynamical mass factors, plotted in \fref{figure.cluster.models.Gamma}, show the ratio of the Newtonian dynamical mass to the observed ICM gas mass, including the best-fits resulting from Moffat's STVG and MSTG theories, Milgrom's MOND theory with variable \(a_0\), and the best-fit core-modified dark matter theory including the X-ray gas mass component, where the results of \tref{table.cluster.models.bestfit} were used, respectively.

\newcommand{\subclusterGamma}{\small The dynamic mass factors, \(\Gamma(r)\), vs. orbital distance, \(r\) in kpc}
\begin{figure}[ht]\index{Dynamic mass factor, \(\Gamma\)|(}
\begin{picture}(460,185)(0,0)
\put(0,40){\includegraphics[width=0.48\textwidth]{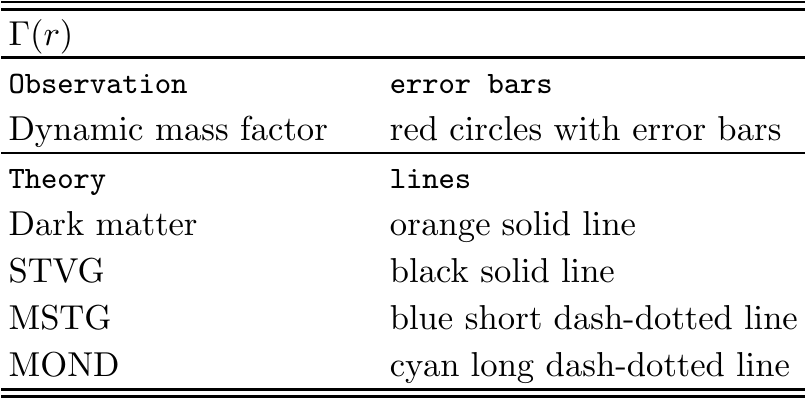}}
\put(225,0){\includegraphics[width=0.5\textwidth]{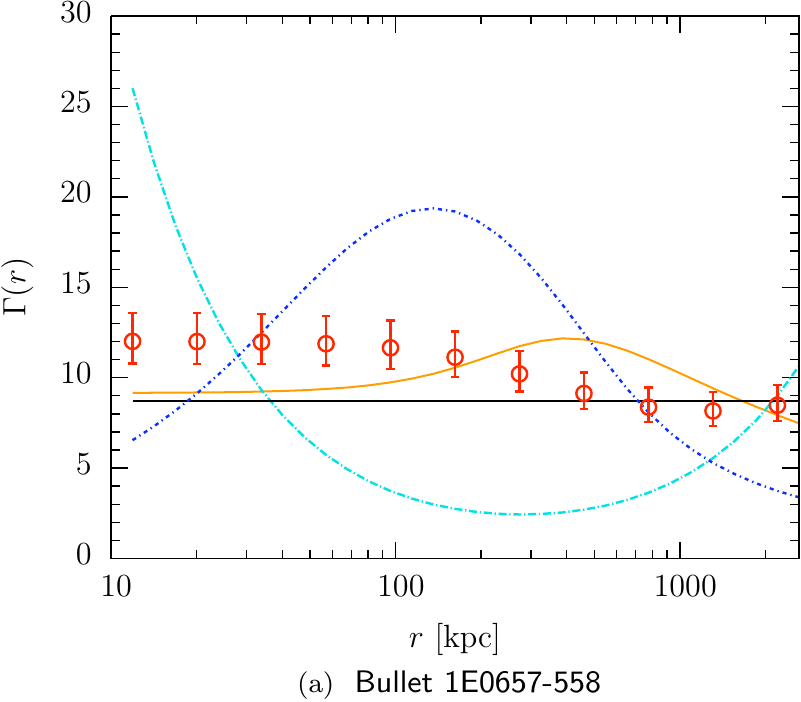}}
\end{picture}
\caption[Dynamic mass factors]{\label{figure.cluster.models.Gamma} {{\sf\small X-ray clusters of galaxies -- Dynamic mass factors.}}\break\break{\subclusterGamma} for a sample of  X-ray clusters.  The dynamic data consist of the ratio of the Newtonian dynamic mass of \eref{eqn.cluster.xraymass.isothermalNewtonsMass}, due to the measured isothermal temperature, to the integrated X-ray gas mass, derived from \erefs{eqn.cluster.xraymass.isothermal.betaRhoModel}{eqn.cluster.xraymass.isothermal.massProfile} using the best-fit King \(\beta\)-model parameters listed in \tref{table.cluster.sample}.  The computed best-fitted results are plotted for Moffat's STVG and MSTG theories and Milgrom's MOND theory with variable parameters.  Results are plotted for the best-fit core-modified dark matter theory including the X-ray gas mass component.  The reduced-\(\chi^2\) statistic is included in \tref{table.cluster.models.bestfit}.  {\it The figure is continued.}}
\end{figure}

\begin{figure}
\begin{picture}(460,450)(82,190)
\put(30,12){\includegraphics[width=1.28\textwidth]{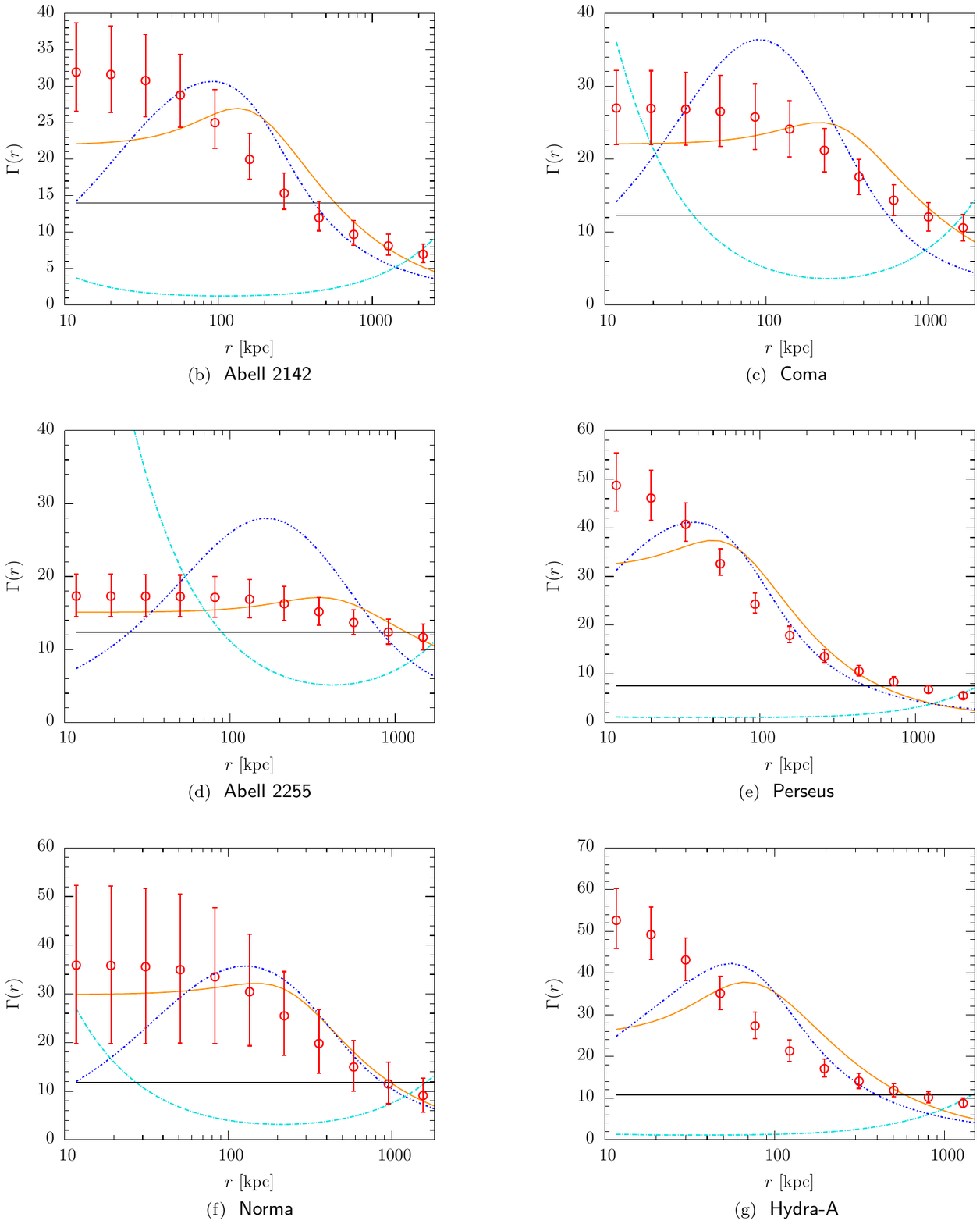}}
\put(82,45){\includegraphics[width=0.98\textwidth]{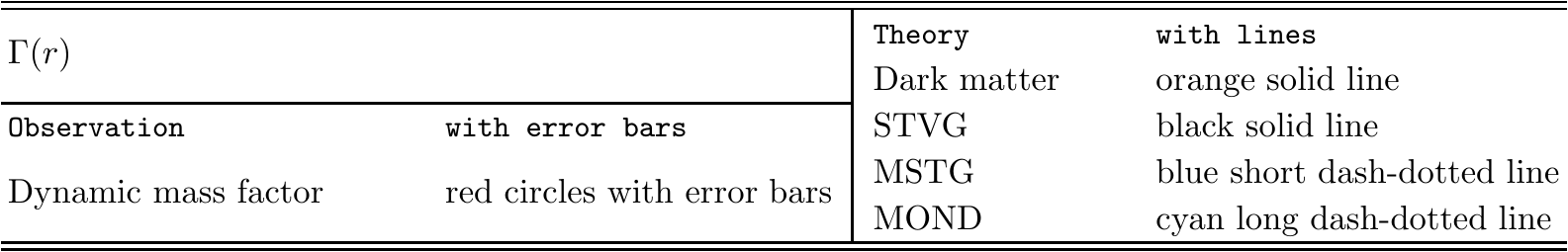}}
\end{picture}
\fcont{figure.cluster.models.Gamma}{\sf\small X-ray clusters of galaxies -- Mass profiles.}
{\subclusterGamma}.
\end{figure}
\begin{figure}
\begin{picture}(460,290)(82,335) 
\put(30,12){\includegraphics[width=1.28\textwidth]{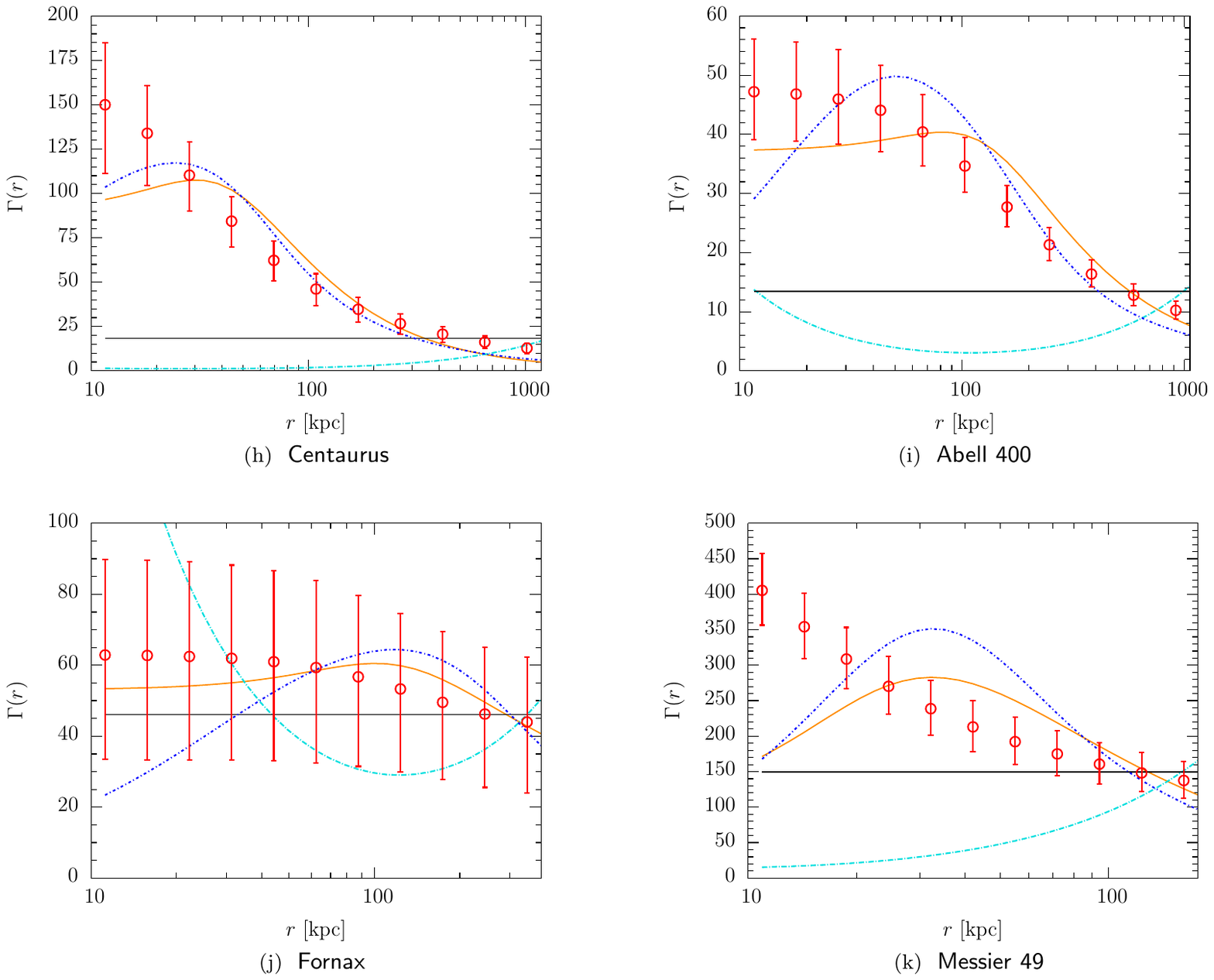}}
\put(82,45){\includegraphics[width=0.98\textwidth]{figure/cluster_models_Gamma_legend}}
\end{picture}
\fcont{figure.cluster.models.Gamma}{\sf\small X-ray clusters of galaxies -- Dynamic mass factors.\break}{\subclusterGamma} for a sample of  X-ray clusters.  The dynamic data consist of the ratio of the Newtonian dynamic mass of \eref{eqn.cluster.xraymass.isothermalNewtonsMass}, due to the measured isothermal temperature, to the integrated X-ray gas mass, derived from \erefs{eqn.cluster.xraymass.isothermal.betaRhoModel}{eqn.cluster.xraymass.isothermal.massProfile} using the best-fit King \(\beta\)-model parameters listed in \tref{table.cluster.sample}.  The computed best-fitted results are plotted for Moffat's STVG and MSTG theories and Milgrom's MOND theory with variable parameters.  Results are plotted for the best-fit core-modified dark matter theory including the X-ray gas mass component.  The reduced-\(\chi^2\) statistic is included in \tref{table.cluster.models.bestfit}.\index{Dynamic mass factor, \(\Gamma\)|)}
\end{figure}

For each cluster, the substitute of missing mass in MOND is the wrong shape, with only the correct total mass predicted due to a variable, best-fit MOND acceleration. For \(r<r_{\rm out}\), the dynamic mass factor predicted by MOND is much smaller than observed leading to too great a predicted gas mass in these regions.  For some of the clusters such as the {\bc}, Abell 2255 and Fornax, this trend is suddenly reversed for \(r<100\) kpc, where MOND predicts a dynamic mass factor which diverges strongly (as does the cuspy NFW profile not shown), but is not actually observed in the data even though the coolest of the clusters such as Messier 49 show a dynamic mass factor as large as \(\Gamma \rightarrow 400\) as \(r \rightarrow 0\).\index{Dark matter!Cusp problem}

Unlike the NFW fitting formula of \citet{Navarro.APJ.1996.462,Navarro.APJ.1997.490}, the core-modified dark matter halos provide the means to fit X-ray masses with constant density cores.  This solution provides missing mass in line with the observations plotted in \fref{figure.cluster.models.Gamma} at all radial positions.

Moffat's MOG theories provide a remarkable picture of the missing mass problem, even though the galactic mass components have been neglected, which are dynamically important in MOG due to the absence of dominant dark matter and the increased weight due to the larger than Newtonian gravitational coupling.  These MOG effects due to  visible baryons are explored in greater detail in \sref{section.cluster.bullet.baryon} as part of the analysis of the strong and weak lensing map of the {\bc}, presented in \sref{section.cluster.bullet}.  

\subsubsection{\label{subsection.cluster.models.mass.observations}Observations}

In the case of the Ursa Major sample of high and low surface brightness galaxies, the dynamic mass factors of \sref{section.galaxy.uma.Gamma} 
\begin{equation}
\label{eqn.cluster.models.mass.Gamma} \Gamma(r) = \frac{G(r)}{G_N}
\end{equation}
are monotonically increasing, nearly linear functions, plotted in \fref{figure.galaxy.Gamma}.  This is a prediction of the modified dynamics at small accelerations, of \sref{section.mog.mond.dynamic}, where the slope is determined by \eref{eqn.mog.mond.aether.dynamic.Gamma.deep} to be the inverse of the transition radius,\index{Gravitational lensing!Modified dynamics}
\begin{equation}
\label{eqn.cluster.models.mass.Gamma.slope.galaxy} \frac{d\Gamma(r)}{dr} = {r_t}^{-1} = \sqrt{\frac{a_0}{G_N M}},
\end{equation}
where \(a_0\) is the transition acceleration.

For the best-fit cluster models of \sref{section.cluster.models}, the dynamic mass factors plotted in \fref{figure.cluster.models.Gamma}, show very different trends, never showing a monotonically linear rise as in \eref{eqn.cluster.models.mass.Gamma.slope.galaxy}.

For each of the clusters of galaxies in the sample, \(\Gamma(r) \gg r\) for all \(r\), having the greatest magnitude in the cores of the smaller (cooler) clusters, in particular Messier 49.  The slope 
\begin{equation}
\label{eqn.cluster.bullet.kappa.Gamma.slope.cluster} \frac{d\Gamma(r)}{dr} \lesssim 0,
\end{equation}
is close to flat for the larger (hotter) clusters, in particular the {\bc} and Abell 2255, but generally having the greatest magnitude outside the cores of the smaller (cooler) clusters.  All of the cluster cores have a particularly slow varying slope.  Therefore \eref{eqn.cluster.bullet.kappa.Gamma.slope.cluster} suggests that clusters of galaxies are observationally inconsistent with singular (cuspy) models.  For the {\bc}, the relationship between the X-ray observed \map{\Sigma} and the gravitational lensing \map{\kappa} is discussed in \sref{section.cluster.bullet.kappa}.\index{Dark matter!Missing mass problem|)}

\section{\label{section.cluster.bullet}{\bc}}

\advancecontents{lof}\addtocontentsheading{lof}{\bc}
\begin{SCfigure}[0.9][h]
\includegraphics[width=0.5\textwidth]{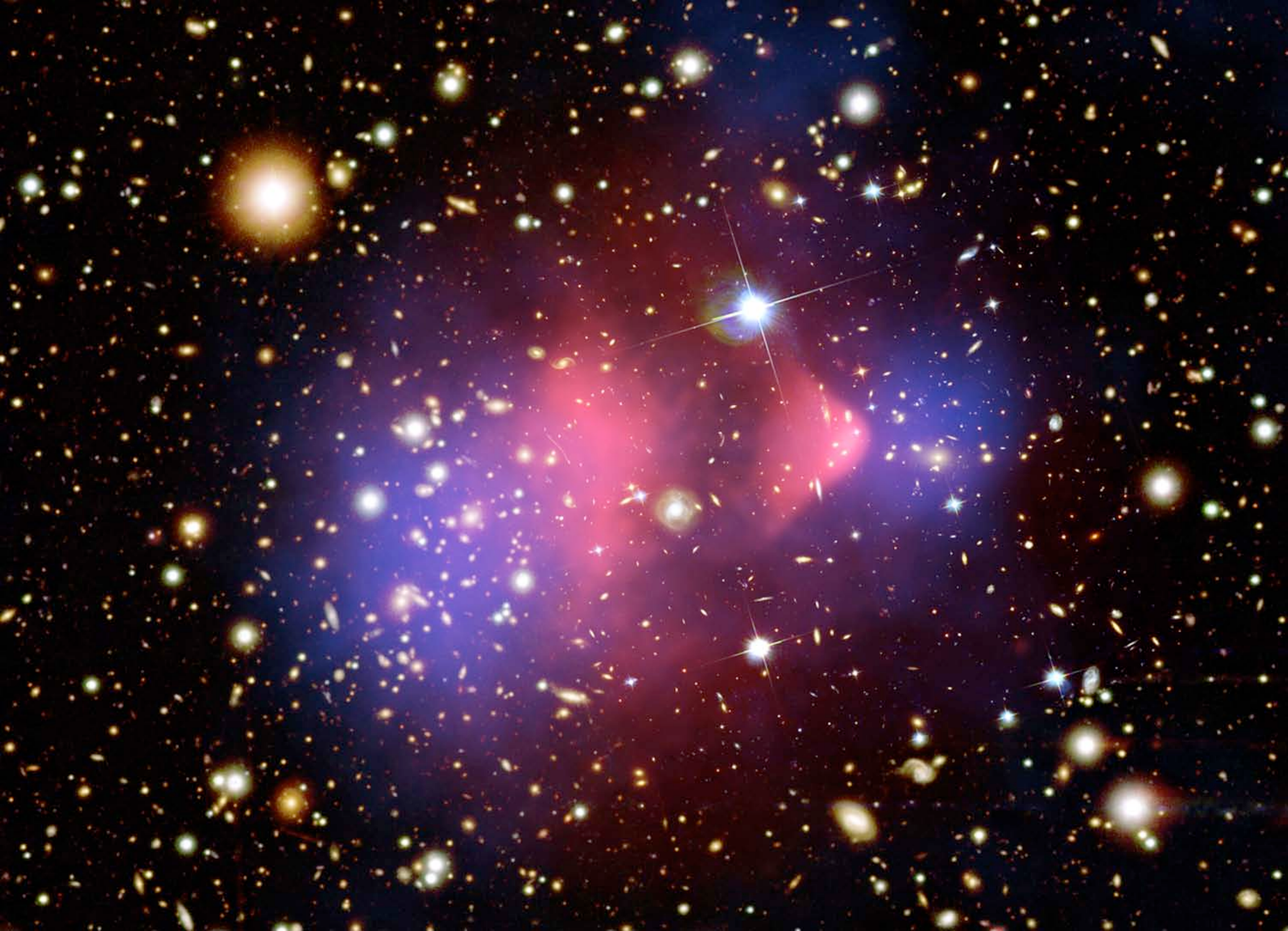}
\caption[False colour composite image]{\label{figure.cluster.bullet.1e0657} {{\sf\small {\bc}\break \phantom{wword}False colour composite image.\phantom{wword}}}\break\break The surface density \map{\Sigma} peaks reconstructed from X-ray imaging observations are shown in red and the convergence \map{\kappa} peaks as reconstructed from strong and weak gravitational lensing observations are shown in blue.  Image provided courtesy of \href{http://chandra.harvard.edu/photo/2006/1e0657/}{Chandra X-ray Observatory}.}
\end{SCfigure}

The {\em Chandra} Peer Review has declared the {\bc} to be the most interesting cluster in the sky.  This system, located at a redshift $z=0.296$ has the highest X-ray luminosity and temperature ($T = 14.1 \pm 0.2\ \mbox{keV} \sim 1.65 \times 10^{8}\ \mbox{K}$), and demonstrates a spectacular merger in the plane of the sky exhibiting a supersonic shock front, with Mach number as high as $3.0 \pm 0.4$~\protect\citep{Markevitch:2006}.  The {\bc} has provided a rich dataset in the X-ray spectrum which has been modelled to high precision.  From the extra-long $5.2 \times 10^{5}\ \mbox{s}$ {\em Chandra} space satellite X-ray image, the surface mass density, $\Sigma(x,y)$, was reconstructed providing a high resolution map of the ICM gas~\protect\citep{Clowe.NPBPS.2007.173}.  The \map{\Sigma}, shown in a false colour composite map (in red) in \SCfref{figure.cluster.bullet.1e0657}{False colour composite image} is the result of a normalized geometric mass model  based upon a $16^{\prime} \times 16^{\prime}$ field in the plane of the sky that covers the entire cluster and is composed of a square grid of $185 \times 185$ pixels ($\sim 8000$ data-points).

Based on observations made with the NASA/ESA Hubble Space Telescope, the Spitzer Space Telescope and with the 6.5 meter Magellan Telescopes, \protect\citet{Clowe.APJL.2006.648,Bradac.APJ.2006.652,Clowe.NPBPS.2007.173} reported on a combined strong and weak gravitational lensing survey used to reconstruct a high-resolution, absolutely calibrated convergence \map{\kappa} of the region of sky surrounding \bc, without assumptions on the underlying gravitational potential.  The \map{\kappa} is shown in the false colour composite map (in blue) in \SCfref{figure.cluster.bullet.1e0657}{False colour composite image}.  The gravitational lensing reconstruction of the convergence map is a remarkable result, considering it is based on a catalogue of strong and weak lensing events and relies upon a thorough understanding of the distances involved  -- ranging from the redshift of the {\bc} ($z = 0.296 $) which puts it at a distance of the order of one million parsecs away.  Additionally, the typical angular diameter distances to the lensing event sources ($z \sim 0.8$ to $z\sim 1.0$) are several million parsecs distant.  

In most observable systems, gravity creates a central potential, where the baryon density peaks.  As exhibited in \SCfref{figure.cluster.bullet.1e0657}{False colour composite image}, the latest results from the {\bc} show, beyond a shadow of doubt, that the \map{\Sigma}, which is a direct measure of the hot ICM gas, is offset from the \map{\kappa}, which is a direct measure of the curvature (convergence) of space-time.  The fact that the \map{\kappa} is centered on the galaxies, and not on the ICM gas mass is certainly either evidence of ``missing mass'', as in the case of the dark matter paradigm, or evidence of a stronger gravitational coupling due to a modification to gravity, as supported by \citet{Brownstein.MNRAS.2007.382}.   \citet{Clowe.NPBPS.2007.173} stated 
\begin{quotation}
One would expect that this (the offset $\Sigma$- and $\kappa$-peaks) indicates that dark matter must be present regardless of the gravitational force law, but in some alternative gravity models, the multiple peaks can alter the lensing surface potential so that the strength of the peaks is no longer directly related to the matter density in them. As such, all of the alternative gravity models have to be tested individually against the observations.
\end{quotation}
\citet{Clowe.NPBPS.2007.173} described this as a degeneracy between whether gravity comes from dark matter, or from the observed baryonic mass of the hot ICM and visible galaxies where the excess gravity is due to a fifth force modification to the potential.  This degeneracy may be split by examining a system that is out of steady state, where there is spatial separation between the hot ICM and visible galaxies.  This is precisely the case in galaxy cluster mergers such as the {\bc}, since the galaxies will experience a different gravitational potential created by the hot ICM than if they were concentrated at the center of the ICM.

The data from the {\bc} provides a laboratory of the greatest scale, where the degeneracy between   ``missing mass'' and  ``extra gravity'' may be distinguished.  We are fortunate, indeed, that the {\bc} is not only one of the hottest, most supersonic, most massive cluster mergers seen, but the plane of the merger is aligned with our sky!  \citet{Brownstein.MNRAS.2007.382} addressed the full-sky data product~\protect\citep{Clowe:dataProduct} for the \bc, and provide first published results for the King $\beta$-model  of the \map{\Sigma}, detailed in \sref{section.cluster.bullet.Sigma}.  \citet{Brownstein.MNRAS.2007.382} utilized the metric skew-tensor gravity model of \sref{section.mog.mstg} to compute component mass profiles, and account for all of the baryons in each of the main and subclusters, including all of the galaxies in the regions near the main central dominant (cD) and the subcluster's brightest central galaxy (BCG), without non-baryonic dark matter, to fit the gravitational lensing convergence map, as in \sref{section.cluster.bullet.kappa}.  The results of the analysis include a map of the visible baryon distribution, as in \sref{section.cluster.bullet.baryon}, and the dark matter distribution, as in \sref{section.cluster.bullet.darkmatter}.
\begin{figure}[ht]
\begin{picture}(460,270)(0,0)
\put(5,0){\includegraphics[width=0.98\textwidth]{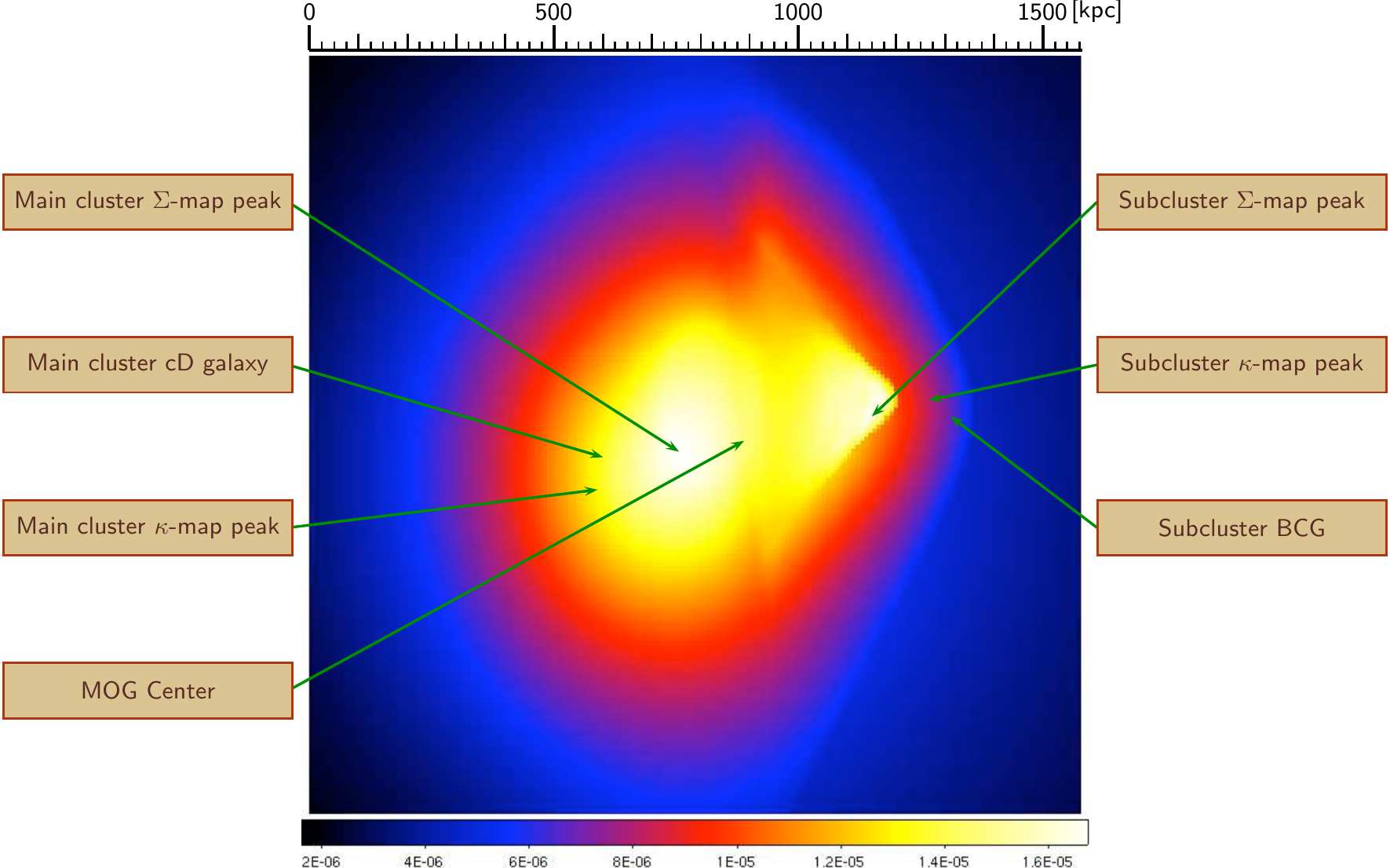}}
\end{picture}
\caption[X-ray gas surface density map]{\label{figure.cluster.bullet.Sigma} {{\sf\small {\bc} -- X-ray gas surface density map.}}\break\break Data reconstructed from X-ray imaging observations of the {\bc}, November 15, 2006 data release~\protect\citep{Clowe:dataProduct}, showing \map{\Sigma} observed peaks (local maxima) and \map{\kappa} observed peaks.  The central dominant (cD) galaxy of the main cluster, the brightest cluster galaxy (BCG) of the subcluster, and the MOG predicted gravitational center are shown.  The colourscale is shown at the bottom, in units of $10^{15}\ M_{\solar}/\mbox{pixel}^{2}$.  The resolution of the \map{\Sigma} is 8.5 kpc/pixel, based upon the measured redshift distance $\sim$ 260.0 kpc/arcminute~\protect\citep{Bradac.APJ.2006.652}.  The scale in kpc is shown at the top of the figure.  J2000 and map (x,y) coordinates are listed in \tref{table.cluster.bullet.coords}.}
\end{figure}

\subsection{\label{section.cluster.bullet.Sigma}X-ray gas map}

With an advance of the \citet{Clowe:dataProduct} November 15, 2006 data release,  \citet{Brownstein.MNRAS.2007.382} performed a precision analysis to model the gross features of the surface density \map{\Sigma} data in order to gain insight into the three-dimensional matter distribution, $\rho(r)$, and to separate the components into a model representing the main cluster and the subcluster -- the remainder after subtraction. 

\advancecontents{lot}\addtocontentsheading{lot}{{\bc}}
\begin{table}[h]
\caption[J2000 sky coordinates]{\label{table.cluster.bullet.coords}{\sf J2000 sky coordinates of the {\bc}}}
\begin{center} \begin{tabular}{c|cc|c|c} \\ \hline
{\sc Observation} & \multicolumn{2}{c|}{{\sc J2000 Coordinates}} & {\sc\map{\Sigma}} & {\sc\map{\kappa}}\\
& {\sc RA} & {\sc Dec} & { $(x,\,y)$} & {$(x,\,y)$} \\ 
(1)&(2)&(3)&(4)&(5) \\ \hline \hline
Main cluster \map{\Sigma} peak & 06 : 58 : 31.1\quad & \quad-55 : 56 : 53.6 & $(89,\,89)$ & $(340,\,321)$ \\
Subcluster \map{\Sigma} peak & 06 : 58 : 20.4\quad & \quad-55 : 56 : 35.9 & $(135,\,98)$ & $(365,\,326)$ \\
Main cluster \map{\kappa} peak & 06 : 58 : 35.6\quad & \quad-55 : 57 : 10.8 & (70,\,80) & (329,\,317)\\
Subcluster \map{\kappa} peak & 06 : 58 : 17.0\quad & \quad-55 : 56 : 27.6 & (149,\,102) & (374,\,327) \\
Main cluster cD & 06 : 58 : 35.3\quad & \quad-55 : 56 : 56.3 & (71,\,88) & (330,\,320) \\
Subcluster BCG & 06 : 58 : 16.0\quad & \quad-55 : 56 : 35.1 & (154,\,98) & (375,\,326) \\
MOG Center & 06 : 58 : 27.6\quad & \quad-55 : 56 : 49.4 & (105,\,92) & (348,\,322) \\ 
\hline \multicolumn{5}{c}{}
\end{tabular} \end{center}
\parbox{6.375in}{\small Notes. --- November 15, 2006 data release~\protect\citep{Clowe:dataProduct}:  Column (1) provide the primary observational features.  Columns (2) and (3) list the J2000 right ascension (RA) and declination (Dec) for each feature.  Columns (4) and (5) provide the \map{\Sigma} and \map{\kappa} \((x,y)\)  coordinates using a resolution of 8.5 kpc/pixel, and 15.4 kpc/pixel, respectively, based upon the measured redshift distance $\sim$ 260.0 kpc/arcminute of the {\bc}~\protect\citep{Bradac.APJ.2006.652}.}
 \end{table}

The \map{\Sigma} is shown in false colour in \fref{figure.cluster.bullet.Sigma}.  There are two distinct peaks in the surface density \map{\Sigma} -- the primary peak centered at the main cluster, and the secondary peak centered at the subcluster. The main cluster gas is the brightly glowing (yellow) region to the left of the subcluster gas, which is the nearly equally bright shockwave region (arrowhead shape to the right).
The \map{\kappa} observed peaks, the central dominant (cD) galaxy of the main cluster, the brightest cluster galaxy (BCG) of the subcluster, and the MOG predicted gravitational center are shown in \fref{figure.cluster.bullet.Sigma} for comparison.  J2000 and map (x,y) coordinates are listed in \tref{table.cluster.bullet.coords}.

Since there is a multitude of source galaxies in a range of redshifts ($z=0.85\pm 0.15$), the source distances, $D_{\rm s}$, may be averaged.  For the {\bc}, \citet{Clowe.APJ.2004.604} used
\begin{equation}
\label{eqn.cluster.bullet.SigmaC.Deff}
\frac{D_{\rm l}D_{\rm ls}}{D_{\rm s}} \approx 540\ \mbox{kpc},
\end{equation}
and the Newtonian critical surface mass density (with vanishing shear) of \eref{eqn.galaxy.uma.SigmaC.Newton},
\begin{equation}
\label{eqn.cluster.bullet.SigmaC.Newton}
\Sigma_{c} = \frac{c^{2}}{4\pi G_{N}} \frac{D_{\rm s}}{D_{\rm l}D_{\rm ls}} \approx 3.1 \times 10^{9}\
M_{\solar}/\mbox{kpc}^{2}
\end{equation}
is effectively constant. 

\begin{SCfigure}[0.9][h]
\includegraphics[width=0.5\textwidth]{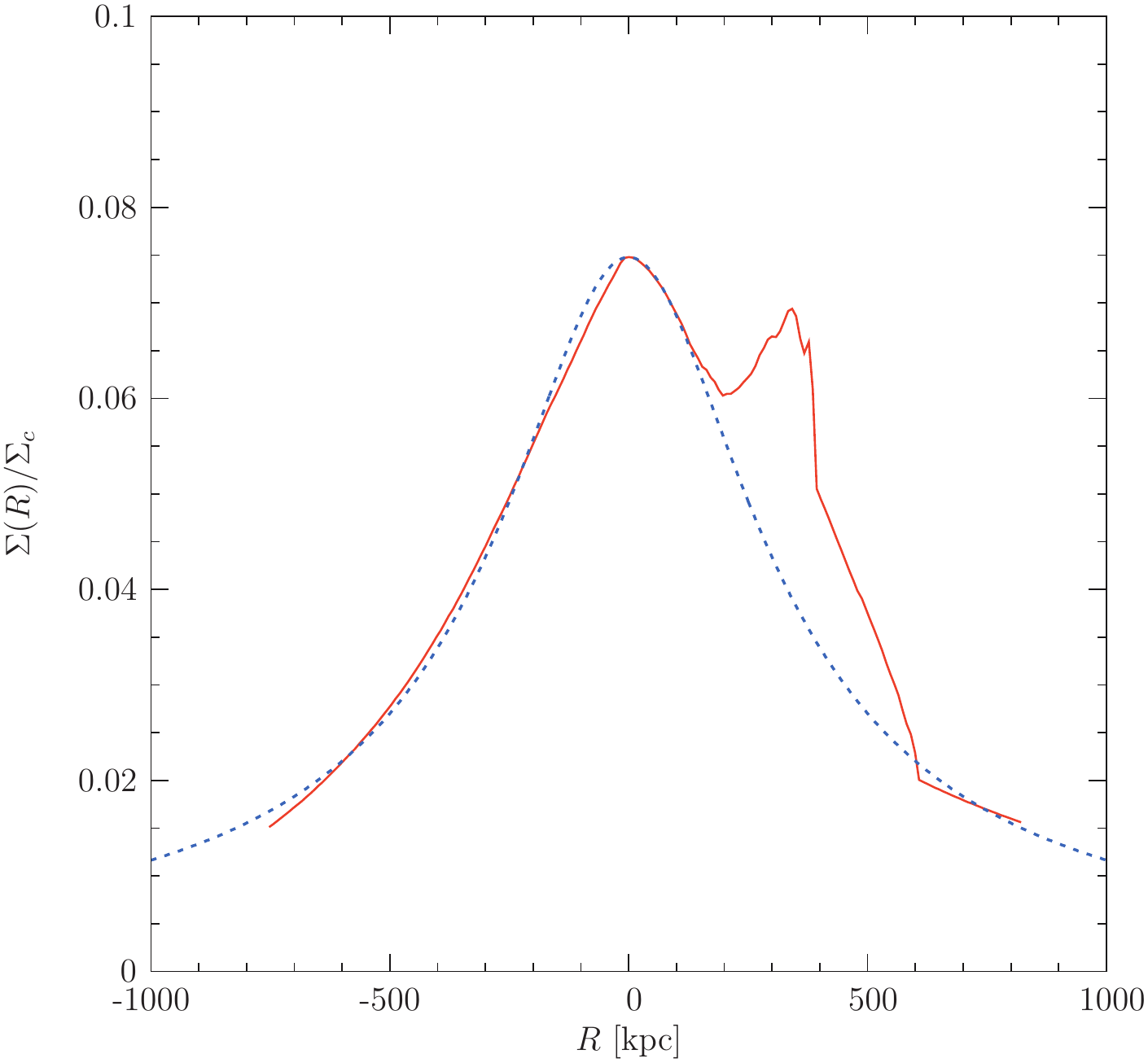}
\caption[King $\beta$-model fit to scaled \map{\Sigma}]{\label{figure.cluster.bullet.SigmaFit} {{\sf\small {\bc}\break \phantom{word}King $\beta$-model fit to scaled \map{\Sigma}.\phantom{word}}}\break\break  A cross-section of the \map{\Sigma} of \fref{figure.cluster.bullet.Sigma} reconstructed from X-ray imaging observations~\protect\citep{Clowe:dataProduct}, shown in solid red, on a straight-line connecting the main X-ray cluster peak to the main central dominant (cD) galaxy. The King $\beta$-model (neglecting the subcluster) of \eref{eqn.cluster.xraymass.Sigma.surfaceMassDensity} is shown in short-dashed blue, best-fit by \erefss{eqn.cluster.bullet.beta}{eqn.cluster.bullet.rc}{eqn.cluster.bullet.Sigma0}.  The unmodeled peak (at $R \sim 300\ \mbox{kpc}$) is due to the subcluster.} \end{SCfigure}

\subsubsection{\label{subsection.cluster.bullet.Sigma.king}King \(\beta\)-model of the main cluster}

To calculate the best-fit parameters, $\beta$, $r_{c}$ and $\rho_{0}$ of the King $\beta$-model of
\erefs{eqn.cluster.xraymass.Sigma.surfaceMassDensity0}{eqn.cluster.xraymass.Sigma.surfaceMassDensity}, \citet{Brownstein.MNRAS.2007.382} applied a nonlinear least-squares fitting routine (including estimated errors) to the {\map{\Sigma}} on a straight-line connecting the main cluster \map{\Sigma} peak ($R\equiv 0$) to the main cD, and then extrapolated the fit to the entire map.  This reduces the complexity of the calculation to a simple algorithm, but is not guaranteed to yield a global best-fit.  However, the approximation provides a very low reduced \(\chi^2\) everywhere on the full sky map, except at the subcluster (which is masked for the best-fit). The X-ray gas surface density \map{\Sigma} data, and the King $\beta$-model of \eref{eqn.cluster.xraymass.Sigma.surfaceMassDensity}, best-fit to the scaled \map{\Sigma}, are shown in \SCfref{figure.cluster.bullet.SigmaFit}{King $\beta$-model fit to scaled \map{\Sigma}}, with the best-fit parameters,
\begin{eqnarray}
\label{eqn.cluster.bullet.beta} \beta &=& 0.803\pm0.013,\\
\label{eqn.cluster.bullet.rc} r_{c} &=& 278.0\pm6.8\ \mbox{kpc},
\end{eqnarray}
where the value of the \map{\Sigma} at the main cluster peak is constrained to the observed value, 
\begin{equation} 
\label{eqn.cluster.bullet.Sigma0} \Sigma_{0} = 1.6859\times 10^{10} \frac{M_{\solar}}{\mbox{pixel}^{2}} \left(\frac{1\ \mbox{pixel}}{8.528\
\mbox{kpc}}\right)^{2}=2.3181\times 10^{8} M_{\solar}/\mbox{kpc}^{2},
\end{equation}
scaled by \(\Sigma_{c}\) of \eref{eqn.cluster.bullet.SigmaC.Newton}.  Solving \eref{eqn.cluster.xraymass.Sigma.surfaceMassDensity0} for the central density of the main cluster,
\begin{equation}
\label{eqn.cluster.bullet.rho0}
\rho_{0} =  \frac{\Sigma_{0}}{\sqrt{\pi}  r_{c}} 
\frac{\Gamma\left(\frac{3}{2}\beta\right)}{\Gamma\left(\frac{3\beta-1}{2}\right)} = 0.334 \times 10^{6}\
M_{\solar}/\mbox{kpc}^{3},
\end{equation}
which is between one and two orders of magnitude less than the dark matter central densities listed in  \tref{table.galaxy.darkmatter} derived from the galaxy rotation curves of \sref{section.galaxy.uma.velocity}, proving that cluster scale dark matter does not affect the dynamics of galaxy rotation curves.  The set of parameters, $\beta$, $r_{c}$ and $\rho_{0}$,  completely determines the isotropic isothermal King $\beta$-model for the density, $\rho(r)$, of \eref{eqn.cluster.xraymass.isothermal.betaRhoModel} of the main cluster X-ray gas, and the isotropic isothermal model of \sref{section.cluster.xraymass.isothermal} may be applied to measure the mass-luminosity relation in the main cluster and compute the ratio of the Newtonian dynamic mass to the  X-ray gas (baryon) mass, per gravity theory.

\citet{Brownstein.MNRAS.2007.382} computed the Newtonian dynamic mass profile for the main cluster of the {\bc}, and determined the MSTG mass profile according to \erefs{eqn.mog.mstg.mass}{eqn.mog.mstg.mass.mstg}, finding that the modified gravity mass profile is an excellent fit to the measured X-ray (baryon) mass profile, as shown in Panel (a) of \fref{figure.cluster.models.mass}.  Across the full range of the $r$-axis, and throughout the radial extent of the {\bc}, the $1\sigma$ correlation between the gas mass, $M(r)$ and the MOG dynamical mass, $M_{\rm MSTG}(r)$, provides excellent agreement between theory and experiment. 

Substituting \erefss{eqn.cluster.bullet.beta}{eqn.cluster.bullet.rc}{eqn.cluster.bullet.rho0} into \eref{eqn.cluster.xraymass.isothermal.rout.0}, we obtain the main cluster outer radial extent,
\begin{equation}
\label{eqn.cluster.bullet.rout}
r_{\rm out} = 2620\ \mbox{kpc},
\end{equation}
the distance at which the density, $\rho(r_{\rm out})$, drops to $\approx 10^{-28}\,\mbox{g/cm}^{3}$, or 250
times the mean cosmological density of baryons.  The total mass of the main cluster may be calculated by 
substituting \erefss{eqn.cluster.bullet.beta}{eqn.cluster.bullet.rc}{eqn.cluster.bullet.rho0} into \eref{eqn.cluster.xraymass.isothermal.Mgas.0}:
\begin{equation}
\label{eqn.cluster.bullet.Mgas}
M_{\rm gas}  = 3.87  \times 10^{14}\ M_{\solar},\qquad\mbox{\tt main cluster.}
\end{equation}

The MOG temperature prediction, from the MSTG best-fit, is increasingly consistent with updated experimental values, shown in \tref{table.cluster.bullet.temp}.

\begin{table}[h]
\caption[Isothermal temperature of the main cluster]{\label{table.cluster.bullet.temp} {\sf Isothermal
temperature of the main cluster}}
\begin{center}\begin{tabular}{c|l|cr} \\ \hline 
{\sc Year} & {\sc Source - Theory or Experiment} & {\sc $T\ (\mbox{keV})$} & {\sc \% error} \\ \hline\hline
2007 &  Computed value  & $15.5\pm3.9$& \\
2002 & accepted experimental value  &$14.8^{+1.7}_{-1.2}$ & $\phantom{X1}4.5$ \\
1999 & ASCA+ROSAT fit  & $14.5^{+2.0}_{1.7}$ &$\phantom{X1}6.5$\\
1998 & ASCA fit & $17.4\pm2.5$ & $\phantom{X}12.3$ \\ \hline
\hline \multicolumn{4}{c}{}
\end{tabular}\end{center}
\parbox{6.375in}{\small Notes. ---  The computed isothermal temperature
 is consistent with the experimental values for the main cluster~\protect\citep{Markevitch:ApJL:2002}.: Column (1) and (2) list the year and source of the temperature result, respectively.  Column (3) provides the temperature in keV, and Column (4) provides the percent error between the computed and experimental values.}
\end{table}
\subsubsection{\label{subsection.cluster.bullet.Sigma.subcluster}Model of the bullet subcluster}

Although the X-ray morphology of the main cluster is very regular, and well described by the King \(\beta\)-model of the main cluster, \citet{Liang.APJ.2000.544} reported on a diffuse radio halo, which requires the acceleration of thermal electrons to ultra-relativistic energies, enhanced at the main X-ray gas peak and more focused at the densest part of the optical galaxy distribution.  Since galaxies are collisionless, at the \(\sim 1\) Mpc cluster scale, a merger with the subcluster -- the bullet in the X-ray gas surface density map of \fref{figure.cluster.bullet.Sigma} -- allows the galaxies to stream through the X-ray gas and generate the radio halo.  

\citet{Markevitch:ApJL:2002} reported on Chandra observations,  providing evidence that the merger is ongoing and the subcluster is in a perturbed state far from hydrostatic equilibrium leading to an apparent increase in the X-ray temperature, 150 million years after its collision with the main cluster core.  \citet{Barrena.AAP.2002.386} studied the dynamics of the collision, and determined that the subcluster is the remnant core of a moderate mass cluster of galaxies, whose properties have been strongly affected.  \citet{Randall.APJ.2008.679} studied the prominent bow shock, estimating the supersonic merger velocity of 4700 km s\(^{-1}\), and concluded that the subcluster X-ray gas mass distribution is significantly more peaked than a King profile.

\begin{figure}[h]
\begin{picture}(460,215)(50,445)
\put(0,0){\includegraphics[width=1.2\textwidth]{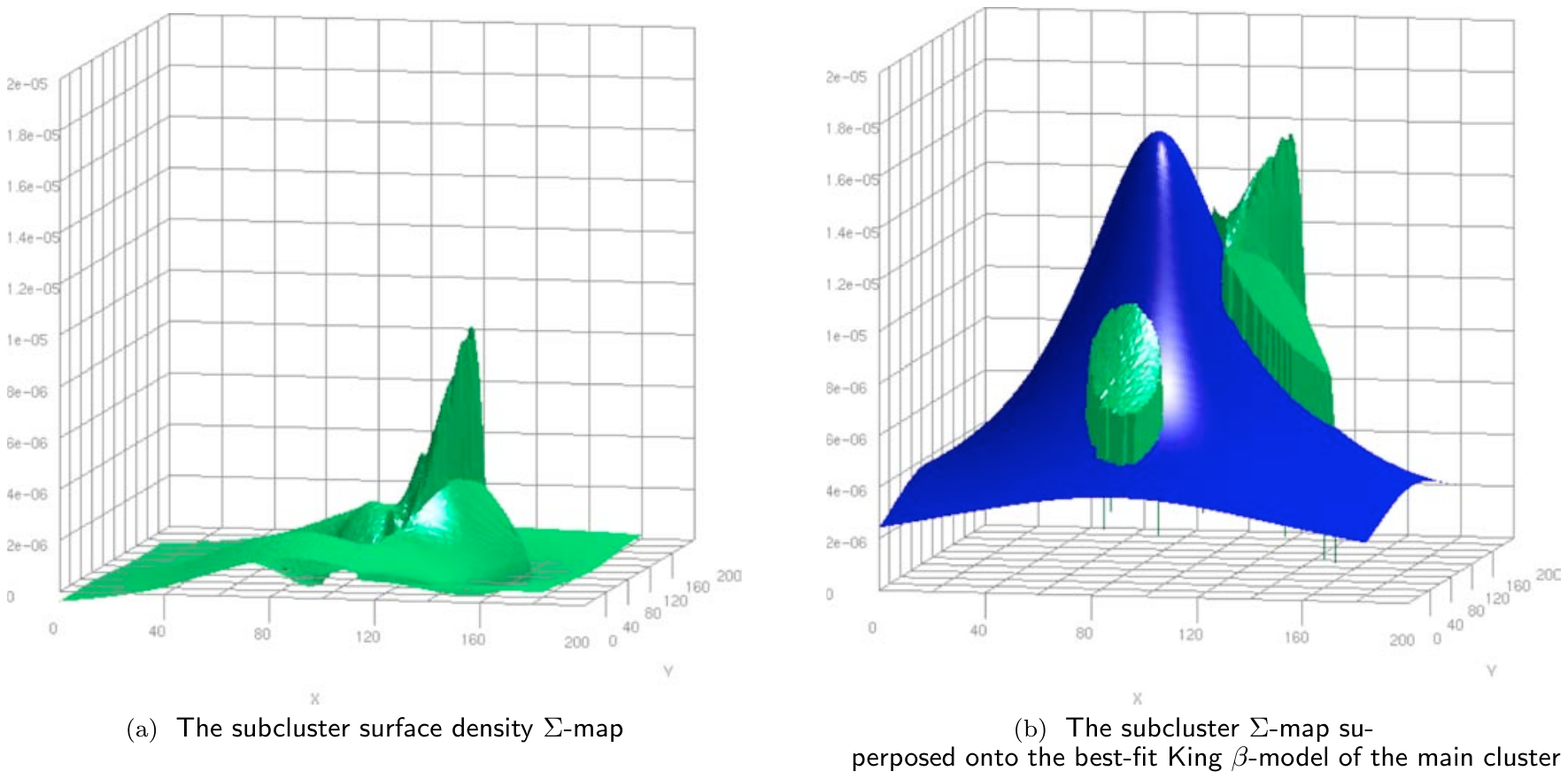}}
\end{picture}
\caption[X-ray gas surface density model]{\label{figure.cluster.bullet.SigmaModel} {{\sf\small {\bc} -- X-ray gas surface density model.}}\break\break  (a) Subtracted subcluster X-ray gas surface density map is shown in green.  (b)  X-ray gas surface density model. The blue surface represents the \map{\Sigma} due to the integrated (line-of-sight) King $\beta$-model fit to main cluster.  The green surface is the contribution to the \map{\Sigma} from the subcluster.}
\end{figure}

\citet{Brownstein.MNRAS.2007.382} computed the surface mass density of the subcluster by subtracting the best-fit ($\chi^{2}<0.2$) King $\beta$-model to the main cluster -- which agreed with the main cluster surface mass \map{\Sigma} (data) within 1\% everywhere -- from the total X-ray surface mass density of \fref{figure.cluster.bullet.Sigma}. The subcluster subtraction is accurate down to $\rho = 10^{-28}\ \mbox{g}/\mbox{cm}^{3}\sim 563.2\ M_{\solar}/\mbox{pc}^{3}$ baryonic background density.  After subtraction, the subcluster \map{\Sigma} peak takes a value of $1.30 \times 10^{8}\ M_{\solar}/\mbox{kpc}^{2}$, whereas the full \map{\Sigma} has a value of  $2.32\times 10^{8}\ M_{\solar}/\mbox{kpc}^{2}$ at the subcluster \map{\Sigma} peak.  Thus the subcluster (at its most dense position) provides only $\approx 56\%$ of the X-ray ICM, the rest is due to the extended distribution of the main cluster.

\fref{figure.cluster.bullet.SigmaModel} is a stereogram of the subcluster subtracted surface density
\map{\Sigma} and the subcluster superposed onto the surface density \map{\Sigma} of the best-fit King
$\beta$-model to the main cluster.

Since the outer radial extent of the subcluster gas is less than 400 kpc, the \map{\Sigma} completely contains all of the subcluster gas mass.  By summing the subcluster subtracted \map{\Sigma} pixel-by-pixel over the entire \map{\Sigma} peak, one is performing an integration of the surface density, yielding the total subcluster mass.  \citet{Brownstein.MNRAS.2007.382} performed such a sum over the subcluster subtracted \map{\Sigma} data, obtaining
\begin{equation}
\label{eqn.cluster.bullet.Mgas.subcluster}
M_{\rm gas}  = 2.58  \times 10^{13}\ M_{\solar},\qquad\mbox{\tt  subcluster.}
\end{equation}
for the mass of the subcluster gas, which is less than 6.7\% of the mass of main cluster gas of \eref{eqn.cluster.bullet.Mgas}.
This justifies the initial assumption that the subcluster may be treated as a perturbation in
order to fit the main cluster to the King $\beta$-model.  The subsequent analysis of the thermal profile confirms that the main cluster X-ray temperature is nearly isothermal, lending further support to the validity of the King $\beta$-model and the reliability of the isothermal temperatures of \tref{table.cluster.bullet.temp}.

\subsection{\label{section.cluster.bullet.kappa}Gravitational lensing convergence map}\index{Gravitational lensing!\map{\kappa}|(}

\begin{figure}[ht]
\begin{picture}(460,270)(0,0)
\put(5,0){\includegraphics[width=0.98\textwidth]{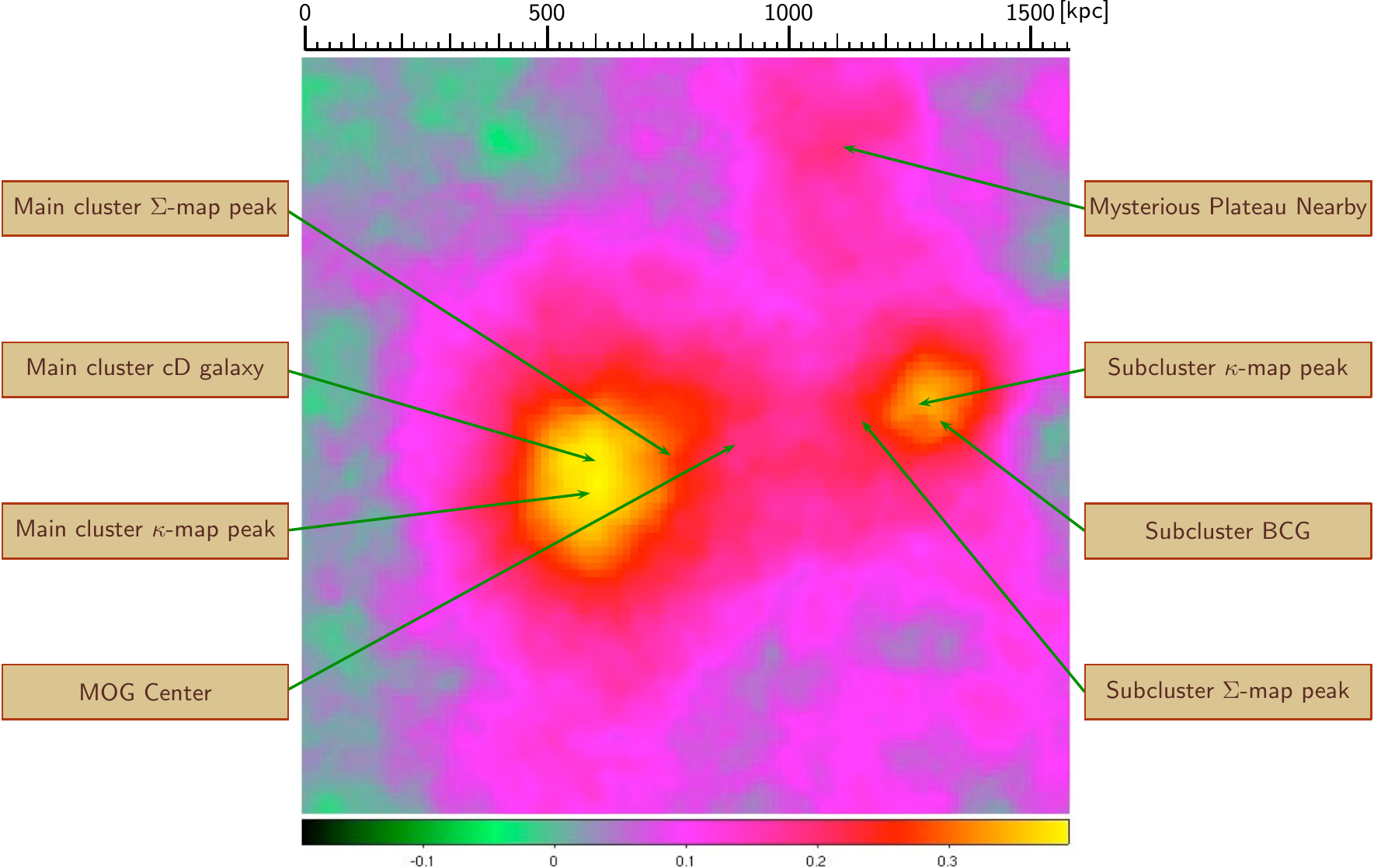}}
\end{picture}
\caption[Gravitational lensing convergence map]{\label{figure.cluster.bullet.kappa} {{\sf\small {\bc} -- Gravitational lensing convergence map.}}\break\break Data reconstructed from strong and weak gravitational lensing of the {\bc}, November 15, 2006 data release~\protect\citep{Clowe:dataProduct}, showing convergence \map{\kappa} observed peaks (local maxima) and \map{\Sigma} observed peaks.  The central dominant (cD) galaxy of the main cluster, the brightest cluster galaxy (BCG) of the subcluster, and the MOG predicted gravitational center are shown.  The scale in kpc is shown at the top of the figure.  J2000 and map (x,y) coordinates are listed in \tref{table.cluster.bullet.coords}.}
\end{figure}

The convergence \map{\kappa} of \fref{figure.cluster.bullet.kappa} is a false colour image of the  strong
and weak gravitational lensing reconstruction~\protect\citep{Clowe.APJL.2006.648,Bradac.APJ.2006.652,Clowe.NPBPS.2007.173} of all of the bending of light over the entire distance from the lensing event source toward the Hubble Space Telescope.  The source of the \map{\kappa} is  $\propto\int G_{N}\rho(r)$, along the line-of-site, as in the Newtonian case of \eref{eqn.galaxy.uma.kappa},  but $\propto\int G(r)\rho(r)$ as in \eref{eqn.galaxy.uma.scaledSurfaceMassDensity.MOG} of modified gravity with a spatially varying gravitational coupling.

\subsubsection{\label{subsection.cluster.bullet.kappa.solution}Modified gravity solution}\index{Gravitational lensing!MOG|(}

\begin{figure}[ht]
\begin{picture}(460,215)(50,445)
\put(0,0){\includegraphics[width=1.2\textwidth]{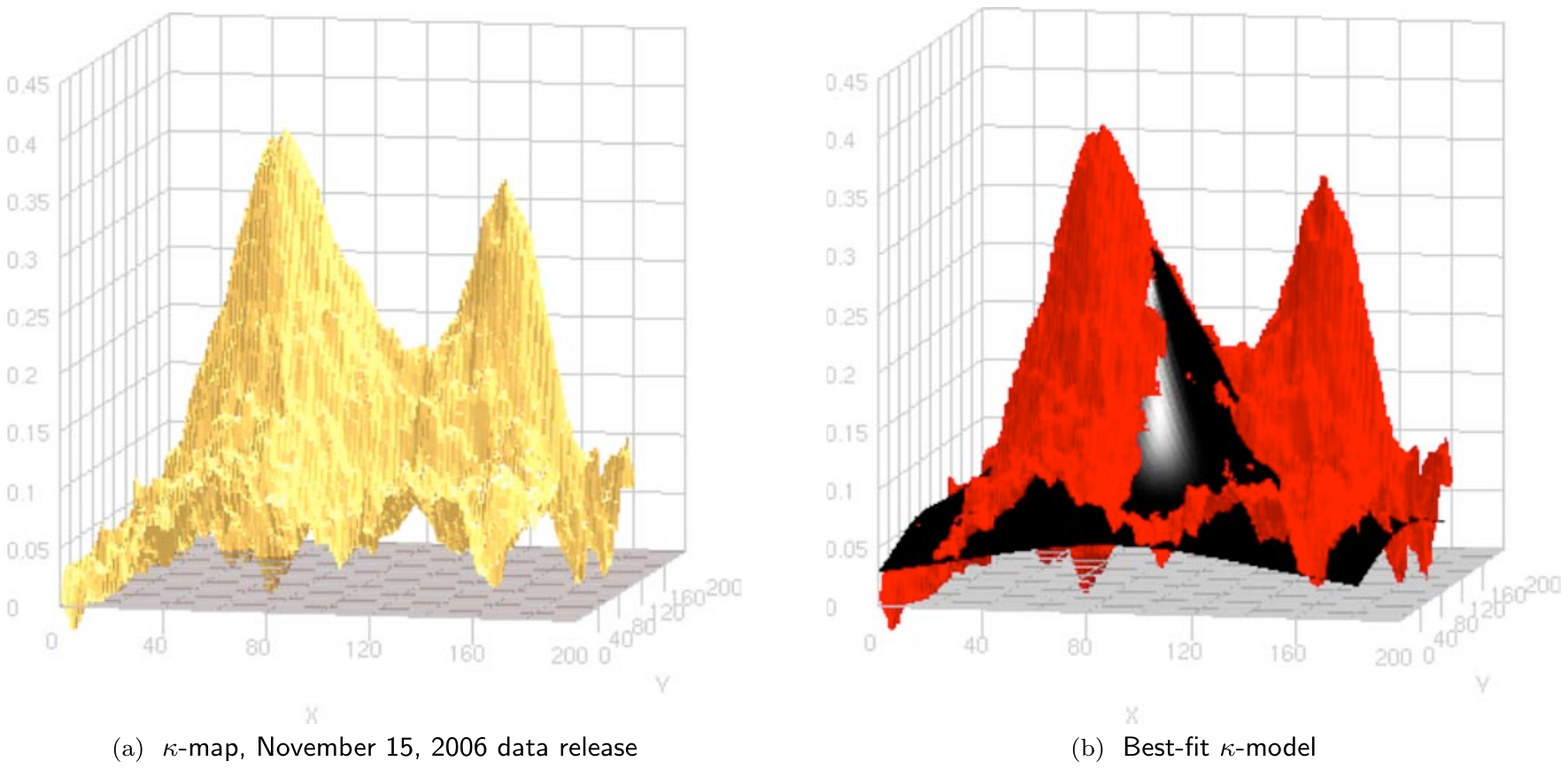}}
\end{picture}
\caption[Best-fit model to the gravitational lensing convergence map]{\label{figure.cluster.bullet.kappaModel} {{\sf\small {\bc} -- Best-fit model to the gravitational lensing convergence map.}}\break\break (a) Gravitational lensing convergence \map{\kappa} November 15, 2006 data release~\protect\citep{Clowe:dataProduct} is shown in gold.  The twin peaks are due to the main and subcluster, respectively.  (b) Gravitational lensing convergence model.  The black surface is the best-fit \map{\kappa} to the main cluster X-ray gas component. The red surface represents the excess \map{\kappa} of the galactic component beyond the best-fit X-ray gas component.}
\end{figure}

The lack of spherical symmetry in the \map{\kappa}, shown in \fref{figure.cluster.bullet.kappa}, is better visualized in Panel (a) of \fref{figure.cluster.bullet.kappaModel}, which demonstrates the importance of the subcluster's dynamic mass.

\citet{Brownstein.MNRAS.2007.382} utilized the metric skew-tensor gravity model of \sref{section.mog.mstg} to compute the weighted surface mass density, $\bar \Sigma$ of \eref{eqn.cluster.bullet.kappa.Sigma.MOG}, of the X-ray gas mass of the main cluster using the King \(\beta\)-model of \eref{eqn.cluster.xraymass.isothermal.betaRhoModel} with the best-fit parameters of \erefss{eqn.cluster.bullet.beta}{eqn.cluster.bullet.rc}{eqn.cluster.bullet.rho0}.  This is shown as \({\bar \Sigma}(r)/\Sigma_{c}\) by the black surface of the \(\kappa\)-model of Panel (b) of \fref{figure.cluster.bullet.kappaModel}.  Including the galaxies is accomplished by \eref{eqn.cluster.bullet.kappa.model} which is shown by the red surface of the \(\kappa\)-model.

\citet{Brownstein.MNRAS.2007.382} proceeded to account for the spherical symmetry breaking effect of the subcluster on the dynamic mass of \eref{eqn.mog.mstg.mass.mstg}:  Remarkably, as the MOG center was separated from the main cluster \map{\Sigma} peak, due to the gravitational effect of the subcluster,  the centroid naturally shifted toward the \map{\kappa} peak, and the predicted height of the \map{\kappa} decreased, flattening the peak and dimpling the core and skewing the distribution in the direction opposite to the shift in the MOG center.  Although \citet{Moffat.ArXiv.astroph:0608675} demonstrated that the integration of the \map{\kappa}, assuming a constant surface mass density for the galaxies, produced a peak offset from the X-ray peak, the effect alone was insufficient to fit the {\bc} \map{\kappa} data. However, the difference can be entirely accounted for by including the  surface mass density of the galaxies, \({\bar \Sigma}_{\rm galax}(r)/\Sigma_{c}\), as indicated by the red surface of the best-fit \(\kappa\) model of \fref{figure.cluster.bullet.kappaModel}.  Combining the black surface and the red surface, we obtain the best-fit model,
\begin{equation}
\label{eqn.cluster.bullet.kappa.model}\kappa(r) = \frac{{\bar \Sigma}(r) + {\bar \Sigma}_{\rm galax}(r)}{\Sigma_{c}},
\end{equation}
which is equivalent to the \map{\kappa} data illustrated by the gold surface on the left hand side of \fref{figure.cluster.bullet.kappaModel}.

As introduced in \sref{section.galaxy.uma.Sigma}, predictions for the \map{\kappa} of high resolution sub-kiloparsec galaxy-galaxy lensing, plotted in \fref{figure.galaxy.Sigma}, are computed by 
\begin{equation}
\label{eqn.cluster.bullet.kappa.scaledSurfaceMassDensity.MOG}
\kappa(r) =  \int \frac{4\pi G(r)}{c^{2}} \frac{D_{\rm l}D_{\rm ls}}{D_{\rm s}} \rho(r) dz  \equiv \frac{\bar \Sigma(r)}{\Sigma_{c}},
\end{equation}
where
\begin{equation}
\label{eqn.cluster.bullet.kappa.Sigma.MOG} {\bar \Sigma}(r) = \int \frac{G(r)}{G_N}\rho(r) dz,
\end{equation}
is the weighted surface mass density.  For the multiple source {\bc} reconstruction, \citet{Clowe.APJ.2004.604} used \(\Sigma_{c} \approx 3.1 \times 10^{9}\,M_{\solar}/\mbox{kpc}^{2}\), without estimate of the uncertainty.  The precision of the \map{\kappa} depends on the validity of the assumption of small variation \(\Delta \Sigma_{c}\) across the lens, which depends on the variation in the ratio, \(D_{\rm ls}/D_{\rm s}\).

Substituting \eref{eqn.cluster.models.mass.Gamma} into \eref{eqn.cluster.bullet.kappa.Sigma.MOG},\index{Dynamic mass factor, \(\Gamma\)|(}
\begin{equation}
\label{eqn.cluster.bullet.kappa.Sigmabar} {\bar \Sigma}(r) = \int \Gamma(r)\rho(r) dz.
\end{equation}
In the Newtonian limit, \(G(r) \rightarrow G_N\), and therefore the factor \(\Gamma(r) \rightarrow 1\) is removed from the integral, and \(\bar \Sigma \rightarrow \Sigma\), recovering the \map{\Sigma} of \eref{eqn.cluster.xraymass.Sigma.surfaceMassDensity.1}.

Whereas \(\Gamma(r)\) is a maximum value for the outer radial positions of galaxies, contributing more weight to the integral of \eref{eqn.cluster.bullet.kappa.Sigmabar} in the galactic halo, \(\Gamma(r)\) contributes more weight to the cores of spherically symmetric clusters of galaxies.\index{Dynamic mass factor, \(\Gamma\)|)}\index{Gravitational lensing!MOG|)}\index{Gravitational lensing!\map{\kappa}|)}
\begin{figure}[ht]
\begin{picture}(460,270)(0,0)
\put(5,0){\includegraphics[width=0.98\textwidth]{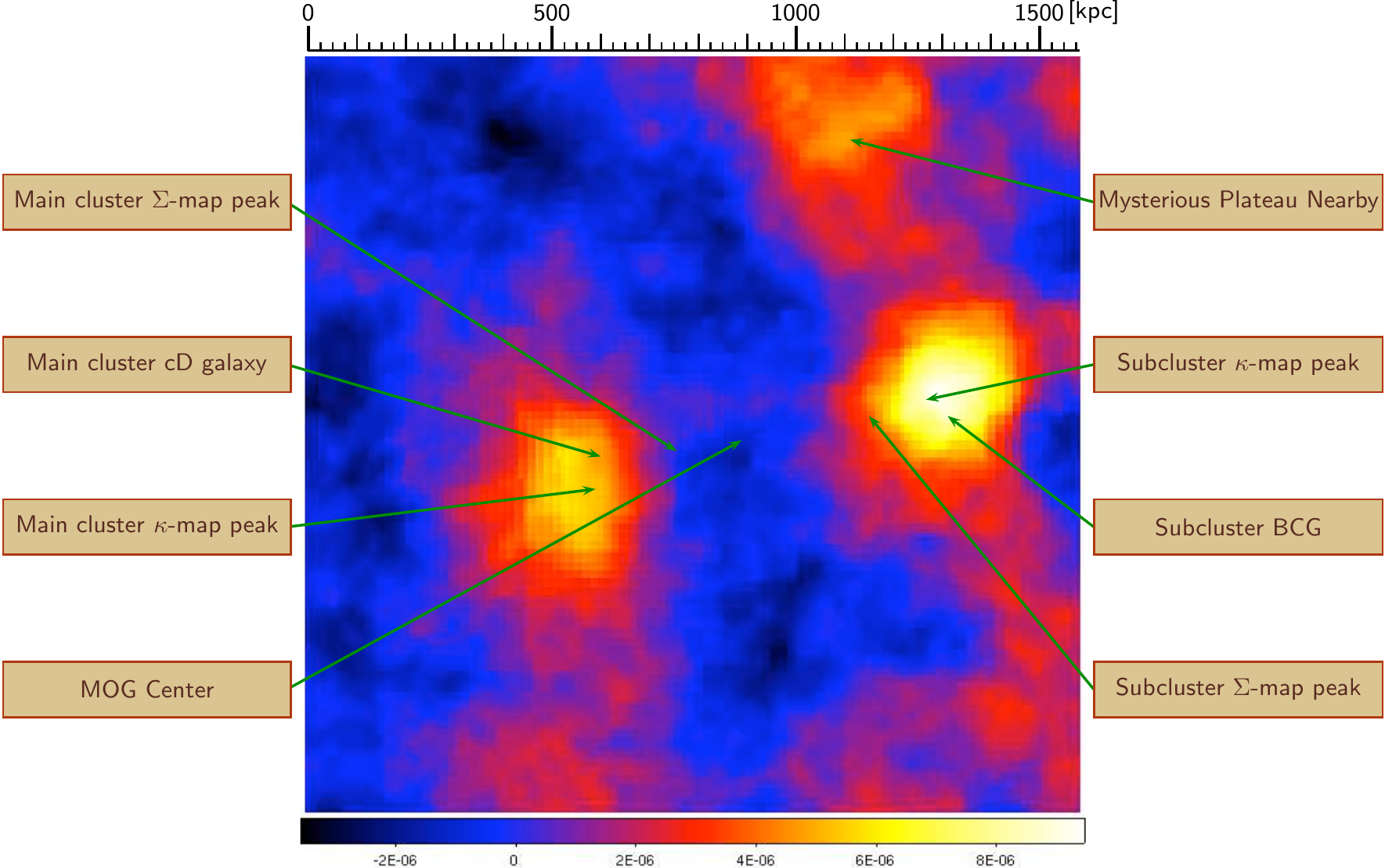}}
\end{picture}
\caption[Galactic surface mass density map]{\label{figure.cluster.bullet.galaxy} {{\sf\small {\bc} -- Galactic surface density map.}}\break\break The predicted \map{\Sigma} due to the galaxies as computed by the difference between the \map{\kappa} and the MOG $\kappa$-model, scaled as surface mass density according to \eref{eqn.cluster.bullet.sigma.galaxy}.  \map{\Sigma} and \map{\kappa} observed peaks are shown for comparison.  The central dominant (cD) galaxy of the main cluster, the brightest cluster galaxy (BCG) of the subcluster, and the MOG predicted gravitational center are shown.  J2000 and map (x,y) coordinates are listed in \tref{table.cluster.bullet.coords}. Component masses (integrated within a 100 kpc radius aperture) for the main and subcluster, the MOG center and the total predicted baryonic mass, $M_{\rm baryon}$, for the {\bc} are shown in \tref{table.cluster.bullet.mass}.}
\end{figure}

\subsection{\label{section.cluster.bullet.baryon}Visible baryon distribution}\index{Modified gravity!Mass profile|(}

\begin{figure}[ht]
\begin{picture}(460,235)(0,0)
\put(5,0){\includegraphics[width=0.98\textwidth]{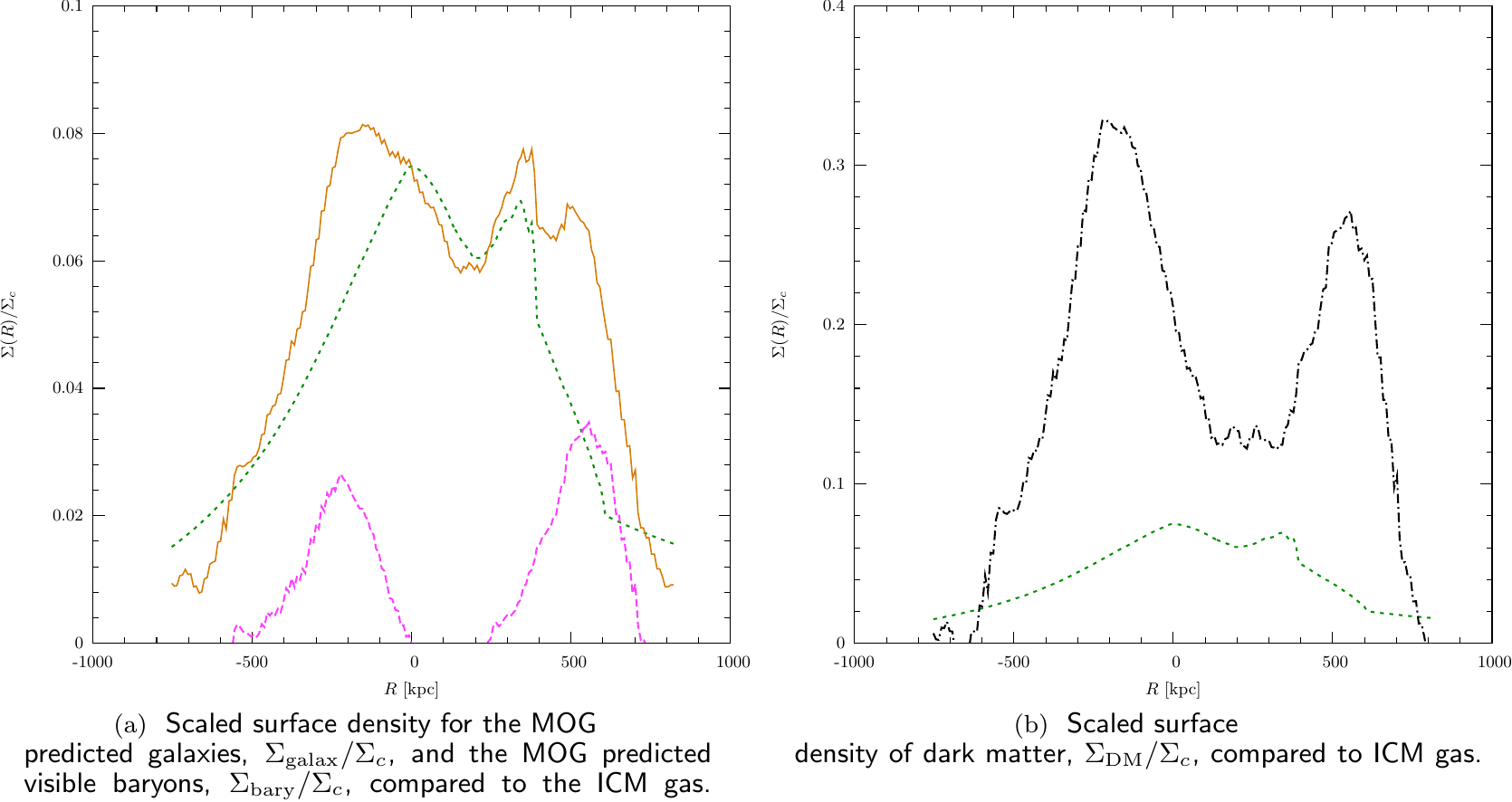}}
\end{picture}
\caption[Distribution of visible and dark matter]{\label{figure.cluster.bullet.distribution} {{\sf\small {\bc} -- Distribution of visible and dark matter}}\break\break Plot of the scaled surface density, 
$\Sigma(r)/\Sigma_{c}$, of the ICM gas is shown in short-dashed green on each
plot along the line connecting the main cluster \map{\Sigma} peak with the main cD.  In the left panel, the MOG prediction for the galaxies of \eref{eqn.cluster.bullet.sigma.galaxy} is shown in long-dashed magenta, and the prediction of the visible baryonic mass due to the combined ICM gas mass and galaxies is shown in solid brown.  The calculation of \eref{eqn.cluster.bullet.sigma.darkmatter} for dark matter is plotted in the right panel in dash-dot black.}
\end{figure}

The galaxies contribute a ``measurable'' surface mass density,
\begin{equation}
\label{eqn.cluster.bullet.sigma.galaxy}\Sigma_{\rm galax}(r) \approx \frac{G_{N}}{G(r)}\left(\kappa(r) \Sigma_{c}-{\bar \Sigma}(r)\right),
\end{equation}
which we may interpret as the difference between the \map{\kappa} and the scaled contribution from the weighted surface density of the ICM gas.  The result of the galaxy subtraction of \eref{eqn.cluster.bullet.sigma.galaxy} is shown as the galactic surface mass density map, in \fref{figure.cluster.bullet.galaxy}.  The surface mass density of the visible baryons is taken to be the sum of the ICM gas component and the galaxies, as shown in the left panel of \fref{figure.cluster.bullet.distribution}.

The total mass of the galaxies is determined by integrating over the \map{\Sigma},
\begin{equation}
\label{eqn.cluster.bullet.mass.galaxies} M_{\rm galax} = \int \Sigma_{\rm galax}(r) dx dy.
\end{equation}

\citet{Brownstein.MNRAS.2007.382} performed the integration within a 100 kpc radius aperture about the main cluster cD and subcluster
BCG, separately, the results of which are listed in \tref{table.cluster.bullet.mass}, where they are compared with the upper limits on
galaxy masses set by HST observations.  If the hypothesis that the predicted $M_{\rm galax}$ is below the bound set by HST observations is true, then it follows that 
\begin{equation}
\label{eqn.cluster.bullet.mass.bary} M_{\rm bary} = M_{\rm gas} + M_{\rm galax},
\end{equation}
requires no addition of non-baryonic dark matter.  The results of our best-fit for $M_{\rm gas}$, $M_{\rm galax}$ and $M_{\rm
bary}$ of \eref{eqn.cluster.bullet.mass.bary} are listed in  \tref{table.cluster.bullet.mass}.  
The result of $M_{\rm galax}/M_{\rm
gas} \approx 0.4\%$ in the central ICM is due to the excellent fit in MOG across the hundreds of kpc separating the main
and subcluster.\index{Modified gravity!Mass profile|)}
\subsection{\label{section.cluster.bullet.darkmatter}Dark matter distribution}\index{Dark matter!Missing mass problem}\index{Dark matter!Mass profile|(}

In the absence of modified gravity, dark matter is hypothesized to account for all of the ``missing mass'' which results
in applying Newton/Einstein gravity.  This means, for the November 15, 2006 data
release~\protect\citep{Clowe.APJL.2006.648,Bradac.APJ.2006.652,Clowe.NPBPS.2007.173,Clowe:dataProduct}, 
that the ``detected''  dark matter must contribute a surface mass density,
\begin{equation}
\label{eqn.cluster.bullet.sigma.darkmatter}\Sigma_{\rm DM}(x,y) \approx  \kappa(x,y)  \Sigma_{c} - \Sigma(x,y), 
\end{equation}
and is plotted in the right panel of \fref{figure.cluster.bullet.distribution}.  

The total mass of the dark matter distribution
with an associated total mass,
\begin{equation}
\label{eqn.cluster.bullet.mass.darkmatter.0} M_{\rm DM} = \int \Sigma_{\rm DM}(r)
dx dy.
\end{equation}
Upon substitution of \eref{eqn.cluster.bullet.sigma.darkmatter}, the integral of \eref{eqn.cluster.bullet.mass.darkmatter.0} becomes:
\begin{equation}
\label{eqn.cluster.bullet.mass.darkmatter} M_{\rm DM} = \Sigma_{c} \int \kappa(r) dx dy - \int \Sigma(r) dx dy
\end{equation}
where we have neglected $M_{\rm galax}$ in \eref{eqn.cluster.bullet.mass.darkmatter}, because the contribution from
the
galaxies is $\le 1$ -- $4\%$ of $M_{\rm total}$ due to dark matter dominance.

\citet{Brownstein.MNRAS.2007.382} computed $M_{\rm DM}$ in
\eref{eqn.cluster.bullet.mass.darkmatter} by a pixel-by-pixel sum over the convergence \map{\kappa} data and surface density
\map{\Sigma}
data, within a 100 kpc radius aperture around the main and subcluster \map{\kappa} peaks, respectively.  The
result of the computation, including the mass ratios, $M_{\rm galax}/M_{\rm gas}$, for the
main and subcluster and central ICM are provided in \tref{table.cluster.bullet.mass}.  

\begin{table}[h]
\caption[Component mass predictions.]{\label{table.cluster.bullet.mass}{\sf Component mass predictions.}}
\begin{center} \begin{tabular}{c||c|c|c||c} \multicolumn{5}{c}{}\\ \hline
{\sc Component}& {\sc Main cluster} & {\sc Subcluster} & {\sc Central ICM} & {\sc Total}  \\ 
(1)&(2)&(3)&(4)&(5) \\ \hline \hline
{$M_{\rm gas}$} & $7.0  \times 10^{12}\ M_{\solar}$ & $5.8  \times 10^{12}\ M_{\solar}$ &$6.3  \times 10^{12}\ M_{\solar}$& $2.2 \times 10^{14}\ M_{\solar}$ \\
{$M_{\rm galax}$} & $1.8  \times 10^{12}\ M_{\solar}$ & $3.1  \times 10^{12}\ M_{\solar}$ &$2.4  \times 10^{10}\ M_{\solar}$&  $3.8  \times 10^{13}\ M_{\solar}$ \\ \hline
{$M_{\rm bary}$} & $8.8\times 10^{12}\ M_{\solar}$ & $9.0\times 10^{12}\ M_{\solar}$ & $4.9 \times 10^{12}\ M_{\solar}$ & $2.6 \times 10^{14}\ M_{\solar}$ \\ \hline\hline
{$M_{\rm DM}$} & $2.1\times 10^{13}\ M_{\solar}$ & $1.7\times 10^{13}\ M_{\solar}$ & $1.4\times 10^{13}\ M_{\solar}$ & $6.8\times 10^{14}\ M_{\solar}$\\ \hline\hline
{$M_{\rm galax}/M_{\rm gas}$} & $26\%$ & $53\%$ &$0.4\%$&  $17\%$ \\ 
{$M_{\rm gas}/M_{\rm DM}$} & $33\%$ & $34\%$ &$45\%$&  $32\%$ \\ 
\hline \multicolumn{5}{c}{}
\end{tabular} \end{center}
\parbox{6.375in}{\small Notes. --- Column (1) specifies the component masses and mass fractions. Columns (2) and (3) list the component masses integrated within a 100 kpc radius aperture for the main and subcluster, respectively.  Columns (4) lists the component mass integrated within a 100 kpc radius aperture for the central ICM located at the MOG center.  Column (5) lists the total of each component masses integrated over the full \map{\Sigma}.}
 \end{table}
The dark matter result of $M_{\rm gas}/M_{\rm
DM} \approx 45\%$ in the central ICM implies that the evolutionary scenario does not lead to a spatial dissociation
between the dark matter and the ICM gas, which confirms that the merger is ongoing.  In contrast, the MOG result shows a true
dissociation between the galaxies and the ICM gas as required by the evolutionary scenario.  The baryon to dark matter fraction over the full \map{\Sigma} is $32\%$, which is significantly higher than the $\Lambda$-CDM cosmological baryon mass-fraction of $17^{+1.9}_{-1.2}\%$~\protect\citep{Spergel:ApJS:2007}.\index{Dark matter!Mass profile|)}

\subsection{\label{section.cluster.bullet.neutrino}Neutrino halos}\index{MOND!Neutrino halos|(}

\citet{Sanders.MNRAS.2003.342} postulated a two component model for the Coma cluster, adding a nonluminous  rigid sphere to include the contribution of finite mass neutrinos -- a candidate for hot non-baryonic dark matter -- with a constant density core,
\begin{equation}\label{eqn.cluster.bullet.neutrino.density}
\rho_{\nu} < 4.8 \times 10^{-27} \mbox{g cm}^{-3} \left(\frac{m_{\nu}}{2 \mbox{eV}}\right)^4 \left(\frac{T}{\mbox{keV}}\right)^{\frac{3}{2}}.
\end{equation}
\citet{Sanders.MNRAS.2003.342} assumed that the constant density cores have finite radii that scale as
\begin{equation}\label{eqn.cluster.bullet.neutrino.rnu}
r_{\nu} = 2 r_{c},
\end{equation}
where \(r_{c}\) is the gas core radius of the isothermal King \(\beta\)-model of \eref{eqn.cluster.xraymass.isothermal.betaRhoModel}.  For a 2 eV neutrino and the accepted experimental value of \tref{table.cluster.bullet.temp} suggest
\begin{equation}\label{eqn.cluster.bullet.neutrino.main.density}
\rho_{\nu} < 3.9^{+0.7}_{-0.5} \times 10^{6} M_{\solar}\mbox{kpc}^{-3}.
\end{equation}
Substituting \eref{eqn.cluster.bullet.rc} into \eref{eqn.cluster.bullet.neutrino.rnu} gives the constant density neutrino core radius,
\begin{equation}\label{eqn.cluster.bullet.neutrino.main.rnu}
r_{\nu} = 556 \pm 14 \mbox{kpc},
\end{equation}
which has an integrated mass within an aperture of 100 kpc of 
\begin{equation}\label{eqn.cluster.bullet.neutrino.main.mass.aperture}
M_{100} = 1.6^{+0.3}_{-0.2} \times 10^{13} M_{\solar},
\end{equation}
which is within 30\% of the required value according to \tref{table.cluster.bullet.mass}.  However, the total integrated core mass of 
\begin{equation}\label{eqn.cluster.bullet.neutrino.main.mass.total}
M_{\nu} = 2.8^{+0.5}_{-0.3} \times 10^{15} M_{\solar},
\end{equation}
exceeds the mass of the {\bc}  by a factor of three, implying that while \eref{eqn.cluster.bullet.neutrino.main.density} is reasonable, \erefs{eqn.cluster.bullet.neutrino.rnu}{eqn.cluster.bullet.neutrino.main.rnu} may be overestimating the extent of the neutrino halos by a factor of two.  

\citet{Sanders:2007MNRAS.380..331S} elaborated on the MOND neutrino-baryon model of clusters,  confirming the result of \sref{section.cluster.models.mass} that the need for dark matter appears to decrease with increasing temperature, suggesting \eref{eqn.cluster.bullet.neutrino.density} is opposite to observation, but argued that the observed trend is caused by the cooling and inflow of baryons.  \citet{Sanders:2007MNRAS.380..331S} provided an improved neutrino halo model for clusters, with the constant density core of \eref{eqn.cluster.bullet.neutrino.density}, and included a theoretically derived scaling relation instead of \eref{eqn.cluster.bullet.neutrino.rnu},
\begin{equation}\label{eqn.cluster.bullet.neutrino.rnu.mond}
r_{\rm \nu} = \left\{ \begin{array}{ll}
0.7\,\mbox{Mpc} \left(\frac{m_{\nu}}{2 {\rm eV}}\right)^{-4/3} \left(\frac{T}{{\rm keV}}\right)^{\frac{1}{6}} & \mbox{for}\,T \le 3 \,\mbox{keV} \left(\frac{m_{\nu}}{2 {\rm eV}}\right)^{-8/5}\\
1.1\,\mbox{Mpc} \left(\frac{m_{\nu}}{2 {\rm eV}}\right)^{-2} \left(\frac{T}{{\rm keV}}\right)^{-\frac{1}{4}} & \mbox{otherwise},\end{array}\right.
\end{equation}
which assumes three flavours of neutrinos, each of which have comparable velocity dispersions to the baryons and maintain their cosmological density ratio, 
\begin{equation}\label{eqn.cluster.bullet.neutrino.cosmo}
\Omega_{\nu}/\Omega_{\rm baryon} = 2.8 \left(\frac{m_{\nu}}{2 \mbox{eV}}\right).
\end{equation}
Now, substituting the accepted experimental value of \tref{table.cluster.bullet.temp} of \(T = 14.5^{+2.0}_{-1.7}\) keV into \eref{eqn.cluster.bullet.neutrino.rnu.mond} gives the constant density 2 eV neutrino core radius,
\begin{equation}\label{eqn.cluster.bullet.neutrino.main.rnu.improved}
r_{\nu} = 575^{+20}_{-17}\ \mbox{kpc},
\end{equation}
which is consistent with \eref{eqn.cluster.bullet.neutrino.main.rnu} and therefore too large by a factor of 2 to explain the total mass of the {\bc}.

\citet{Angus.APJL.2007.654} confirmed that a simple model of 4 dominant constant density cores of 2 eV neutrinos can supply the missing mass in the peaks of the gravitational lensing convergence \map{\kappa} of the {\bc}, provided the neutrino cores have radii \(r_{\nu} \lesssim 50\) kpc.  \citet{Brownstein.MNRAS.2007.382} provided a comparison of the surface density of dark matter to the surface density of the X-ray emitting ICM gas, shown in Panel (b) of \fref{figure.cluster.bullet.distribution}, consistent with 2 extended, overlapping halos centered at the galactic regions, which may have constant density cores in the inner 50 kpc, but then declining more rapidly.

In the absence of scaling relations, such as those investigated by \citet{Sanders.MNRAS.2003.342}, each neutrino halo requires additional free parameters to specify the shape of the density profile, which may be better described by other possibilities including the King \(\beta\)-model, or the core-modified dark matter profile of \sref{section.cluster.models.darkmatter} which fits clusters of galaxies without the necessity of MOND.   The core-modified dark matter fit to the main cluster of the {\bc} from \sref{section.cluster.models}, with best-fit parameters listed in the top row of Panel (a) of \tref{table.cluster.models.bestfit}, provides
\begin{equation}\label{eqn.cluster.bullet.neutrino.coremodified.density}
\rho_{0} = (2.73 \pm 0.47) \times 10^{6} M_{\solar}\mbox{kpc}^{-3},
\end{equation}
\begin{equation}\label{eqn.cluster.bullet.neutrino.coremodified.rs}
r_{s} = 328.1 \pm 26.7\,\mbox{kpc},
\end{equation}
Identifying \eref{eqn.cluster.bullet.neutrino.coremodified.density} with \eref{eqn.cluster.bullet.neutrino.density}, we may solve for the upper limit on the neutrino mass, 
\begin{equation}\label{eqn.cluster.bullet.neutrino.upperlimit}
m_{\nu} < 1.8 \pm 0.1\,\mbox{eV},
\end{equation}
which is below the Mainz/Troitsk experimental limit on the electron neutrino, \(m_{\nu,e}<2.2\) eV, but is falsifiable in the near future.

\citet{Angus.MNRAS.2008.387} decomposed the mass profiles of 26 X-ray systems according to MOND, with temperatures ranging from 0.5 keV to 9 keV, and concluded that whatever the equilibrium distribution, 2 eV neutrino halos cannot explain the inner 100 to 150 kiloparsecs of clusters within MOND.  This issue is seen in the dynamic mass factors plotted in \fref{figure.cluster.models.Gamma}, since each plot is maximized in the inner region of every cluster in the sample, where the missing mass problem is most pronounced.\index{MOND!Neutrino halos|)}

\chapterquote{I was like a boy playing on the sea-shore, and diverting myself now and then finding a smoother pebble or a prettier shell than ordinary, whilst the great ocean of truth lay all undiscovered before me.}{Sir Isaac Newton}
\chapter{\label{chapter.solar}Solar system}\index{Newton's central potential!Kepler's three laws}

The motion of the planets and planetoids, their satellites, and the chunks of matter that comprise the asteroids and the comets are along paths derived from matter's response to gravity.  The opportunity to discover new celestial physics in the solar system provides a challenge to form deeper understandings of Kepler's eponymous laws, from which Newton's theory of universal gravitation is founded.  Precise observation of orbits of the many bodies in the solar system suggest Kepler's three laws require subtle corrections:

\begin{description}
\item[\sc Kepler's first law:] The path of planets and bodies about the sun are {\it near} elliptic in shape, with a focus {\it near} the center of the sun, but changing in time under the influence of Jupiter and the other solar bodies.
\item[\sc Kepler's second law:] An imaginary line drawn from the center of a body to the center of a body in orbit will sweep out {\it nearly} equal changing areas in equal intervals of time, where the change in area slightly increases if orbital angular momentum is transferred to the orbiting body from the spin of the central body, and decreases if angular momentum is transferred in the opposite direction.
\item[\sc Kepler's third law:] The ratio of the squares of the periods of any two planets is {\it nearly} equal to the ratio of the cubes of their average distances from the sun, where the difference in this {\it near equality} is most significant at the orbit of Jupiter. 
\end{description}

Because the sun is not the only source of gravity in the solar system, and since so few solutions to Einstein's gravity theory are known, modelling gravity in the solar system is a managed process, such as the astronomer's ephemerides, which are datacentric solutions without the elegance and utility of a theoretical prediction, which does not need daily updates to correct for unmodelled physics, deemed unnecessary. Jupiter adds a significant source of gravity to the solar system, with mass \(M_{\jupiter} = 0.0095 M_{\solar}\), which is more than twice the total mass of all the other smaller bodies, combined.

New physics beyond the orbit of Jupiter must be {\it nearly} consistent with Kepler's three laws, and should make quantitative predictions of the necessary amendments.  However, all terrestrial and solar system attempts to falsify Moffat's nonsymmetric gravity theory (NGT) have led only to upper bounds on the possible strength of the modified gravity fifth force, including predictions for the Gravity Probe B experiment.  \citet{Moffat.PRD.1990.41} considered spinning test particles and the motion of a gyroscope, finding that the difference between the NGT correction to the gyroscope precession, and the Einstein correction, would be smaller than the Gravity Probe B experiment could detect in orbit about Earth.

\citet{Brownstein:CQG:2006} considered the motion of the Pioneer 10 and 11 spacecraft in the metric skew-tensor gravity theory, as in \sref{section.mog.mstg}, proving that the unexpected sunward acceleration can be explained by modified gravity without leading to disagreement between the predicted and actual orbits of the outermost planets.  According to the Pioneer Explorer Collaboration, the most likely explanation is that there is a systematic origin to the effect, such as a thermal recoil force investigated by \citet{Toth.PRD.2009}, using a simulated Pioneer 10 data set, but neither has NASA ruled out the modified gravity solution, presented in \sref{section.solar.pioneer.solution}.
\addtocontentsheading{lof}{Solar System}
\section{\label{section.solar.pioneer}Pioneer 10/11 Anomaly}

The radio tracking data from the Pioneer 10/11 spacecraft during their travel to the outer parts of the solar system have revealed a possible anomalous acceleration. The Doppler data obtained at distances \(r\) from the Sun between \(20\) and \(70\) astronomical units (AU) showed the anomaly as a deviation from Newton's and Einstein's gravitational theories.  At this time, NASA continues to support the search for a gravitational solution, as in \sref{section.solar.pioneer.solution}, but the Pioneer Explorer Collaboration may eventually be able to rule out  modified gravity as the origin of the effect, once the recovered data sets have been formatted and a comprehensive model can be applied, as progressing according to \citet{Toth.PRD.2009}.

\citet{Brownstein:CQG:2006} applied the metric skew-tensor gravity theory of \sref{section.mog.mstg}, in which Einstein gravity is coupled to a Kalb-Ramond-Proca field, as in \sref{section.mog.mstg.action}, and provided a fit to the available anomalous acceleration data or the Pioneer 10/11 spacecraft consistent with all current satellite, laser ranging and observations for the inner planets.

The Pioneer anomalous acceleration observations are described in \sref{section.solar.pioneer.anomaly}, and the fit is presented in  \sref{section.solar.pioneer.solution}. The effect of modified gravity in the solar system on Kepler's law of motion and the planetary ephemerides are explored in \sref{section.solar.pioneer.kepler} and \sref{section.solar.pioneer.ephemeris}, respectively, and the constraints set by observations of the anomalous perihelion advance are identified in \sref{section.solar.pioneer.perihelion}.

\subsection{\label{section.solar.pioneer.anomaly}Pioneer anomalous acceleration}

\citet{Anderson.PRL.1998.81,Anderson.PRD.2002.65,Turyshev.gr-qc/0510081} observed the Doppler residuals data as the differences of the observed Doppler velocity from the modelled Doppler velocity, and computed the  anomalous acceleration directed towards the Sun, with an {\it approximately constant} amplitude over the range of distance, \(20\au < r < 70\au\):
\begin{equation} \label{aP}
a_P=(8.74\pm 1.33)\times 10^{-8}\,{\rm cm}\,s^{-2}.
\end{equation}

After a determined attempt to account for all {\it known} sources of systematic errors, \citet{Anderson.PRL.1998.81,Anderson.PRD.2002.65,Turyshev.gr-qc/0510081} reached the conclusion that the Pioneer anomalous acceleration towards the Sun could be a real physical effect that requires a physical explanation. \citet{Turyshev.IJMPD.2006.15} reviewed NASA's efforts to recover the extended Pioneer doppler data set,  emphasizing  that the apparent difficulty to explain the anomaly within standard physics is a motivation to look for new physics, including the model of \citet{Brownstein:CQG:2006}.  

In NASA's official statement, \citet{Turyshev.2007.update} reported,
\begin{quotation}``As of March 2007, the existence of the anomaly is confirmed by seven independent investigations using different navigational codes -- the signal is present in the Doppler data received from both Pioneers 10 and 11. The most important question now is to identify the cause of this anomalous frequency drift discovered in the Pioneer data.

``\ldots Our thermal modelling of the Pioneer vehicles is progressing very well. We finished the development of the geometric mathematical models of the spacecraft that include geometry and properties of most of the important spacecraft components and surfaces. We are now working on the thermo-dynamical model of the vehicles. At this stage, we have a very good understanding of heat re-distribution within the craft and soon will be ready to compute the heat flow to the outside of the craft. Soon, we will be able to tell whether or not heat contributes to the formation of the anomaly.''\end{quotation}

MOND is not considered a viable mechanism because the value of the MOND universal acceleration of \eref{eqn.galaxy.mond.a0} that provides good fits to galaxy rotation curves, as in \cref{chapter.galaxy}, is orders of magnitude smaller than the acceleration of the Pioneer satellites, \(a_{P}\), until the satellite reaches the MOND transition radius of \eref{eqn.mog.mond.aether.dynamic.rt},
\begin{equation}\label{eqn.solar.pioneer.anomaly.rt}
r_{t} = \sqrt{\frac{G_{N}M_{\solar}}{a_0}} \sim 7700 \au,
\end{equation}
and is not likely to be observed on the scales of the solar system.

Galaxy scale dark matter cannot affect the solar system, since the density of the Milky Way dark matter halo in the vicinity of the solar system is 
\begin{equation} \label{eqn.solar.pioneer.darkmatter.density}
\rho \sim 5.0 \times 10^{-19} M_{\solar}/\au^3,
\end{equation}
and therefore a galaxy scale dark matter globe, equivalent in mass to Earth, would have a radius of greater than 10,000 \au.

\subsection{\label{section.solar.pioneer.solution}Gravitational solution}

The acceleration law of \erefs{eqn.mog.mstg.mog.Gforce}{eqn.mog.mstg.mog.fullGspherical}, derived from the metric skew-tensor gravity theory of \sref{section.mog.mstg.mog}, can be written
\begin{equation}
\label{eqn.solar.pioneer.accelerationGrun} a(r)=-\frac{G(r)M}{r^2},
\end{equation}
where
\begin{equation}
\label{eqn.solar.pioneer.runningNewton} G(r)=G_{N}\biggl[1+\alpha(r)\biggl(1-\exp(-r/\lambda(r))
\biggl(1+\frac{r}{\lambda(r)}\biggr)\biggr)\biggr].
\end{equation}
\citet{Brownstein:CQG:2006} postulated a gravitational solution that the Pioneer 10/11 anomaly is caused by the difference between
the running \(G(r)\) of \eref{eqn.solar.pioneer.runningNewton} and the bare value, \(G_{N}\). So the
Pioneer anomalous acceleration directed towards the center of the Sun is given by
\begin{equation}\label{eqn.solar.pioneer.aP}
a_P=-\frac{\delta G(r)M_{\odot}}{r^2},
\end{equation}
where
\begin{equation} \label{eqn.solar.pioneer.deltaG}
\delta G(r)=G(r) - G_{N} = G_{N}\alpha(r)\biggl[1-\exp(-r/\lambda(r))
\biggl(1+\frac{r}{\lambda(r)}\biggr)\biggr].
\end{equation}
The dynamic mass factor is defined as
\begin{eqnarray} \label{eqn.solar.pioneer.Gamma}
\Gamma(r) &=& G(r)/G_{N} = 1 + \frac{\delta G(r)}{G_N}\\
\label{eqn.solar.pioneer.Gamma.mog} &=& G(r)/G_{N} = 1 + \alpha(r)\biggl[1-\exp(-r/\lambda(r))
\biggl(1+\frac{r}{\lambda(r)}\biggr)\biggr],
\end{eqnarray}
and measures the degree to which the observed acceleration of the Pioneer satellite differs from the Newtonian acceleration,
\begin{equation} \label{eqn.solar.pioneer.Gamma.aP}
a_P(r) = (\Gamma(r) - 1)a_N(r),
\end{equation}
where the Newtonian acceleration is 
\begin{equation} \label{eqn.solar.pioneer.Gamma.Newtonian}
a_N(r) = -\frac{G_N M}{r^2}.
\end{equation}
Therefore, a measurement of \(\Gamma(r) \sim 1\) dismisses the Pioneer anomaly, whereas modified gravity predicts a monotonically increasing \(\Gamma(r)\) due to \eref{eqn.solar.pioneer.Gamma.mog}. 

\citet{Brownstein:CQG:2006} proposed the following parametric representations of the running of \(\alpha(r)\) and \(\lambda(r)\):
\begin{equation}
\label{eqn.solar.pioneer.alpha} \alpha(r) =\alpha_\infty(1-\exp(-r/{\bar r}))^{b/2},
\end{equation}
\begin{equation}
\label{eqn.solar.pioneer.lambda} \lambda(r)=\frac{\lambda_\infty}{{(1-\exp(-r/{\bar r}))^b}}.
\end{equation}
Here, \({\bar r}\) is a non-running distance scale parameter and \(b\) is a constant.

\newcommand{\subanomaly}{\small Pioneer 10/11 anomalous acceleration, \(a_{P}(r)\) in \(10^{-8}\) cm s\(^{-1}\), vs. orbital distance, \(r\) in \(\au\)}
\begin{figure}
\begin{picture}(460,175)(0,0)
\put(0,0){\includegraphics[width=1.0\textwidth]{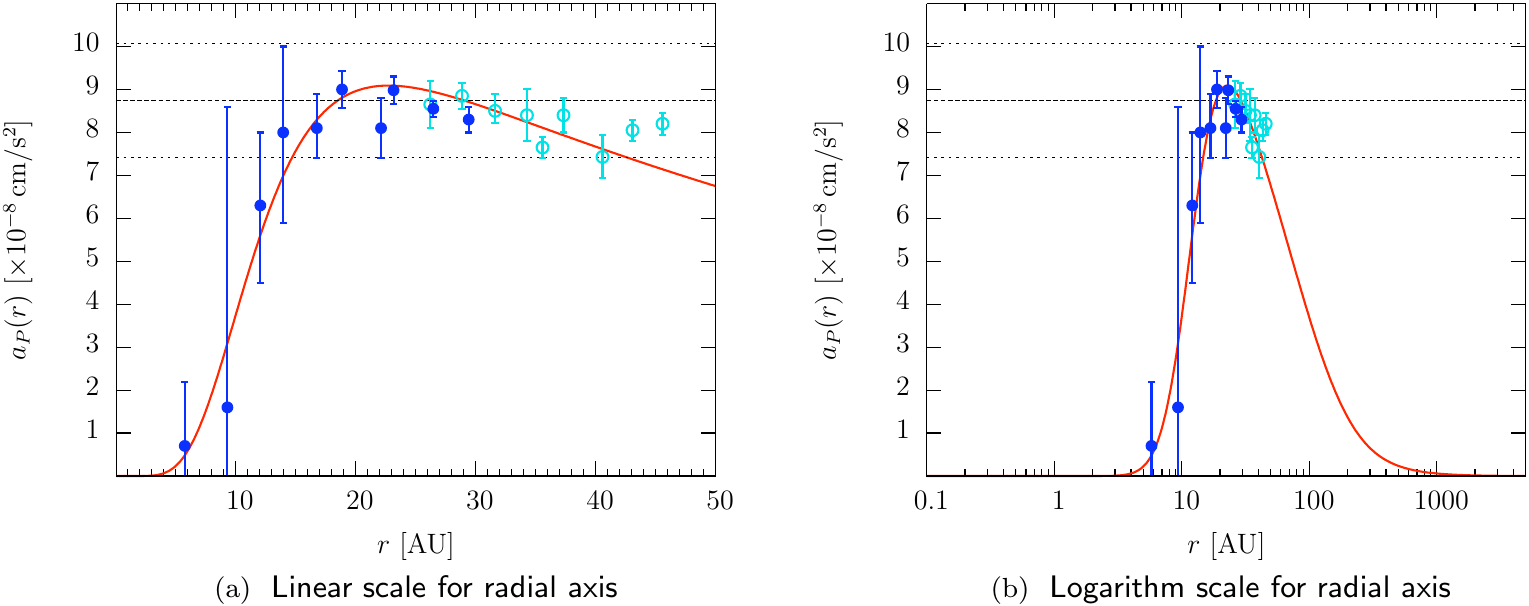}}
\end{picture}
\caption[Pioneer 10/11 anomalous acceleration]{\label{figure.solar.pioneer.anomaly}{{\sf\small Pioneer 10/11 anomalous acceleration}\break\break{\subanomaly}, of \eref{eqn.solar.pioneer.aP}, is plotted on a linear scale out to \(r=50 \au\), in the left panel, and on a logarithmic scale out to \(r = 5,000 \au\), in the right panel.}}
\begin{picture}(460,55)(0,0)
\put(0,0){\includegraphics[width=1.0\textwidth]{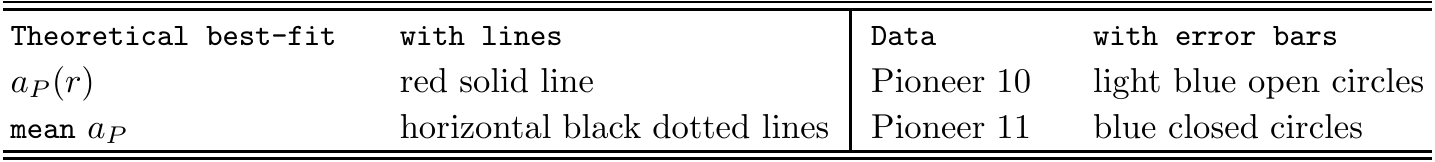}}
\end{picture}
\end{figure}

In \fref{figure.solar.pioneer.anomaly}, we display a best-fit to the Pioneer 10/11 anomalous acceleration data from \citet[Figure 4]{Nieto.CQG.2005.22}
obtained using a nonlinear least-squares fitting routine including estimated errors from the Doppler shift
observations~\protect\citep{Anderson.PRD.2002.65}.

The best-fit parameters are:
\begin{eqnarray}
\nonumber \alpha_\infty &=& (1.00\pm0.02)\times 10^{-3},\\
\nonumber \lambda_\infty &=& 47\pm 1\au ,\\
\nonumber {\bar r} &=& 4.6\pm 0.2\au ,\\
\label{eqn.solar.pioneer.bestparameters} b &=& 4.0.
\end{eqnarray}
The small uncertainties in the best-fit parameters are due to the remarkably low variance of residuals corresponding 
to a reduced \(\chi^{2}\) per degree of freedom of 0.42 signalling a good fit.

\newcommand{\subPioneerGamma}{\small Pioneer 10/11 anomaly dynamic mass factor, \(\Gamma(r)\), vs. orbital distance, \(r\) in \(\au\)}
\begin{figure}
\begin{picture}(460,175)(0,0)
\put(0,0){\includegraphics[width=1.0\textwidth]{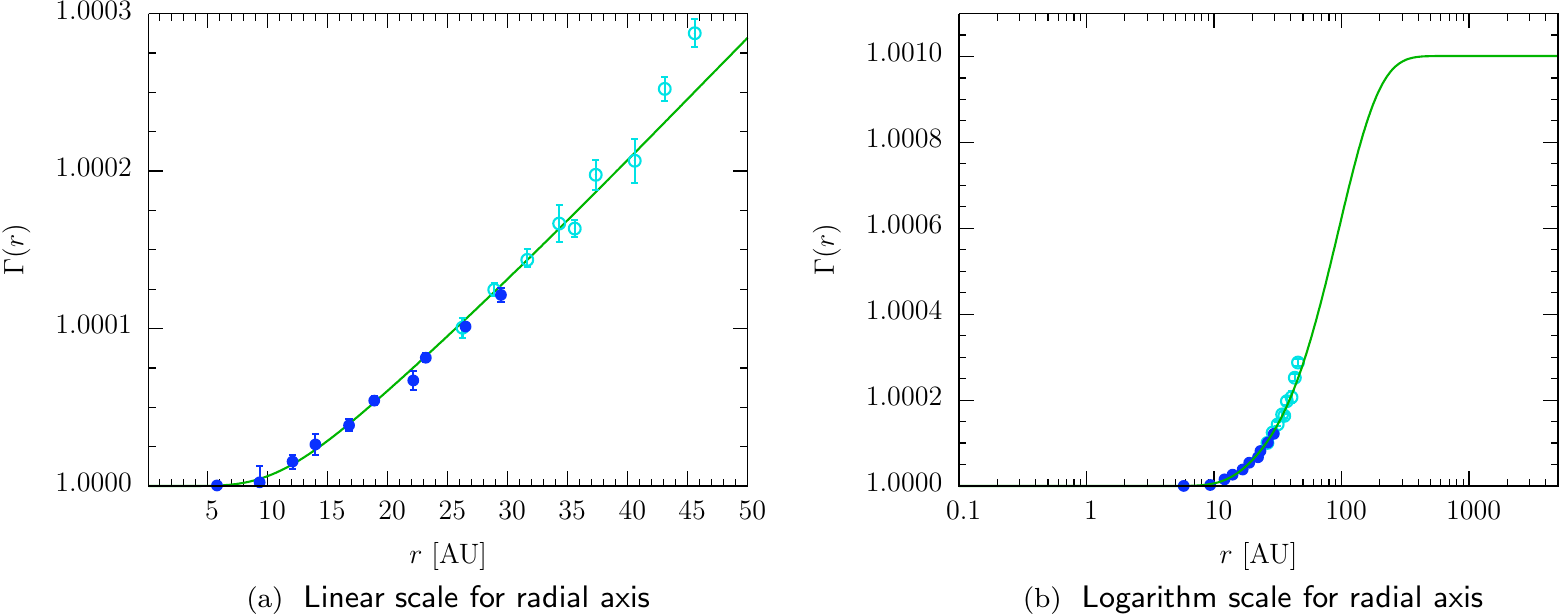}}
\end{picture}
\caption[Dynamic mass factor]{\label{figure.solar.pioneer.Gamma}{{\sf\small Dynamic mass factor in the Solar system}\break\break{\subPioneerGamma}, of \eref{eqn.solar.pioneer.Gamma}, is plotted on a linear scale out to \(r=50 \au\), in the left panel, and on a logarithmic scale out to \(r = 5,000 \au\), in the right panel.}}
\begin{picture}(460,55)(0,0)
\put(0,0){\includegraphics[width=1.0\textwidth]{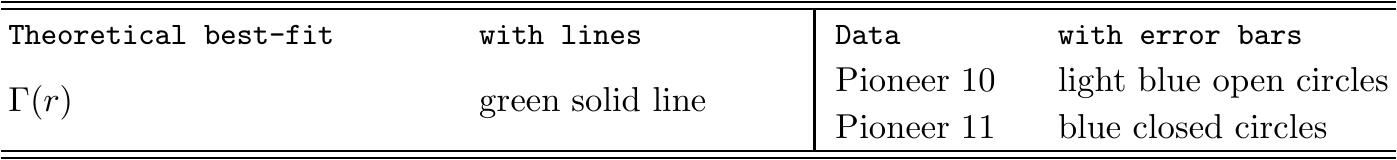}}
\end{picture}
\end{figure}

In \fref{figure.solar.pioneer.Gamma}, we display the Pioneer 10/11 data in the form of dynamic mass factors: 
\begin{equation} \label{eqn.solar.pioneer.Gamma.data}
\Gamma(r) = \frac{a_N(r) + a_P(r)}{a_N(r)},
\end{equation}
which is a rearrangement of \eref{eqn.solar.pioneer.Gamma.aP}, and compare to the MOG prediction of \eref{eqn.solar.pioneer.Gamma.mog}  for
the parametric values of
\(\alpha(r)\) and \(\lambda(r)\) of \eref{eqn.solar.pioneer.alpha} and \eref{eqn.solar.pioneer.lambda}, respectively, using the best-fit values for the parameters given in
\eref{eqn.solar.pioneer.bestparameters}.
The behaviour of \(G(r)/G_{N}\) is closely constrained to unity over the inner planets until beyond the orbit
of Saturn (\(r \gtrsim 10 \au\)) where the deviation in Newton's constant increases to an asymptotic value of
\(G_{\infty}/G_{N} \rightarrow 1.001\) over a distance of hundreds of \(\au\).

Although MOND is not expected to provide a viable solution to the Pioneer 10/11 anomaly because of \eref{eqn.solar.pioneer.anomaly.rt}, the variation in the dynamic mass factor, \(\Gamma(r)\), shown in Panel (a) of \fref{figure.solar.pioneer.Gamma}, is consistent with the deep MOND linear relation of \eref{eqn.mog.mond.aether.dynamic.Gamma.deep} with a best-fit MOND acceleration of 
\begin{equation} \label{eqn.solar.pioneer.mondacceleration}
a_0 = (3.0 \pm 0.3) \times 10^{-11}\,\mbox{cm s}^{-2},
\end{equation}
provided the MOND interpolating function is so gentle that the onset of the deep MOND regime occurs at 
\begin{equation} \label{eqn.solar.pioneer.mondacceleration.newregime}
r = 10.5 \pm 0.5 \au,
\end{equation}
instead of \(r_t\), which is not consistent with MOND, and improbable to explain using any generalized theory involving a preferred frame, as in \sref{section.mog.mond.aether}, including Bekenstein's TEVES theory of \sref{section.mog.mond.aether.bekenstein}.

Since the density of the Milky Way dark matter in the vicinity of the solar system, according to \eref{eqn.solar.pioneer.darkmatter.density}, is at least 10 orders of magnitude too small to affect the acceleration of spacecraft, then the Solar System must have its own halo for dark matter to provide a viable solution to the Pioneer 10/11 anomaly.  \citet{Frere.PRD.2008} calculated the bound on the dark matter density of a spherical halo centered about the sun from high precision Solar System measurements, finding that a dark matter halo around the Solar System may be as much as 5 to 6 orders of magnitude more dense than the Milky Way's dark matter halo, but this is still at least 4 orders of magnitude too low to affect the acceleration of spacecraft in the Solar System. 
\newcommand{\subKepler}{\small The cube of the orbital distance, \(a_{PL}^3\) in \(\au^3\), vs. the square of the orbital period, \(T_{PL}^2\) in day\(^2\)}
\begin{figure}
\begin{picture}(460,185)(0,0)
\put(0,40){\includegraphics[width=0.48\textwidth]{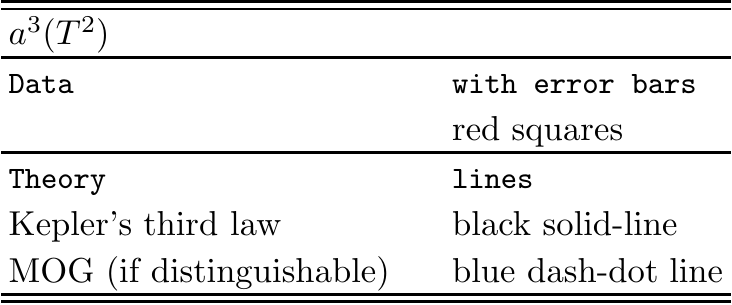}}
\put(225,0){\includegraphics[width=0.5\textwidth]{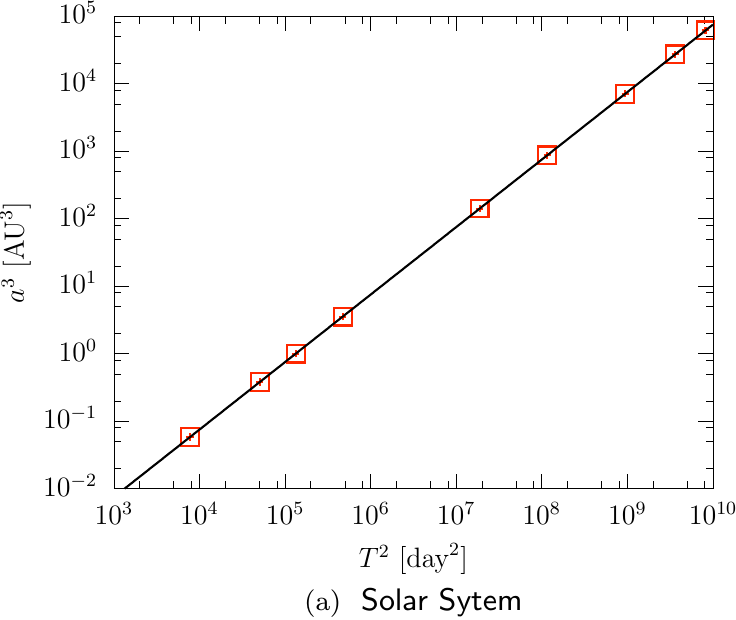}}
\end{picture}
\caption[Kepler's third law]{\label{figure.solar.pioneer.kepler} {{\sf\small  Solar system --- Kepler's third law.}}\break\break{\subKepler} for the Solar system.  The orbital data consist of the measured semi-major axis of the planateary orbit, \(a_{PL}\) and the measured planetary sidereal orbital period, \(T_{PL}\), listed in \tref{table.solar.pioneer.ephemerides}, respectively. Corrections due to MOG in Kepler's third law to \eref{eqn.solar.pioneer.Gamma} is plotted using the result for \(\Gamma(r)\) of \fref{figure.solar.pioneer.Gamma}. {\it The figure is continued.}}
\end{figure}

\begin{figure}
\begin{picture}(460,375)(82,265) 
\put(30,12){\includegraphics[width=1.28\textwidth]{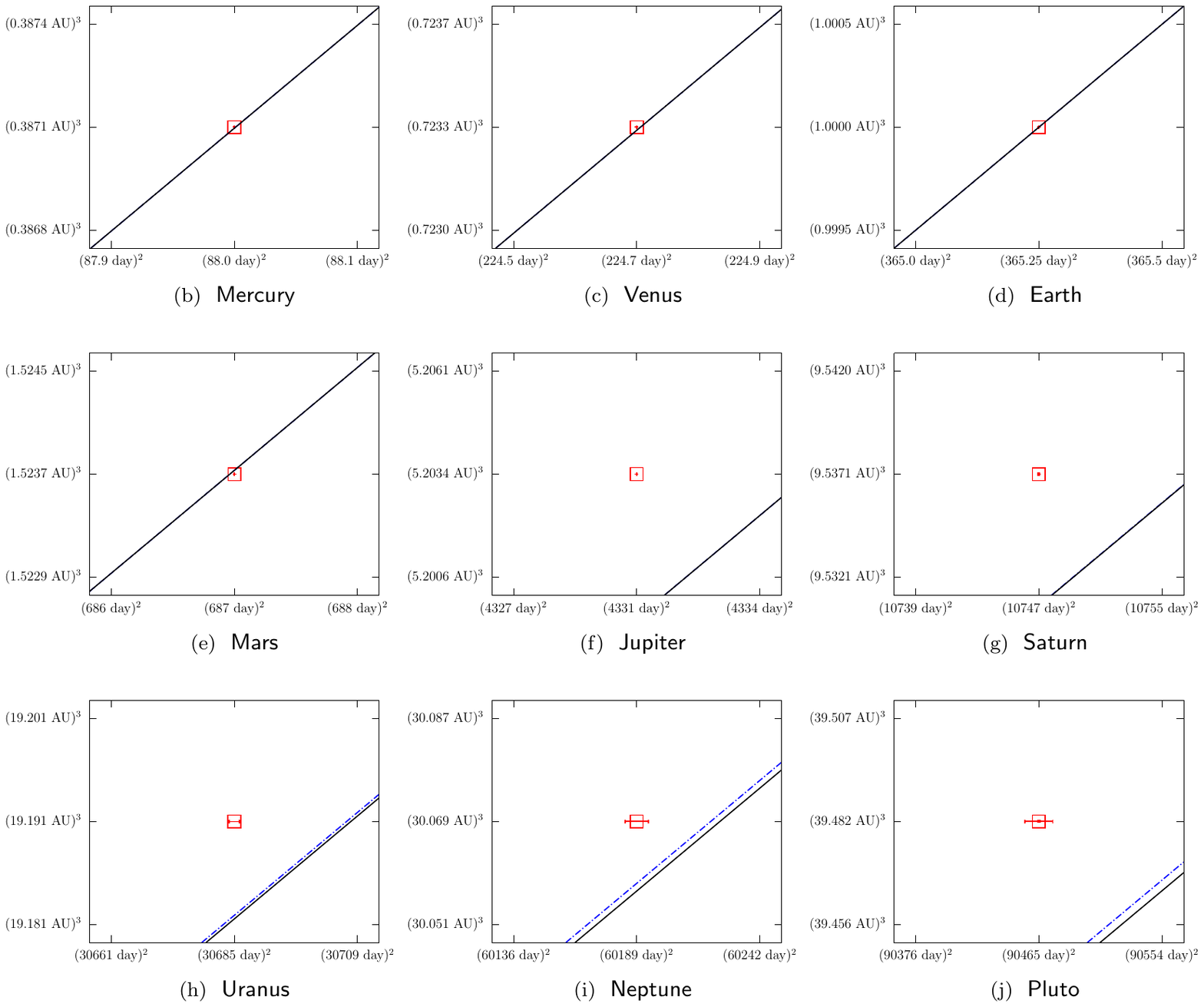}}
\put(82,50){\includegraphics[width=0.98\textwidth]{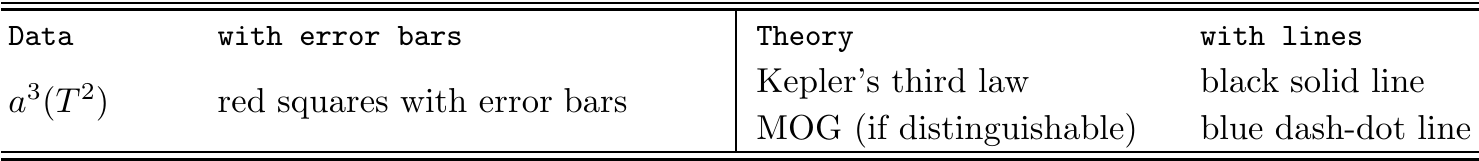}}
\end{picture}
\fcont{figure.solar.pioneer.kepler}{\sf\small Solar system --- Kepler's third law.\break}
{\subKepler} for the planets of the Solar system.  The orbital data consist of the measured semi-major axis of the planateary orbit, \(a_{PL}\) and the measured planetary sidereal orbital period, \(T_{PL}\), listed in \tref{table.solar.pioneer.ephemerides}, respectively. Corrections due to MOG in Kepler's third law to \eref{eqn.solar.pioneer.Gamma} is plotted using the result for \(\Gamma(r)\) of \fref{figure.solar.pioneer.Gamma}. 
\end{figure}

\subsection{\label{section.solar.pioneer.kepler}Kepler's laws of motion}

A consequence of a variation of \(G\) and \(GM_\odot\) for the solar system is a modification of Kepler's third law:
\begin{equation}\label{eqn.solar.pioneer.Kepler}
a_{PL}^3=G(a_{PL})M_\odot\biggl(\frac{T_{PL}}{2\pi}\biggr)^2,
\end{equation}
where \(T_{PL}\) is the planetary sidereal orbital period and \(a_{PL}\) is the physically measured semi-major axis of the planetary orbit. For given
values of \(a_{PL}\) and \(T_{PL}\), \eref{eqn.solar.pioneer.Kepler} can be used to determine \(G(r)M_\odot\). The standard method is to use
astrometric data to define \(GM_\odot\) for a constant value, 
\begin{equation} \label{eqn.solar.pioneer.gauss0}
G(r)M_\odot=G(a_{\oplus})M_\odot=\kappa^2,
\end{equation}
where \(a_{\oplus}\) is the semi-major axis for Earth's orbit about the Sun, and \(\kappa\) is the Gaussian gravitational constant given
by\footnote[1]{\url{http://ssd.jpl.nasa.gov/?constants}}
\begin{equation} \label{eqn.solar.pioneer.gauss}
\kappa=0.01720209895 \au^{3/2}/{\rm day}.
\end{equation}
We obtain the standard semi-major axis value at 1 AU:
\begin{equation} \label{eqn.solar.pioneer.abar}
{\bar a}_{PL}^3 =G(a_\oplus)M_\odot\biggl(\frac{T_{PL}}{2\pi}\biggr)^2.
\end{equation}
For several planets such as Mercury, Venus, Mars and Jupiter there are planetary ranging data, spacecraft tracking data and
radiotechnical flyby observations available, and it is possible to measure \(a_{PL}\) directly. For a distance varying \(GM_\odot\) we
derive~\protect\citep{Fischbach.1999,Talmadge.PRL.1988.61}:
\begin{equation}
\label{eqn.solar.pioneer.eta0}
\biggl(\frac{a_{PL}}{{\bar a}_{PL}}\biggr)=1+\eta_{PL} =\biggl[\frac{G(a_{PL})M_\odot}{\kappa^2}\biggr]^{1/3}.
\end{equation}
Here, it is assumed that \(GM_\odot\) varies with distance such that \(\eta_{PL}\) can be treated as a constant for the
orbit of a planet.  We may substitute the Gaussian gravitational constant of \eref{eqn.solar.pioneer.gauss0} into \eref{eqn.solar.pioneer.eta0} and obtain
\begin{equation}
\label{eqn.solar.pioneer.eta}
\eta_{PL} =\left[\frac{G(a_{PL})}{G(a_{\oplus})}\right]^{1/3}-1.
\end{equation}

\subsection{\label{section.solar.pioneer.ephemeris}Planetary Ephemerides}

For the nine planets, we obtain the values of \(\eta_{PL}\) shown in \tref{table.solar.pioneer.predictions}. We see that we are able to
obtain agreement well within the bounds of possible variation of \(GM_\odot\) consistent with the data~\protect\citep{Fischbach.1999,Talmadge.PRL.1988.61}
for Mercury, Venus, Mars and Jupiter.  No observational limit on \(\eta_{PL}\) for Saturn or the outer planets has yet
been established; but this is precisely where the deviation \(\delta G(r)/G_{N}\) leads to a sizeable contribution in the
theoretical prediction for \(\eta_{PL}\).

\addtocontentsheading{lot}{Solar System}

\begin{table}[ht]
\caption[Planetary predictions and observational limits]{\label{table.solar.pioneer.predictions}\sf Planetary predictions and observational limits}
\begin{center}
\begin{tabular}{@{}l|cccc}  \multicolumn{5}{c}{}\\ \hline
{\sc Planet} & \(r\) & {\sc Prediction} & {\sc Observational Limit} & \(a_{P}\) \\
&(AU)&\(\eta_{PL}\ (10^{-10})\) &\(\eta_{PL}\ (10^{-10})\) & \((10^{-8}\ \mbox{cm/s}^{2})\)\\
\phantom{Pl}(1)&(2)&(3)&(4)&(5)\\ \hline\hline
Mercury & 0.38 & \(-6.55 \times 10^{-5}\) & \(+40\pm50\) & \(1.41 \times 10^{-10}\)\\
Venus & 0.72 & \(-6.44 \times 10^{-5}\) & \(-55\pm35\) & \(5.82 \times 10^{-8}\)\\
Earth & 1.00 & \(0.00\times 10^{0}\) & \(0\) & \(1.16 \times 10^{-6}\)\\
Mars & 1.52 & \(4.93 \times 10^{-3}\) & \(-0.9\pm2.1\) & \(4.42 \times 10^{-5}\) \\
Jupiter & 5.20 & \(4.19 \times 10^{2}\) & \(+200\pm400\) & \(2.76 \times 10^{-1}\)\\
Saturn & 9.54 & \(1.67 \times 10^{4}\) & \ldots & \(3.27 \times 10^{0}\) \\ 
Uranus & 19.22 & \(1.84 \times 10^{5}\) & \ldots & \(8.86 \times 10^{0}\) \\
Neptune & 30.06 & \(4.39 \times 10^{5}\) & \ldots & \(8.65 \times 10^{0}\) \\
Pluto & 39.52 & \(6.77 \times 10^{5}\) & \ldots & \(7.72 \times 10^{0}\) \\ 
\hline \multicolumn{5}{c}{}
\end{tabular} \end{center}
\parbox{6.375in}{\small Notes. --- Theoretical predictions of the values of \(\eta_{PL}\)  and the best-fit theoretical predictions for the Pioneer Anomaly, \(a_{P}\), for the planetary bodies and observational limits.  Planetary bodies are listed in Column (1), with their mean distance, \(r\), from the Sun shown in Column (2).   Column (3) is the theoretical prediction of \(\eta_{PL}\)  of \eref{eqn.solar.pioneer.eta}, and may be compared to Column (4) for the observational limits taken from \citet{Talmadge.PRL.1988.61}.  No observational limits were computed beyond Saturn in \citet{Talmadge.PRL.1988.61} due to uncertainty in opical data. Beyond the outer planets, the theoretical predictions for \(\eta(r)\) approaches the asymptotic value \(\eta_{\infty} = 3.34 \times 10^{-4}\).  Column (5) lists the anomalous accelerations, at the planetary positions, predicted by the best-fit to the Pioneer 10/11 anomaly. }
\end{table}

The reason for the uncertainty beyond the orbit of Saturn and the lack of
observational limits on \(\eta_{PL}\) is that the ephemerides for the outer planets is based on optical measurements. 
Even in the context of Newton's theory, the extrapolation of Kepler's third law of \eref{eqn.solar.pioneer.Kepler} using the Gaussian
gravitational constant of \eref{eqn.solar.pioneer.gauss} which fits the inner planets missestimates the semi-major axis, \(a_{PL}\), or the
orbital period, \(T_{PL}\), of the outer planets resulting from Newtonian perturbations due to Jupiter and the gas giants
and their satellites, the Kuiper belt and hundreds of asteroids.  The latest version of the planetary part of the numerical ephemerides is a
numerical integration of the post-Newtonian metric.  It attempts to account for these perturbations from Kepler's law
beyond Saturn by a least squares adjustment to all the available observations including the CCD optical astrometric
observations of the outer planets.  These values (without uncertainty) are
available from the Solar System Dynamics Group (SSD) of the Jet Propulsion Laboratory (JPL) through the Horizon's
ephermeris DE410 online\footnote[2]{http://ssd.jpl.nasa.gov/horizons.html}. The Russian Academy of Sciences has also
placed their latest values known as
EPM2004 online\footnote[3]{ftp://quasar.ipa.nw.ru/incoming/EPM2004}.  Because the perturbations change daily due to the
motion within the
solar system, the planetary ephemerides quoted values for \(a_{PL}\) and \(T_{PL}\) change daily.  In order to compute
deviations from Kepler's third law for the outer planets, we
have listed today's best known values in \tref{table.solar.pioneer.ephemerides}.  

The uncertainty in the
EPM2004 deduced values for the semi-major axes of the planets, \(\Delta a_{PL}\), have been studied in \citet{Pitjeva.SSR.2005.39}
and the quoted values are listed in \tref{table.solar.pioneer.ephemerides}.  \citet{Pitjeva.SSR.2005.39} warned that the real errors
may be larger by an order of magnitude.  The uncertainty in the periods for the outer planets are not quoted in either
EPM2004 or DE410, and so we have assumed small uncertainties based on the precision provided by the JPL Horizon's
online ephemeris.

\begin{landscape}
\begin{table}
\caption[Mean ephemerides of planetary orbits]{\label{table.solar.pioneer.ephemerides}\sf Mean ephemerides of planetary orbits} 
\begin{center}
\begin{tabular}{@{}l|cccccc}  \multicolumn{7}{c}{}\\ \hline
{\sc Planet} & \(a_{PL}\) & \(\Delta a_{PL}\) & \(e_{PL}\) & \(T_{PL}\) & \(\Delta T_{PL}\) & \(\Delta \eta_{PL}\)  \\
&(AU)&(AU)&(days)&(days)&\((10^{-10})\)\\
\phantom{Pl}(1)&(2)&(3)&(4)&(5)&(6)&(7)\\ \hline\hline
Mercury & 0.38709893 & \(7.02 \times 10^{-13}\) & 0.206 & 87.968435 & \(5.0 \times 10^{-7}\) & \(3.79\times 10^{1}\) \\
Venus & 0.72333199 & \(2.20 \times 10^{-12}\) & 0.007 & 224.695434 & \(5.0 \times 10^{-7}\) & \(1.48\times 10^{1}\) \\
Earth & 1.00000011 & \(9.76 \times 10^{-13}\) & 0.017 & 365.256363051 & \(5.0 \times 10^{-10}\) & \(1.33\times 10^{-2}\) \\
Mars & 1.52366231 & \(4.39 \times 10^{-12}\) & 0.093 & 686.980 & \(5.0 \times 10^{-4}\) & \(4.85\times 10^{3}\) \\
Jupiter & 5.20336301 & \(4.27 \times 10^{-9}\) & 0.048 & 4330.595 & \(5.0 \times 10^{-4}\) & \(7.68\times 10^{2}\) \\
Saturn & 9.53707032 & \(2.82 \times 10^{-8}\) & 0.056 & 10746.94 & \(5.0 \times 10^{-2}\) & \(3.10\times 10^{4}\) \\ 
Uranus & 19.19126393 & \(2.57 \times 10^{-7}\) & 0.047 & 30685.4 & \(1.0 \times 10^{0}\)& \(2.17\times 10^{5}\) \\
Neptune & 30.06896348 & \(3.20 \times 10^{-6}\) & 0.009 & 60189. & \(5.0 \times 10^{0}\) & \(5.54\times 10^{5}\) \\
Pluto & 39.48168677 & \(2.32 \times 10^{-5}\) & 9.250 & 90465. & \(1.0 \times 10^{1}\) &\(7.37\times 10^{5}\) \\
\hline \multicolumn{7}{c}{}
\end{tabular}
 \end{center}
\parbox{1.0in}{\phantom{Notes.}}
\parbox{7.5in}{\small Notes. ---  Mean ephemerides for the planets' semi-major axes, orbital eccentricities, and the sidereal orbital periods.  Planetary bodies are listed in Column (1), with their semi-major axes (J2000), \(a_{PL}\), shown in Column (2).   The errors in the semi-major axes, listed in Column (3), are deduced from Table 4 of \citet{Pitjeva.SSR.2005.39} with 1 AU = \((149597870696.0 \pm 0.1)\) m.  Column (4) shows the planetary orbital eccentricities.  Column (5) and Column (6) show the sidereal orbital periods and uncertainties (JPL Horizon's online ephemeris), respectively.  Column (7) tabulates \(\Delta \eta_{PL}\) according to the propagation of the uncertainties of \eref{eqn.solar.pioneer.Deta}.}
\end{table}
\end{landscape}

We may calculate the uncertainty, \(\Delta \eta_{PL}\),
by propagating the errors \(\Delta a_{PL}\) and \(\Delta T_{PL}\) according to Equations \eref{eqn.solar.pioneer.abar} and \eref{eqn.solar.pioneer.eta0},
neglecting any uncertainty in the Gaussian gravitational constant of \eref{eqn.solar.pioneer.gauss}:
\begin{equation}
\label{eqn.solar.pioneer.Deta}
\Delta \eta_{PL} = \sqrt{\left(\frac{\Delta a_{PL}}{\bar a_{PL}}\right)^{2}+\left(\frac{2}{3}\frac{a_{PL}}{\bar a_{PL}}\frac{\Delta
T_{PL}}{T_{PL}}\right)^{2}}.
\end{equation}

Although according to \tref{table.solar.pioneer.predictions} we are consistent with the observational limits of  \(\eta_{PL}\) for the inner
planets to Jupiter, the computation of \citet{Talmadge.PRL.1988.61} attempted
to set model-independent constraints on the possible modifications of Newtonian gravity. 
The procedure was to run the planetary ephemerides numerical integration with the addition of \(\eta_{PL}\) as free
parameters.  Because there was one additional parameter for each planet, they were only able to find observational limits
for the inner planets including Jupiter.  In order to compute the observational limit for \(\eta_{PL}\) for the outer planets, it would be necessary to
compute the planetary ephemerides using the modified acceleration law of Equations \eref{eqn.solar.pioneer.accelerationGrun} and \eref{eqn.solar.pioneer.runningNewton}. 
Although this is beyond the scope of the current investigation, we may approximate here the observational limit of
\(\eta_{PL}\) for the outer planets as the uncertainty \(\Delta \eta_{PL}\) from \eref{eqn.solar.pioneer.Deta}, for the perturbations
of \fref{figure.solar.pioneer.Gamma}, \(\delta G(r)/G_{N}\), are small compared to the Newtonian perturbations
acting on the outer planets. The results for \(\Delta \eta_{PL}\) due to the uncertainty in the planetary ephemerides are presented in \tref{table.solar.pioneer.ephemerides} for the nine planets and exceed 
the predictions, \(\eta_{PL}\), of \tref{table.solar.pioneer.predictions}.

Modified gravity can explain the Pioneer anomalous acceleration data and still be consistent
with the accurate equivalence principle, lunar laser ranging and satellite data for the inner solar system as well as
the outer solar system planets including Pluto at a distance of $r= 39.52 \au =5.91\times 10^{12}$ meters. The ephemerides
for the outer planets are not as well know as the inner planets due to their large distances from the Sun.

The orbital data for Pluto only correspond to the planetoid having gone round 1/3 of its orbit. It is important that the
distance range parameter lies in the region $47 \au <
\lambda(r) < \infty$ for the best-fit to the Pioneer acceleration data, for the range in the modified Yukawa correction
to Newtonian gravity lies in a distance range beyond Pluto.  Further investigation of fifth force bounds obtained by an
analysis of the planetary data for the outer planets, based on the modified gravity theory is required.  We are
predicting that measurements of a fifth force in the solar system will become measurable at distances $r \gtrsim  10
\au$ from the Sun where as shown in \fref{figure.solar.pioneer.Gamma}, $\delta G(r)/G_0$ (and $\eta_{PL}$) become potentially
measurable.  The likely possibility that the Pioneer 10/11 anomaly is caused by thermal effects would cause these predictions to be treated as bounds on the effects of MOG in the solar system.  \citet{Moffat.CQG.2009.26} give good agreement with solar system bounds using the scalar-tensor-vector modified acceleration law of \sref{section.mog.stvg.yukawa}.

\subsection{\label{section.solar.pioneer.perihelion}Anomalous perihelion advance}

The relativistic equation of motion for a test particle in our gravitational theory may be solved
perturbatively in a weak field approximation for 
the anomalous perihelion advance of a planetary orbit:
\begin{equation}
\label{eqn.solar.pioneer.perihelionAdvance}\Delta\omega_{PL}=\frac{6\pi G_{N}M_\odot}{c^2a_{PL}(1-e_{PL}^2)}(1-\alpha_{PL}),
\end{equation}
where we have assumed as with Kepler's third law that \(GM_\odot\) and \(\alpha\) vary with distance such that they can
be
treated as constants for the orbit of a planet, where we have made use of the approximation \(G(r) \approx
G_{N}\)~\protect\citet{Moffat.JCAP03.2006}, which is the case from the fit to the Pioneer 10/11 anomalous acceleration data.
We may rewrite
\eref{eqn.solar.pioneer.perihelionAdvance} as the perihelion advance in arcseconds per century:
\begin{equation}
\label{eqn.solar.pioneer.perihelionAdvancedot}{\dot\omega_{PL}}=\frac{\Delta\omega_{PL}}{2\pi T_{PL}} = \frac{3 G_{N}M_\odot}{c^2a_{PL}(1-e_{PL}^2) T_{PL}}(1-\alpha_{PL}),
\end{equation}
where \(T_{PL}\) is the planetary orbital period, and \(e_{PL}\) is the planetary orbital eccentricity.  
We may separate
\eref{eqn.solar.pioneer.perihelionAdvancedot} into the usual Einstein anomalous perihelion advance, and a prediction of the correction
to the anomalous perihelion advance:
 \begin{equation}
\label{eqn.solar.pioneer.perihelionAdvance.split}{\dot \omega_{PL}}={\dot \omega_{0}} +{\dot \omega_{1}},
\end{equation}
where \begin{equation}
\label{eqn.solar.pioneer.perihelionAdvance.Einstein}{\dot \omega_{0}} = \frac{3 G_{N}M_\odot}{c^2a_{PL}(1-e_{PL}^2)T_{PL}},
\end{equation} 
\begin{equation}
\label{eqn.solar.pioneer.perihelionAdvance.Retrograde}{\dot \omega_{1}} = -\alpha_{PL} {\dot \omega_{0}},
\end{equation}

\begin{landscape}
\begin{table}
\caption[Planetary perihelion advance]{\label{table.solar.pioneer.perihelion}\sf Planetary perihelion advance}
\begin{center}
\begin{tabular}{@{}l|ccc|ccc} \multicolumn{7}{c}{}\\ \hline
&&&&\multicolumn{3}{|c}{ ${\dot \omega}\ (^{\prime\prime}/$century)}  \\
{\sc Planet} & $\alpha_{PL}$ & $\lambda_{PL}\ (\au)$ & $\delta G_{PL}/G_{0}$ & {\sc Einstein} & {\sc Retrograde} & {\sc Ephemeris}  \\
\phantom{Pl}(1)&(2)&(3)&(4)&(5)&(6)&(7)\\ \hline\hline
Mercury & $6.51 \times 10^{-6}$ & $1.11 \times 10^6$ & $3.98 \times 10^{-19}$ & $42.99$& $-2.80 \times\ 10^{-4}$
&$-0.0336\pm0.0050$ \\
Venus & $2.12 \times 10^{-5}$ & $1.05 \times 10^5$ & $5.04 \times 10^{-16}$ & $8.63$ & $-1.83 \times\ 10^{-4}$  & \ldots\\
Earth & $3.82 \times 10^{-5}$ & $3.23 \times 10^4$ & $1.84 \times 10^{-14}$ & $3.84$ & $-1.47 \times\ 10^{-4}$  &$-0.0002\pm 0.0004$\\
Mars & $7.95 \times 10^{-5}$ & $7.44 \times 10^3$ & $1.67 \times 10^{-12}$ & $1.35$ & $-1.07 \times\ 10^{-4}$  &$0.0001\pm 0.0005$\\
Jupiter & $4.59 \times 10^{-4}$ & $2.23 \times 10^2$ & $1.23 \times 10^{-7}$ & $0.0624$ & $-2.86 \times\
10^{-5}$  &$0.0062\pm 0.036$ \\
Saturn & $7.64 \times 10^{-4}$ & $8.05 \times 10^1$ & $4.96 \times 10^{-6}$ & $0.0137$ & $-1.05 \times\
10^{-5}$  &$-0.92\pm 2.9$\\
Uranus & $9.69 \times 10^{-4}$ & $5.00 \times 10^1$ & $5.55 \times 10^{-5}$  & $0.00239$ & $-2.31 \times\
10^{-6}$  &$0.57\pm 13.0$\\
Neptune & $9.97 \times 10^{-4}$ & $4.73 \times 10^1$ & $1.34 \times 10^{-4}$ & $0.00078$ & $-7.73 \times\
10^{-7}$  &
\ldots\\
Pluto & $1.00 \times 10^{-3}$ & $4.70 \times 10^1$ & $2.05 \times 10^{-4}$ & $0.00042$ & $-4.18 \times\ 10^{-7}$  & \ldots\\
\hline \multicolumn{7}{c}{}
\end{tabular} \end{center}
\parbox{1.0in}{\phantom{Notes.}}
\parbox{7.5in}{\small Notes. --- The values of the running parameters, $\alpha(r)$ of \eref{eqn.solar.pioneer.alpha} and $\lambda(r)$ of \eref{eqn.solar.pioneer.lambda} and the deviation in the dimensionless gravitational constant, $\delta G(r)/G_{0}$ of \eref{eqn.solar.pioneer.runningNewton}, calculated for each planet.  
Included on the right of the table is the theoretical (Einstein) perihelion advance of \eref{eqn.solar.pioneer.perihelionAdvance.Einstein},
and the predicted retrograde of \eref{eqn.solar.pioneer.perihelionAdvance.Retrograde} for the planets, and the limits set by the ephemeris~\protect\citet{Pitjeva.SSR.2005.39}.  Planetary bodies are listed in Column (1), with their mean distance, \(r\), from the Sun shown in Column (2).   Column (3) is the theoretical prediction of \(\eta_{PL}\)  of \eref{eqn.solar.pioneer.eta}, and may be compared to Column (4) for the observational limits taken from \citet{Talmadge.PRL.1988.61}.  No observational limits were computed beyond Saturn in \citet{Talmadge.PRL.1988.61} due to uncertainty in opical data. Beyond the outer planets, the theoretical predictions for \(\eta(r)\) approaches the asymptotic value \(\eta_{\infty} = 3.34 \times 10^{-4}\).  Column (5) lists the anomalous accelerations, at the planetary positions, predicted by the best-fit to the Pioneer 10/11 anomaly. }
\end{table}
\end{landscape}
\noindent are the Einstein anomalous perihelion advance, and the predicted retrograde, respectively.  Note the {\it minus-sign} in the predicted retrograde of \eref{eqn.solar.pioneer.perihelionAdvance.Retrograde} as compared to the Einstein anomalous perihelion advance of \eref{eqn.solar.pioneer.perihelionAdvance.Einstein}. The measured perihelion precession is
best known for the inner planets  (for Mercury the precession obtained from ranging data is known to \(0.5\%\)~\protect\citep{Will:LLR:2006}).  For each of the planets in the solar system, we find that
\(\alpha_{PL} << 1\), so that our fit to the Pioneer anomalous acceleration is in agreement with the relativistic
precession data.  The results for the Einstein perihelion advance, and our predicted retrograde for each planet, and the
observational limits set by the recent ultra-high precision ephemeris are listed in \tref{table.solar.pioneer.perihelion}.

The validity of the bounds on a possible fifth force obtained from the ephemerides of the outer planets Uranus, Neptune
and Pluto are critical in the exclusion of a parameter space for our fits to the Pioneer anomaly acceleration. Beyond
the outer planets, the theoretical prediction for \(\eta(r)\) approaches an asymptotic value:
\begin{equation} \label{eqn.solar.pioneer.etalim}
\eta_{\infty} \equiv \lim_{r \to \infty} \eta(r)= 3.34 \times 10^{-4}.
\end{equation}
We
see that the variations (running) of \(\alpha(r)\) and \(\lambda(r)\) with distance play an important role in
interpreting the data for the fifth force bounds. This is in contrast to the standard non-modified Yukawa correction to the Newtonian force law with fixed universal values of \(\alpha\) and \(\lambda\) and for the range of values \(0 < \lambda < \infty\), for which the equivalence principle and lunar laser ranging and radar ranging data to planetary probes exclude the possibility of a gravitational and fifth force explanation for the Pioneer anomaly.

Perhaps, a future deep space probe can produce data that can check the predictions obtained for the Pioneer anomaly from  modified gravity theory. Or perhaps utilizing Mars or Jupiter may clarify whether the Pioneer
anomaly is caused by the gravitational field.~\protect\citep{Page.APJ.2006.642}.  An
analysis of anomalous acceleration data obtained from earlier Doppler shift data
retrieval will clarify in better detail the apparent onset of the anomalous acceleration, or support the thermal recoil explanation of \citet{Toth.PRD.2009}, perhaps to as low a \(\chi^2\) as the modified gravity solution.

\part{\label{part.conclusions}Conclusions}		
\chapterquote{Any intelligent fool can make things bigger and more complex and more violent. It takes a touch of genius -- and a lot of courage to move in the opposite direction.}{Albert Einstein}
\chapter{\label{chapter.summary}Summary}

The mysteries of the gravitational field continue to challenge mankind as our physical models of the universe evolve.  Isaac Newton's great contribution was to deduce the analytical form of the force of gravity exerted by an isolated object.  However, the principle of superposition is not exact, as Newton assumed, because the gravitational field is non-linear, and models of the gravity internal to astrophysical matter distributions are Newtonian approximations.  Albert Einstein's great contribution was to deduce the geometric form of the relativity principle, reinterpreting the force of gravity as a geometric distortion of space and time, but this sets the gravitational field apart from the three other known forces.  Gravity is the only force that couples universally to matter and energy.  However, the strong equivalence principle, which Einstein first assumed, does not hold for stable gravity theories that include scalar, vector and tensor modifications to the metric with associated couplings, even when additional charged quantum numbers associated with new symmetries are suppressed.  Such modified gravity theories suggest the presence of a fifth force, which is assumed to couple universally to matter and energy, and gains in strength at astrophysical scales to become the dominant force.  This dominant force, if neglected by means of a Newton-Einstein approximation, emerges as the phantom of dark matter in galaxies and clusters of galaxies.\index{Equivalence principle!Universality of free fall}

\index{Dark matter!Missing mass problem}\index{Modified gravity!Phantom of dark matter}To address the hypothesis, stated in \sref{chapter.introduction.objective.theory}, to the missing mass problem in galaxy rotation curves and clusters of galaxies, the following theories were studied:
\begin{enumerate}
\item Cold non-baryonic dark matter (CDM),
\item Milgrom's modified Newtonian dynamics (MOND),\index{MOND}
\item Moffat's metric skew-tensor gravity theory (MSTG),\index{Modified gravity!Metric skew-tensor gravity}
\item Moffat's scalar tensor vector gravity theory (STVG).\index{Modified gravity!Scalar-tensor-vector gravity}
\end{enumerate}
Conclusions drawn upon identification of the missing mass as CDM is presented in \sref{section.summary.darkmatter}.  Some common lessons learned from the modified gravity theories are summarized in \sref{section.summary.theory}.  Corresponding to each of the astrophysical scales in \tref{table.introduction.organization.astroph},  conclusions based upon galactic-scale and cluster-scale astrophysics are summarized in \sref{section.summary.galaxy} and \sref{section.summary.cluster}, respectively.  Some possible directions for future astrophysical tests are suggested in \cref{chapter.future}.

\section{\label{section.summary.lessons}Lessons learned}

Whether identified as dark matter halos with profiles defined in \cref{chapter.darkmatter}, or the massive fifth-force fields of a modified gravity theory with acceleration laws derived in \cref{chapter.mog},  conclusions are drawn in \sref{section.summary.darkmatter} and \sref{section.summary.theory}, respectively.

\subsection{\label{section.summary.darkmatter}CDM halos}\index{Dark matter!Core-modified|(}

According to \tref{table.galaxy.darkmatter} from the Ursa Major sample of galaxies of \sref{section.galaxy.uma}, every one of the galaxy rotation curves presented in \fref{figure.galaxy.velocity} have excellent fits, within Einstein-Newton gravity including a non-baryonic dark matter halo density described by the core-modified dark matter profile of \eref{eqn.galaxy.dynamics.dm.coremodified}, with reasonable mean stellar mass to light ratios \(\Upsilon\), as compared to baryon suppressed fits with \(\Upsilon=0\).  

Provided baryons are included in the core-modified dark matter halo, the total mass with dark matter vs.\ velocity relation, plotted in \fref{figure.galaxy.halos.tfr.darkmatter}, showed the least scatter of any of the Tully-Fisher relations, even when compared to the modified gravity alternatives

Shown in the same table, in the leftmost columns, the Navarro-Frenk-White (NFW) profile of \eref{eqn.galaxy.dynamics.dm.nfw} showed low \(\chi^2\) when including the visible baryons, but because the cusp problem occurs precisely where baryons are important, there were exceptions in the sample of \cref{chapter.galaxy}, involving high and low surface dwarf galaxies, that could not be \(\chi^2\) best-fit for any non-zero value of \(\Upsilon\).  For those galaxies that the NFW formula produced excellent best-fits, the halo mass function overcompensated for the baryons by suppressing the best-fit stellar mass to light ratio, \(\Upsilon < 1\). \index{Dark matter!NFW formula}\index{Dark matter!Cusp problem}

The dynamic mass factors show no indication of a cusp at small \(r\), but unlike for clusters of galaxies, take on a global minimum, with \(\lim_{r \rightarrow 0} \Gamma(r) = 1\), whereas the core-modified dark matter formula provides excellent fits to the dynamic mass factors, \(\Gamma(r)\) for all positions, \(r\), plotted in \fref{figure.galaxy.Gamma}, for every galaxy in the sample including the dwarfs.\index{Dark matter!Dynamic mass factor \(\Gamma\)}

In the CDM hierarchical structure formation scenario, galactic halos are considered subhalos to the larger structure which is their cluster (or filament).  These subhalos are self-similar to the cluster halo if they are describable by similar fitting formulae.  Shown in Panel (a) of \tref{table.cluster.models.bestfit},  the core-modified dark matter profile of \eref{eqn.galaxy.dynamics.dm.coremodified} showed the lowest \(\chi^2\) best-fit cluster model parameters for the sample of \cref{chapter.cluster}, including X-ray clusters of varying mass, scale radius, and central temperature, as compared to the modified gravity alternatives.

The NFW profile could not be \(\chi^2\) best-fit to the sample of X-ray clusters because the data does not exhibit the cusp in the core, as shown by the variation in the dynamic mass factor, plotted in \fref{figure.cluster.models.Gamma}.  

Therefore, the core-modified dark matter formula describes a self-similar halo profile for both high and low surface brightness galaxies, including the dwarfs, and also for the clusters of galaxies, including the dwarf clusters and the {\bc}, provided the baryons are not suppressed in the fits.\index{Dark matter!Core-modified|)}
 
\subsection{\label{section.summary.theory}Modified gravity theories}\index{Modified gravity!Phantom of dark matter|(}

Unlike the core-modified dark matter or NFW profiles, each of the modified gravity theories produced low  \(\chi^2\) fits to the galaxy rotation curves of \cref{chapter.galaxy} with universal parameters, averaged over the sample, shown in \tref{table.galaxy.mond} for MOND, \tref{table.galaxy.mstg} for MSTG, and \tref{table.galaxy.stvg} for STVG.  The monotonic (near-linear) rise of the dynamic mass factor, plotted in  \fref{figure.galaxy.Gamma} across all gravity theories, suggests that the missing mass problem is most pronounced at the edge of the luminous disk.  

Only the STVG acceleration law of \sref{section.mog.stvg.yukawa} is derived without any phenomenological input from the Tully-Fisher relation.  The Tully-Fisher relations are compared in \tref{table.galaxy.halos.tfr}, for each of the gravity theories, and plotted in \fref{figure.galaxy.halos.tfr.diskbary}.

For the sample of X-ray clusters of galaxies of \cref{chapter.cluster}, it was confirmed that MOND provides poor best-fits to X-ray cluster mass profiles, as plotted in \fref{figure.cluster.models.mass}, even with a varying MOND acceleration, \(a_0\), as shown in Panel (b) of \tref{table.cluster.models.bestfit}.\index{MOND!Mass profile}  The possibility, within MOND, of including a non-luminous component, such as neutrino halos, was considered in the case of the {\bc} in \sref{section.cluster.bullet.neutrino}, although the models of \citet{Sanders.MNRAS.2003.342,Sanders:2007MNRAS.380..331S} may be overestimating the extent of the neutrino halos by a factor of two to provide a universally consistent explanation.\index{MOND!Neutrino halos}

Unlike MOND, both MSTG and STVG theories provided excellent best fits to the X-ray cluster mass profiles of \fref{figure.cluster.models.mass}, as shown in Panels (c) and (d) of \tref{table.cluster.models.bestfit}, respectively.  Whereas the STVG theory produced a constant dynamic mass factor, which is an approximate feature present in the data, the MSTG produced dynamic mass factors which mimicked the core-modified dark matter result and produced excellent fits, as shown in \fref{figure.cluster.models.Gamma}, from the smallest of the clusters to the largest {\bc}.  

Effectively, the common feature of the modified gravity paradigm is given by \eref{eqn.mog.mstg.mass}, in which 
\begin{equation}\label{eqn.conclusions.mog.mstg.mass}
M(r)  = \frac{G_{N}M_{N}(r)}{G(r)},
\end{equation}
where \(G(r)\) is the spatially varying gravitational coupling.  The modified gravity hypothesis of phantom dark matter suggests that, for sufficiently large \(r\), \(G(r)/G_{N} \gg 1\) so that the luminous, baryonic mass, \(M(r)\), is less than the observed Newtonian dynamic mass, \(M_{N}(r)\), by the same factor.

The metric skew-tensor gravity (MSTG) theory,\index{Modified gravity!Metric skew-tensor gravity|(} presented in \sref{section.mog.mstg}, identifies the phantom of dark matter with the massive skewon of the Kalb-Ramond-Proca field, as described in \sref{section.mog.mstg.action}, with a Yukawa interaction that leads to motion under the fifth force, as in \sref{section.mog.mstg.eom}.  For MSTG astrophysical predictions, the gravitational coupling, \(G(r)\),  of \eref{eqn.mog.mstg.mog.fullGspherical}, was substituted into \eref{eqn.conclusions.mog.mstg.mass} leading to a nonlinear equation because \(G(r)\) itself is a function of \(M(r)\), which was solved exactly in \erefs{eqn.mog.mstg.mass.mstg.soln}{eqn.mog.mstg.mass.mstg.xi}.  This solution permits the analytic computation of baryon masses according to MSTG from dynamic measurements such as velocity rotation curves or X-ray temperature distributions, without approximation.\index{Fifth force!Yukawa meson}

Using a simpler field structure, the MSTG skewon can be replaced by a massive vector phion in the scalar-tensor-vector gravity (STVG) theory,\index{Modified gravity!Scalar-tensor-vector gravity|(} presented in \sref{section.mog.stvg}, which then identifies the Maxwell-Proca field , as in \sref{section.mog.stvg.action}, as the phantom of dark matter.  The STVG action includes three dynamical scalar fields which lead to gravitationally strong interactions, as shown in the Yukawa phion theory of  \sref{section.mog.stvg.yukawa}, which leads to the modified central force law of \sref{section.mog.stvg.mog}.  For STVG astrophysical predictions, the gravitational coupling, \(G(r)\),  of \eref{eqn.mog.stvg.yukawa.Geff}, was substituted into \eref{eqn.mog.mstg.mass} leading to \eref{eqn.mog.mstg.mass.stvg}, where the Yukawa strength and phion mass, \(\alpha\) and \(\mu\), are given by \erefs{eqn.mog.stvg.mog.alpha}{eqn.mog.stvg.mog.mu}, respectively. \index{Modified gravity!Scalar-tensor-vector gravity|)}\index{Modified gravity!Metric skew-tensor gravity|)}

To compare the predictions of the gravity theories relevant to astrophysical scales, \eref{eqn.conclusions.mog.mstg.mass} can be written in terms of the dynamic mass factor, of \eref{eqn.darkmatter.astrophGamma.mog},
\begin{equation}\label{eqn.conclusions.mog.mass} 
M(r)  = \frac{M_{N}(r)}{\Gamma(r)}.
\end{equation}
It was shown that MOND may be the weak-field limit of certain Lorentz violating theories, including the family of theories discussed in \sref{section.mog.mond.aether}, which have non-metric field structure that are subject to violations of the strong equivalence principle, as described in \sref{section.mog.equivalence.violations}, and may thereby provide the phantom of dark matter detected in galaxy rotation curves and X-ray cluster masses.  In \sref{section.mog.mond.dynamic} it was shown that the dynamic mass factor due to modified dynamics at small accelerations, is precisely the inverse of the MOND interpolating function, according to \eref{eqn.mog.mond.aether.dynamic.Gamma},\index{Equivalence principle!Violations}\index{MOND!Interpolating function, \(\mu\)}
\begin{equation}\label{eqn.conclusions.qg.aether.dynamic.Gamma}
\Gamma(r) = 1/\mu(x(r)).
\end{equation}
At sufficiently small accelerations, the dynamic mass factor is a linear function in \(r\) according to \eref{eqn.mog.mond.aether.dynamic.Gamma.deep}, for all MOND interpolating functions, and is proportional to \(\sqrt{a_0}\).  Whereas this fits the observations in galaxy rotation curves, it does not correspond to the observations in X-ray cluster masses, even with larger values of the MOND universal acceleration.

The dynamic mass factor,\index{Dynamic mass factor, \(\Gamma\)}
\begin{equation}\label{eqn.conclusions.contributions.astroph.Gamma} 
\Gamma(r)  = \frac{M_{N}(r)}{M(r)},
\end{equation}
is the ratio of the Newtonian dynamic mass to the observed, baryonic mass.  For CDM, \eref{eqn.conclusions.contributions.astroph.Gamma} is constrained by the particular choice of the dark matter  fitting formula.

\index{Modified gravity!Phantom of dark matter|)}
\section{\label{section.summary.galaxy}Galactic astrophysics}\index{Dark matter!NFW formula}\index{Dark matter!Core-modified|(}

In \cref{chapter.galaxy}, a core-modified fitting formula was derived in \ssref{section.galaxy.dynamics}{subsection.galaxy.dynamics.coremodified}, and found to fit the sample of high and low surface brightness galaxies, including all of the dwarfs.  The NFW fitting formula led either to fits with very poor \(\chi^2\), if baryons were omitted, or to a suppression of the stellar mass-to-light ratio if photometric data was included, particularly for the dwarf galaxies.  The worst of these dwarf galaxies could not be fitted using the NFW profile with any non-zero value of the stellar mass-to-light ratio, as discussed in \ssref{section.galaxy.dynamics.dm}{subsection.newton.darkmatter.observations}.\index{Dark matter!Observations}

The core-modified fitting formula of \eref{eqn.newton.darkmatter.coremodified} produced the lowest reduced-\(\chi^2\) best-fits to the galaxy rotation curves for the sample of \sref{section.galaxy.uma}, plotted in \fref{figure.galaxy.velocity}, with two parameters, \(\rho_{0}\) and \(r_s\), which varied across the sample, tabulated in \tref{table.galaxy.darkmatter}.  The surface mass densities, plotted in \fref{figure.galaxy.Sigma}, show the baryon dominated cores transitioning to dark matter dominated halos.  The best-fitting mass profiles are plotted in \fref{figure.galaxy.mass}, showing the halo component negligible in the galaxy cores, but adding up to the dominant mass at the outermost radial points.  At large distances from the center of each galaxy in the sample, the density profile of the dark matter halo is well described by a steep power-law, with power-law index \(\gamma \rightarrow 3\), whereas at distances toward the center of the galaxy an increasingly shallow power-law is observed, as plotted in \fref{figure.galaxy.powerlaw}.\index{Dark matter!Power law, \(\gamma\)|(}  For distances less than the dark matter halo core radius, \(r < r_s\),the total density profile including baryons shows a universal \(\gamma \rightarrow 1\) power-law index, and the density profile of the dark matter component alone approaches a rarified, constant density core.\index{Dark matter!Core-modified|)}

\ssref{section.galaxy.dynamics.mond}{subsection.mog.mond.observations} discusses the MOND best-fits to the galaxy rotation curves for the sample of \sref{section.galaxy.uma}, plotted in \fref{figure.galaxy.velocity}, with universal acceleration, \(a_{0}\), tabulated in \tref{table.galaxy.mond}.\index{MOND!Observations}

\ssref{section.galaxy.dynamics.mog}{subsection.galaxy.dynamics.mog.observations} discusses the MSTG and STVG best-fits to the galaxy rotation curves for the sample of \sref{section.galaxy.uma}, plotted in \fref{figure.galaxy.velocity}, with universal parameters depending on the theory, tabulated in \tref{table.galaxy.mstg} and \tref{table.galaxy.stvg}, for MSTG and STVG respectively.\index{Modified gravity!Observations}

All of best-fits showed a central disk dominated by the Newtonian potential, where \(\Gamma(r) \sim 1\), outside of which, the dynamical mass factor increased approximately linearly with distance, plotted in \fref{figure.galaxy.Gamma}.  The results, plotted (with lines) are nearly equivalent for all of the gravity theories studied. However each theory's data (with error bars) has a dependence on the result of the best-fit stellar mass-to-light ratio, \(\Upsilon\), provided in \tref{table.galaxy.darkmatter} for the NFW profile and the core-modified dark matter profile, \tref{table.galaxy.mond} for MOND, \tref{table.galaxy.mstg} for MSTG, and \tref{table.galaxy.stvg} for STVG.\index{Dynamic mass factor, \(\Gamma\)}

The analysis of the Ursa Major sample of \sref{section.galaxy.uma} involved a series of calculations using a variety of computational resources.  The error analysis was a fruitful exercise in measuring the properties of the variation of the stellar mass-to-light ratio within high or low surface brightness galaxies, and the quality of the best-fit.  The stellar mass-to-light ratio, \(\Upsilon(r)\), varied strongly within every galaxy, in comparison to the variation across different galaxies of the best-fit stellar mass-to-light ratio.\index{Stellar mass-to-light ratio, \(\Upsilon\)|(}

The Newtonian core was calculated from the plot of the stellar mass-to-light ratio, \(\Upsilon(r)\) of \fref{figure.galaxy.masslight} (vertical lines), at the position, \(r_{c}\), where the Newtonian dynamics induce a rapid increase in the slope of \(\Upsilon(r)\), for every galaxy including the dwarfs. The best-fit stellar mass-to-light ratio for the Newtonian core is plotted (horizontal line) to \(r=r_{c}\). Within the luminous disk, the stellar mass-to-light ratio never exceeds a value of ten in any galaxy.  This naturally constrains the total amount of dark matter required and allows the best-fit dark matter theory, with two parameters \(\rho_0\) and \(r_s\), and the best-fit stellar mass-to-light ratio, \(\Upsilon\), to be simultaneously varied toward minimum \(\chi^2\).  

The stellar mass-to-light ratios, \(\Upsilon(r)\), plotted in \fref{figure.galaxy.masslight} (with lines) are nearly equivalent for MOND, MSTG and STVG, with similar mean values, near \(\Upsilon = 1\), with larger values for the best-fit Newtonian core model, in all galaxies.\index{Newton's central potential!Galaxy core}  The best-fit values, determined by a non-linear least squares fitting algorithm, are plotted (with horizontal lines).  Since the galaxies of Ursa Major are at a common distance from the Milky Way, the variation in the actual stellar mass-to-light ratio from galaxy to galaxy is not expected to be large.  The results for the best-fit stellar mass-to-light ratio, with a core-modified dark matter halo, are plotted (with horizontal lines), with mean values close to predicted values according to MOND and the MOG theories.  This is an example of the importance of the luminous baryons in the computation.  It is the variable stellar mass-to-light ratio which includes the data, by force, and allows the ultra-low reduced \(\chi^2\) test for the core-modified dark matter model.  The computational results are provided in \tref{table.galaxy.darkmatter} for the NFW profile and the core-modified dark matter profile, \tref{table.galaxy.mond} for MOND, \tref{table.galaxy.mstg} for MSTG, and \tref{table.galaxy.stvg} for STVG.\index{Stellar mass-to-light ratio, \(\Upsilon\)|)}

Qualitative assessment of each theory's predictions for the Ursa Major galaxies is provided in \sref{section.galaxy.halos}, and includes the predictions of the best-fit Newtonian core model of \sref{section.galaxy.halos.core}, which is a base-line for any improvement.  Whereas repatriating orphan features, as described in \sref{section.galaxy.halos.orphans}, provides a reasonable test for theories which fit galaxy rotation curves, particularly the core-modified dark matter halos which properly include the luminous baryons, as in \sref{section.galaxy.halos.coremodified}. The power-law profile, for either the NFW or core-modified models is derived as the logarithm slope of \eref{eqn.galaxy.uma.powerlaw.gamma}, which relates\index{Dark matter!Power law, \(\gamma\)|)}
\begin{equation}\label{eqn.conclusions.contributions.astroph}
\gamma(r) + 2 = - \frac{d \ln M(r)}{d \ln r},
\end{equation}
which depends on the baryon distribution through the mass-to-light ratio, according to \erefs{eqn.galaxy.uma.masstolight}{eqn.galaxy.uma.Gamma.darkmatter.mass}.  The halo component is computed by substituting \(M = M_{\rm halo}\) into \eref{eqn.conclusions.contributions.astroph} and the dark matter logarithm slope is computed by substituting \(M = \frac{4}{3}M_{\rm HI} + \Upsilon M_{\rm disk} + M_{\rm halo}\), thereby including the gaseous and luminous stellar disks into the Newtonian dynamic mass as discussed in \ssref{section.galaxy.uma.powerlaw}{subsection.galaxy.uma.powerlaw.cuspproblem}.\index{Dark matter!Cusp problem}

The implication of Occam's razor,\index{Dark matter!Occam's razor} that the total mass of a galaxy should be less in a theory without non-baryonic dark matter, depends on how large the halo is taken to be, beyond the luminous disk, where there is data.  The results for the variation in the total integrated mass, \(M(r)\), to the outermost radial position, \(r_{\rm out}\), are plotted in \fref{figure.galaxy.mass} (with lines), per theory.  Components are plotted for the actual HI gas, and the stellar disk is normalized with \(\Upsilon=1\) for relevance across theories, each with a best-fit \(\Upsilon\).  The dark matter halo component is plotted, which is a small part of the total mass in the dark matter model in the core of every galaxy studied.  However the dark matter halo component is the fastest rising mass in the galaxy because of the spherical distribution, compared to both the exponentially-thin gaseous disk, and the best-fit luminous disk, becoming dominant outside the Newtonian core, \(r_c\), per galaxy.

The Tully-Fisher relation, as in \sref{section.galaxy.tullyfisher}, confronts the dynamical importance of the luminous baryons compared to the ordinary Tully-Fisher relation, plotted in \SCfref{figure.galaxy.halos.tfr.kband}{Empirical K-band Tully-Fisher relation} (blue) with power-law index of \(a=4.1 \pm 0.4\) according to the best-fit of \eref{eqn.galaxy.halos.tullyfisher.index.kband}.  This result, although familiar, is today considered too large with larger samples providing a K-band Tully Fisher relation power-law index of \(a=3.4 \pm 0.1\)~\protect\citep{Tully.APJ.2000.533}.  \tref{table.galaxy.halos.tfr} provides the best-fit logarithm slopes and intercepts for six relations of the form \(M \propto v^{a}\), where either \(M = \Upsilon M_{\rm disk}\) for the stellar Tully-Fisher relation, or \(M = \frac{4}{3}M_{\rm HI} + \Upsilon M_{\rm disk}\) for the baryonic Tully-Fisher relation, or \(M = \frac{4}{3}M_{\rm HI} + \Upsilon M_{\rm disk} + M_{\rm halo}\) for the total mass-velocity relation, including the best-fit core-modified dark matter halo, as in \SCfref{figure.galaxy.halos.tfr.darkmatter}{Dark matter total mass vs.\,velocity}.  The best-fits (with lines) to the theoretical Tully-Fisher relations are plotted (with error bars) for the two stellar and two baryonic relations in \fref{figure.galaxy.halos.tfr.diskbary}, per gravity theory.\index{Galaxy rotation, \(v\)!Tully-Fisher relation}

In MOND, the relation with the least scatter is the baryonic mass to outermost velocity, \(v_{\rm out}\), with a logarithm slope of \(a=3.3\pm0.3\) whereas the ordinary, \(v_{\rm max}\), stellar relation has a logarithm slope of \(a=4.1\pm0.4\).  However, the STVG, MSTG and core-modified dark matter theories show less scatter for the ordinary baryonic relation with smaller logarithm slopes of \(a=2.6\pm0.2\), \(a=2.5\pm0.2\), and \(a=2.5\pm0.3\), respectively.  Whereas overall, the relation which shows the least scatter is that of the total mass, including luminous baryons and core-modified dark matter, vs.\ velocity as plotted in \SCfref{figure.galaxy.halos.tfr.darkmatter}{Dark matter total mass vs.\,velocity}, with \(a=2.9\pm0.2\) for the ordinary case.  This restores the Tully-Fisher relation to the dark matter solution, provided the baryons are included, and dismisses the notion that the Tully-Fisher relation is unnatural due to dark matter dominance.\index{MOND}

Some possible directions for future galactic astrophysical tests are presented in \sref{section.future.galaxy}.
\section{\label{section.summary.cluster}Cluster-scale astrophysics}

In \cref{chapter.cluster}, the missing problem is studied with X-ray clusters, as described in \sref{section.cluster.xraymass}, using the astrophysical sample of \sref{section.cluster.xraymass.astroph}.  The King \(\beta\)-model of the X-ray gas distribution of \eref{eqn.cluster.xraymass.isothermal.betaRhoModel} is presented in \sref{section.cluster.xraymass.isothermal}, and the collisionless Boltzmann equations are derived, and the solution is shown in \eref{eqn.cluster.xraymass.isothermal.isothermalAccelerationProfile}.  In \sref{section.cluster.xraymass.Sigma}, a computation expresses the surface mass density map, \(\Sigma(x,y)\), as the simple analytical result of \erefs{eqn.cluster.xraymass.Sigma.surfaceMassDensity0}{eqn.cluster.xraymass.Sigma.surfaceMassDensity}, in terms of the King \(\beta\)-model best-fit values (\(\rho_{0},~\beta,~r_{c}\)).  This was used in the analysis of the {\bc}, in \sref{section.cluster.bullet}, as the initial study of the X-ray gas map, described in \sref{section.cluster.bullet.Sigma}.  The Newtonian dynamical mass of \eref{eqn.cluster.xraymass.isothermalNewtonsMass} is derived in \sref{section.cluster.xraymass.Gamma}.\index{Surface mass, \(\Sigma\)}

Each of the theories that were tested at galactic-scale using galaxy rotation curves are studied at cluster-scale  using the best-fit cluster models of \sref{section.cluster.models}.  The NFW fitting formula of \eref{eqn.newton.darkmatter.nfw} generated large uncertainties due to a parameter degeneracy between the central density parameter, \(\rho_{0}\), and the scale radius, \(r_{s}\), and could not be \(\chi^2\)-fitted.  Without numerical convergence, the NFW results either over-predicted the density at the core or under-predicted the total mass.  Alternatively, the core-modified dark matter model of \sref{section.cluster.models.darkmatter} provided excellent fits using the fitting formula of  \eref{eqn.newton.darkmatter.coremodified}, with results provided in Panel (a) of  \tref{table.cluster.models.bestfit}.\index{Dark matter!Observations}

Although MOND does not fit X-ray cluster masses, there have been studies that claim improvements using a larger value of the MOND universal acceleration, \(a_{0}\), or to include a non-luminous component.  Both avenues were considered in this thesis.  In \sref{section.cluster.models.mond}, the MOND universal acceleration was treated as a variable parameter, and found to lead to very poor fits with larger than galaxy-scale values of \(a_0\), as shown in Panel (b) of \tref{table.cluster.models.bestfit}.  Furthermore, the problem of too large a dynamic mass factor, for small \(r\), and too small a dynamic factor for intermediate \(r\), was not corrected although the increase in the value of \(a_0\) did lead to the correct dynamic mass factor for \(r \sim r_{\rm out}\).  This means that the best-fit MOND solution in clusters of galaxies without dark matter does not fit the shape of the X-ray mass profile, except at \(r_{\rm out}\) so that the total mass is corrected.\index{MOND!Observations}

Both MSTG and STVG theories, as discussed in \sref{section.cluster.models.mog}, provided excellent fits to the X-ray gas masses, with results provided in Panels (c) and (d), respectively, of \tref{table.cluster.models.bestfit}.\index{Modified gravity!Observations}

The missing mass problem at cluster-scale, presented in \sref{section.cluster.models.mass}, is best demonstrated by the dynamic mass factor of \eref{eqn.conclusions.contributions.astroph.Gamma}, which is plotted (with lines) in \fref{figure.cluster.models.Gamma}, per gravity theory.  The observations are plotted (with error bars) as the ratio of the  Newtonian dynamical mass, \(M_N(r)\) of \eref{eqn.cluster.xraymass.isothermalNewtonsMass}, to the best-fit King \(\beta\)-model to the gas mass, \(M(r)\), of \erefs{eqn.cluster.xraymass.isothermal.betaRhoModel}{eqn.cluster.xraymass.isothermal.massProfile}, and is therefore a ratio of two very large masses, increasing with separation, \(r\), as plotted (with red crosses, and green circles, including error bars).

The study of the {\bc}, in \sref{section.cluster.bullet},  includes a detailed analysis of the X-ray gas map, of \sref{section.cluster.bullet.Sigma}, with the subcluster masked-out, and a best-fit to the King \(\beta\)-model for the main cluster is derived in \erefss{eqn.cluster.bullet.beta}{eqn.cluster.bullet.rc}{eqn.cluster.bullet.rho0}.  The subcluster subtracted from the X-ray data was added to the best-fit King \(\beta\) model of the main cluster, as shown in \fref{figure.cluster.bullet.SigmaModel}.  The study of the gravitational lensing convergence map, according to \sref{section.cluster.bullet.kappa}, applies the derivation of \erefs{eqn.galaxy.uma.scaledSurfaceMassDensity.MOG}{eqn.galaxy.uma.Sigma.MOG} with running gravitational coupling, \(G(r)\), leading to the best-fit \(\kappa\)-model of \fref{figure.cluster.bullet.kappaModel}.  The best-fit MOG model is used to compute the visible baryon surface mass profile in \sref{section.cluster.bullet.baryon}, and the result of the galaxy subtraction of \eref{eqn.cluster.bullet.sigma.galaxy} is shown in \fref{figure.cluster.bullet.galaxy}.  Predictions for the component masses of the main cluster, subcluster and central ICM are listed in \tref{table.cluster.bullet.mass}.  The dark matter distribution is computed as the difference between the \map{\kappa} and the scaled \map{\Sigma} by a pixel by pixel subtraction in \sref{section.cluster.bullet.darkmatter}.  The distribution of visible and dark matter, plotted in \fref{figure.cluster.bullet.distribution}, provides a comparison of the distribution of galaxies, gas and total baryons according to MOG in Panel (a), as compared to the distribution of gas and dark matter in Panel (b).  Although the NFW profile does not fit the main cluster, the excellent fit to the core-modified profile is shown in Panel (a) of \fref{figure.cluster.models.mass}.

Some possible directions for future cluster-scale astrophysical tests are presented in \sref{section.future.cluster}.
\chapterquote{If I have seen further than others, it is by standing upon the shoulders of giants.}{Sir Isaac Newton}
\chapter{\label{chapter.future}Future astrophysical tests}
The work of \pref{part.astroph} explored the importance of directly measuring the dynamics of the gravitational field and comparing to the observed galactic and gaseous components, within the gravity theories of \pref{part.theory}.  The results show a strong interplay between current astrophysical measurements and the resulting surface density maps predicted by gravity theory, with predictions which are testable and falsifiable.  Some possible directions in future space observations are presented in \sref{section.future.galaxy} for galactic-scale astrophysics, and in \sref{section.future.cluster} for cluster-scale astrophysics.

\section{\label{section.future.galaxy}Galactic astrophysics}

Testing whether galaxy dynamics are dominated by a distribution of cold non-baryonic dark matter, like the models of \cref{chapter.darkmatter}, or whether galaxy dynamics are dominated by a modified gravity theory which violates the strong equivalence principle, such as any of the candidates of \cref{chapter.mog}, is possible through a combination of sub-kiloparsec resolution luminous disk observations, galaxy rotation curve measurements, and future directions in galaxy-galaxy lensing, depending on the next generation of space observatories.  Future directions for rotation curve methods, and galaxy-galaxy lensing are explored in \sref{section.future.galaxy.grc}, and \sref{section.future.galaxy.lensing}, respectively.\index{Equivalence principle!Violations}

\subsection{\label{section.future.galaxy.grc}Galaxy rotation curves}

The rotation curves for the Ursa Major filament of galaxies, studied in \sref{section.galaxy.uma}, are already sufficiently detailed to provide features challenging to any candidate theory's best-fit.  As the state of the art of computational models of gravity theories continues to improve, astrophysical observations of nearby filaments and clusters of the Local Group offer the best possible future laboratory.\index{Galaxy rotation, \(v\)}

Within the Local Group, the Milky Way is the most well studied galaxy, but the available rotation curve remains poorly known.  \citet{Brownstein:ApJ:2006} performed a MOND and MSTG best-fit to the Milky Way galaxy rotation curve supplied by \citet{Sofue:ApJ.1996.458} using a parametric model for the surface mass density, independent of photometric observations, whereas the method using photometry, as described in \sref{section.galaxy.uma.photometry}, for the surface mass computation of \sref{section.galaxy.uma.Sigma}, and best-fitting by means of the stellar mass to light ratio, as described in \sref{section.galaxy.uma.masslight}, is preferred.  According to our best-fit parametric model, the total baryonic mass of the Milky Way was determined to be
\begin{equation}\label{eqn.future.galaxy.milkyway}
M_{\rm Milky\,Way} = \left\{
\begin{array}{rc} 9.12 \pm0.28\ M_{\odot} & {\rm MSTG}, \\ 10.60 \pm0.37\ M_{\odot} & {\rm MOND}.
\end{array}\right.
\end{equation}

A new study of the rotation curve for the Milky Way should include better high-resolution rotation curve measurements, particularly for orbital distances greater than the Solar system's, and an updated model of the luminous stellar and gaseous components using available photometry, providing a better answer of the mass of our galaxy as determined by each gravity theory.  The 21-cm line emission of neutral hydrogen, as traced by the Leiden/Argentine/Bonn (LAB) galactic HI survey~\citep{Kalberla.AA.2005.440}, is a full sky map which may be used to provide the {\it gas} component in \eref{eqn.galaxy.uma.mass.gas}, assuming a big-bang nucleosynthesis relation such as \eref{eqn.galaxy.uma.rotmodGas}.  Spitzer Space Telescope's encompassing infrared view of the plane of the Milky Way consists of nearly one million images which have been composed into a \(120^{\circ}\) mosaic, and may be used to compute the vertical scale height, \(z_0\), and provide the stellar {\it disk} component in \eref{eqn.galaxy.uma.mass.disk}, assuming a Van der Kruit and Searle law such as \eref{eqn.galaxy.uma.rotmodDisk}.

The Milky Way has three large dwarf satellite galaxies.  Sagittarius is a 6 kpc dwarf spheroidal, located just below the galactic plane, twice as close as the Large and Small Magellanic Clouds, which are between 4 and 8 kpc in diameter.  In addition, the Milky Way has sixteen dwarf spheroidal galaxies each less than 2 kpc in diameter.  These satellite dwarf galaxies are an important part of the future laboratory for rotation curve astrophysics.

\subsection{\label{section.future.galaxy.lensing}Galaxy-Galaxy lensing}\index{Gravitational lensing!MOG}

Although galaxy rotation curves, as in \fref{figure.galaxy.velocity}, are the primary research tool for measuring dark matter or predicting modified gravity line-of-sight surface mass densities, as in \fref{figure.galaxy.Sigma}, and their galactic mass profiles, as in \fref{figure.galaxy.mass}, it is the  dynamic mass factor of \fref{figure.galaxy.Gamma} which best shows where the {\it missing mass} according to Einstein-Newton gravity is distributed.  For every high and low surface brightness galaxy, including the dwarf galaxies, the dynamic mass factor increases monotonically with distance from the galactic center.  It is this factor which amplifies the different predictions between cold non-baryonic dark matter and each of the modified gravity solutions.   \(\Sigma\)-maps, as in \fref{figure.galaxy.Sigma} for the Ursa Major filament of galaxies,  are predictions for future galaxy-galaxy lensing measurements.\index{Surface mass, \(\Sigma\)}\index{Gravitational lensing!\map{\Sigma}}

Galaxy-galaxy lensing is a unique method to probe the dynamic mass of the foreground (lens) galaxy out to large radii.  Strong galaxy-galaxy lensing, in which multiple background (source) galaxies appear in the image, allow direct measurements of the dynamic mass on scales \(\lesssim 10\) kpc.  The Center for Astrophysics at Harvard -- Arizona Space Telescope Lens Survey (CASTLES)  has identified \((\sim 100)\) strong gravitational lens galaxies from Hubble Space Telescope images.  However, in order to directly measure the line-of-sight surface mass distribution across the foreground galaxy out to \(\sim 100\) kpc, precision weak lensing is used, whereby the slight, coherent, gravitational lensing induced shape correlations can be averaged over multiple \((\sim 1000)\) source galaxies.  The Cosmological Evolution Survey (COSMOS) has observed a 2 square degree equatorial field with the Advanced Camera for Surveys aboard the Hubble Space Telescope, and \citet{Leauthaud.ApJS.2007.172} report on the source catalogue constructed from COSMOS, containing almost 400,000 galaxies with shape measurements and uncertainties.  \citet{Dune.2008SPIE.7010E..38R} report on plans for the Dark Universe Explorer (DUNE) wide-field mission concept, consisting of a 1.2\,m telescope designed to carry out an all-sky survey in one visible and three near infrared bands, optimised for weak gravitational lensing.  With the future direction of galaxy-galaxy lensing measurements, sub-kiloparsec distributions of the dynamic and visible components are attainable, providing high resolution images of the distribution of matter within galaxies.

Whether identified as CDM or the massive fifth-force fields of a modified gravity theory, the solutions to the missing mass problem are testable and falsifiable by current galactic observations, as in \cref{chapter.galaxy}, and through future directions in measuring galaxy rotation curves, with infrared photometry on the stellar and gaseous disks, as in \sref{section.future.galaxy.grc}, and through techniques such as gravitational lensing for directly measuring the dynamic mass, as in \sref{section.future.galaxy.lensing}.			
\section{\label{section.future.cluster}Cluster-scale astrophysics}

Some tests involving X-ray space observatories for measuring the intercluster medium are introduced in \sref{section.future.cluster.icm}, which should be expected to generate higher resolution and greater bandwidth \(\Sigma\)-maps, which are correlated to the ionized electron density, and thereby a measurement of the visible baryon distribution.  Some directions involving the Hubble and Spitzer space telescopes for the future of gravitational lensing measurements are introduced in \sref{section.future.cluster.lensing}, which provide \(\Gamma\)-maps correlated to the missing mass, whether identified as dark matter, or the massive fifth-force fields of a modified gravity theory, according to the lessons learned in \sref{section.summary.lessons}.

\subsection{\label{section.future.cluster.icm}Intercluster medium}

The intercluster medium (ICM) is X-ray measurable with temperatures as high as \(15.5 \pm 3.9\)~keV, from our 2007 value shown in \tref{table.cluster.bullet.temp} for the main cluster of the {\bc}~\protect\citep{Brownstein:MNRAS:2006}, based on the Chandra observatory's very high resolution surface mass density \map{\Sigma}, shown in \fref{figure.cluster.bullet.Sigma}. However, Chandra's bandwidth limitation prevents the {\bc} being identified as the {\it hottest} cluster in the sky, until the next generation of X-ray space observatories, and a combination of the high bandwidth, but low resolution, data available from methods involving the Sunyaev-Zeldovich effect.  New \map{\Sigma}s of the ICM will therefore be of even greater precision and resolution, and more complete at the hottest concentrations of ionized electrons, where the baryon densities are the greatest.\index{Surface mass, \(\Sigma\)}\index{Gravitational lensing!\map{\Sigma}}

Unlike the high precision, high resolution data analysis for the {\bc} of \sref{section.cluster.bullet}, the X-ray mass profiles of \fref{figure.cluster.models.mass} are based on the high precision, but low resolution, data analysis of \citet{Reiprich:2001,Reiprich:2002}, and therefore updating the results with Chandra data is an important future effort, particularly as the next generation of {\it high resolution} Sunyaev-Zeldovich imaging also becomes available.

\subsection{\label{section.future.cluster.lensing}Gravitational lensing}\index{Gravitational lensing!\map{\kappa}}

Weak and strong gravitation lensing surveys of large clusters, such as presented in \sref{section.cluster.bullet} for the {\bc}, may be used to map the scaled surface mass density by means of the convergence \map{\kappa}, discussed in \sref{section.cluster.bullet.kappa}.  The November 15, 2006 data release~\protect\citep{Clowe:dataProduct}, shown in \fref{figure.cluster.bullet.kappa}, is a computational reconstruction based an a survey of both strong and weak gravitational lensing, using images provided by the Advanced Camera for Surveys aboard the Hubble Space Telescope, providing a much more complete convergence \map{\kappa} than was previously known.  Gravitational lensing surveys, particularly the CASTLES and COSMOS surveys discussed in \sref{section.future.cluster.lensing}, have provided catalogues of lensing events, which may lead to as yet undiscovered distributions of mass, whether it be non-baryonic dark matter or the massive fifth-force fields of a modified gravity theory.

The solutions to the missing mass problem in clusters of galaxies are testable and falsifiable by current observations, as in \cref{chapter.cluster}, and through increasingly complete, more precise, and higher resolution  measurements of the intercluster medium will lead to a better identification of the dynamic mass measured by gravitational lensing reconstructions.\index{Gravitational lensing!MOG}			
\bibliographystyle{thesis} 
\bibliography{thesis}
\printindex
\end{document}